%% file: VVA.tex
\documentclass[a4paper, 10pt]{article}
\pdfoutput=1

\usepackage{jheppub}

\usepackage{amsthm}

\usepackage[numbers,sort&compress]{natbib}
\usepackage{dsfont}
\usepackage{slashed}
\usepackage{color}
\usepackage{empheq}

\usepackage[export]{adjustbox}

\usepackage{bibgerm}

\usepackage{soul}

\usepackage[english]{babel}

\usepackage[utf8x]{inputenc}

\usepackage{graphicx,epsfig}
\usepackage{pstricks}

\usepackage{tikz}

\usepackage{bashful}

\usepackage{setspace}

\usepackage{booktabs}

\usepackage{float}

\usepackage{nextpage}

\usepackage{pdfpages}

\usepackage{hypcap}

\usepackage[small,bf]{caption}
\usepackage{subcaption}

\usepackage{longtable}

\usepackage{fancyhdr}

\usepackage[makeroom]{cancel}

\usepackage{microtype}

\usepackage{xparse}
\usepackage{etoolbox}

\usepackage[subfigure]{tocloft}

\usepackage{listings}

\usepackage{ifmtarg}

\usepackage{multirow}

\usepackage{braket}

\usepackage{siunitx}

\newcommand{\tr}{\ensuremath{\operatorname{Tr}}}
\newcommand*\dif{d}
\newcommand{\imag}{\ensuremath{i}}
\newcommand{\T}{\ensuremath{T}}
\newcommand{\LQCD}{\ensuremath{\Lambda_\text{QCD}}}
\newcommand{\Order}{\ensuremath{\mathcal{O}}}
\newcommand{\M}{\ensuremath{\mathcal{M}}}

\renewcommand{\O}{\mathcal{O}}
\newcommand{\Q}{\mathcal{Q}}
\newcommand{\W}{\mathcal{W}}
\newcommand{\A}{\mathcal{A}}
\newcommand{\C}{\mathcal{C}}
\newcommand{\F}{\mathcal{F}}
\newcommand{\U}{\mathcal{U}}
\newcommand{\V}{\mathcal{V}}
\renewcommand{\Re}{\ensuremath{\operatorname{Re}}}
\renewcommand{\Im}{\ensuremath{\operatorname{Im}}}
\newcommand{\nn}{\nonumber\\}
\newcommand{\chpt}{$\chi$PT}
\newcommand{\GeV}{\,\mathrm{GeV}}
\newcommand{\MeV}{\,\mathrm{MeV}}
\newcommand{\p}{\partial}
\newcommand{\msbar}{\ensuremath{\overline{\text{MS}}}}

\newcommand{\<}{\langle}
\renewcommand{\>}{\rangle}

\newcommand{\minidiagSize}[2]{\begin{minipage}{#2} \includegraphics[width=#2]{images/#1} \end{minipage}}

\makeatletter
  \def\my@tag@font{\normalsize}
  \def\maketag@@@#1{\hbox{\m@th\normalfont\my@tag@font#1}}
  \let\amsmath@eqref\eqref
  \renewcommand\eqref[1]{{\let\my@tag@font\relax\amsmath@eqref{#1}}}
\makeatother

\makeatletter
\renewcommand\paragraph{\@startsection{paragraph}{4}{\z@}%
  {-3.25ex\@plus -1ex \@minus -.2ex}%
  {1.5ex \@plus .2ex}%
  {\normalfont\normalsize\bfseries}}
\makeatother

\cftsetindents{section}{0em}{2em}
\cftsetindents{subsection}{2em}{3em}
\cftsetindents{subsubsection}{5em}{4em}

\allowdisplaybreaks[1]
\title{\boldmath Dispersion relations for the hadronic VVA correlator}

\author[a]{Jan Lüdtke,}
\author[a]{Massimiliano Procura,}
\author[b,c]{Peter Stoffer}

\affiliation[a]{Faculty of Physics, University of Vienna, Boltzmanngasse 5, 1090 Vienna, Austria}
\affiliation[b]{Physik-Institut, Universit\"at Z\"urich, Winterthurerstrasse 190, 8057 Z\"urich, Switzerland}
\affiliation[c]{PSI Center for Neutron and Muon Sciences, 5232 Villigen PSI, Switzerland}

\abstract{
We derive two types of dispersion relations for the hadronic vector--vector--axial-vector (VVA) correlator: one in generic three-point kinematics with fixed photon virtualities, the second in the kinematic limit of one soft photon. The VVA correlator enters in the electroweak contribution to the muon anomalous magnetic moment and it also emerges as the leading term in the operator-product expansion for hadronic light-by-light scattering in the limit of two large photon momenta. Previously, this correlator was only described in terms of hadronic models. Our new dispersive treatments are in analogy to the established and the newly proposed dispersive approaches to hadronic light-by-light. The VVA correlator allows us to investigate the relation between the two types of dispersion relations in a simpler example and to elucidate the reshuffling of different hadronic contributions in this comparison. As a byproduct, we reduce the theoretical uncertainties on the first-family VVA contribution to the muon anomalous magnetic moment by combining the dispersive representation with both asymptotic and low-energy constraints.
}

\numberwithin{equation}{section}

\begin{document}

\preprint{
\mbox{}\hfill{} PSI-PR-24-18 \\
\mbox{}\hfill{} UWThPh 2024-20 \\
\mbox{}\hfill{} ZU-TH 49/24
}
	\maketitle


	\input{sections/Intro}
	
	\input{sections/Lorentz}

	\input{sections/DRfixed}

	\input{sections/DRgm2}

	\input{sections/Comparison}

	\input{sections/OPE}

	\input{sections/Numerics}

	\input{sections/Results}

	\input{sections/Conclusions}

	
	\section*{Acknowledgements}
	\addcontentsline{toc}{section}{\numberline{}Acknowledgements}

	We thank Gilberto Colangelo, Martin Hoferichter, Bai-Long Hoid, Bastian Kubis, Pere Masjuan, Pablo Roig, and Pablo S\'anchez-Puertas for valuable discussions and Martin Hoferichter for useful comments on the manuscript.
	Financial support by the Swiss National Science Foundation (Project No.~PCEFP2\_194272) is gratefully acknowledged. 	J.L. is supported by the FWF-DACH Grant I~3845-N27 and by the FWF doctoral program Particles and Interactions, project no. W1252-N27.



	\appendix

	\input{sections/Nomenclature}

	\input{sections/TFFs}

	\input{sections/VgammaA}

	\input{sections/AgammaPiPi}
	
	\clearpage
	
	\phantomsection
	\addcontentsline{toc}{section}{\numberline{}References}
	\bibliographystyle{utphysmod}
	\bibliography{Literature}
	
\end{document}

%% file: sections/Intro.tex

\section{Introduction}
\label{sec:Intro}

The study of the correlator of one axial and two vector currents, along with its contributions from fermion triangle diagrams, played a crucial role in establishing several important findings in quantum field theory. These include the Adler--Bell--Jackiw axial-current anomaly~\cite{Adler:1969gk,Bell:1969ts} and a set of non-renormalization theorems for the coefficients of the Lorentz structures in the covariant decomposition of the correlator~\cite{Adler:1969er,Vainshtein:2002nv,Knecht:2003xy}, which prompted also dedicated high-precision perturbative QCD calculations~\cite{Jegerlehner:2005fs,Melnikov:2006qb,Mondejar:2012sz}.

One particularly interesting phenomenological application of the VVA correlator is its role in the precision calculation of the muon anomalous magnetic moment~\cite{Aoyama:2020ynm}, $a_\mu$. In particular, the hadronic VVA correlator enters the theory prediction of $a_\mu$ in two significant ways: it is part of the electroweak (EW) contribution~\cite{Czarnecki:2002nt,Gnendiger:2013pva} and it also appears in the leading term of the operator product expansion (OPE) for hadronic light-by-light (HLbL) scattering in the limit of two large (Euclidean) photon momenta~\cite{Melnikov:2003xd,Colangelo:2019uex,Ludtke:2020moa,Bijnens:2022itw}. In both cases, the VVA correlator enters in the kinematic limit where one of the two vector currents represents the static external electromagnetic field with vanishing momentum. The hadronic VVA contribution to the EW part of $a_\mu$ has so far been estimated only through models~\cite{Czarnecki:2002nt}. While its size and uncertainty are not critical in view of much larger uncertainties in the hadronic contributions to $a_\mu$, a new analysis is timely: a better understanding of the hadronic VVA correlator informs the data-driven evaluation of HLbL~\cite{Melnikov:2003xd,Pauk:2014rta,Colangelo:2014dfa,Colangelo:2014qya,Colangelo:2014pva,Colangelo:2015ama,Danilkin:2016hnh,Masjuan:2017tvw,Colangelo:2017qdm,Colangelo:2017fiz,Jegerlehner:2017gek,Knecht:2018sci,Hoferichter:2018kwz,Gerardin:2019vio,Hoferichter:2019nlq,Bijnens:2019ghy,Roig:2019reh,Eichmann:2019bqf,Colangelo:2019lpu,Colangelo:2019uex,Danilkin:2021icn,Colangelo:2021nkr,Stamen:2022uqh,Ludtke:2023hvz,Stoffer:2023gba,Hoferichter:2024fsj}, both as an input to the matching to short-distance constraints (SDCs)~\cite{Colangelo:2019lpu,Colangelo:2019uex,Ludtke:2020moa,Hoferichter:2024bae}, but also by providing a better understanding of the dispersive approach, which is crucial to obtain robust uncertainty estimates that match the precision of the forthcoming final results of the Muon $g-2$ Experiment at Fermilab~\cite{Muong-2:2021ojo,Muong-2:2023cdq,Muong-2:2024hpx}.\footnote{Lattice-QCD evaluations of HLbL are seeing similar improvements~\cite{Blum:2019ugy,Chao:2021tvp,Blum:2023vlm} and lead to results compatible with the data-driven approach.}

To achieve these goals, we adopt a data-driven approach based on dispersion relations, similar to the method applied to HLbL, both in general~\cite{Colangelo:2014dfa,Colangelo:2014pva,Colangelo:2015ama,Colangelo:2017qdm,Colangelo:2017fiz} and three-point kinematics~\cite{Ludtke:2023hvz}. The latter approach to HLbL has been recently introduced to fully remove the spurious kinematic singularities from the contribution of tensor-meson resonances or two-particle $D$- and higher partial waves that affect the dispersive framework at fixed photon virtualities~\cite{Hoferichter:2024fsj}. This enables the inclusion in a controlled fashion of resonances beyond scalar and axial-vector mesons, which presently is not possible in the dispersive approach in general kinematics. Furthermore, the novel dispersive approach offers new insights into the optimal strategy for incorporating SDCs in the data-driven determination of HLbL.

In this paper, we correspondingly derive two types of dispersion relations for the hadronic VVA correlator: one for general kinematics with fixed photon virtualities, and another formulated directly in the limit of a soft external photon. Following the necessary truncation of the sum over individual intermediate states, there is a reshuffling between included and neglected contributions in the two approaches, as for HLbL~\cite{Ludtke:2023hvz}. In the present study, we perform the first analytic comparison between the two types of dispersion relations, investigate the reshuffling of intermediate-state contributions, including pseudoscalars, axial-vector, and vector mesons as well as two pions, and we achieve an improved understanding of the impact of neglected contributions, all in a simpler context compared to HLbL. In order to perform the analytic comparison, we also derive a dispersive representation of axial-vector meson transition form factors (TFFs), a useful input in the HLbL evaluation. Our analysis paves the way for an optimal strategy to combine the two complementary dispersive approaches to achieve improved data-driven determinations of VVA and HLbL, with more precise uncertainty estimates. We also perform a numerical study of both dispersive representations of the hadronic VVA correlator. Our results lead to a reduction of the theoretical uncertainty on the first-family VVA contribution to $a_\mu$ compared to the analysis in Ref.~\cite{Czarnecki:2002nt}. This is obtained by combining the dispersive representation with both low-energy and asymptotic constraints, from Chiral Perturbation Theory (\chpt{}) and the OPE for large space-like momenta, respectively.

The paper is structured as follows. We review some results for the VVA correlator and its role in the evaluation of $a_\mu$ in Sect.~\ref{sec:VVA}. In Sect.~\ref{sec:DRFixedVirtualities}, we introduce the dispersive approach to VVA in general kinematics with fixed photon virtualities. The alternative dispersive approach, formulated directly in the kinematic limit of one soft photon, is introduced in Sect.~\ref{sec:DRgm2}. In Sect.~\ref{sec:Comparison}, we perform a detailed analytic comparison of the two approaches, which sheds light on the reshuffling of intermediate-state contributions and the effect of the truncation of hadronic intermediate states. In Sect.~\ref{sec:OPEandChPTConstraints}, we review constraints at low and high energies, before we explain our numerical implementation in Sect.~\ref{sec:Numerics}. A numerical analysis of our results is presented in Sect.~\ref{sec:Results}, including a detailed comparison of the two dispersive approaches and the respective matching to SDCs, and we provide an improved prediction for the first-family VVA contribution to $a_\mu$. We conclude in Sect.~\ref{sec:Conclusions} and collect the details on the hadronic sub-amplitudes in the appendices, including a dispersive analysis of axial-vector meson TFFs.

%% file: sections/Lorentz.tex

\section{\boldmath The VVA Green's function and applications in the context of \texorpdfstring{$a_\mu$}{a mu}}
\label{sec:VVA}

In this section, we review some important results for the VVA correlator, which can be found, e.g., in Ref.~\cite{Knecht:2020xyr}. An overview of our conventions and nomenclature is provided in App.~\ref{sec:Conventions}.

\subsection{Kinematics and matrix element}
\label{sec:VVA_Kin}

The light-quark hadronic VVA correlator is defined as the three-point function
\begin{equation}
	\label{eq:defW}
	\W_{\mu\nu\rho}(q_1, q_2) = \imag \int \dif^4 x \int \dif^4 y \, e^{\imag (q_1\cdot x + q_2\cdot y)} \< 0| \T\{j_\mu^\mathrm{em}(x) j_\nu^\mathrm{em}(y) j_\rho^5(0)\} |0 \>
\end{equation}
of two insertions of the electromagnetic current $j_\mathrm{em}^\mu(x) = \bar q(x) \Q \gamma^\mu q(x)$ and an axial-vector current
\begin{equation}
	\label{eq:VVA_axialCurrent}
	j_\mu^5(x) = \bar q(x) \Q_5 \gamma_\mu \gamma_5 q(x) \,,
\end{equation}
where $\Q_5$ is a matrix in flavor space. In the context of the Melnikov--Vainshtein (MV) SDC for HLbL scattering~\cite{Melnikov:2003xd}, the flavor matrix is given by $\Q_5 = \Q^2$, see Sect.~\ref{sec:MVSDC}. In the context of the EW contribution to $g-2$, where the axial current $j_\mu^{5,\mathrm{EW}}$ couples to a $Z$ boson, the flavor matrix is $\Q_5^\mathrm{EW} = 2 I_3 = \operatorname{diag}(1, -1, -1)$ with the weak isospin $I_3$, see Sect.~\ref{sec:electroweak}.
In both cases, the axial current can be decomposed into its flavor components according to
\begin{equation}
	j_\mu^5(x) = \sum_{\mathclap{a=3,8,0}} \, 2 C_a \, j_\mu^{5(a)}(x)\,, \qquad j_\mu^{5(a)}(x) = \bar q(x) \frac{\lambda^a}{2} \gamma_\mu \gamma_5 q(x) \,,
\end{equation}
where $\lambda^a$ are the Gell-Mann matrices and $\lambda^0 = \sqrt{2/3} \, \mathds{1}$.
The coefficients $C_a$ are given by $C_a = \tr(\Q_5 \lambda^a) / 2$, explicitly
\begin{equation}
	\label{eq:Ca}
	C_3 = \frac{1}{6} \,, \qquad C_8 = \frac{1}{6\sqrt{3}} \,,  \qquad C_0 = \frac{2}{3\sqrt{6}}
\end{equation}
for the SDC for HLbL and
\begin{equation}
	C_3^\mathrm{EW} = 1 \,,  \qquad C_8^\mathrm{EW} = \frac{1}{\sqrt{3}}\,,  \qquad C_0^\mathrm{EW} = -\frac{1}{\sqrt{6}}
\end{equation}
for the EW contribution to $a_\mu$. We define the flavor components of the VVA correlator as\footnote{This differs by a factor of $2C_a$ from the convention in Ref.~\cite{Colangelo:2019uex}.}
\begin{equation}
	\W_{\mu\nu\rho}^{(a)}(q_1, q_2) = \imag \int \dif^4 x \int \dif^4 y \, e^{\imag (q_1\cdot x + q_2\cdot y)} \< 0 | \T\{j^\mathrm{em}_\mu(x) j^\mathrm{em}_\nu(y) j_\rho^{5(a)}(0)\} | 0\> \,.
\end{equation}

The Ward identities for the VVA correlator read
\begin{equation}
	\label{eq:WardIdentities}
	\{q_1^\mu, q_2^\nu \}\W_{\mu\nu\rho}^{(a)}(q_1, q_2) = \{0,0\}\,, \qquad (q_1 + q_2)^\rho \W_{\mu\nu\rho}^{(a)}(q_1, q_2) = \A^{(a)} \epsilon_{\mu\nu\alpha\beta} q_1^\alpha q_2^\beta\,,
\end{equation}
where $\A^{(a)} = N_c \tr(\Q^2 \lambda^a) / (4\pi^2)$ according to the QED axial anomaly\footnote{Eq.~\eqref{eq:Gamma5LCConvention} explains the different sign in $\A$ compared to Ref.~\cite{Knecht:2020xyr}.} and the second identity receives corrections from quark masses and the gluon anomaly. By summing over the flavor components, one obtains the Ward identities for the full correlator, leading to $\A = \sum_a 2 C_a \A^{(a)} = 1 / (3\pi^2)$ in the anomalous one. The anomalous Ward identity has been checked up to three loops in perturbative QCD~\cite{Mondejar:2012sz}.

\subsection{Tensor decomposition}
\label{sec:VVA_Ten}
In Ref.~\cite{Knecht:2020xyr}, a decomposition of $\W_{\mu\nu\rho}(q_1, q_2)$ into independent Lorentz structures free of kinematic singularities and zeros has been given:
\begin{align}
	\label{eq:TensorDecompVVA}
	\W^{\mu\nu\rho}(q_1, q_2) &= \sum_{i=0}^{3} \W_i(q_1^2, q_2^2, (q_1 + q_2)^2) \tau_i^{\mu\nu\rho}(q_1, q_2)\,,\nn
	\tau_0^{\mu\nu\rho}(q_1, q_2) &= \epsilon^{\mu\nu\alpha\beta} q_{1\alpha} q_{2\beta} (q_1 + q_2)^\rho\,,\nn
	\tau_1^{\mu\nu\rho}(q_1, q_2) &= \epsilon^{\mu\nu\alpha\beta} q_{1\alpha} q_{2\beta} (q_1 - q_2)^\rho\,,\nn
	\tau_2^{\mu\nu\rho}(q_1, q_2) &= q_1^\nu \epsilon^{\mu\rho\alpha\beta} q_{1\alpha} q_{2\beta} - q_2^\mu \epsilon^{\nu\rho\alpha\beta} q_{1\alpha} q_{2\beta} - (q_1 \cdot q_2) \epsilon^{\mu\nu\rho\alpha} (q_1 - q_2)_\alpha\,,\nn
	\tau_3^{\mu\nu\rho}(q_1, q_2) &= q_1^\mu \epsilon^{\nu\rho\alpha\beta} q_{1\alpha} q_{2\beta} + q_2^\nu \epsilon^{\mu\rho\alpha\beta} q_{1\alpha} q_{2\beta} - q_1^2 \epsilon^{\mu\nu\rho\alpha} q_{2\alpha} - q_2^2 \epsilon^{\mu\nu\rho\alpha} q_{1\alpha}\,.
\end{align}
The Lorentz structures fulfill the crossing relations
\begin{equation}
	\C_{12}[\tau_0^{\mu\nu\rho}] = \tau_0^{\mu\nu\rho}\,, \quad \C_{12}[\tau_1^{\mu\nu\rho}] = -\tau_1^{\mu\nu\rho}\,,  \quad \C_{12}[\tau_2^{\mu\nu\rho}] = \tau_2^{\mu\nu\rho}\,,  \quad \C_{12}[\tau_3^{\mu\nu\rho}] = -\tau_3^{\mu\nu\rho}\,,
\end{equation}
where the crossing operator is defined by $\C_{12}[f] := f(\mu\leftrightarrow\nu, q_1\leftrightarrow q_2)$.
An equivalent decomposition has been obtained in Ref.~\cite{Hoferichter:2020lap} in the context of the axial-vector meson TFFs by using the recipe by Bardeen, Tung~\cite{Bardeen:1968ebo}, and Tarrach~\cite{Tarrach:1975tu} (BTT). These structures are linearly related to the ones of Ref.~\cite{Knecht:2020xyr} by
\begin{align}
	\label{eq:AxialTFFTensors}
	\tau_0^{\mu\nu\rho}(q_1, q_2) &= \bar T_3^{\mu\nu\rho}(q_1, q_2) \,, \nn
	\tau_1^{\mu\nu\rho}(q_1, q_2) &= T_1^{\mu\nu\rho}(q_1, q_2) \,, \nn
	\tau_2^{\mu\nu\rho}(q_1, q_2) &= \bar T_3^{\mu\nu\rho}(q_1, q_2) - T_2^{\mu\nu\rho}(q_1, q_2) + T_3^{\mu\nu\rho}(q_1, q_2) \,, \nn
	\tau_3^{\mu\nu\rho}(q_1, q_2) &= -T_2^{\mu\nu\rho}(q_1, q_2) - T_3^{\mu\nu\rho}(q_1, q_2) \,,
\end{align}
where we made use of the Schouten identity.

For both the MV constraint in HLbL and the EW contribution to $a_\mu$, we need the VVA correlator for small $q_2$ up to linear order. Expanding the Lorentz structures gives $\tau_i^{\mu\nu\rho}(q_1, q_2) = q_{2\sigma}\tau_i^{\mu\nu\rho\sigma}(q_1) + \Order(q_2^2)$ with the Lorentz tensors
\begin{align}
	\tau_0^{\mu\nu\rho\sigma}(q_1) &= \tau_1^{\mu\nu\rho\sigma}(q_1) = \epsilon^{\mu\nu\alpha\sigma} q_{1\alpha} q_1^\rho\,,\nn
	\tau_2^{\mu\nu\rho\sigma}(q_1) &= \epsilon^{\mu\rho\alpha\sigma} q_1^\nu q_{1\alpha} - \epsilon^{\mu\nu\rho\alpha} q_1^\sigma q_{1\alpha}\,,\nn
	\tau_3^{\mu\nu\rho\sigma}(q_1) &= \epsilon^{\nu\rho\alpha\sigma} q_1^\mu q_{1\alpha} - q_1^2 \epsilon^{\mu\nu\rho\sigma}\,.
\end{align}
Since $\tau_0^{\mu\nu\rho\sigma}(q_1) = \tau_1^{\mu\nu\rho\sigma}(q_1)$ and $\tau_3^{\mu\nu\rho\sigma}(q_1) = \tau_2^{\mu\nu\rho\sigma}(q_1) - \tau_1^{\mu\nu\rho\sigma}(q_1)$ (again due to the Schouten identity), in this kinematic limit there are only two independent Lorentz structures. Thus, we can write
\begin{equation}
	\label{eq:decompositiongm2}
	\W^{\mu\nu\rho}(q_1, q_2) = \sum_{i=1}^{2} \tilde{\W}_i(q_1^2) q_{2\sigma} \tau_i^{\mu\nu\rho\sigma}(q_1) + \Order(q_2^2)
\end{equation}
and the scalar functions
\begin{equation}
	\label{eq:W_Wtilde}
	\tilde{\W}_1(q^2) = \tilde{\W}_1(q^2, 0, q^2)\,, \qquad
	\tilde{\W}_2(q^2) = \tilde{\W}_2(q^2, 0, q^2)
\end{equation}
are related to the ones in Eq.~\eqref{eq:TensorDecompVVA} by
\begin{align}
	\label{eq:W_WtildeFull}
	\tilde{\W}_1(q_1^2, q_2^2, q_3^2) &:= \W_0(q_1^2, q_2^2, q_3^2) + \W_1(q_1^2, q_2^2, q_3^2) - \W_3(q_1^2, q_2^2, q_3^2)\,,\nn
	\tilde{\W}_2(q_1^2, q_2^2, q_3^2) &:= \W_2(q_1^2, q_2^2, q_3^2) + \W_3(q_1^2, q_2^2, q_3^2) \,.
\end{align}
Due to their properties with respect to the axial-vector current,
\begin{equation}
	(q_{1\rho} q_{1\rho'} - q_1^2 g_{\rho\rho'} ) \tau_1^{\mu\nu\rho\sigma}(q_1) = 0 \,, \qquad q_{1\rho} \tau_2^{\mu\nu\rho\sigma}(q_1) = 0 \,,
\end{equation}
the function $\tilde\W_1$ is called the longitudinal and $\tilde\W_2$ the transverse component.

The anomalous Ward identity in Eq.~(\ref{eq:WardIdentities}) fixes the longitudinal function in the chiral limit and when the gluon anomaly is neglected:
\begin{equation}
	\label{eq:WardConstraint}
	\tilde{\W}_1(q^2) = \frac{\A}{q^2} \, .
\end{equation}
We see that it contains a massless pole in $q^2$, which has to be a dynamical singularity since we started from a basis free of kinematic singularities. The only massless states in the chiral limit and without the gluon anomaly are the Goldstone bosons and the $\eta'$. Due to the diagonal flavor structure only $\pi^0$, $\eta$, and $\eta'$ can contribute, each according to the coefficients $C_a$ if mixing is neglected.

The non-renormalization theorem for the transverse component derived in Ref.~\cite{Vainshtein:2002nv} translates into
\begin{equation}
	\label{eq:NonRenTheorem}
	\tilde{\W}_2(q^2) = -\frac{\A}{2 q^2} \,.
\end{equation}
However, in contrast to Eq.~(\ref{eq:WardConstraint}), this equation receives non-perturbative corrections even in the chiral limit and hence is valid only for $-q^2 \gg \Lambda_\mathrm{QCD}^2$, where also quark-mass effects are suppressed, see Sect.~\ref{sec:VVA_OPE}. Perturbative corrections away from the chiral limit were calculated in Ref.~\cite{Melnikov:2006qb}.

\subsection{Connection with HLbL in the MV regime}
\label{sec:MVSDC}

The VVA correlator is related to the HLbL tensor in the regime where two virtualities are comparable and much larger than both the third virtuality and $\LQCD^2$~\cite{Melnikov:2003xd, Colangelo:2019uex, Bijnens:2022itw}.
We consider the HLbL tensor
\begin{equation}
	\label{eq:HLbLTensor}
	\Pi^{\mu\nu\lambda\sigma}(q_1,q_2,q_3) = -\imag \int \dif^4 x\, \dif^4 y \,\dif^4 z \, e^{-\imag(q_1\cdot x + q_2 \cdot y + q_3 \cdot z)} \< 0| \T\{j_\mathrm{em}^\mu (x) j_\mathrm{em}^\nu (y) j_\mathrm{em}^\lambda (z) j_\mathrm{em}^\sigma (0) \} |0\>
\end{equation}
after the Wick rotation has been performed, i.e., for Euclidean momenta, in the regime where $\hat{q} \equiv (q_1 - q_2)/2$ is large. The leading contribution to the integral in Eq.~\eqref{eq:HLbLTensor} then comes from the region $x\approx y$ and the product of electromagnetic currents in the matrix element can be expanded around $x=y$. We consider the tensor operator
\begin{equation}
	\label{eq:Pimunu}
	\Pi^{\mu\nu}(q_1, q_2) = \imag \int \dif^4 x\, \dif^4 y \,e^{-\imag(q_1\cdot x + q_2 \cdot y)}\T\{j_\mathrm{em}^\mu (x) j_\mathrm{em}^\nu (y)\}\,,
\end{equation}
which we want to expand into a series of local operators. Operators with the lowest mass dimension are multiplied by the largest power of $\hat{q}$ and thus dominate at large $\hat{q}$. Neglecting the unit operator, which would lead to a disconnected contribution to HLbL, the OPE reads (see, e.g., Refs.~\cite{Melnikov:2003xd, Colangelo:2019uex})
\begin{equation}
	\label{eq:OPE}
	\Pi^{\mu\nu}(q_1, q_2) = \int \dif^4 z\, e^{-\imag(q_1 + q_2) \cdot z} \left(-\frac{2\imag}{\hat{q}^2}\epsilon^{\mu\nu\alpha\beta} \hat{q}_\alpha j^5_{\beta}(z)\right) + \dots\,.
\end{equation}
The flavor matrix $\Q_5$ in $j^5_{\beta}$ is given by $\Q^2$ due to the two vector currents in Eq.~\eqref{eq:Pimunu}. The ellipsis denotes sub-leading terms suppressed by powers of $\{|q_1+q_2|/|\hat{q}|,\,\allowbreak\LQCD/|\hat{q}|\}$. This result implies that, at leading order in the OPE and at leading order in $\alpha_s$, the HLbL tensor can be expressed in terms of the correlator of two vector currents and an axial current,
\begin{align}
	\label{eq:HLbLVVA}
	\Pi_{\mu\nu\lambda\sigma}^\mathrm{MV}(q_1,q_2,q_3) &= \frac{2}{\hat{q}^2} \epsilon_{\mu\nu\alpha\beta} \hat{q}^\alpha \W_{\lambda \sigma}^{\ \ \,\beta}(-q_3, q_4) = \frac{4}{\hat{q}^2} \epsilon_{\mu\nu\alpha\beta} \hat{q}^\alpha \sum_{\mathclap{a={3,8,0}}} C_a \W^{(a)\beta}_{\lambda\sigma}(-q_3, q_4) \, ,
\end{align}
for $Q_1^2 \sim Q_2^2 \gg \{Q_3^2,Q_4^2,\LQCD^2\}$ and where we defined $q_4 = q_1 + q_2 + q_3$. The leading perturbative correction to this OPE was studied in Ref.~\cite{Ludtke:2020moa} and Ref.~\cite{Bijnens:2022itw} dealt with non-perturbative corrections.

The anomaly result and the non-renormalization theorem for VVA thus imply SDCs on the BTT functions of HLbL~\cite{Colangelo:2019uex}. The saturation of these constraints has been studied in a simple hadronic model~\cite{Melnikov:2003xd, Melnikov:2019xkq}, in holographic QCD~\cite{Leutgeb:2019gbz, Cappiello:2019hwh, Leutgeb:2021mpu, Leutgeb:2022lqw}, in a large-$N_c$ Regge model~\cite{Colangelo:2019lpu, Colangelo:2019uex}, and using general interpolants~\cite{Ludtke:2020moa}.

\subsection[Connection with the hadronic EW contribution to \texorpdfstring{$a_\mu$}{amu}]{\boldmath Connection with the hadronic EW contribution to \texorpdfstring{$a_\mu$}{amu}}
\label{sec:electroweak}

\begin{figure}
	\centering
	\includegraphics{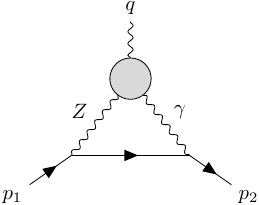} \hspace{1.5cm} \includegraphics{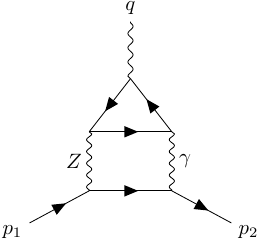}
	\caption{Left: EW contribution to $a_\mu$ due to the hadronic VVA correlator. Right: contribution to $a_\mu$ due to a charged-fermion loop.}
	\label{Fig:EW_VVA}
\end{figure}

In the context of $a_\mu$, the VVA correlator is not only relevant due to the connection with HLbL via the OPE, but it also directly enters as an input for a part of the EW contribution, given by the diagram on the left-hand side of Fig.~\ref{Fig:EW_VVA} (and the diagram with reversed fermion flow). Due to Furry's theorem~\cite{Furry:1937zz}, only the axial component of the $Z$-boson contributes and as for HLbL, the hadronic correlator is only needed for a soft external photon.

This contribution can be reduced to an integral over space-like values for the arguments of the scalar functions~\cite{Jegerlehner:2017gek}
\begin{equation}
	\label{eq:VVAMaster}
	a_\mu^\text{VVA} = \int_0^{\Lambda^2} \dif Q^2 \left[ K_1(Q^2) \tilde{\W}_1(-Q^2) + K_2(Q^2) \tilde{\W}_2(-Q^2) \right] \,,
\end{equation}
where the cutoff $\Lambda^2$ should be taken to infinity in the end. The form of the kernel functions $K_i(Q^2)$ depends on the gauge chosen for the EW part of the diagram since only the sum of all diagrams at a given order in the EW couplings is gauge invariant. In the unitary gauge, the kernel functions read~\cite{Jegerlehner:2017gek}
\begin{align}
	K_1(Q^2) &= -\frac{G_\mu \alpha Q^2}{12 \sqrt{2} \pi} \left[ \left( \frac{Q^2}{m_\mu^2} - 2 \right) \left( 1 - \sqrt{1 + \frac{4 m_\mu^2}{Q^2}} \right) + 2 \right] \,, \nn
	K_2(Q^2) &= -\frac{G_\mu \alpha Q^2}{12 \sqrt{2} \pi} \frac{M_Z^2}{M_Z^2 + Q^2} \left[ \left( \frac{Q^2}{m_\mu^2} + 4 \right) \left( 1 - \sqrt{1 + \frac{4 m_\mu^2}{Q^2}} \right) + 2 \right] \,.
\end{align}
Since the axial part of the $Z$ boson couples to fermions according to their weak isospin, the functions $\tilde{\W}_i$ are needed here for the axial current in Eq.~\eqref{eq:VVA_axialCurrent} with $\Q_5 = 2 I_3$ as already mentioned in Sect.~\ref{sec:VVA_Kin}.

Both kernel functions go to constants for large $Q^2$ and the OPE predicts that the scalar functions $\tilde{\W}_i$ asymptotically behave as $1/Q^2$, see Sect.~\ref{sec:VVA_OPE}. Thus, the integrals in Eq.~\eqref{eq:VVAMaster} diverge logarithmically in the limit $\Lambda^2 \to \infty$. However, if we consider a complete family of fermions, the leading term in the $1 / Q^2$ expansion drops out due to the gauge-anomaly cancellation condition
\begin{equation}
	\sum_f N_f I_3^f \Q_f^2 = 0 \,,
\end{equation}
where the sum runs over all quarks and leptons of one family, $N_f = N_c = 3$ for quarks and $N_f = 1$ for leptons, $I_3^f$ is the fermion's weak isospin and $\Q_f$ its charge. This makes $a_\mu^\text{VVA}$ finite in the limit $\Lambda \to \infty$. For this reason, it is crucial that the leading OPE constraint on the VVA correlator be implemented exactly.

The contribution from a charged lepton is given, to leading order in $\alpha$, by the loop diagram on the right-hand side of Fig.~\ref{Fig:EW_VVA} (and diagrams with reversed fermion flow), which yields
\begin{equation}
	\label{eq:VVAFermionLoop}
	\tilde{\W}_1^l(q^2) = -2 \tilde{\W}_2^l(q^2) = -\frac{1}{2 \pi^2} \int_0^1 \dif x \frac{x(1 - x)}{x(1 - x)q^2 - m_l^2} \,.
\end{equation}

In our numerical study in Sect.~\ref{sec:Numerics}, we will focus on the first-family contribution to $a_\mu^\text{VVA}$. Since $2 I_3$ restricted to the first family of quarks coincides with the third Pauli matrix, this only requires the isovector part of the hadronic VVA correlator, together with the electron-loop contribution.

\subsection{Hadronic models for the VVA correlator}
\label{sec:VVAModels}

So far, the VVA correlator and the corresponding contribution to $a_\mu$ have only been calculated using hadronic models. For the compilation of the White Paper~\cite{Aoyama:2020ynm}, the model of Ref.~\cite{Czarnecki:2002nt} was used, which contains only pseudoscalar, axial-vector, and vector poles. For the isovector component, it takes the form
\begin{align}
	\label{eq:VVA_CMVModel}
	\tilde{\W}_{1, \text{CMV}}^{(3)}(q^2) &= \frac{1}{4 \pi^2 (q^2 - m_\pi^2)} \,, \nn
	\tilde{\W}_{2, \text{CMV}}^{(3)}(q^2) &= \frac{1}{8 \pi^2} \frac{1}{m_{a_1}^2 - m_\rho^2} \left( \frac{m_\rho^2 - m_\pi^2}{q^2 - m_{a_1}^2} - \frac{m_{a_1}^2 - m_\pi^2}{q^2 - m_\rho^2} \right) \,.
\end{align}
The residues of the poles were fixed from the OPE constraints that will be discussed in Sect.~\ref{sec:VVA_OPE}. For the contribution to $a_\mu$ from the first family of quarks and leptons, the model leads to\footnote{In Ref.~\cite{Czarnecki:2002nt}, the electron mass as well as terms suppressed by $m_\mu^2 / m_\rho^2$ have been neglected in the $Q^2$ integral. This explains the difference to the numbers quoted there, $a_{\mu, \text{CMV}}^{\text{VVA}, L}[u,d,e] = \num{-0.96e-11}$ and $a_{\mu, \text{CMV}}^{\text{VVA}, T}[u,d,e] = \num{-1.32e-11}$.}
\begin{equation}
	\label{eq:VVA_res_CMV}
	a_{\mu, \text{CMV}}^{\text{VVA}, L}[u,d,e] = \num{-0.94e-11} \,, \qquad a_{\mu, \text{CMV}}^{\text{VVA}, T}[u,d,e] = \num{-1.33e-11} \,,
\end{equation}
where the two results correspond to the longitudinal and transverse contribution, i.e., the contribution from $\tilde{\W}_1$ and $\tilde{\W}_2$, respectively. Reference~\cite{Czarnecki:2002nt} quotes an uncertainty estimate of $\pm \num{0.2e-11}$ for the sum, based on a generous variation of the $\rho$ and $a_1$ masses.

References~\cite{Leutgeb:2019gbz, Cappiello:2019hwh} discuss holographic models for the axial-vector meson contributions to $a_\mu^\text{HLbL}$. We consider here the model introduced in Ref.~\cite{Hirn:2005nr} and called HW2 in Ref.~\cite{Leutgeb:2019gbz}. In order to obtain the correct asymptotic limit, we use the anomaly, $N_c = 3$, and $F_\pi$ to fix the parameters, which is called HW2(UV-fit) in Ref.~\cite{Leutgeb:2019gbz} and set 2 in Ref.~\cite{Cappiello:2019hwh}. Formulae for the singly-virtual VVA scalar functions can be found in Ref.~\cite{Cappiello:2019hwh}
\begin{align}
	\tilde{\W}_{1, \text{hQCD}}^{(3)}(-Q^2) &= -\frac{1}{4 \pi^2 Q^2} \left( 1 - \frac{m_\pi^2}{Q^2 + m_\pi^2} \tilde{F}_{\pi^0 \gamma^* \gamma^*}(-Q^2, 0)  \right) \,, \nn
	\tilde{\W}_{2, \text{hQCD}}^{(3)}(-Q^2) &= \frac{1}{8 \pi^2 Q^2} - \frac{z_0^2}{16 \pi^2} \left( \frac{K_0(\sqrt{Q^2} z_0)}{I_0(\sqrt{Q^2} z_0)} + \frac{K_1(\sqrt{Q^2} z_0)}{I_1(\sqrt{Q^2} z_0)} \right) \, ,
\end{align}
with the Bessel functions $K_i$ and $I_i$, the size of the holographic dimension $z_0 \approx \SI{2.42}{\GeV^{-1}}$ and the normalized pion TFF
\begin{equation}
	\tilde{F}_{\pi^0 \gamma^* \gamma^*}(-Q^2, 0) = \int_0^{z_0} \dif z \frac{2 z}{z_0^2} \sqrt{Q^2} z \left( K_1(\sqrt{Q^2} z) + \frac{K_0(\sqrt{Q^2} z_0)}{I_0(\sqrt{Q^2} z_0)} I_1(\sqrt{Q^2} z) \right) \,.
\end{equation}
These functions have the correct leading asymptotic behavior due to the anomaly and the non-renormalization theorem, but do not include the effects of non-vanishing condensates~\cite{Cappiello:2019hwh}. This model leads to
\begin{equation}
	\label{eq:VVA_res_hQCD}
	a_{\mu\text{, hQCD}}^{\text{VVA}, L}[u, d, e] = \num{-0.90e-11} \,, \qquad a_{\mu\text{, hQCD}}^{\text{VVA}, T}[u, d, e] = \num{-1.29e-11} \,.
\end{equation}
Given the large hadron masses in the holographic model~\cite{Leutgeb:2019gbz} and the very different analytic form compared to the model~\eqref{eq:VVA_CMVModel}, these values are surprisingly close to the ones of Ref.~\cite{Czarnecki:2002nt}.

In the time-like region, the holographic functions exhibit a series of poles on the real axis since the model is constructed in the large-$N_c$ limit. $\tilde{\W}_1^{(3)}$ contains a pion pole and a tower of vector-meson poles, whereas $\tilde{\W}_2^{(3)}$ contains towers of vector-meson and axial-vector meson poles. The masses of these states are, however, $SU(3)$ symmetric and typically too large~\cite{Leutgeb:2019gbz}.

\begin{table}
	\centering
	\begin{tabular}{c c c c c}
		\toprule
		 & \multicolumn{2}{c}{$4 \pi^2 \operatorname{Res} \left( \tilde{\W}_{1, \text{hQCD}}^{(3)}, m_i^2 \right)$} & \multicolumn{2}{c}{$8 \pi^2 \operatorname{Res} \left( \tilde{\W}_{2, \text{hQCD}}^{(3)}, m_i^2 \right)$} \\
		\midrule
		$n$ & $\pi$ & $\rho$ & $a_1$ & $\rho$ \\
		\midrule
		0 & \num{1.0204} & \num{-0.0208} & \num{6.16} & \num{-3.71} \\
		1 & --- & \num{4.9e-4} & \num{11.10} & \num{-8.64} \\
		2 & --- & \num{-6.5e-5} & \num{16.04} & \num{-13.57} \\
		3 & --- & \num{1.6e-5} & \num{20.97} & \num{-18.51} \\
		\bottomrule
	\end{tabular}
\caption[Residues of the lightest poles in the two isovector scalar functions in hQCD]{Residues of the lightest poles in the two isoscalar scalar functions in hQCD. $n$ is the radial excitation and $\rho$ stands for all vector mesons from the multiplet.}
\label{Tab:VVA_HolographyResidues}
\end{table}
The residues of the lightest poles can be found in Tab.~\ref{Tab:VVA_HolographyResidues}. The pion pole completely dominates the first scalar function and the vector-meson poles only give small corrections to it. This is consistent with the pion pole being the only contribution in the chiral limit. Conversely, in the second scalar function the whole tower of vector and axial-vector poles is important and the residues even increase with $n$. Thus, an intricate cancellation between the infinite towers of axial-vector and vector poles is needed to reach the space-like asymptotic value in Eq.~\eqref{eq:NonRenTheorem}. A similar---but simpler---cancellation is also built into the model in Eq.~\eqref{eq:VVA_CMVModel}, where both the $a_1$ and the $\rho$ pole have large residues with opposite sign.

In Ref.~\cite{Masjuan:2020jsf}, a propagator model is used to describe pseudoscalar and axial contributions both to HLbL and the VVA correlator. An infinite tower of axial-vector mesons is constructed in a Regge-like approach, which allows to satisfy the anomaly constraint.

The Regge model introduced in Refs.~\cite{Colangelo:2019lpu, Colangelo:2019uex} to saturate the longitudinal MV SDC to HLbL within the established dispersive approach uses an infinite tower of pseudoscalar mesons. Although not designed for this purpose, this model also leads to a prediction for the longitudinal component of the VVA correlator~\cite{Colangelo:2021nkr}.

%% file: sections/DRfixed.tex

\section{Dispersion relations for fixed photon virtualities}
\label{sec:DRFixedVirtualities}

By construction, the scalar functions $\tilde \W_i$ in the tensor decomposition of the VVA correlator are free from kinematic singularities. Therefore, their singularities are of purely dynamical origin and dictated by unitarity. Based on the principles of unitarity and analyticity, dispersion relations provide a means to reconstruct the scalar functions non-perturbatively and, a priori, in a model-independent way. This approach is only limited by the need to truncate the infinite sum over intermediate states in the unitarity relation and by the knowledge about the hadronic sub-processes.
Previously, dispersion relations were applied to the VVA correlator in the context of perturbative calculations~\cite{Horejsi:1985qu,Horejsi:1994aj,Pasechnik:2005ae}, but they have not been used to calculate the VVA correlator at the non-perturbative level.

In the following, we develop two complementary dispersive approaches that can be used to evaluate the VVA scalar functions. In this section, we derive dispersion relations in the axial-current invariant mass squared $q_3^2$, keeping the two photon virtualities $q_1^2$ and $q_2^2$ as fixed external variables. These dispersion relations are valid in the absence of anomalous thresholds, in particular for arbitrary space-like photon virtualities. They can also be evaluated in $g-2$ kinematics by taking the limit $q_2 \to 0$ in the very end. This is in analogy to the dispersive approach to HLbL developed in Refs.~\cite{Colangelo:2015ama, Colangelo:2017fiz}. In Sect.~\ref{sec:DRgm2}, we derive alternative dispersion relations by starting directly in $g-2$ kinematics, in analogy to the novel approach to HLbL proposed in Ref.~\cite{Ludtke:2023hvz}. A detailed comparison of the results of the two formalisms will be provided in Sect.~\ref{sec:Comparison}.

In general kinematics, the VVA correlator $\W^{\mu\nu\rho}(q_1, q_2)$ can be decomposed into four scalar functions according to Eq.~\eqref{eq:TensorDecompVVA}. These depend on $q_1^2$, $q_2^2$ and $q_3^2 = (q_1 + q_2)^2$ and a dispersion relation in $q_3^2$ holds for fixed space-like photon virtualities $q_1^2$ and $q_2^2$:
\begin{equation}
	\label{eq:DispRelq3}
	\W_i(q_1^2, q_2^2, q_3^2) = \frac{1}{\pi} \int \dif s \frac{\Delta_3 \W_i(q_1^2, q_2^2, s)}{s - q_3^2 -\imag \epsilon}\,,
\end{equation}
where the limit $\epsilon\to0$ is always implicit and the discontinuity is given by
\begin{equation}
	\label{eq:DiscontinuityDefinition}
	\Delta_3 \W_i(q_1^2, q_2^2, q_3^2) = \frac{\W_i(q_1^2 + i\epsilon, q_2^2 + i\epsilon, q_3^2 + i\epsilon) - \W_i(q_1^2 + i\epsilon, q_2^2 + i\epsilon, q_3^2 - i\epsilon)}{2i} \, ,
\end{equation}
in contrast to the imaginary part
\begin{equation}
	\label{eq:ImaginaryPartDefinition}
	\Im \W_i(q_1^2, q_2^2, q_3^2) = \frac{\W_i(q_1^2 + i\epsilon, q_2^2 + i\epsilon, q_3^2 + i\epsilon) - \W_i(q_1^2 - i\epsilon, q_2^2 - i\epsilon, q_3^2 - i\epsilon)}{2i} \, ,
\end{equation}
where we made use of the multivariate version of the Schwarz reflection principle. In the present situation of space-like $q_1^2$ and $q_2^2$, the two expressions~\eqref{eq:DiscontinuityDefinition} and \eqref{eq:ImaginaryPartDefinition} are identical. In general, the discontinuity entering a single-variable dispersion relation is obtained by considering the imaginary part below the thresholds of the fixed external variables, which can be obtained from the appropriate unitarity relation. If the discontinuity is needed for fixed external variables above the threshold of their cuts, one has to perform a subsequent analytic continuation in these variables to the upper rim of the respective cuts.

\begin{figure}
	\begin{align*}
		\includegraphics[valign=c]{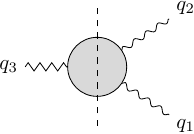}
		\quad = \quad 
		\includegraphics[valign=c]{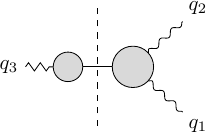}
		\quad + \quad
		\includegraphics[valign=c]{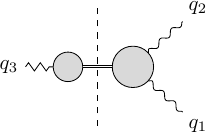}
		\quad + \quad \ldots
	\end{align*}
	\caption{Pseudoscalar and axial-vector poles due to intermediate states in the unitarity relation Eq.~\eqref{eq:Unitarity3}. The normal line denotes a pseudoscalar meson, the double line an axial-vector resonance and the zigzag line the axial current.}
	\label{fig:VVACutFixedPhotonVirtualities}
\end{figure}

In order to determine the discontinuity in Eq.~\eqref{eq:DispRelq3}, we consider the off-shell amplitude for $\gamma^*(-q_1) \gamma^*(-q_2) \to a^*(q_3)$, where $a$ denotes an auxiliary axial-vector particle coupling to the axial current with coupling constant $g_A$, see Eq.~\eqref{eq:LSZAuxiliaryAxial}. The unitarity relation for this process takes the form
\begin{align}
	\label{eq:Unitarity3}
	&\Delta_3 \left( -\imag \braket{a^*(q_3, \lambda_a) | \gamma^*(-q_1, \lambda_1) \gamma^*(-q_2, \lambda_2)} \right)\nn
	&\quad= \sum_n \frac{1}{2S_n} \left(\prod_{i=1}^{N_n} \int \widetilde{\dif p_i}\right) \braket{\gamma^*(-q_1, \lambda_1)  \gamma^*(-q_2, \lambda_2)|n;\{p_i\}}^* \braket{a^*(q_3, \lambda_a)|n;\{p_i\}} \,,
\end{align}
where $S_n$ is the symmetry factor of the state $\ket{n;\{p_i\}}$, $N_n$ its multiplicity, and $\widetilde{\dif p_i} = \frac{\dif^3 p_i}{ (2 \pi)^3 2 p_i^0}$ with $p_i^0 = (\vec{p}_i^{\,2} + m_i^2)^{1/2}$. Inserted into the dispersion relation~\eqref{eq:DispRelq3}, the unitarity relation allows us to express the four scalar functions as sums over intermediate states, as illustrated in Fig.~\ref{fig:VVACutFixedPhotonVirtualities}.

The lightest intermediate state is a single pseudoscalar and will be discussed in Sect.~\ref{sec:PseudoscalarPoleFixedPhotonVirtualities}.
Two-pion intermediate states cannot appear in $\Delta_3 \W^{\mu\nu\rho}(q_1, q_2)$ due to angular-momentum conservation and parity conservation of the strong interaction (the axial current has parity $P=1$ and angular momentum $J=1$, whereas the two-pion state has parity $P = P_\pi^2 P_J = (-1)^2 (-1)^J = -1$).
A full treatment of three-pion intermediate states would be very challenging, but we expect them to be important only when resonantly enhanced. In that case we can employ a narrow-width approximation (NWA) and include axial-vector mesons in the sum, as discussed in Sect.~\ref{sec:AxialPoleFixedPhotonVirtualities}. Neither scalar mesons nor $J^{PC} = 2^{++}$ tensor mesons couple to the axial current. We neglect further resonances with heavier masses, in particular $J^{PC} = 2^{-+}$ tensor mesons that only show up in the $1.6\GeV$ region.

\subsection{Pseudoscalar pole}
\label{sec:PseudoscalarPoleFixedPhotonVirtualities}

Selecting a single-pseudoscalar state $P$ from the sum in the unitarity relation in Eq.~\eqref{eq:Unitarity3} gives
\begin{align}
	&\Delta_3^P \left( -\imag \< a^*(q_3, \lambda_a) | \gamma^*(-q_1, \lambda_1) \gamma^*(-q_2, \lambda_2) \> \right)\nn
	&\quad= \frac{1}{2} \int \widetilde{\dif p} \, \< \gamma^*(-q_1, \lambda_1) \gamma^*(-q_2, \lambda_2)|P(p) \>^* \< a^*(q_3, \lambda_a)|P(p)\> \,.
	\label{eq:Unitarity3Pi}
\end{align}
We relate the two matrix elements on the right-hand side to the pseudoscalar TFF and decay constant, respectively,
\begin{align}
	\< \gamma^*(-q_1, \lambda_1) & \gamma^*(-q_2, \lambda_2)|P(p)\> \nn
	&= \imag e^2 \epsilon_\mu^{\lambda_1}(-q_1)^* \epsilon_\nu^{\lambda_2}(-q_2)^* (2\pi)^4 \delta^{(4)}(q_1 + q_2 + p) \epsilon^{\mu\nu\alpha\beta} q_{1\alpha} q_{2\beta} F_{P \gamma^* \gamma^*}(q_1^2, q_2^2) \, ,
\end{align}
with the TFF defined as
\begin{equation}
	\label{eq:pionTFF}
	\imag \int \dif^4 x\, e^{\,\imag q_1\cdot x} \< 0 | \T\{j_\mu^\mathrm{em}(x) j_\nu^\mathrm{em}(0)\} | P(q_1+q_2) \> = \epsilon_{\mu\nu\alpha\beta} q_1^\alpha q_2^\beta F_{P \gamma^* \gamma^*}(q_1^2,q_2^2) \, ,
\end{equation}
and
\begin{align}
	\label{eq:AxialPion}
	\< a^*(q_3, \lambda_a)|P(p) \> &= - \imag g_A \epsilon_\rho^{\lambda_a}(q_3)^* \int \dif^4 z \, e^{\imag q_3 \cdot z} \< 0 | j^\rho_5(z) | P(p) \> \nn
	&= - \imag g_A \epsilon_\rho^{\lambda_a}(q_3)^* (2\pi)^4 \delta^{(4)}(q_3 - p) \< 0 | j^\rho_5(0) | P(p) \> \nn
	&= 2 g_A \epsilon_\rho^{\lambda_a}(q_3)^* (2\pi)^4 \delta^{(4)}(q_3 - p) p^\rho \sum_a C_a F_P^a \,,
\end{align}
where the factor $2 C_a$ is due to the flavor decomposition of the axial current. Inserting these matrix elements into Eq.~\eqref{eq:Unitarity3Pi} and performing the phase-space integral gives
\begin{equation}
	\label{eq:Delta3PS}
	\Delta_3^P \W^{\mu\nu\rho} (q_1, q_2) = 2 \epsilon^{\mu\nu\alpha\beta} q_{1\alpha} q_{2\beta} F_{P \gamma^* \gamma^*}(q_1^2, q_2^2) q_3^\rho \sum_a C_a F_P^a \pi \delta(q_3^2 - m_P^2)\,.
\end{equation}
The projection onto the tensor decomposition leads to
\begin{equation}
	\Delta_3^P \W_0(q_1^2, q_2^2, q_3^2) = - 2 \pi \delta(q_3^2 - m_P^2) F_{P\gamma^*\gamma^*}(q_1^2, q_2^2) \sum_a C_a F_P^a
\end{equation}
for the first scalar function, whereas the other three do not receive a contribution from pseudoscalar single-particle intermediate states. Plugging this discontinuity into the dispersion relation, we obtain
\begin{equation}
	\label{eq:PSPoleFixedVirtGeneralKin}
	\W_i^\text{$P$-pole}(q_1^2, q_2^2, q_3^2) = 2 \sum_a C_a F_P^a \frac{F_{P\gamma^*\gamma^*}(q_1^2, q_2^2)}{q_3^2 - m_P^2} \delta_{i0}\,.
\end{equation}
In the $g-2$ limit, using Eq.~\eqref{eq:W_Wtilde}, this leads to
\begin{equation}
	\label{eq:PSPoleFixedVirt}
	\tilde{\W}_1^\text{$P$-pole}(q^2) = 2 \sum_a C_a F_P^a \frac{F_{P \gamma^* \gamma^*}(q^2, 0)}{q^2 - m_P^2} \, , \qquad \tilde{\W}_2^\text{$P$-pole}(q^2) = 0 \,.
\end{equation}

\subsection{Axial-vector pole}
\label{sec:AxialPoleFixedPhotonVirtualities}

Similarly to the pseudoscalar pole, we consider axial-vector poles in the $q_3^2$-channel, see Fig.~\ref{fig:VVACutFixedPhotonVirtualities}. For the contribution of a narrow axial-vector meson $A$, treated as an asymptotic state, we find
\begin{align}
	&\Delta_3^A \left( -\imag \braket{a^*(q_3, \lambda_a) | \gamma^*(-q_1, \lambda_1) \gamma^*(-q_2, \lambda_2)} \right)\nn
	&\quad = \frac{1}{2} \int \widetilde{\dif p} \sum_{\lambda_A}  \braket{\gamma^*(-q_1, \lambda_1) \gamma^*(-q_2, \lambda_2)|A(p, \lambda_A)}^* \braket{a^*(q_3, \lambda_a)|A(p, \lambda_A)}\,.
\end{align}
The two matrix elements evaluate to
\begin{align}
	\label{eq:axialGammaGamma}
	\< \gamma^*(-q_1, \lambda_1) & \gamma^*(-q_2, \lambda_2)|A(p, \lambda_A) \> \nn
	&\quad = \imag e^2 \epsilon_\mu^{\lambda_1}(-q_1)^* \epsilon_\nu^{\lambda_2}(-q_2)^* \epsilon_\alpha^{\lambda_A}(p) (2\pi)^4 \delta^{(4)}(q_1 + q_2 + p) \M^{\mu\nu\alpha}(-q_1, -q_2)
\end{align}
and
\begin{equation}
	\label{eq:AxialMatEl}
	\braket{a^*(q_3, \lambda_a)|A(p, \lambda_A)} = - 2 \imag g_A \epsilon_\rho^{\lambda_a}(q_3)^* (2\pi)^4 \delta^{(4)}(q_3 - p) \sum_a C_a F_A^a m_A \epsilon^\rho_{\lambda_A}(p)
\end{equation}
in the notation of Ref.~\cite{Hoferichter:2020lap}.

The polarization sum for the axial meson is given by
\begin{equation}
	\label{eq:polSumVec}
	\sum_{\lambda_A} \epsilon_\alpha^{\lambda_A}(p)^* \epsilon_\beta^{\lambda_A}(p) = -g_{\alpha \beta} + \frac{p_\alpha p_\beta}{m_A^2}\,,
\end{equation}
which leads to
\begin{align}
	\label{eq:Delta3Apole}
	\Delta_3^{A} \W^{\mu\nu\rho}(q_1, q_2) &= - 2 \imag \pi \delta\left( (q_1+q_2)^2 - m_A^2 \right) \sum_a C_a F_A^a m_A \nn
	&\quad \times \left( -{g_\alpha}^\rho + \frac{(q_1 + q_2)_\alpha (q_1 + q_2)^\rho}{m_A^2} \right) \M^{\mu\nu\alpha}(-q_1, -q_2)^*\,.
\end{align}
Using the decomposition of $\M^{\mu\nu\alpha}(q_1, q_2)$ in terms of TFFs~\cite{Hoferichter:2020lap} (see also Refs.~\cite{Kuhn:1979bb,Rudenko:2017bel,Roig:2019reh}),
\begin{equation}
	\label{eq:AxialTFFDefinition}
	\M^{\mu\nu\alpha}(q_1, q_2) = \frac{i}{m_A^2} \sum_{i=1}^3 T_i^{\mu\nu\alpha} \F_i^A(q_1^2,q_2^2) \, ,
\end{equation}
and projecting onto the tensor structures gives
\begin{align}
	\Delta_3^A \W_0(q_1^2, q_2^2, q_3^2) &= \frac{\pi}{m_A^3} \delta(q_3^2 - m_A^2) \sum_a C_a F_A^a \Big[ 2 (q_1^2 - q_2^2) \F_1^A(q_1^2, q_2^2) \nn
		&\qquad - (m_A^2 - 2 q_1^2) \F_2^A(q_1^2, q_2^2) + (m_A^2 - 2 q_2^2)\F_3^A(q_1^2, q_2^2) \Big] \,, \nn
	\Delta_3^A \W_1(q_1^2, q_2^2, q_3^2) &= - 2 \frac{\pi}{m_A} \delta(q_3^2 - m_A^2) \sum_a C_a F_A^a \F_1^A(q_1^2, q_2^2) \,, \nn
	\Delta_3^A \W_2(q_1^2, q_2^2, q_3^2) &= \frac{\pi}{m_A} \delta(q_3^2 - m_A^2) \sum_a C_a F_A^a \left[\F_2^A(q_1^2, q_2^2) - \F_3^A(q_1^2, q_2^2) \right] \,, \nn
	\Delta_3^A \W_3(q_1^2, q_2^2, q_3^2) &= \frac{\pi}{m_A} \delta(q_3^2 - m_A^2) \sum_a C_a F_A^a \left[\F_2^A(q_1^2, q_2^2) + \F_3^A(q_1^2, q_2^2) \right] \,,
\end{align}
which leads to the following contributions to the scalar functions
\begin{align}
	\W_0^\text{$A$-pole}(q_1^2, q_2^2, q_3^2) &= -\frac{1}{m_A^3} \sum_a C_a F_A^a \frac{1}{q_3^2 - m_A^2} \Big[ 2 (q_1^2 - q_2^2) \F_1^A(q_1^2, q_2^2) \nn
		&\qquad - (m_A^2 - 2 q_1^2)\F_2^A(q_1^2, q_2^2) + (m_A^2 - 2 q_2^2)\F_3^A(q_1^2, q_2^2) \Big] \,, \nn
	\W_1^\text{$A$-pole}(q_1^2, q_2^2, q_3^2) &= \frac{2}{m_A} \sum_a C_a F_A^a \frac{\F_1^A(q_1^2, q_2^2)}{q_3^2 - m_A^2} \,, \nn
	\W_2^\text{$A$-pole}(q_1^2, q_2^2, q_3^2) &= -\frac{1}{m_A} \sum_a C_a F_A^a \frac{\F_2^A(q_1^2, q_2^2) - \F_3^A(q_1^2, q_2^2)}{q_3^2 - m_A^2} \,, \nn
	\W_3^\text{$A$-pole}(q_1^2, q_2^2, q_3^2) &= -\frac{1}{m_A} \sum_a C_a F_A^a \frac{\F_2^A(q_1^2, q_2^2) + \F_3^A(q_1^2, q_2^2)}{q_3^2 - m_A^2} \,.
\end{align}
Similar findings were obtained in Ref.~\cite{Masjuan:2020jsf} based on a propagator model, which however led to a dependence on a fourth unphysical axial-vector TFF that should appear only in the case of an off-shell axial current. This TFF was then set to zero based on the anomaly constraint. Since only on-shell intermediate states enter the unitarity relation, the dispersive result does not contain any dependence on unphysical off-shell quantities, distinguishing it from a model calculation. In $g-2$ kinematics, it reduces to
\begin{align}
	\label{eq:APoleFixedVirt}
	\tilde{\W}_1^\text{$A$-pole}(q^2) &= -\frac{2}{m_A^3} \sum_a C_a F_A^a \left[ \F_1^A(q^2, 0) + \F_2^A(q^2, 0) \right] \,, \nn
	\tilde{\W}_2^\text{$A$-pole}(q^2) &= -\frac{2}{m_A} \sum_a C_a F_A^a \frac{\F_2^A(q^2, 0)}{q^2 - m_A^2} \,.
\end{align}
We note that in this limit, the dependence on the TFF $\F_3^A(q^2, 0)$ drops out: the corresponding tensor structure
\begin{equation}
	T_3^{\mu\nu\alpha}(q_1,q_2) = \epsilon^{\alpha\mu\beta\gamma} {q_1}_\beta {q_2}_\gamma q_2^\nu + \epsilon^{\alpha\mu\nu\beta} {q_1}_\beta q_2^2
\end{equation}
becomes unphysical in the singly-virtual limit with $q_2^2 = 0$ and $q_2 \cdot \epsilon^{\lambda_A}(q_2) = 0$, hence $\F_3^A(q^2, 0)$ cannot contribute to observables.

\subsection{Results for fixed photon virtualities}

In summary, if we take pseudoscalar mesons as well as axial-vector mesons as intermediate states into account, the dispersion relation for fixed photon virtualities leads to the following results for the longitudinal and transverse scalar functions in the soft-photon limit:
\begin{align}
	\label{eq:ResultScalarFunctionsFixedVirt}
	\tilde{\W}_1(q^2) &= 2 \sum_P \sum_a C_a F_P^a \frac{F_{P \gamma^* \gamma^*}(q^2, 0)}{q^2 - m_P^2} - \sum_A\sum_a \frac{2}{m_A^3} C_a F_A^a \left[ \F_1^A(q^2, 0) + \F_2^A(q^2, 0) \right] + \ldots \,, \nn 
	\tilde{\W}_2(q^2) &= - \sum_A\sum_a \frac{2}{m_A} C_a F_A^a \frac{\F_2^A(q^2, 0)}{q^2 - m_A^2} + \ldots \,,
\end{align}
where the ellipses denote neglected intermediate states. In analogy to HLbL~\cite{Colangelo:2021nkr}, this dispersion relation leads to a non-vanishing but regular (non-pole) contribution of axial-vector mesons to the longitudinal scalar function $\tilde\W_1$.

The results for the VVA correlator in the soft-photon limit obtained here depend on the singly-virtual pseudoscalar and axial-vector TFFs. In the pseudoscalar case, precise results from dispersive analyses exist~\cite{Hoferichter:2014vra, Hoferichter:2018dmo, Hoferichter:2018kwz, Hanhart:2013vba, Holz:2022hwz, Holz:2015tcg, Kubis:2018bej, Holz:2024diw}. The axial TFFs have been studied in the literature using vector-meson-dominance parametrizations~\cite{Zanke:2021wiq,Hoferichter:2023tgp} as well as holographic models~\cite{Leutgeb:2019gbz}, but so far no dispersive reconstruction is available. In order to facilitate the comparison with the complementary dispersive approach in Sect.~\ref{sec:Comparison}, we introduce dispersive reconstructions of the TFFs in App.~\ref{sec:TFFs}.

The asymptotic behavior of the scalar functions $\tilde\W_i(q_1^2)$ for large space-like momenta $q^2 = -Q^2$ follows from the OPE, see Sect.~\ref{sec:VVA_OPE}. The leading term predicts
\begin{equation}
	\label{eq:VVALeadingOPE}
	\tilde\W_i(-Q^2) \propto \frac{1}{Q^2} \left[ 1 + \Order\left(  \frac{m_q^2}{Q^2}, \frac{\LQCD^2}{Q^2} \right) \right]
\end{equation}
for both functions. The expressions~\eqref{eq:ResultScalarFunctionsFixedVirt} show that the contribution of a finite number of pseudoscalar or axial-vector intermediate states falls off too fast to saturate the asymptotic constraints on the scalar functions $\tilde\W_i$. We will come back to this aspect in Sect.~\ref{sec:OPEandChPTConstraints}.

%% file: sections/DRgm2.tex

\section{\boldmath Dispersion relations in \texorpdfstring{$g-2$}{g-2} kinematics}
\label{sec:DRgm2}

We now turn to dispersion relations directly in $g-2$ kinematics, in analogy to the formalism for HLbL introduced in Ref.~\cite{Ludtke:2023hvz}.

We assume that the falloff~\eqref{eq:VVALeadingOPE} predicted by the OPE remains true in the time-like region, which allows us to write an unsubtracted dispersion relation for the scalar functions $\tilde{\W}_i(q_1^2)$ in $g-2$ kinematics:
\begin{equation}
	\label{eq:DispRel}
	\tilde{\W}_i(q^2) = \frac{1}{\pi} \int \dif s \frac{\Delta \tilde{\W}_i(s)}{s - q^2 -\imag \epsilon}\,.
\end{equation}
The discontinuity of the scalar functions in $g-2$ kinematics $\Delta \tilde{\W}_i(s)$ can be expressed in terms of the discontinuities of the scalar functions $\tilde{\W}_i(q_1^2, q_2^2, q_3^2)$ in different physical regions, in complete analogy to the triangle-kinematics discontinuity in HLbL~\cite{Ludtke:2023hvz},
\begin{align}
	\label{eq:DiscWtildei}
	\Delta \tilde{\W}_i(q_1^2) &= \frac{1}{2\imag} \Big(\tilde{\W}_i(q_1^2 + \imag \epsilon, 0, q_1^2 + \imag \epsilon) - \tilde{\W}_i(q_1^2 - \imag \epsilon, 0, q_1^2 - \imag \epsilon) \Big) \nn
	&= \lim_{q_3^2 \to q_1^2} \frac{1}{2\imag} \Big(\tilde{\W}_i(q_1^2 + \imag \epsilon, 0, q_3^2 + \imag \epsilon) - \tilde{\W}_i(q_1^2 - \imag \epsilon, 0, q_3^2 - \imag \epsilon) \Big) \nn
	&= \lim_{q_3^2 \to q_1^2} \frac{1}{2\imag} \Big(\tilde{\W}_i(q_1^2 + \imag \epsilon, 0, q_3^2 + \imag \epsilon) - \tilde{\W}_i(q_1^2 + \imag \epsilon, 0, q_3^2 - \imag \epsilon) \nn
	&\qquad\qquad + \tilde{\W}_i(q_1^2 + \imag \epsilon, 0, q_3^2 - \imag \epsilon) - \tilde{\W}_i(q_1^2 - \imag \epsilon, 0, q_3^2 - \imag \epsilon) \Big) \nn
	&= \lim_{q_3^2 \to q_1^2} \bigg[\frac{1}{2\imag} \Big(\tilde{\W}_i(q_1^2 + \imag \epsilon, 0, q_3^2 + \imag \epsilon) - \tilde{\W}_i(q_1^2 + \imag \epsilon, 0, q_3^2 - \imag \epsilon)\Big) \nn
	&\qquad\quad + \bigg(\frac{1}{2\imag}\Big(\tilde{\W}_i(q_1^2 + \imag \epsilon, 0, q_3^2 + \imag \epsilon) - \tilde{\W}_i(q_1^2 - \imag \epsilon, 0, q_3^2 + \imag \epsilon)\Big)\bigg)^* \bigg] \nn
	&= \lim_{q_3^2 \to q_1^2} \bigg[\Delta_3 \tilde{\W}_i(q_1^2 + \imag \epsilon, 0, q_3^2) + \left(\Delta_1 \tilde{\W}_i(q_1^2, 0, q_3^2 + \imag \epsilon)\right)^*\bigg]\,.
\end{align}
In this equation, $\Delta_{1,3}$ denote the discontinuities in the $q_{1,3}^2$-channels, respectively. As mentioned in Sect.~\ref{sec:DRFixedVirtualities}, these discontinuities can be obtained by performing the analytic continuation of imaginary parts that follow from the unitarity relations: e.g., in order to obtain the discontinuity $\Delta_1\tilde\W_i$, we consider the unitarity relation in the $q_1^2$-channel, evaluated for $q_3^2$ below the $q_3^2$-channel thresholds. This unitarity relation provides the imaginary part $\Im\tilde\W_i$, which is a real quantity and needs to be continued analytically to $q_3^2 + \imag \epsilon$, resulting in $\Delta_1\tilde\W_i$. Due to this analytic continuation, the discontinuities are in general complex quantities.

\begin{figure}
	\centering
	\begin{align*}
		\Delta \left[ \includegraphics[valign=c]{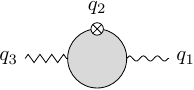} \right]
		\quad = \quad 
		\left( \includegraphics[valign=c]{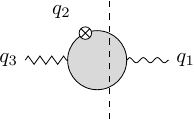} \right)^*
		\quad + \quad
		\includegraphics[valign=c]{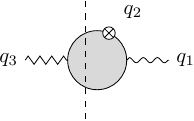}
	\end{align*}
	\caption{Unitarity cuts contributing to the discontinuity of the VVA function in the limit of $g-2$ kinematics. The axial current is denoted by a zigzag line, the off-shell photon by a wavy line, and the static external electromagnetic field by a crossed circle.}
	\label{fig:VVATwoCuts}
\end{figure}

In the third line of Eq.~\eqref{eq:DiscWtildei}, we could equally well have added and subtracted the function $\tilde{\W}_i(q_1^2 - \imag \epsilon, 0, q_3^2 + \imag \epsilon)$, which would have resulted in having the complex conjugation on $\Delta_3\tilde\W_i$ instead of $\Delta_1\tilde\W_i$. This is equivalent, because the final discontinuity $\Delta \tilde{\W}_i(s) = \Im \tilde{\W}_i(s)$ is real. In general, soft-photon singularities cancel between the two cuts such that the limit of the sum exists~\cite{Ludtke:2023hvz}. However, all contributions discussed in the following are free of these singularities so that the limit of each term can be taken individually. Thus, the imaginary part in degenerate kinematics can be obtained as the sum of two cuts in general kinematics as illustrated in Fig.~\ref{fig:VVATwoCuts}. This can be understood from the fact that the two cuts become kinematically indistinguishable in the limit $q_2 \to 0$.

\begin{figure}
	\begin{align*}
		\includegraphics[valign=c]{images/q3-cut}
		\quad = \quad 
		\includegraphics[valign=c]{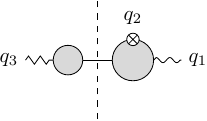}
		\quad + \quad
		\includegraphics[valign=c]{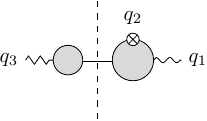}
		\quad + \quad \ldots
	\end{align*}
	\caption{Pseudoscalar and axial-vector poles due to intermediate states in the unitarity relation Eq.~\eqref{eq:Unitarity3}, evaluated in the soft-photon limit. This corresponds to Fig.~\ref{fig:VVACutFixedPhotonVirtualities} but is evaluated in a different kinematic configuration.}
	\label{fig:VVACut3}
\end{figure}

\begin{figure}
	\begin{align*}
		\includegraphics[valign=c]{images/q1-cut}
		\quad = \quad 
		\includegraphics[valign=c]{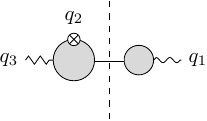} 
		\quad + \quad
		\includegraphics[valign=c]{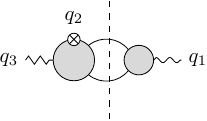}
		\quad + \quad \ldots
	\end{align*}
	\caption{Vector-meson poles and two-pion cuts due to intermediate states in the unitarity relation Eq.~\eqref{eq:Unitarity1}.}
	\label{fig:VVACut1}
\end{figure}

In addition to the $q_3^2$-channel unitarity relation in Eq.~\eqref{eq:Unitarity3}, we thus need the unitarity relation in the $q_1^2$-channel
\begin{align}
	\label{eq:Unitarity1}
	&\Delta_1 \left( -\imag \braket{\gamma^*(q_2, \lambda_2) a^*(q_3, \lambda_a) | \gamma^*(-q_1, \lambda_1)} \right)\nn
	&\quad= \sum_n \frac{1}{2S_n}\left(\prod_{i=1}^{N_n} \int \widetilde{\dif p_i}\right)\braket{\gamma^*(-q_1, \lambda_1) |n;\{p_i\}}^* \braket{\gamma^*(q_2, \lambda_2) a^*(q_3, \lambda_a)|n;\{p_i\}} \,,
\end{align}
continued analytically to $q_3^2 + i \epsilon$. The two unitarity relations Eqs.~\eqref{eq:Unitarity3} and \eqref{eq:Unitarity1} allow us to describe the two contributions to the discontinuity in the limit of $g-2$ kinematics in terms of a sum over intermediate states. As before, we consider pseudoscalar and axial-vector mesons as intermediate states in the $q_3^2$-channel, shown in Fig.~\ref{fig:VVACut3}. In the $q_1^2$-channel, we take vector-meson as well as two-pion intermediate states into account, as shown in Fig.~\ref{fig:VVACut1}. Scalar mesons or tensor mesons with $J^{PC} = 2^{++}$ do not couple to vector currents and thus do not contribute to the $q_1^2$-channel unitarity relation.

\subsection{Pseudoscalar pole}

A single pseudoscalar meson does not couple to the electromagnetic current and thus only appears in the $q_3^2$-channel unitarity relation, shown in Fig.~\ref{fig:VVACut3}. The single-pseudoscalar contribution to the unitarity relation is given in Eq.~\eqref{eq:Delta3PS}. We expand this in $q_2$ to linear order,
\begin{align}
	\Delta_3^P \W^{\mu\nu\rho} (q_1, q_2) &= - 2 \pi \epsilon^{\mu\nu\alpha\beta} q_{1\alpha} q_{2\beta} q_1^\rho F_{P \gamma^* \gamma^*}(m_P^2, 0) \sum_a C_a F_P^a \delta(q_1^2 - m_P^2) + \Order(q_2^2)\nn
	&= - 2 \pi \tau_1^{\mu\nu\rho\sigma} q_{2\sigma} F_{P \gamma^* \gamma^*}(m_P^2, 0) \sum_a C_a F_P^a \delta(q_1^2 - m_P^2) + \Order(q_2^2) \, ,
\end{align}
and obtain through projection onto the Lorentz structures in the limit $q_2 \to 0$
\begin{equation}
	\label{eq:Delta3PWtildei}
	\Delta^P \tilde{\W}_i (q^2) = - 2 \pi F_{P \gamma^* \gamma^*}(m_P^2, 0) \sum_a C_a F_P^a \delta(q^2 - m_P^2) \delta_{i1}\,.
\end{equation}
This can be inserted into the dispersion relation Eq.~\eqref{eq:DispRel}, from which we obtain a contribution of a single-pseudoscalar intermediate state only to the longitudinal scalar function
\begin{equation}
	\label{eq:PSpolesg-2kin}
	\tilde{\W}_1^P(q^2) = 2 \sum_a C_a F_P^a \frac{F_{P \gamma^* \gamma^*}(m_P^2, 0)}{q^2 - m_P^2} \, , \qquad \tilde{\W}_2^P(q^2) = 0 \,.
\end{equation}

In the chiral limit (and neglecting the gluon anomaly in the singlet case), we have $m_P = 0$ for the Goldstone bosons and $\sum\limits_P F_P^a F_{P \gamma^* \gamma^*} (0, 0) = \frac{N_c}{4\pi^2} \tr(\Q^2 \lambda^a)$, which leads to
\begin{equation}
	\sum_P \tilde{\W}_1^P(q^2) = \frac{N_c \sum_a C_a \tr(\Q^2 \lambda^a)}{2 \pi^2 q^2} = \frac{2 \sum_a C_a \mathcal{A}^{(a)}}{q^2} = \frac{\mathcal{A}}{q^2} \, ,
\end{equation}
in agreement with Eq.~\eqref{eq:WardConstraint}. This shows that the poles due to the lightest pseudoscalars defined by this dispersion relation saturate $\tilde{\W}_1(q^2)$ in the chiral limit (and if the gluon anomaly is neglected) for arbitrary $q^2$ and the sum over all other contributions to $\tilde{\W}_1(q^2)$ has to vanish in this limit.

The result obtained here differs from the pseudoscalar-pole contribution in the dispersion relation with fixed photon virtualities by non-pole contributions: while the result~\eqref{eq:PSpolesg-2kin} is given by a pure pole in $q^2$, taking the limit $q_2 \to 0$ only after writing dispersion relations gives rise to the additional non-pole piece in Eq.~\eqref{eq:PSPoleFixedVirt}, in analogy to the situation in HLbL~\cite{Colangelo:2019uex,Ludtke:2023hvz}. This will be discussed in more detail in Sect.~\ref{sec:Comparison}.

\subsection{Axial-vector pole}

Similarly to the pseudoscalar pole, the axial-vector pole can only appear in the discontinuity in $q_3^2$, see Fig.~\ref{fig:VVACut3}. Again, the discontinuity can be taken from the dispersive approach in general kinematics, Eq.~\eqref{eq:Delta3Apole}. Using the decomposition of $\M^{\mu\nu\alpha}(q_1, q_2)$ in terms of TFFs~\cite{Hoferichter:2020lap}, taking into account the analytic continuation to $q_1^2 + \imag \epsilon$, which removes the complex conjugation from the TFFs, and expanding in $q_2$, we arrive at
\begin{equation}
	\label{eq:Delta3MWtildei}
	\Delta^A \tilde{\W}_i(q^2) = 2 \pi \delta(q^2 - m_A^2) \sum_a \frac{C_a F_A^a}{m_A} \F_2^A(m_A^2, 0) \delta_{i2}\,.
\end{equation}
Inserting this into the dispersion relation leads to
\begin{equation}
	\label{eq:Apole_g-2kin}
	\tilde{\W}_1^A (q^2) = 0 \, , \qquad \tilde{\W}_2^A (q^2) = -2 \sum_a \frac{C_a F_A^a}{m_A} \frac{\F_2^A(m_A^2, 0)}{q^2 - m_A^2}\,.
\end{equation}
As opposed to the pseudoscalar poles, the axial poles only contribute to the transverse scalar function. Analogously to the case of pseudoscalars, the difference to the dispersion relations for fixed photon virtualities only consists of contributions without an axial-vector meson pole.

\subsection{Vector-meson poles}
Vector-meson poles can only contribute to the discontinuity in $q_1^2$, shown in Fig.~\ref{fig:VVACut1}. While the contribution of the $\rho$-meson will be part of the two-pion intermediate state in Sect.~\ref{sec:VVA_2pi}, the very narrow $\omega$ and $\phi$ mesons can be described in terms of a NWA. Together with the axial-vector poles, this is expected to capture the most important effects of three-pion intermediate states in VVA.

The contribution of a vector meson $V$ to the sum over intermediate states in Eq.~\eqref{eq:Unitarity1} is given by
\begin{align}
	&\Delta_1^V \left( -\imag \braket{\gamma^*(q_2, \lambda_2) a^*(q_3, \lambda_A) | \gamma^*(-q_1, \lambda_1)} \right)\nn
	&\qquad= \frac{1}{2} \int \widetilde{\dif p} \sum_{\lambda_V} \braket{\gamma^*(-q_1, \lambda_1) | V(p, \lambda_V)}^* \braket{\gamma^*(q_2, \lambda_2) a^*(q_3, \lambda_a) | V(p, \lambda_V)}\,.
\end{align}
The first matrix element on the right-hand side evaluates to
\begin{equation}
	\label{eq:vectorGammaMatrixElement}
	\braket{\gamma^*(-q_1, \lambda_1) | V(p, \lambda_V)} = - \imag e \left(\epsilon^{\lambda_1}(-q_1)^* \cdot \epsilon^{\lambda_V}(p) \right) (2\pi)^4 \delta^{(4)}(q_1 + p) m_V f_V\,,
\end{equation}
where the vector decay constant $f_V$ is defined by
\begin{equation}
	\label{eq:VectorDecayConst}
	\bra{0} j^\mu_\mathrm{em}(0) \ket{V(p, \lambda_V)} = m_V f_V \epsilon^\mu_{\lambda_V}(p)\,.
\end{equation}
The second matrix element can be written as
\begin{align}
	\label{eq:gammaAVMatrixElement}
	\< \gamma^*(q_2, \lambda_2) & a^*(q_3, \lambda_a) | V(p, \lambda_V) \> \nn
	&\quad = -e g_A \epsilon_\nu^{\lambda_2}(q_2)^* \epsilon_\rho^{\lambda_a}(q_3)^* (2\pi)^4 \delta^{(4)}(q_2 + q_3 - p) \epsilon_\lambda^{\lambda_V}(p) \V^{\lambda \nu \rho}(-p, q_2) \, ,
\end{align}
with
\begin{equation}
	\label{eq:vectorMesonTFF}
	\epsilon_\lambda^{\lambda_V}(p) \V^{\lambda \nu \rho}(-p, q_2) = \int \dif^4 x e^{\imag q_2 \cdot x} \bra{0} T\{j^\nu_\mathrm{em}(x) j^\rho_5(0)\} \ket{V(p, \lambda_V)}\,.
\end{equation}
The BTT decomposition of the function $\V^{\lambda \rho \nu}(-p, q_2)$ leads to the result that the same tensor structures as for the VVA correlator can be used without introducing kinematic singularities:
\begin{equation}
	\label{eq:tensorDecompositionVAVectorMeson}
	\V^{\lambda\nu\rho}(-p, q_2) = \sum_{i=0}^3 \tau^{\lambda\nu\rho}_i(-p, q_2) \V_i(q_2^2, (p-q_2)^2) \, .
\end{equation}
Putting everything together, we find
\begin{align}
	\Delta_1^V \W^{\mu\nu\rho}(q_1, q_2) &= \pi \delta(q_1^2 - m_V^2) m_V f_V \V^{\lambda \nu \rho}(q_1, q_2) \left(-g_\lambda^\mu + \frac{q_1^\mu q_{1\lambda}}{m_V^2}\right) \nn
	&= -\pi \delta(q_1^2 - m_V^2) m_V f_V \V^{\mu \nu \rho}(q_1, q_2)\,,
\end{align}
where we have used the polarization sum similar to Eq.~\eqref{eq:polSumVec}. In $g-2$ kinematics, we obtain
\begin{equation}
	\label{eq:Delta1VWtildei}
	\Delta^V \tilde{\W}_i(q^2) = -\pi \delta^{(4)}(q^2 - m_V^2) m_V f_V \tilde{\V}_i(m_V^2) \, ,
\end{equation}
with
\begin{align}
	\label{eq:VtildeDef}
	\tilde{\V}_1(q^2) &= \V_0(0, q^2) + \V_1(0, q^2) - \V_3(0, q^2) \,, \nn*
	\tilde{\V}_2(q^2) &= \V_2(0, q^2) + \V_3(0, q^2) \,.
\end{align}
Plugging this into the dispersion relation and assuming that both $f_V$ and $\tilde{\V}_i(m_V^2)$ are real leads to
\begin{align}
	\tilde{\W}_i^V(q^2) = \frac{m_V f_V \tilde{\V}_i(m_V^2)}{q^2 - m_V^2}\,.
	\label{eq:resVPole}
\end{align}

In analogy to the other contributions, we can also perform a flavor decomposition of the axial current in $\V^{\lambda \nu \rho}(-p, q_2)$, which gives
\begin{align}
	\tilde{\W}_i^V(q^2) = 2 \sum_{a=3,8,0} \frac{m_V f_V C_a \tilde{\V}_i^{(a)}(m_V^2)}{q^2 - m_V^2}\,.
\end{align}
The determination of the constants $\tilde{\V}_i^{(a)}(m_V^2)$ is discussed in App.~\ref{sec:VectorMesonTransitionAmplitude}.

\subsection{Two-pion intermediate state}
\label{sec:VVA_2pi}

The two-pion intermediate state contributes to $\Delta_1 \W^{\mu\nu\rho}(q_1, q_2)$, as illustrated in Fig.~\ref{fig:VVACut1}. The unitarity relation Eq.~\eqref{eq:Unitarity1} gives
\begin{align}
	\label{eq:Unitarity1PiPi}
	&\Delta_1^{\pi\pi} \left( -\imag \braket{\gamma^*(q_2, \lambda_2) a^*(q_3, \lambda_a) | \gamma^*(-q_1, \lambda_1)} \right) \nn
	&\quad = \frac{1}{2} \int \widetilde{\dif p_1} \widetilde{\dif p_2} \braket{\gamma^*(-q_1, \lambda_1)| \pi^+(p_1) \pi^-(p_2)}^* \braket{\gamma^*(q_2, \lambda_2) a^*(q_3, \lambda_a)| \pi^+(p_1) \pi^-(p_2)} \, .
\end{align}
Since the photon only couples to a pure isospin $I=1$ two-pion state, there is no $\pi^0 \pi^0$ contribution.

The first matrix element on the right-hand side of Eq.~\eqref{eq:Unitarity1PiPi} can be reduced to the pion vector form factor (VFF)
\begin{equation}
	\label{eq:pionVFF}
	\bra{\pi^+(-p_2)} j^\mu(0) \ket{\pi^+(p_1)} = (p_1 - p_2)^\mu F_\pi^V \left((p_1 + p_2)^2\right)\,,
\end{equation}
see Refs.~\cite{Ananthanarayan:2012tn,Ananthanarayan:2012tt,Ananthanarayan:2013zua,Ananthanarayan:2016mns,Ananthanarayan:2017efc,Colangelo:2018mtw,Ananthanarayan:2018nyx,Ananthanarayan:2020vum,Colangelo:2020lcg,Colangelo:2022prz,Stoffer:2023gba,Simula:2023ujs} for recent dispersive analyses.
With the isospin decomposition of the two-pion state
\begin{equation}
	\ket{\pi^+ \pi^-} = \frac{1}{\sqrt{6}} \ket{2,0} + \frac{1}{\sqrt{2}} \ket{1, 0} + \frac{1}{\sqrt{3}} \ket{0, 0} \, ,
\end{equation}
we define the isospin $I=1$ projection
\begin{equation}
	\label{eq:IsospinProj}
	\ket{\pi^+ \pi^-}_{I=1} = \ket{1, 0} \braket{1, 0 | \pi^+ \pi^-} = \frac{1}{\sqrt{2}} \ket{1, 0} \, ,
\end{equation}
which is odd under pion crossing $p_1 \leftrightarrow p_2$. We write the second matrix element in Eq.~\eqref{eq:Unitarity1PiPi} as
\begin{align}
	\label{eq:VApipiReduction}
	&\braket{\gamma^*(q_2, \lambda_2) a^*(q_3, \lambda_a)| \pi^+(p_1) \pi^-(p_2)}_{I=1} \nn
	&\qquad = -e g_A \epsilon_\nu^{\lambda_2}(q_2)^* \epsilon_\rho^{\lambda_a}(q_3)^* \int \dif^4 x \dif^4 y \, e^{\imag q_2 \cdot x} e^{\imag q_3 \cdot y} \bra{0} \T\{j_\mathrm{em}^\nu(x) j_5^\rho(y)\} \ket{\pi^+(p_1) \pi^-(p_2)}_{I=1} \nn
	&\qquad = - e g_A \epsilon_\nu^{\lambda_2}(q_2)^* \epsilon_\rho^{\lambda_a}(q_3)^* (2\pi)^4 \delta^{(4)}(q_2 + q_3 - p_1 - p_2) \U^{\nu \rho}(q_2, p_1, p_2) \,,
\end{align}
where in the last line we have defined
\begin{equation}
	\label{eq:defU}
	\U^{\nu \rho}(q_2, p_1, p_2) = \int \dif^4 x \, e^{\imag q_2 \cdot x} \bra{0} \T\{j_\mathrm{em}^\nu(x) j_5^\rho(0)\} \ket{\pi^+(p_1) \pi^-(p_2)}_{I=1} \,.
\end{equation}
To our knowledge, the matrix element $\U^{\nu\rho}$ has not yet been discussed in the literature. In App.~\ref{sec:AxialCurrentGammaPiPi}, we present its tensor decomposition and the dispersive reconstruction of the corresponding scalar coefficient functions.
Inserting the two matrix elements into Eq.~\eqref{eq:Unitarity1PiPi}, we obtain
\begin{align}
	\label{eq:2PiDisc}
	&\Delta_1^{\pi\pi} \W^{\mu\nu\rho}(q_1, q_2) \nn
	&\qquad = -\frac{1}{2}\int \widetilde{\dif p}_1 \widetilde{\dif p}_2 (2\pi)^4 \delta^{(4)}(q_1 + p_1 + p_2) (p_1 - p_2)^\mu F_\pi^V (q_1^2)^* \,  \U^{\nu\rho}(q_2, p_1, p_2)\,,
\end{align}
where we have taken into account that unitarity is used in the $q_1^2$-channel and thus no analytic continuation of the pion VFF is needed. In the limit of $g-2$ kinematics, we find that the two-pion contribution to the VVA scalar functions can be written as
\begin{align}
	\label{eq:WipipiDisc}
	\Delta^{\pi\pi} \tilde{\W}_1 (q^2) &= -\frac{1}{96\pi} \sigma_\pi^3(q^2) F_\pi^V(q^2) \left[2 \tilde\U_1(q^2)^* + q^2 \tilde\U_2(q^2)^* \right] \,, \nn
	\Delta^{\pi\pi} \tilde{\W}_2 (q^2) &= \frac{1}{48\pi} \sigma_\pi^3(q^2) F_\pi^V(q^2) \tilde\U_1(q^2)^* \,,
\end{align}
where the two scalar functions $\tilde\U_{1,2}$ arise in the tensor decomposition of $\U^{\nu\rho}$ and are free of kinematic singularities, see App.~\ref{sec:AxialCurrentGammaPiPiTensorDecomposition}.
The dispersion relation for VVA leads to the two-pion contribution
\begin{align}
	\label{eq:Wipipi}
	\tilde\W_1^{\pi\pi}(q^2) &= - \frac{1}{96\pi^2} \int_{4m_\pi^2}^\infty ds \frac{\sigma_\pi^3(s) F_\pi^V(s) \bigl[2 \tilde\U_1(s)^* + s \tilde\U_2(s)^* \bigr]}{s-q^2 - i \epsilon} \,, \nn
	\tilde\W_2^{\pi\pi}(q^2) &= \frac{1}{48\pi^2} \int_{4m_\pi^2}^\infty ds \frac{\sigma_\pi^3(s) F_\pi^V(s) \, \tilde\U_1(s)^*}{s-q^2 - i \epsilon} \,.
\end{align}

\subsection[Results in $g-2$ kinematics]{\boldmath Results in $g-2$ kinematics}
\label{sec:DRgm2LimitResults}

Collecting the contributions of pseudoscalar, axial-vector, vector, and two-pion intermediate states, we find that the VVA scalar functions are given by
\begin{align}
	\label{eq:ResultScalarFunctionsgm2}
	\tilde{\W}_1(q^2) &= 2 \sum_P \sum_a C_a F_P^a \frac{F_{P \gamma^* \gamma^*}(m_P^2, 0)}{q^2 - m_P^2} + \sum_V \frac{m_V f_V \tilde\V_1(m_V^2)}{q^2 - m_V^2} \nn
		&\quad - \frac{1}{96 \pi^2} \int_{4m_\pi^2}^\infty \dif s \frac{\sigma_\pi^3(s) F_\pi^V(s) \bigl[2 \tilde\U_1(s)^* + s \tilde\U_2(s)^* \bigr]}{s - q^2 - \imag \epsilon} + \ldots \,,\nn
	\tilde{\W}_2(q^2) &= - 2 \sum_A \sum_a \frac{C_a F_A^a}{m_A} \frac{\F_2^A(m_A^2, 0)}{q^2 - m_A^2} + \sum_V \frac{m_V f_V \tilde\V_2(m_V^2)}{q^2 - m_V^2} \nn
		&\quad + \frac{1}{48 \pi^2} \int_{4m_\pi^2}^\infty \dif s \frac{\sigma_\pi^3(s) F_\pi^V(s) \tilde\U_1(s)^*}{s - q^2 - \imag \epsilon} + \ldots \,,
\end{align}
where the ellipses denote the contribution of neglected higher intermediate states.

The anomaly constraint for $\tilde{\W}_1$ in Eq.~\eqref{eq:WardConstraint} is saturated by the first term in Eq.~\eqref{eq:ResultScalarFunctionsgm2}, which implies that in the chiral limit and neglecting the gluon anomaly the other contributions have to drop out. As shown in App.~\ref{sec:TwoPionResultsInsertedIntoVVA}, this indeed happens for the two-pion contribution. The vector-meson contributions to $\tilde{\W}_1$ have poles at finite $q^2$, which cannot cancel against another contribution, hence their residues have to vanish in the chiral limit. The implications of the non-renormalization constraint Eq.~\eqref{eq:NonRenTheorem} will be discussed in Sect.~\ref{sec:VVA_OPE}.

In order to compare in more detail the results of the two dispersion relations, Eqs.~\eqref{eq:ResultScalarFunctionsFixedVirt} and~\eqref{eq:ResultScalarFunctionsgm2}, we need to analyze the sub-processes: changing the dispersion relation corresponds to a reshuffling of unitarity cuts between the VVA correlator and its sub-processes. Therefore, in App.~\ref{sec:TFFs} we consider dispersive representations of the pseudoscalar and axial-meson TFFs. The analysis of the additional vector-meson and two-pion sub-amplitudes $\tilde\V_i$ and $\tilde\U_i$ can be found in App.~\ref{sec:VectorMesonTransitionAmplitude} and~\ref{sec:AxialCurrentGammaPiPi}. Inserting the dispersive representation for the two-pion sub-amplitudes $\tilde\U_i(s)$, one obtains the following results for the VVA scalar functions:
\begin{align}
	\label{eq:ResultScalarFunctionsgm2TwoPionExplicit}
	\tilde{\W}_1(q^2) &= 2 \sum_P \sum_a C_a F_P^a \frac{F_{P \gamma^* \gamma^*}(m_P^2, 0)}{q^2 - m_P^2} + \sum_V \frac{m_V f_V \tilde{\V}_1(m_V^2)}{q^2 - m_V^2} \nn
		&\quad - \frac{1}{48 \pi^2} \int_{4m_\pi^2}^\infty \dif s \frac{\sigma_\pi^3(s)  F_\pi^V(s) \Omega_1^1(s)^* \, \tilde\U_1(0)}{s - q^2 - \imag \epsilon} \nn
		&\quad + \frac{1}{48\pi^2} \sum_P \int_{4m_\pi^2}^\infty \dif s \frac{s \sigma_\pi^3(s)  F_\pi^V(s) \Omega_1^1(s)^*}{(s - q^2 - \imag \epsilon)(s - m_P^2 - \imag \epsilon)}  \sum_a \frac{C_a F_P^a \F_{P}}{\Omega_1^1(m_P^2)} + \ldots \,, \nn
	\tilde{\W}_2(q^2) &= - 2 \sum_A \sum_a \frac{C_a F_A^a}{m_A} \frac{\F_2^A(m_A^2, 0)}{q^2 - m_A^2} + \sum_V \frac{m_V f_V \tilde{\V}_2(m_V^2)}{q^2 - m_V^2} \nn
		&\quad + \frac{1}{48 \pi^2} \int_{4m_\pi^2}^\infty \dif s \frac{\sigma_\pi^3(s) F_\pi^V(s) \Omega_1^1(s)^* \, \tilde\U_1(0)}{s - q^2 - \imag \epsilon} \nn
		&\quad + \frac{1}{24\pi^2} \sum_A \int_{4m_\pi^2}^\infty \dif s \frac{s \sigma_\pi^3(s) F_\pi^V(s) \Omega_1^1(s)^*}{(s - q^2 - \imag \epsilon)(s - m_A^2 - \imag \epsilon)} \sum_a \frac{C_a F_A^a}{m_A} \frac{\tilde\F_{A,1}}{\Omega_1^1(m_A^2)} + \ldots \,,
\end{align}
where the Omn\`es function is
\begin{equation}
	\label{eq:Omnes}
	\Omega_1^1(s) = \exp\left\{ \frac{s}{\pi} \int_{4m_\pi^2}^\infty ds' \frac{\delta_1^1(s')}{s'(s'-s-i\epsilon)} \right\} \, ,
\end{equation}
with the elastic isospin $I=1$ $\pi\pi$-scattering  $P$-wave phase shift $\delta_1^1$. Through the two-pion sub-process, the longitudinal function $\tilde\W_1$ depends on $\F_P = \F_P(m_P^2, m_\pi^2, m_\pi^2;0)$, which describes $\gamma\to\pi^+\pi^- P$ in the soft-photon limit, see Eq.~\eqref{eq:gammaPiPiPScFun}. The transverse function $\tilde\W_2$ depends on $\tilde\F_{A,1} = \tilde\F_{A,1}(m_A^2, m_\pi^2, m_\pi^2)$, which describes the soft-photon limit of $\gamma\to\pi^+\pi^- A$.

The scalar functions $\tilde\W_i$ need to be real on the real axis apart from poles and cuts due to the considered intermediate states. In Eq.~\eqref{eq:ResultScalarFunctionsgm2TwoPionExplicit}, this is not manifest if $m_{P,A}^2 > 4 m_\pi^2$. In this case, both the TFFs and the two-pion contributions exhibit additional imaginary parts. However, with the explicit form of the pseudoscalar and axial-vector TFFs derived in App.~\ref{sec:DispForPSTFFs} and \ref{sec:axialTFF}, one indeed finds that all unphysical imaginary parts cancel, up to effects due to neglected higher intermediate states. In particular, the pion VFF is given by $F_\pi^V(s) = \Omega_1^1(s)$ if inelastic effects are neglected.

%% file: sections/Comparison.tex

\section{Analytic comparison of the two dispersive approaches to VVA}
\label{sec:Comparison}

\begin{table}[t]
	\centering
	\begin{tabular}{ c | c c c }
	\toprule
	& \multicolumn{3}{c}{DR for fixed photon virtualities} \\
	\cmidrule{2-4}
	$g-2$ DR & $\pi^0, \eta, \eta'$ & $A$ & $\ldots$ \\
	\midrule
	\midrule
	$\pi^0, \eta, \eta'$
		& $\includegraphics[height=1cm,valign=c]{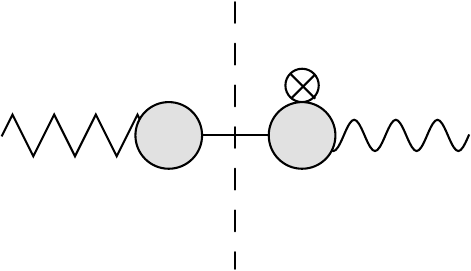}$
		& $\times$
		& $\times$
		\\
	\midrule
	$2\pi$
		& \includegraphics[height=1cm,valign=c]{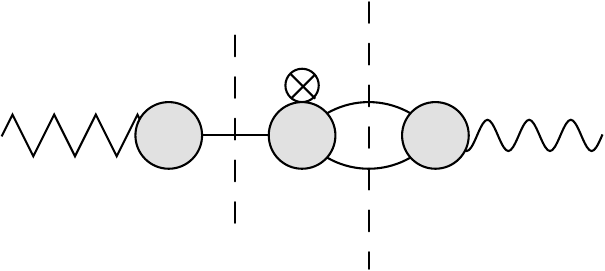}
		& \includegraphics[height=1cm,valign=c]{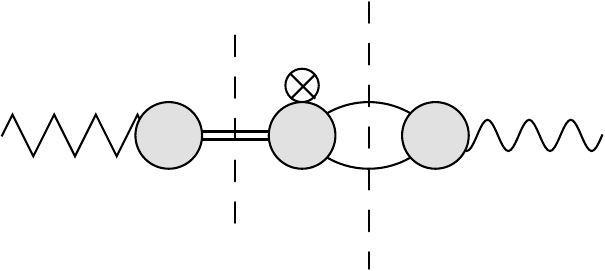}
		& \includegraphics[height=1cm,valign=c]{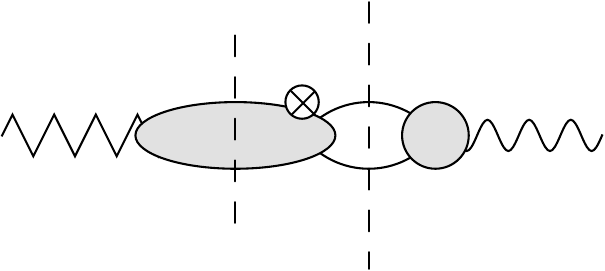}
		\\
	\midrule
	$V$
		&	\includegraphics[height=1cm,valign=c]{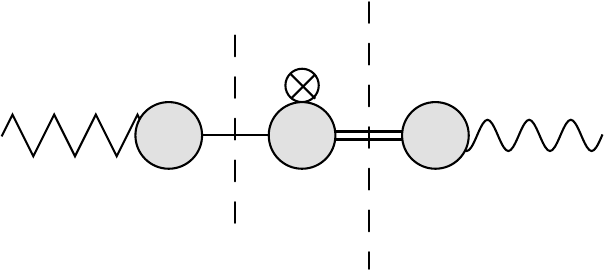}
		&	\includegraphics[height=1cm,valign=c]{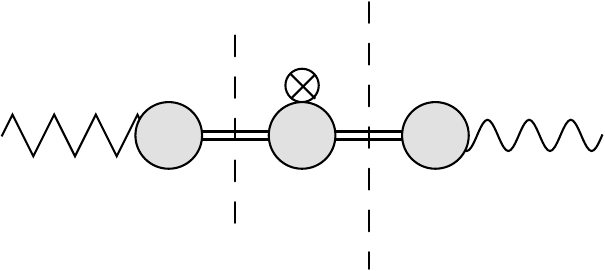}
		&	\includegraphics[height=1cm,valign=c]{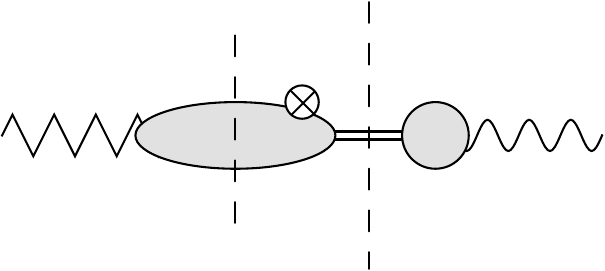}
		\\
	\midrule
	$A$
		& $\times$
		& \includegraphics[height=1cm,valign=c]{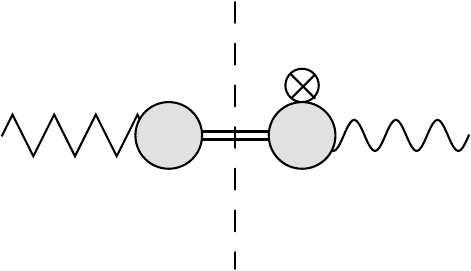}
		& $\times$
		\\
	\midrule
	$\ldots$
		& \includegraphics[height=1cm,valign=c]{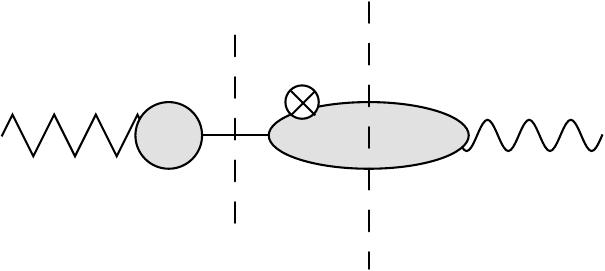}
		& \includegraphics[height=1cm,valign=c]{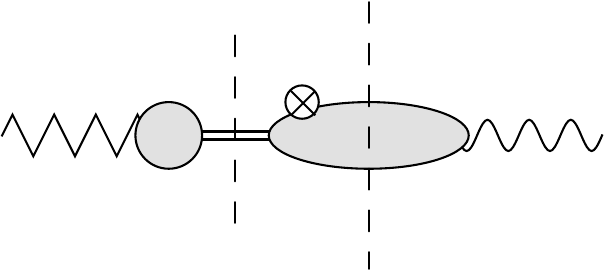}
		& $\ldots$
		\\
	\bottomrule
	\end{tabular}
	\caption{Comparison of different unitarity contributions in the two dispersive approaches, in analogy to Ref.~\cite{Ludtke:2023hvz}. The soft external photon is denoted by a crossed circle. The longer dashed line is the primary cut in the dispersion relations in $g-2$ kinematics. Cuts through gray blobs denote even higher intermediate states (denoted by ellipses) that need to be covered via the implementation of asymptotic constraints. The dispersive reconstruction of the sub-process $V\to\gamma a^*$ in the $V$ row needs to be performed away from the soft-photon limit. A cross~$\times$ means that no such contribution exists.}
	\label{tab:Reshuffling}
\end{table}

In the previous two sections, we have worked out two dispersive representations for the VVA correlator in the limit where one of the photons is soft.

In the formalism discussed in Sect.~\ref{sec:DRFixedVirtualities}, we derived dispersion relations in general kinematics and took the soft-photon limit only at the end. In this way, we only had to deal with one unitarity cut in VVA. The results depend on singly-virtual pseudoscalar and axial TFFs. Two-pion and vector-meson intermediate states arise in these sub-processes, which in turn can be reconstructed dispersively.

In contrast, in Sect.~\ref{sec:DRgm2} we set the soft-photon momentum to zero at the beginning. In this limit, the imaginary part needed for the dispersion relation resulted from a sum of two cuts in general kinematics, which become indistinguishable in the limit of vanishing photon momentum. For that reason, we had to not only consider pseudoscalar and axial-vector intermediate states in the axial-current channel, but also two-pion and vector-meson intermediate states in the vector-current channel.

As the two approaches rely on different types of dispersion relations, it is clear that the contributions from individual intermediate states do not agree. However, in order for the results to be consistent, the infinite sum over all contributions has to be the same. Therefore, a reshuffling is taking place between the contribution of different intermediate states. This is illustrated in Tab.~\ref{tab:Reshuffling}: the different intermediate states in the dispersion relation in general kinematics correspond to the different columns, while the intermediate states in the dispersion relation in the soft-photon limit correspond to the rows of Tab.~\ref{tab:Reshuffling}. In practice, the infinite sum over intermediate states needs to be truncated: we explicitly include pseudoscalars, axial-vector mesons, vector mesons, as well as two-pion intermediate states.

Since the reshuffling can involve terms that are neglected in one of the two approaches, the truncated results are not guaranteed to agree exactly. As we will check below, the difference between the results of the two approaches can be attributed to neglected intermediate states, hence it can also be used to estimate the uncertainty due to the truncation.

The dispersion relation for fixed photon virtualities includes pseudoscalar and axial-vector poles. The result of the dispersion relation in $g-2$ kinematics also contains these two classes of contributions, with the poles appearing either directly in VVA or in one of the sub-processes. In addition, the dispersive representation in $g-2$ kinematics contains contributions without any pseudoscalar or axial-vector poles. In the following, we compare these different contributions separately.

\subsection{Contributions with a pseudoscalar pole}
\label{sec:ComparisonPSPole}

The pseudoscalar pole in the dispersion relations for fixed photon virtualities Eq.~\eqref{eq:PSPoleFixedVirt} contains the TFF evaluated at $q^2$. It can be separated into a pure pole and a part regular at $q^2 \to m_P^2$~\cite{Colangelo:2019uex}
\begin{align}
	\label{eq:PSPoleFixedVirtSeparated}
	\tilde{\W}_1^\text{$P$-pole}(q^2) = 2 \sum_a C_a F_P^a \left(\frac{F_{P \gamma^* \gamma^*}(m_P, 0)}{q^2 - m_P^2} + \frac{F_{P \gamma^* \gamma^*}(q^2, 0) - F_{P \gamma^* \gamma^*}(m_P^2, 0)}{q^2 - m_P^2}\right)\,.
\end{align}
In the dispersion relation in $g-2$ kinematics, pseudoscalar poles contribute to the main process, Eq.~\eqref{eq:PSpolesg-2kin}, and this precisely reproduces the pure pole part of the previous equation. This contribution is shown in the first row of the first column of Tab.~\ref{tab:Reshuffling}. The non-pole part corresponds to the remaining rows of the first column. It has to be compared with the pseudoscalar poles in the sub-processes entering the dispersion relation in soft-photon kinematics: these are $V \to a^* \gamma$, appearing in the vector-meson pole contribution, as well as the $\gamma P \to \pi^+ \pi^-$ sub-process of the two-pion contribution. To this end, we employ the TFF representation in Eq.~\eqref{eq:PTFFsinglyVirtual} for the non-pole part in Eq.~\eqref{eq:PSPoleFixedVirtSeparated}, which leads to
\begin{align}
	\label{eq:PpoleFPVminusgm2}
	\tilde{\W}_1^\text{$P$-pole}(q^2) - \tilde{\W}_1^P(q^2) &= \sum_a C_a F_P^a \frac{1}{48\pi^2} \int_{4 m_\pi^2}^\infty \dif s \frac{s \sigma_\pi^3(s) F_\pi^V(s)^* f_P^1(s, 0)}{(s - m_P^2)(s - q^2)} \nn
	& \quad - 2 \sum_a C_a F_P^a \sum_V \frac{m_V f_V F_{VP}}{(q^2 - m_V^2)(m_V^2 - m_P^2)}\,.
\end{align}
The second line of this equation exactly agrees with the insertion of the pseudoscalar pole Eq.~\eqref{eq:VtildeiPpole} into the vector-meson pole contribution Eq.~\eqref{eq:resVPole} to the dispersion relation in $g-2$ kinematics. In Tab.~\ref{tab:Reshuffling}, this contribution is shown in the `$V$' row of the first column. The first line in Eq.~\eqref{eq:PpoleFPVminusgm2} should be compared with the two-pion contribution with a pseudoscalar pole in the sub-process, given in the third line of Eq.~\eqref{eq:ResultScalarFunctionsgm2TwoPionExplicit} and shown in the $2\pi$ row of the first column of Tab.~\ref{tab:Reshuffling}. The difference between the two expressions is
\begin{align}
	\label{eq:diffPS}
	\tilde{\W}_1^\text{$P$-pole}&(q^2) - \left( \tilde{\W}_1^P(q^2) + \tilde{\W}_1^{V,P}(q^2) + \tilde{\W}_1^{\pi^+\pi^-,P}(q^2) \right) \nn
		&= \sum_a \frac{C_a F_P^a}{48\pi^2} \int_{4 m_\pi^2}^\infty \dif s \frac{s \sigma_\pi^3(s)}{(s - m_P^2)(s - q^2)} \left( F_\pi^V(s)^* f_P^1(s, 0) - \frac{F_\pi^V(s) \Omega_1^1(s)^* f_P^1(m_P^2, 0)}{\Omega_1^1(m_P^2)} \right) \nn
		&= \sum_a \frac{C_a F_P^a}{48\pi^2} \int_{4 m_\pi^2}^\infty \dif s \frac{s \sigma_\pi^3(s) |\Omega_1^1(s)|^2}{s - q^2} \left( \frac{\frac{f_P^1(s, 0)}{\Omega_1^1(s)} - \frac{f_P^1(m_P^2, 0)}{\Omega_1^1(m_P^2)}}{s - m_P^2} \right) \,,
\end{align}
where we have used that $F_\pi^V(s) = \Omega_1^1(s)$ if inelastic contributions are neglected. The pole of the integrand at $s = m_P^2$, corresponding to a pseudoscalar pole in the sub-process, cancels as required. The expression would vanish if we took $f_P^1(s,0) = \mathrm{const} \times \Omega_1^1(s)$, which corresponds to an approximation for $f_P^1(s,0)$ that neglects the effects of crossed-channel rescattering. These effects are described by the inhomogeneity in the Omn\`es solution to the Khuri--Treiman equations~\cite{Khuri:1960zz} for $\gamma\to3\pi$~\cite{Niecknig:2012sj, Hoferichter:2018kwz,Niehus:2021iin}.

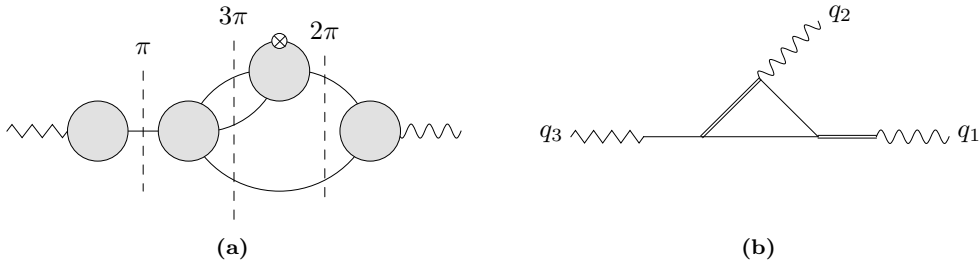
\begin{figure}
	\centering
	\begin{subfigure}[t]{0.45\textwidth}
		\centering
		\setlength{\unitlength}{1cm}
		\begin{picture}(6,3)
			\put(1.7,2.2){$\pi$}
			\put(2.8,2.6){$3\pi$}
			\put(4.0,2.4){$2\pi$}
			\includegraphics[width=6cm]{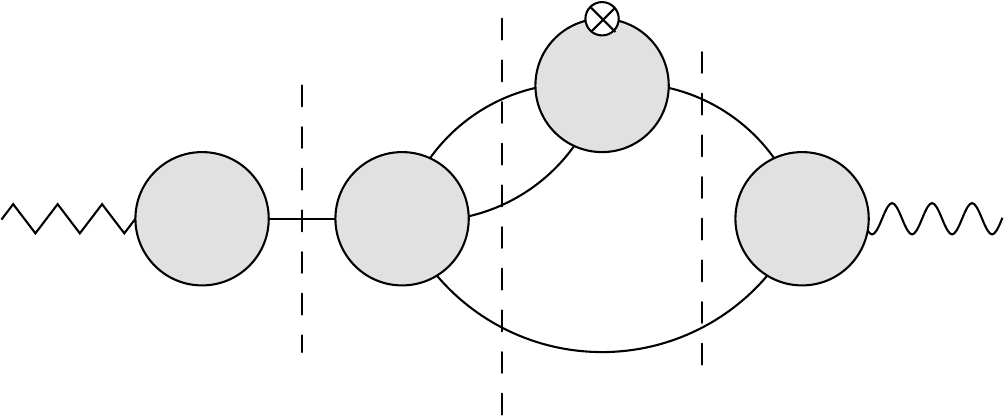}
		\end{picture}
		\caption{}
		\label{fig:ThreeCuts}
	\end{subfigure}
	\begin{subfigure}[t]{0.45\textwidth}
		\centering
		\setlength{\unitlength}{1cm}
		\begin{picture}(5,3)
			\put(-0.4,1.1){$q_3$}
			\put(3.4,2.7){$q_2$}
			\put(5.1,1.1){$q_1$}
			\put(0,1){\includegraphics[width=5cm]{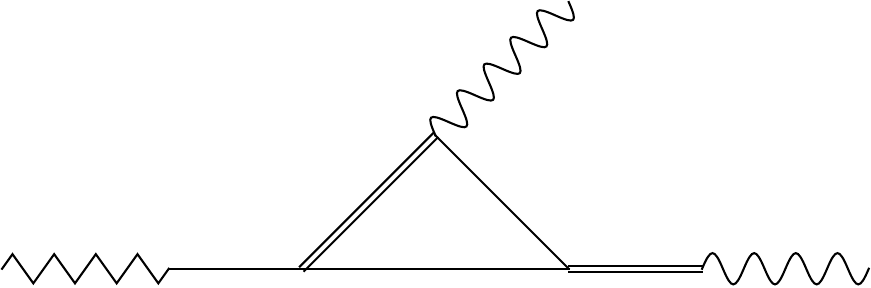}}
		\end{picture}
		\caption{}
		\label{fig:ToyDiagram}
	\end{subfigure}
	\caption{{\bf (a)} Topology with a single-pion, a two-pion, and a three-pion intermediate state, which is only partially included in the two dispersive approaches. {\bf (b)} Toy-model diagram with an internal $\rho$ resonance, used to illustrate the different truncation schemes of the two dispersive approaches to VVA.}
\end{figure}

With the Khuri--Treiman solution for $f_P^1(s,0)$, the remainder in Eq.~\eqref{eq:diffPS} does not vanish and has a two-pion cut. At first sight, it might seem surprising that we find a difference in an intermediate state that is included in both dispersive approaches. However, the difference in fact corresponds to a three-pion intermediate state in a sub-process: Fig.~\ref{fig:ThreeCuts} shows the leading topology that has a single-pion, a two-pion, and a three-pion intermediate state. The effect of the three-pion intermediate state (and even higher intermediate states) is only partially captured in the two dispersive approaches in terms of the crossed-channel rescattering of the inhomogeneous Omn\`es solution to the Khuri--Treiman equations. While it is not obvious if these approximations have a larger effect in one of the two approaches, we note that in the treatment in the soft-photon limit, the three-pion intermediate state of the leading topology shown in Fig.~\ref{fig:ThreeCuts} cannot go on shell once we have cut either the single-pion or the two-pion intermediate state. Therefore, the neglected three-pion cut affects the dispersion relation in $g-2$ kinematics only at higher orders in the two-pion rescattering, suggesting that the approximations have a larger effect on the dispersion relation in general kinematics. We illustrate this in the following within a toy example.

\subsection{Toy example}
\label{sec:VVA_Toy}

In order to better understand the difference between the two formalisms, we consider a contribution of the form of the diagram in Fig.~\ref{fig:ToyDiagram}, where double lines represent $\rho$-meson propagators. This toy diagram therefore is a model for the topology shown in Fig.~\ref{fig:ThreeCuts}. We ignore overall factors of couplings and assume Feynman rules such that the contribution of the diagram to the three-point function is proportional to
\begin{equation}
	\label{eq:ToyResultW}
	\W(q_1^2, q_2^2, q_3^2) = \frac{1}{q_3^2 - m_\pi^2} \frac{m_\rho^2}{q_1^2 - m_\rho^2} C_{00}(q_1^2, q_2^2, q_3^2; m_\pi^2, m_\pi^2, m_\rho^2) \, ,
\end{equation}
where the loop function is defined as
\begin{equation}
	\frac{\mu^{2\epsilon}}{i}\! \int\! \frac{\dif^D \ell}{(2\pi)^D} \frac{\ell^\mu\ell^\nu}{(\ell^2 - m_\pi^2)((\ell - q_1)^2 - m_\pi^2)((\ell + q_3)^2 - m_\rho^2)} = g^{\mu\nu} C_{00} + (\text{terms}\propto q_i^\mu q_j^\nu) \, .
\end{equation}
In the limit of $g-2$ kinematics, $q_2\to0$, this becomes
\begin{equation}
	\label{eq:ToyResultWgm2}
	\tilde\W(q^2) = \frac{1}{q^2 - m_\pi^2} \frac{m_\rho^2}{q^2 - m_\rho^2} C_{00}(q^2, 0, q^2; m_\pi^2, m_\pi^2, m_\rho^2)\,.
\end{equation}
In the dispersive framework in $g-2$ kinematics, we have four contributions to the discontinuity
\begin{align}
	\label{eq:ToyDiscgm2}
	\Delta \tilde{\W}(s) &= \Delta^\pi \tilde{\W}(s) + \Delta^\rho \tilde{\W}(s) + \Delta^{\pi\pi} \tilde{\W}(s) + \Delta^{\rho\pi} \tilde{\W}(s) \, , \nn
	\Delta^{\pi} \tilde{\W}(s) &= -\pi \delta(s - m_\pi^2) \frac{m_\rho^2}{m_\pi^2 - m_\rho^2} C_{00}(m_\pi^2, 0, m_\pi^2; m_\pi^2, m_\pi^2, m_\rho^2) \,,\nn
	\Delta^{\rho} \tilde{\W}(s) &= \pi \delta(s - m_\rho^2) \frac{m_\rho^2}{m_\pi^2 - m_\rho^2} C_{00}(m_\rho^2, 0, m_\rho^2; m_\pi^2, m_\pi^2, m_\rho^2) \,,\nn
	\Delta^{\pi\pi} \tilde{\W}(s) &= \frac{1}{192\pi} \frac{s \sigma_\pi^3(s)m_\rho^2}{(s - m_\pi^2)(s - m_\rho^2)(m_\rho^2 - m_\pi^2)} \theta(s - 4 m_\pi^2) \,,\nn
	\Delta^{\rho\pi} \tilde{\W}(s) &= -\frac{1}{192\pi} \frac{\lambda^{3/2}(s, m_\pi^2, m_\rho^2)m_\rho^2}{s^2 (s - m_\pi^2)(s - m_\rho^2)(m_\rho^2 - m_\pi^2)}  \theta(s - (m_\pi + m_\rho)^2) \,.
\end{align}
The full function fulfills an unsubtracted dispersion relation (the UV divergence is contained in the pion and $\rho$ poles)
\begin{equation}
	\tilde\W(q^2) = \frac{1}{\pi} \int ds \frac{\Delta\tilde\W(s)}{s-q^2-i\epsilon} \, .
\end{equation}
Neglecting the $\rho\pi$-cut amounts to a modification of the discontinuity at $s > (m_\pi + m_\rho)^2$ only.

Next, we analyze this diagram in the dispersive formalism with fixed photon virtualities, where it only has a pion pole and a $\rho\pi$-cut. The corresponding discontinuities read
\begin{align}
	\label{eq:ToyDelta3}
	\Delta_3^{\pi} \W(q_1^2, q_2^2, q_3^2) &= -\pi \delta(q_3^2 - m_\pi^2) \frac{m_\rho^2}{q_1^2 - m_\rho^2} C_{00}(q_1^2, q_2^2, m_\pi^2; m_\pi^2, m_\pi^2, m_\rho^2) \,, \nn
	\Delta_3^{\rho\pi} \W(q_1^2, q_2^2, q_3^2) &= \frac{1}{q_3^2 - m_\pi^2} \frac{m_\rho^2}{q_1^2 - m_\rho^2} \Delta_3^{\rho\pi} C_{00}(q_1^2, q_2^2, q_3^2; m_\pi^2, m_\pi^2, m_\rho^2) \theta(q_3^2 - (m_\pi + m_\rho)^2) \,,
\end{align}
where the $\rho\pi$-cut of the loop function shows up. The full function can be recovered from the unsubtracted dispersion relation Eq.~\eqref{eq:DispRelq3} if the sum of both discontinuities is plugged in. Using only the pion pole in general kinematics, we obtain
\begin{equation}
	\W^\text{$\pi$-pole}(q_1^2, q_2^2, q_3^2) = \frac{1}{q_3^2 - m_\pi^2} \frac{m_\rho^2}{q_1^2 - m_\rho^2} C_{00}(q_1^2, q_2^2, m_\pi^2; m_\pi^2, m_\pi^2, m_\rho^2)\,,
\end{equation}
which in the $g-2$ limit becomes
\begin{equation}
	\label{eq:ToyPionPole}
	\tilde{\W}^\text{$\pi$-pole}(q^2) = \frac{1}{q^2 - m_\pi^2} \frac{m_\rho^2}{q^2 - m_\rho^2} C_{00}(q^2, 0, m_\pi^2; m_\pi^2, m_\pi^2, m_\rho^2) \,.
\end{equation}
Note that compared to the result for the full diagram Eq.~\eqref{eq:ToyResultWgm2}, the third argument of the loop function is replaced by $m_\pi^2$. In the $g-2$ limit, Eq.~\eqref{eq:ToyPionPole} has a pion pole, a $\rho$ pole, and a two-pion cut,
\begin{align}
	\label{eq:ToyDiscPionPole}
	\Delta^{\pi} \tilde{\W}^\text{$\pi$-pole}(s) &= - \pi \delta(s - m_\pi^2) \frac{m_\rho^2}{m_\pi^2 - m_\rho^2} C_{00}(m_\pi^2, 0, m_\pi^2; m_\pi^2, m_\pi^2, m_\rho^2) \,, \nn
	\Delta^{\rho} \tilde{\W}^\text{$\pi$-pole}(s) &= \pi \delta(s - m_\rho^2) \frac{m_\rho^2}{m_\pi^2 - m_\rho^2} C_{00}(m_\rho^2, 0, m_\pi^2; m_\pi^2, m_\pi^2, m_\rho^2) \,, \nn
	\Delta^{\pi\pi} \tilde{\W}^\text{$\pi$-pole}(s) &= \theta(s - 4 m_\pi^2) \frac{1}{s - m_\pi^2} \frac{m_\rho^2}{s - m_\rho^2}\Delta_s^{\pi\pi} C_{00}(s, 0, m_\pi^2; m_\pi^2, m_\pi^2, m_\rho^2) \nn
		&= \theta(s - 4 m_\pi^2) \frac{m_\rho^2}{32 \pi} \Bigg[ \frac{s(s + 2 m_\rho^2 - 3 m_\pi^2) \sigma_\pi(s)}{2(s - m_\pi^2)^3(s - m_\rho^2)} \nn
		&\quad + \frac{m_\pi^6 + m_\rho^2 s (m_\rho^2 + s - 3 m_\pi^2)}{(s - m_\pi^2)^4(s - m_\rho^2)} \log \left(\frac{3 m_\pi^2 - 2 m_\rho^2 - s + \sigma_\pi(s) (s - m_\pi^2)}{3 m_\pi^2 - 2 m_\rho^2 - s - \sigma_\pi(s) (s - m_\pi^2)} \right) \Bigg] \, .
\end{align}
The pion-pole residue agrees with the residue of the full function, but the two-pion discontinuity as well as the $\rho$-pole residue differ from Eq.~\eqref{eq:ToyDiscgm2}. The $\rho\pi$-cut in general kinematics gives precisely the missing contributions to the two-pion discontinuity and $\rho$-pole residue after taking the $g-2$ limit.

In practice, we neglect higher cuts also in the sub-processes, which in the toy model affects the residue of the $\rho$-pole. We write it as
\begin{equation}
	\Delta^{\rho} \tilde{\W}(s) = \pi \delta(s - m_\rho^2) m_\rho^2 \, \tilde\U(s) \,, \quad \tilde\U(s) = \frac{1}{m_\pi^2 - s} C_{00}(s, 0, s; m_\pi^2, m_\pi^2, m_\rho^2) \, ,
\end{equation}
and we reconstruct $\tilde\U(s)$ similarly to the real sub-amplitude for $a^*\gamma\to\pi\pi$ in VVA with a once-subtracted dispersion relation
\begin{equation}
	\tilde\U(s) = \tilde\U(0) + \frac{s}{\pi} \int ds' \frac{\Delta \tilde\U(s')}{(s'-s)s'} \, ,
\end{equation}
where we will apply the same approximation to $\Delta\tilde\U$ as to $\Delta\tilde\W$. E.g., neglecting the $\rho\pi$ discontinuity both in $\Delta\tilde\W$ and $\Delta\tilde\U$ in $g-2$ kinematics amounts to a modification of $\tilde\W$ by
\begin{align}
	\tilde\W^{\rho\pi}(q^2) &+ \tilde\W^{\rho,\rho\pi}(q^2) = \frac{1}{\pi} \int ds \frac{1}{s-q^2} \left( \Delta^{\rho\pi}\tilde\W(s) + \pi\delta(s-m_\rho^2) m_\rho^2 \frac{m_\rho^2}{\pi} \int ds' \frac{\Delta^{\rho\pi}\tilde\U(s')}{(s'-m_\rho^2)s'} \right) \nn
		&= \frac{1}{\pi} \int ds \frac{1}{s-m_\pi^2} \frac{m_\rho^2}{s-m_\rho^2} \left( \frac{1}{s-q^2} - \frac{m_\rho^2}{s(m_\rho^2-q^2)} \right) \Delta^{\rho\pi}C_{00}(s,0,s;m_\pi^2,m_\pi^2,m_\rho^2) \, .
\end{align}
Note that this expression vanishes at $q^2 = 0$.

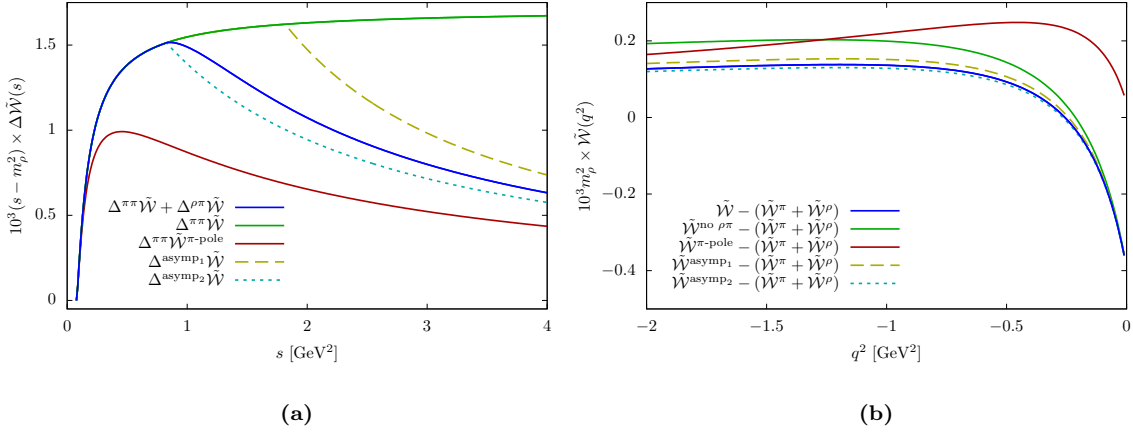
\begin{figure}[t]
	\centering
	\begin{subfigure}[t]{0.45\textwidth}
		\scalebox{0.62}{\input{plots/ToyModelDisc}} \\
		\caption{}
		\label{fig:ToyModelDisc}
	\end{subfigure} \qquad
	\begin{subfigure}[t]{0.45\textwidth}
		\scalebox{0.62}{\input{plots/ToyModel}} \\
		\caption{}
		\label{fig:ToyModel}
	\end{subfigure}
	\caption{{\bf (a)} Comparison of the discontinuities of the toy-model function, multiplied by $(s-m_\rho^2)$: the blue solid line shows the full discontinuity in $g-2$ kinematics, the green solid line only includes the two-pion discontinuity in $g-2$ kinematics, whereas the red line shows the two-pion discontinuity (in the $g-2$ limit) of the pion pole defined in general kinematics. {\bf (b)} Dispersive reconstructions of the toy-model function, with UV-divergent $\pi$- and $\rho$-poles subtracted.}
\end{figure}

It is instructive to numerically compare the different approximations in the toy example. In Fig.~\ref{fig:ToyModelDisc}, we show the discontinuity of $\Delta\tilde\W(s)$, multiplied by $(s-m_\rho^2)$, comparing the full discontinuity in blue with the $\pi\pi$ discontinuity in $g-2$ kinematics (green) as well as the $\pi\pi$ discontinuity of the pion pole in general kinematics (red). Since the $\pi\pi$ discontinuity has a different asymptotic behavior than the total discontinuity, we also use two variants of the discontinuity, where we switch from the $\pi\pi$ discontinuity to an asymptotic form
\begin{equation}
	(s-m_\rho^2) \Delta^{\text{asymp}_1}\tilde\W(s) = \frac{1}{64\pi}\frac{m_\rho^2}{s} \, , \quad
	(s-m_\rho^2) \Delta^{\text{asymp}_2}\tilde\W(s) = \frac{1}{64\pi}\frac{m_\rho^2}{s+b} \, ,
\end{equation}
with the parameter $b$ chosen so that the second asymptotic form matches the $\pi\pi$ discontinuity at the $\rho\pi$ threshold, shown as dotted line in Fig.~\ref{fig:ToyModelDisc}. In Fig.~\ref{fig:ToyModel}, we compare the different dispersive reconstructions in the space-like region. In all cases, we subtract the pion pole and the full $\rho$-pole in $g-2$ kinematics, which both are UV divergent. The blue curve corresponds to the full result. The green curve shows the dispersive result in $g-2$ kinematics, where $\rho\pi$-cuts are neglected both in $\Delta\tilde\W$ and in the residue of the $\rho$-pole. The red curve corresponds to the pion pole defined in general kinematics. We also show the results that one obtains with the discontinuities matched to the asymptotic form.

What we learn from this toy example applies similarly to the VVA topology in Fig.~\ref{fig:ThreeCuts}: neglecting the $3\pi$ cut in the formalism in $g-2$ kinematics leads to missing contributions to the discontinuity only above the three-pion threshold.\footnote{For higher two-pion rescattering contributions, differences appear already above the two-pion threshold, analogous to the modification of the $\rho$-pole residue in the toy model.} Conversely, in the formalism with fixed photon virtualities, cuts in $g-2$ kinematics can emerge from any cut in $q_3^2$ in general kinematics. Therefore, taking into account all cuts below some threshold $s_0$ in general kinematics does not imply that the discontinuity of $\tilde{\W}_i(s)$ in $g-2$ kinematics is correctly reproduced for $s < s_0$. In the toy model, we see that the dispersion relation in $g-2$ kinematics performs much better in reconstructing the low-energy behavior of the full function when $\rho\pi$-cuts are neglected. The sensitivity of the low-energy behavior to neglected higher cuts in general kinematics will become even more problematic if a matching to asymptotic constraints requires an infinite tower of higher states, which introduces a model dependence in the low-energy region.

\subsection{Contributions with an axial-vector pole}

In the dispersion relation for fixed photon virtualities, axial-vector mesons contribute to both scalar VVA functions. We first consider the longitudinal function $\tilde{\W}_1(q^2)$, which does not contain a pole at $q^2 = m_A^2$. This is consistent with the result that in the dispersion relation in the $g-2$ limit, there is no axial-vector pole contribution to this scalar function. Inserting the TFFs from Eqs.~\eqref{eq:AxialTFFPiPi} and \eqref{eq:AxialTFFVpole} gives
\begin{align}
	\tilde{\W}_1^\text{$A$-pole}(q^2) &= -\frac{2}{m_A^3} \sum_a C_a F_A^a \left[ \F_1^A(q^2, 0) + \F_2^A(q^2, 0) \right] \nn
		&= - \frac{1}{12 \pi^2} \sum_a C_a F_A^a \int_{4 m_\pi^2}^\infty \dif s \frac{\sigma_\pi^3(s)}{s - q^2}  F_\pi^V(s)^* \left(\frac{ m_A g_{++}^1(s) - \sqrt{2s} g_{+0}^1(s) }{s - m_A^2}\right) \nn
		&\quad + \frac{1}{12 \pi^2} \sum_a C_a F_A^a \int_{4 m_\pi^2}^\infty \dif s \frac{\sigma_\pi^3(s) F_\pi^V(s)^*}{m_A\sqrt{s}} \left(\frac{\sqrt{s} g_{++}^1(s) - \sqrt{2} m_A g_{+0}^1(s) }{s - m_A^2}\right) \nn
		&\quad + \frac{2}{m_A} \sum_a C_a F_A^a \sum_V \frac{m_V f_V}{q^2 - m_V^2} \left( \frac{q^2}{m_V^2} F_{VA}^1 - F_{VA}^2 - F_{VA}^3 \right) \,,
\end{align}
where the helicity partial waves $g_{\lambda_2\lambda_A}^l(s)$ for $A\gamma\to\pi\pi$ are defined in App.~\ref{sec:AgammaPiPiHelAmps}.
The last line is reproduced in the dispersion relation in $g-2$ kinematics by the insertion of the axial-vector pole Eq.~\eqref{eq:VtildeiApole} into the vector-meson pole contribution Eq.~\eqref{eq:resVPole}, up to a constant term.

The integrand of the two-pion part does not have a pole at $s = m_A^2$ due to Eq.~\eqref{eq:AGammaPiPiConstraint}. This is consistent with the observation that the two-pion contribution to $\W_1$ in the dispersion relation in $g-2$ kinematics in Eq.~\eqref{eq:ResultScalarFunctionsgm2TwoPionExplicit} does not contain an axial-vector contribution. Within the approximation of App.~\ref{sec:AgpipiParametrization} for the $A\gamma\to\pi\pi$ partial waves and an elastic form factor $F_\pi^V(s) = \Omega_1^1(s)$, the difference between the two approaches amounts to
\begin{align}
	\label{eq:axialDifferenceW1}
	\tilde{\W}_1^\text{$A$-pole}&(q^2) - \tilde\W_1^{V,A}(q^2) \nn*
		&= - \frac{1}{12 \pi^2} \sum_a C_a F_A^a \int_{4 m_\pi^2}^\infty \dif s \frac{\sigma_\pi^3(s)}{s - q^2}  F_\pi^V(s)^* \left(\frac{ m_A g_{++}^1(s) - \sqrt{2s} g_{+0}^1(s) }{s - m_A^2}\right) \nn
		&\quad + \frac{1}{12 \pi^2} \sum_a C_a F_A^a \int_{4 m_\pi^2}^\infty \dif s \frac{\sigma_\pi^3(s) F_\pi^V(s)^*}{m_A\sqrt{s}} \left(\frac{\sqrt{s} g_{++}^1(s) - \sqrt{2} m_A g_{+0}^1(s) }{s - m_A^2}\right) \nn
		&\quad + \frac{2}{m_A} \sum_a C_a F_A^a \sum_V \frac{f_V}{m_V} F_{VA}^1 \nn
		&= \frac{1}{24 \pi^2} \sum_a \frac{C_a F_A^a}{m_A} \int_{4 m_\pi^2}^\infty \dif s \frac{\sigma_\pi^3(s) |\Omega_1^1(s)|^2}{s - q^2} \left( \bar\F_{A,1} + s \bar\F_{A,1}' - m_A^2 \bar\F_{A,2} \right) \nn
		&\quad + \frac{1}{24 \pi^2} \sum_a \frac{C_a F_A^a}{m_A} \int_{4 m_\pi^2}^\infty \dif s \, \sigma_\pi^3(s) |\Omega_1^1(s)|^2 \bar\F_{A,2} + \frac{2}{m_A} \sum_a C_a F_A^a \sum_V \frac{f_V}{m_V} F_{VA}^1  \,.
\end{align}
In contrast to the pion-pole contributions, here we find a difference with a two-pion cut that cannot be easily attributed to the effects of neglected higher intermediate states. As there are further two-pion contributions, we will come back to this point below.

In the dispersion relation for fixed photon virtualities, the transverse scalar function $\W_2$ contains an axial-vector pole given in Eq.~\eqref{eq:APoleFixedVirt}, which can be split up in analogy to the pseudoscalar pole as
\begin{equation}
	\tilde{\W}_2^\text{$A$-pole}(q^2) = - 2 \sum_a \frac{C_a F_A^a}{m_A} \frac{\F_2^A(m_A^2, 0)}{q^2 - m_A^2} - 2 \sum_a \frac{C_a F_A^a}{m_A} \frac{\F_2^A(q^2, 0) - \F_2^A(m_A^2, 0)}{q^2 - m_A^2}\,.
\end{equation}
The first term agrees with the axial-vector pole in the formalism in $g-2$ kinematics, Eq.~\eqref{eq:Apole_g-2kin}. It corresponds to the `$A$' row of the `$A$' column in Tab.~\ref{tab:Reshuffling}. For the second term, we again employ the TFF representation derived in Eqs.~\eqref{eq:AxialTFFPiPi} and \eqref{eq:AxialTFFVpole}. This leads to
\begin{align}
	\tilde{\W}_2^\text{$A$-pole}(q^2) - \tilde{\W}_2^A(q^2) &= \frac{m_A}{12 \pi^2} \sum_a C_a F_A^a \int_{4 m_\pi^2}^\infty \dif s \frac{\sigma_\pi^3(s) F_\pi^V(s)^* g_{++}^1(s)}{(s - m_A^2)(s - q^2)} \nn
	&\qquad + 2 \sum_a C_a F_A^a \frac{m_A m_V f_V (F_{VA}^2+F_{VA}^3)}{(m_A^2 - m_V^2)(q^2 - m_V^2)} \, .
\end{align}
The second line of this expression is reproduced in the dispersion relations in $g-2$ kinematics by the insertion of an axial-vector pole Eq.~\eqref{eq:VtildeiApole} into the vector-meson pole contribution Eq.~\eqref{eq:resVPole} to VVA. It corresponds to the `$V$' row of the `$A$' column in Tab.~\ref{tab:Reshuffling}. The remainder has to be compared with the two-pion contribution with an axial-vector pole in the sub-process, given in the last line of Eq.~\eqref{eq:ResultScalarFunctionsgm2TwoPionExplicit}. The difference of the two expressions is
\begin{align}
	\label{eq:axialDifferenceW2}
	\tilde{\W}_2^\text{$A$-pole}(q^2) &- \left( \tilde{\W}_2^A(q^2) + \tilde{\W}_2^{V,A}(q^2) + \tilde{\W}_2^{\pi^+\pi^-,A}(q^2) \right) \nn
		&= \sum_a \frac{m_A C_a F_A^a}{12 \pi^2} \int_{4 m_\pi^2}^\infty \dif s \frac{s \sigma_\pi^3(s) |\Omega_1^1(s)|^2}{s - q^2} \left(\frac{ \frac{g_{++}^1(s)}{s \Omega_1^1(s)} - \frac{g_{++}^1(m_A^2)}{m_A^2\Omega_1^1(m_A^2)} }{s - m_A^2} \right) \nn
		&= - \sum_a \frac{C_a F_A^a}{24 \pi^2} \int_{4 m_\pi^2}^\infty \dif s \frac{\sigma_\pi^3(s) |\Omega_1^1(s)|^2}{s - q^2} \left(\frac{ \bar\F_{A,1} - m_A^2 \bar\F_{A,2} }{m_A} \right) \, ,
\end{align}
where we used $F_\pi^V(s) = \Omega_1^1(s)$ in the elastic approximation, we inserted the threshold constraint~\eqref{eq:AGammaPiPiConstraint}, and the second equation follows in the approximation of App.~\ref{sec:AgpipiParametrization} for the $A\gamma\to\pi\pi$ partial waves.
As expected, the integrand does not have a pole at $s = m_A^2$, hence the obtained difference only has a two-pion discontinuity.

\subsection{Further contributions}

In the dispersive approach in $g-2$ kinematics, there are further two-pion contributions with the subtraction constant $\tilde{\U}_1(0)$ in the sub-process, see Eq.~\eqref{eq:ResultScalarFunctionsgm2TwoPionExplicit}. These only have a two-pion cut with a non-singular discontinuity. As the subtraction constant may capture effects due to higher intermediate states that are not included explicitly, we cannot expect this contribution to arise identically in the dispersion relation in general kinematics.

We noted in Sect.~\ref{sec:ComparisonPSPole} that the difference between the two dispersive approaches regarding the contribution of a pseudoscalar intermediate state can be traced back to higher cuts (due to three and more pions) that are generated by the inhomogeneous Omn\`es solution to the Khuri--Treiman equations for $\gamma\to3\pi$. In a crude approximation that neglects crossed-channel rescattering effects in the Khuri--Treiman equations, the difference in Eq.~\eqref{eq:diffPS} would vanish, depending on the asymptotic behavior of the partial wave. The situation is similar for the axial-vector contributions: in App.~\ref{sec:AgpipiParametrization}, we make an approximation for $A\gamma\to\pi\pi$, which neglects crossed-channel cuts. In this case, we expect the two dispersive approaches to agree, at least if we impose a sufficiently restrictive asymptotic behavior. This requires that the two-pion contributions in the differences~\eqref{eq:axialDifferenceW1} and~\eqref{eq:axialDifferenceW2} correspond to the two-pion contribution with the subtraction constant $\tilde\U_1(0)$ in the dispersion relations in $g-2$ kinematics. Indeed, we find agreement if we identify
\begin{align}
	\label{eq:AxialDRCondition}
	0 &= \sum_A \sum_a \frac{C_a F_A^a}{m_A} \bar\F_{A,1}' \, , \nn
	\tilde\U_1(0) &= - \sum_A \sum_a \frac{2C_a F_A^a}{m_A}  (\bar\F_{A,1} - m_A^2 \bar\F_{A,2} ) \, .
\end{align}
We will come back to these conditions in Sect.~\ref{sec:Numerics}.

%% file: plots/ToyModelDisc.tex
\begingroup
  \makeatletter
  \providecommand\color[2][]{%
    \GenericError{(gnuplot) \space\space\space\@spaces}{%
      Package color not loaded in conjunction with
      terminal option `colourtext'%
    }{See the gnuplot documentation for explanation.%
    }{Either use 'blacktext' in gnuplot or load the package
      color.sty in LaTeX.}%
    \renewcommand\color[2][]{}%
  }%
  \providecommand\includegraphics[2][]{%
    \GenericError{(gnuplot) \space\space\space\@spaces}{%
      Package graphicx or graphics not loaded%
    }{See the gnuplot documentation for explanation.%
    }{The gnuplot epslatex terminal needs graphicx.sty or graphics.sty.}%
    \renewcommand\includegraphics[2][]{}%
  }%
  \providecommand\rotatebox[2]{#2}%
  \@ifundefined{ifGPcolor}{%
    \newif\ifGPcolor
    \GPcolorfalse
  }{}%
  \@ifundefined{ifGPblacktext}{%
    \newif\ifGPblacktext
    \GPblacktexttrue
  }{}%
  \let\gplgaddtomacro\g@addto@macro
  \gdef\gplbacktext{}%
  \gdef\gplfronttext{}%
  \makeatother
  \ifGPblacktext
    \def\colorrgb#1{}%
    \def\colorgray#1{}%
  \else
    \ifGPcolor
      \def\colorrgb#1{\color[rgb]{#1}}%
      \def\colorgray#1{\color[gray]{#1}}%
      \expandafter\def\csname LTw\endcsname{\color{white}}%
      \expandafter\def\csname LTb\endcsname{\color{black}}%
      \expandafter\def\csname LTa\endcsname{\color{black}}%
      \expandafter\def\csname LT0\endcsname{\color[rgb]{1,0,0}}%
      \expandafter\def\csname LT1\endcsname{\color[rgb]{0,1,0}}%
      \expandafter\def\csname LT2\endcsname{\color[rgb]{0,0,1}}%
      \expandafter\def\csname LT3\endcsname{\color[rgb]{1,0,1}}%
      \expandafter\def\csname LT4\endcsname{\color[rgb]{0,1,1}}%
      \expandafter\def\csname LT5\endcsname{\color[rgb]{1,1,0}}%
      \expandafter\def\csname LT6\endcsname{\color[rgb]{0,0,0}}%
      \expandafter\def\csname LT7\endcsname{\color[rgb]{1,0.3,0}}%
      \expandafter\def\csname LT8\endcsname{\color[rgb]{0.5,0.5,0.5}}%
    \else
      \def\colorrgb#1{\color{black}}%
      \def\colorgray#1{\color[gray]{#1}}%
      \expandafter\def\csname LTw\endcsname{\color{white}}%
      \expandafter\def\csname LTb\endcsname{\color{black}}%
      \expandafter\def\csname LTa\endcsname{\color{black}}%
      \expandafter\def\csname LT0\endcsname{\color{black}}%
      \expandafter\def\csname LT1\endcsname{\color{black}}%
      \expandafter\def\csname LT2\endcsname{\color{black}}%
      \expandafter\def\csname LT3\endcsname{\color{black}}%
      \expandafter\def\csname LT4\endcsname{\color{black}}%
      \expandafter\def\csname LT5\endcsname{\color{black}}%
      \expandafter\def\csname LT6\endcsname{\color{black}}%
      \expandafter\def\csname LT7\endcsname{\color{black}}%
      \expandafter\def\csname LT8\endcsname{\color{black}}%
    \fi
  \fi
    \setlength{\unitlength}{0.0500bp}%
    \ifx\gptboxheight\undefined%
      \newlength{\gptboxheight}%
      \newlength{\gptboxwidth}%
      \newsavebox{\gptboxtext}%
    \fi%
    \setlength{\fboxrule}{0.5pt}%
    \setlength{\fboxsep}{1pt}%
    \definecolor{tbcol}{rgb}{1,1,1}%
\begin{picture}(6120.00,4284.00)%
    \gplgaddtomacro\gplbacktext{%
      \csname LTb\endcsname
      \put(198,653){\makebox(0,0)[r]{\strut{}$0$}}%
      \put(198,1684){\makebox(0,0)[r]{\strut{}$0.5$}}%
      \put(198,2715){\makebox(0,0)[r]{\strut{}$1$}}%
      \put(198,3746){\makebox(0,0)[r]{\strut{}$1.5$}}%
      \put(330,330){\makebox(0,0){\strut{}$0$}}%
      \put(1774,330){\makebox(0,0){\strut{}$1$}}%
      \put(3218,330){\makebox(0,0){\strut{}$2$}}%
      \put(4661,330){\makebox(0,0){\strut{}$3$}}%
      \put(6105,330){\makebox(0,0){\strut{}$4$}}%
    }%
    \gplgaddtomacro\gplfronttext{%
      \csname LTb\endcsname
      \put(-275,2405){\rotatebox{-270}{\makebox(0,0){\strut{}$10^3 (s-m_\rho^2) \times \Delta\tilde\W(s)$}}}%
      \put(3217,0){\makebox(0,0){\strut{}$s$ [GeV${}^2$]}}%
      \csname LTb\endcsname
      \put(2218,1780){\makebox(0,0)[r]{\strut{}$\Delta^{\pi\pi}\tilde\W + \Delta^{\rho\pi}\tilde\W$}}%
      \csname LTb\endcsname
      \put(2218,1560){\makebox(0,0)[r]{\strut{}$\Delta^{\pi\pi}\tilde\W$}}%
      \csname LTb\endcsname
      \put(2218,1340){\makebox(0,0)[r]{\strut{}$\Delta^{\pi\pi}\tilde\W^{\pi\text{-pole}}$}}%
      \csname LTb\endcsname
      \put(2218,1120){\makebox(0,0)[r]{\strut{}$\Delta^{\text{asymp}_1}\tilde\W$}}%
      \csname LTb\endcsname
      \put(2218,900){\makebox(0,0)[r]{\strut{}$\Delta^{\text{asymp}_2}\tilde\W$}}%
    }%
    \gplbacktext
    \put(0,0){\includegraphics[width={306.00bp},height={214.20bp}]{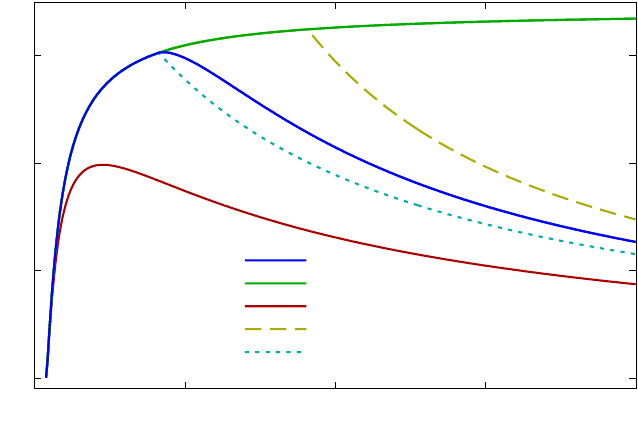}}%
    \gplfronttext
  \end{picture}%
\endgroup

%% file: plots/ToyModel.tex
\begingroup
  \makeatletter
  \providecommand\color[2][]{%
    \GenericError{(gnuplot) \space\space\space\@spaces}{%
      Package color not loaded in conjunction with
      terminal option `colourtext'%
    }{See the gnuplot documentation for explanation.%
    }{Either use 'blacktext' in gnuplot or load the package
      color.sty in LaTeX.}%
    \renewcommand\color[2][]{}%
  }%
  \providecommand\includegraphics[2][]{%
    \GenericError{(gnuplot) \space\space\space\@spaces}{%
      Package graphicx or graphics not loaded%
    }{See the gnuplot documentation for explanation.%
    }{The gnuplot epslatex terminal needs graphicx.sty or graphics.sty.}%
    \renewcommand\includegraphics[2][]{}%
  }%
  \providecommand\rotatebox[2]{#2}%
  \@ifundefined{ifGPcolor}{%
    \newif\ifGPcolor
    \GPcolorfalse
  }{}%
  \@ifundefined{ifGPblacktext}{%
    \newif\ifGPblacktext
    \GPblacktexttrue
  }{}%
  \let\gplgaddtomacro\g@addto@macro
  \gdef\gplbacktext{}%
  \gdef\gplfronttext{}%
  \makeatother
  \ifGPblacktext
    \def\colorrgb#1{}%
    \def\colorgray#1{}%
  \else
    \ifGPcolor
      \def\colorrgb#1{\color[rgb]{#1}}%
      \def\colorgray#1{\color[gray]{#1}}%
      \expandafter\def\csname LTw\endcsname{\color{white}}%
      \expandafter\def\csname LTb\endcsname{\color{black}}%
      \expandafter\def\csname LTa\endcsname{\color{black}}%
      \expandafter\def\csname LT0\endcsname{\color[rgb]{1,0,0}}%
      \expandafter\def\csname LT1\endcsname{\color[rgb]{0,1,0}}%
      \expandafter\def\csname LT2\endcsname{\color[rgb]{0,0,1}}%
      \expandafter\def\csname LT3\endcsname{\color[rgb]{1,0,1}}%
      \expandafter\def\csname LT4\endcsname{\color[rgb]{0,1,1}}%
      \expandafter\def\csname LT5\endcsname{\color[rgb]{1,1,0}}%
      \expandafter\def\csname LT6\endcsname{\color[rgb]{0,0,0}}%
      \expandafter\def\csname LT7\endcsname{\color[rgb]{1,0.3,0}}%
      \expandafter\def\csname LT8\endcsname{\color[rgb]{0.5,0.5,0.5}}%
    \else
      \def\colorrgb#1{\color{black}}%
      \def\colorgray#1{\color[gray]{#1}}%
      \expandafter\def\csname LTw\endcsname{\color{white}}%
      \expandafter\def\csname LTb\endcsname{\color{black}}%
      \expandafter\def\csname LTa\endcsname{\color{black}}%
      \expandafter\def\csname LT0\endcsname{\color{black}}%
      \expandafter\def\csname LT1\endcsname{\color{black}}%
      \expandafter\def\csname LT2\endcsname{\color{black}}%
      \expandafter\def\csname LT3\endcsname{\color{black}}%
      \expandafter\def\csname LT4\endcsname{\color{black}}%
      \expandafter\def\csname LT5\endcsname{\color{black}}%
      \expandafter\def\csname LT6\endcsname{\color{black}}%
      \expandafter\def\csname LT7\endcsname{\color{black}}%
      \expandafter\def\csname LT8\endcsname{\color{black}}%
    \fi
  \fi
    \setlength{\unitlength}{0.0500bp}%
    \ifx\gptboxheight\undefined%
      \newlength{\gptboxheight}%
      \newlength{\gptboxwidth}%
      \newsavebox{\gptboxtext}%
    \fi%
    \setlength{\fboxrule}{0.5pt}%
    \setlength{\fboxsep}{1pt}%
    \definecolor{tbcol}{rgb}{1,1,1}%
\begin{picture}(6120.00,4284.00)%
    \gplgaddtomacro\gplbacktext{%
      \csname LTb\endcsname
      \put(198,1014){\makebox(0,0)[r]{\strut{}$-0.4$}}%
      \put(198,1942){\makebox(0,0)[r]{\strut{}$-0.2$}}%
      \put(198,2869){\makebox(0,0)[r]{\strut{}$0$}}%
      \put(198,3797){\makebox(0,0)[r]{\strut{}$0.2$}}%
      \put(330,330){\makebox(0,0){\strut{}$-2$}}%
      \put(1774,330){\makebox(0,0){\strut{}$-1.5$}}%
      \put(3218,330){\makebox(0,0){\strut{}$-1$}}%
      \put(4661,330){\makebox(0,0){\strut{}$-0.5$}}%
      \put(6105,330){\makebox(0,0){\strut{}$0$}}%
    }%
    \gplgaddtomacro\gplfronttext{%
      \csname LTb\endcsname
      \put(-407,2405){\rotatebox{-270}{\makebox(0,0){\strut{}$10^3 m_\rho^2 \times \tilde\W(q^2)$}}}%
      \put(3217,0){\makebox(0,0){\strut{}$q^2$ [GeV${}^2$]}}%
      \csname LTb\endcsname
      \put(2651,1739){\makebox(0,0)[r]{\strut{}$\tilde\W - (\tilde\W^\pi + \tilde\W^\rho)$}}%
      \csname LTb\endcsname
      \put(2651,1519){\makebox(0,0)[r]{\strut{}$\tilde\W^{\text{no }\rho\pi} - (\tilde\W^\pi + \tilde\W^\rho)$}}%
      \csname LTb\endcsname
      \put(2651,1299){\makebox(0,0)[r]{\strut{}$\tilde\W^{\pi\text{-pole}} - (\tilde\W^\pi + \tilde\W^\rho)$}}%
      \csname LTb\endcsname
      \put(2651,1079){\makebox(0,0)[r]{\strut{}$\tilde\W^{\text{asymp}_1} - (\tilde\W^\pi + \tilde\W^\rho)$}}%
      \csname LTb\endcsname
      \put(2651,859){\makebox(0,0)[r]{\strut{}$\tilde\W^{\text{asymp}_2} - (\tilde\W^\pi + \tilde\W^\rho)$}}%
    }%
    \gplbacktext
    \put(0,0){\includegraphics[width={306.00bp},height={214.20bp}]{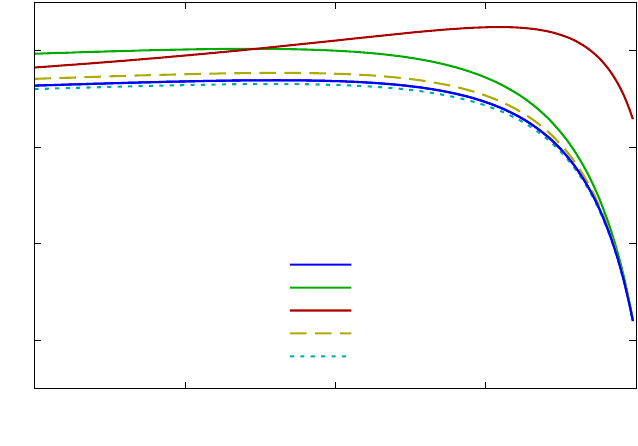}}%
    \gplfronttext
  \end{picture}%
\endgroup

%% file: sections/OPE.tex

\section{\boldmath Constraints from \chpt{} and the OPE}
\label{sec:OPEandChPTConstraints}

In neither of the two dispersive approaches to VVA presented above it is possible to include all intermediate states in a model independent way. Thus, the strategy is to include those contributions explicitly for which we are able to derive dispersive expressions and for which data exist for the required input, and to impose constraints from low and high energies on the contribution from the infinite tower of remaining states. Therefore, we will review low-energy constraints from \chpt{} in Sect.~\ref{sec:ChPTConstraint} and SDCs in Sect.~\ref{sec:VVA_OPE}. In Sect.~\ref{sec:VVA_SDC_Implementation}, we will discuss their implementation in the two dispersive approaches.

\subsection[\chpt{} constraint]{\boldmath \chpt{} constraint}
\label{sec:ChPTConstraint}

In Ref.~\cite{Knecht:2020xyr}, the VVA correlator has been calculated in \chpt. Evaluated at the origin and adapted to our definition of the scalar functions, these read
\begin{align}
	\tilde{\W}_{1,\text{\chpt}}^{(3)}(0) &= \frac{64}{3} C_7^W - \frac{16}{3} C_{22}^W(\mu) + \frac{1}{384 \pi^4 F_\pi^2} \left( \log \frac{m_\pi^2}{\mu^2} + \log \frac{m_K^2}{\mu^2} + 2 \right) - \frac{1}{4 \pi^2 m_\pi^2} \,, \nn
	\tilde{\W}_{2,\text{\chpt}}^{(3)}(0) &= \frac{16}{3} C_{22}^W(\mu) - \frac{1}{384 \pi^4 F_\pi^2} \left( \log \frac{m_\pi^2}{\mu^2} + \log \frac{m_K^2}{\mu^2} + 2 \right) \,.
\end{align}
These expressions depend on two low-energy constants, which have been introduced in Ref.~\cite{Bijnens:2001bb}. They can be fixed from the normalization and slope parameter of the $\pi^0$ TFF. With this replacement, the expressions simplify to (see also Ref.~\cite{Masjuan:2020jsf})
\begin{align}
	\label{eq:VVA_WtiNormalizationChPT}
	\tilde{\W}_{1,\text{\chpt}}^{(3)}(0) &= - \frac{(1 + a_\pi) F_\pi F_{\pi \gamma^* \gamma^*}(0, 0)}{m_\pi^2} \,, \nn
	\tilde{\W}_{2,\text{\chpt}}^{(3)}(0) &= \frac{a_\pi F_\pi F_{\pi \gamma^* \gamma^*}(0, 0)}{m_\pi^2} \,,
\end{align}
where we have used the definition
\begin{equation}
	a_\pi = \frac{m_\pi^2}{F_{\pi \gamma^* \gamma^*}(0, 0)} \frac{\partial}{\partial q^2} F_{\pi \gamma^* \gamma^*}(q^2, 0) \bigg|_{q^2=0} \,.
\end{equation}
Using the experimental value for the normalization~\cite{PrimEx-II:2020jwd} and the dispersive value for the slope~\cite{Hoferichter:2018kwz}, this leads to
\begin{equation}
	\label{eq:NormChPT}
	\tilde{\W}_{1,\text{\chpt}}^{(3)}(0) = \SI{-1.439 \pm 0.011}{\GeV^{-2}} \,, \quad \tilde{\W}_{2,\text{\chpt}}^{(3)}(0) = \SI{0.044 \pm 0.003}{\GeV^{-2}} \,.
\end{equation}
For the longitudinal scalar function, the $1$ in the bracket in Eq.~\eqref{eq:VVA_WtiNormalizationChPT} represents the leading order and the NLO term $a_\pi$ gives a correction of approximately \SI{3}{\percent}. We expect the NNLO correction to be below \SI{5}{\percent} of the NLO term\footnote{Although we have started from $SU(3)$ \chpt, Eq.~\eqref{eq:VVA_WtiNormalizationChPT} only contains $SU(2)$ quantities. Thus, we expect the same expressions to arise if the calculation were done in $SU(2)$ \chpt\ so that higher-order corrections should be small. Indeed, the second relation in Eq.~\eqref{eq:VVA_WtiNormalizationChPT} can be derived in the chiral limit without using \chpt{}~\cite{Masjuan:2020jsf}.} and we added this to the uncertainty coming from the input parameters. In the case of the transverse component, there is just an NLO term and we thus add a \SI{5}{\percent} uncertainty on the total.

\subsection{Review of OPE constraints}
\label{sec:VVA_OPE}

The OPE for the VVA correlator has been discussed in Refs.~\cite{Knecht:2002hr,Czarnecki:2002nt,Vainshtein:2002nv,Knecht:2003xy,Melnikov:2006qb}. Here, we review the results relevant for our analysis. We are interested in the behavior of the operator
\begin{equation}
	\hat{T}_{\mu\nu}(q) = \imag \int \dif^4 x\, e^{\imag q\cdot x} T\{j_\mu^\mathrm{em}(x) j_\nu^5(0) \}
\end{equation}
for a large space-like momentum $q$.\footnote{A recent treatment of the SDCs for the case where all three currents are close to each other can be found in Ref.~\cite{Kadavy:2020hox}.} The matrix element of this operator between the vacuum and a photon state is related to the VVA correlator
\begin{equation}
	\bra{0} \hat{T}^{\mu \rho}(q_1) \ket{\gamma(-q_2, \lambda_2)} = -\imag e \epsilon_\nu^{\lambda_2}(-q_2) \W^{\mu \nu \rho}(q_1, q_2) \,,
\end{equation}
where we restrict ourselves to $q_2^2 = 0$. The OPE can be written as~\cite{Czarnecki:2002nt}
\begin{equation}
	\hat{T}_{\mu \nu}(q) = \sum_i \Bigl[ -c_1^i(q^2) q_\nu q_\alpha g_{\mu\beta} + c_2^i(q^2) \left( -q^2 g_{\mu\alpha} g_{\nu\beta} + q_\mu q_\alpha g_{\nu\beta} - q_\nu q_\alpha g_{\mu\beta} \right) \Bigr] \O_i^{\alpha \beta}(0) \, ,
\end{equation}
with antisymmetric local operators $\O_i^{\alpha \beta}$. Their matrix elements have the form
\begin{equation}
	\bra{0} \O_i^{\alpha \beta}(0) \ket{\gamma(-q_2, \lambda_2)} = - \imag e \epsilon^{\alpha \beta \gamma \delta} q_{2\gamma} \epsilon_\delta^{\lambda_2}(-q_2) \kappa_i \,,
\end{equation}
where $\kappa_i$ are numbers. Comparison with Eq.~\eqref{eq:decompositiongm2} gives
\begin{equation}
	\tilde{\W}_1(q^2) = \sum_i \kappa_i c_1^i(q^2) \, , \quad \tilde{\W}_2(q^2) = \sum_i \kappa_i c_2^i(q^2) \,.
\end{equation}
At large $q^2$, the contribution of higher-dimension operators are suppressed by the associated higher power of $q^{-2}$ in the coefficient.

At mass dimension $2$, there is only one operator that has a non-vanishing matrix element between these states
\begin{equation}
	\O_F^{\alpha \beta} = -e \epsilon^{\alpha \beta \gamma \delta} \partial_\gamma A_\delta\,,
\end{equation}
where the prefactor is chosen such that $\kappa_F = 1$. The renormalized Wilson coefficient to logarithmic accuracy is
\begin{equation}
	\label{eq:OPEWilsonD=2}
	c_1^F(q^2) = -2 c_2^F(q^2) = \frac{N_c}{2 \pi^2 q^2} \left[ \tr(\Q^2 \Q_5) + \frac{2 \tr(\Q^2 \Q_5 \M^2)}{q^2} \log \Bigl( -\frac{q^2}{\bar\mu^2} \Bigr) + \Order \left( \frac{\M^4}{q^4} \right) \right] \, ,
\end{equation}
where the \msbar{} renormalization scale is defined by $\mu^2 = \bar\mu^2 \frac{e^{\gamma_E}}{4\pi}$. Translated to the VVA correlator, the first term in the bracket gives the massless fermion loop and the second one is a chiral correction that corresponds to dimension $4$.

At dimension $3$, there is one operator for each quark flavor
\begin{equation}
	\O_q^{\alpha \beta} = \frac{1}{2} \epsilon^{\alpha \beta \gamma \delta} \bar{q} \sigma_{\gamma \delta} q\,.
\end{equation}
The Wilson coefficient, however, contains an explicit power of the quark mass
\begin{equation}
	c_1^q(q^2) = -2 c_2^q(q^2) = -\frac{4 \Q_5^q \Q^q m_q}{q^4}\,,
\end{equation}
which effectively increases the mass dimension of the operator to $4$. Here, $\Q^q$ and $\Q_5^q$ are the respective entries of the charge matrices. At mass dimension $4$, no further operators appear, as derivative terms can be related to $\O_q^{\alpha\beta}$ by the equations of motion.

Neglecting non-perturbative QCD effects, the matrix elements of these operators are given to logarithmic accuracy by
\begin{equation}
	\kappa_q^\text{pert} = -\frac{\Q^q N_c m_q}{4 \pi^2} \log \frac{\bar\mu^2}{m_q^2}\,.
\end{equation}
This combines with the quark-mass correction in Eq.~\eqref{eq:OPEWilsonD=2} to give the term of order $m_q^2$ in the perturbative quark loop, canceling the dependence on $\bar\mu$. Since the same results also hold for the leptonic contributions to the VVA correlator, this shows the consistency of the OPE with perturbation theory.

For quarks, however, the matrix elements are non-perturbative and the condensates $\kappa_q$ are given by
\begin{equation}
	\kappa_q = -\frac{\Q^q X_q}{e}\,,
\end{equation}
where $X_q$ are so-called tensor coefficients related to the magnetic susceptibility and are known from lattice QCD in the $\overline{\text{MS}}$ scheme at $\bar\mu = \SI{2}{\GeV}$~\cite{Bali:2012jv} (see also Ref.~\cite{Bijnens:2019ghy})
\begin{equation}
	X_u = \SI{40.7\pm1.3}{\MeV}\,, \quad X_d = \SI{39.4\pm1.4}{\MeV}\,, \quad X_s = \SI{53.0\pm7.2}{\MeV} \,,
\end{equation}
whereas Refs.~\cite{Vainshtein:2002nv,Czarnecki:2002nt} use
\begin{equation}
	\label{eq:kappa_qVainshtein}
	\kappa_q^\text{\cite{Czarnecki:2002nt}} = \frac{N_c \Q^q \braket{\bar{q}q}}{4 \pi^2 F_\pi^2}\,,
\end{equation}
which with $\braket{\bar{q}q} = -(\SI{270}{\MeV})^3$ and $F_\pi = \SI{92}{\MeV}$ would give somewhat larger estimates than the lattice-QCD values,\footnote{References~\cite{Vainshtein:2002nv,Czarnecki:2002nt} used $\braket{\bar{q}q} = -(\SI{240}{\MeV})^3$, which gives estimates slightly below the lattice values for up- and down-quark tensor coefficients.}
\begin{equation}
	X_u^\text{\cite{Czarnecki:2002nt}} = X_d^\text{\cite{Czarnecki:2002nt}} = X_s^\text{\cite{Czarnecki:2002nt}} = \SI{53.5}{\MeV}\,.
\end{equation}

The flavor decomposition proceeds according to
\begin{equation}
	\sum_q \kappa_q c_1^q(q^2) = \frac{4 \tr (\Q_5 \Q^2 \M X)}{e q^4} = \sum_a \frac{4 C_a \tr (\Q^2 \M X \lambda^a)}{e q^4}
\end{equation}
and similarly for the dimension $2$ term. Thus, we obtain to order $q^{-4}$
{\small\begin{align}
	\label{eq:OPEresNLOIsovector}
	\tilde{\W}_1^{(3)}(q^2) &= \frac{1}{4 \pi^2 q^2} + \frac{2}{9 e q^4} (4 m_u X_u - m_d X_d) + \Order (q^{-6}) \nn
		&= \frac{1}{4 \pi^2 q^2} + \frac{\hat m (X_u + X_d)}{3 e q^4} + \frac{5\hat m (X_u - X_d)}{9 e q^4} + \frac{(m_u-m_d) (4 X_u + X_d)}{9 e q^4} + \Order (q^{-6}) \nn
		&= \frac{1}{4 \pi^2 q^2} \left[ 1 + \frac{12.0(1.3) \times 10^{-3}\GeV^2}{q^2} + \frac{0.3(5) \times 10^{-3}\GeV^2}{q^2} - \frac{7.4(1.6) \times 10^{-3}\GeV^2}{q^2} \right] + \Order (q^{-6}) \,, \nn
	\tilde{\W}_2^{(3)}(q^2) &= - \frac{1}{2} \tilde{\W}_1^{(3)}(q^2) + \Order (q^{-6})
\end{align}}%
for the isovector component, where $\hat m = (m_u + m_d)/2$ and we neglected terms involving two or more powers of the quark mass. Here, we have used the quark masses in the \msbar{} scheme at $\bar\mu = \SI{2}{\GeV}$~\cite{ParticleDataGroup:2022pth}, consistent with the tensor coefficients given above. The coefficients of the isovector $q^{-4}$ terms are sensitive to isospin-breaking contributions. However, as argued in Ref.~\cite{Vainshtein:2002nv}, they should be disregarded, as they are related to $\pi^0$--$\eta$--$\eta'$ mixing, which we do not take into account. Therefore, we will only use the first of the $q^{-4}$ terms in Eq.~\eqref{eq:OPEresNLOIsovector} and drop the isospin-breaking terms proportional to $m_u-m_d$ and $X_u-X_d$.

For the strange-quark component, we find
\begin{align}
	\label{eq:OPEresNLOstrange}
	\tilde{\W}_1^{(s)}(q^2) &= \frac{1}{\sqrt{6}} \tilde{\W}_1^{(0)}(q^2) - \frac{1}{\sqrt{3}} \tilde{\W}_1^{(8)}(q^2) =  \frac{1}{12 \pi^2 q^2} + \frac{2 m_s X_s}{9 e q^4}  + \Order (q^{-6}) \nn
		&=  \frac{1}{12 \pi^2 q^2} \left[ 1 + \frac{0.43(7)\GeV^2}{q^2} \right] + \Order (q^{-6}) \, , \nn
	\tilde{\W}_2^{(s)}(q^2) &= - \frac{1}{2} \tilde{\W}_1^{(s)}(q^2) + \Order (q^{-6}) \, .
\end{align}
The numerators of the $q^{-4}$ terms in $\tilde{\W}_1^{(3)}$ and $\tilde\W_1^{(s)}$ can be compared to $m_\pi^2 \approx \SI{18.2e-3}{\GeV^2}$ and $2m_{\eta'}^2 - m_{\eta}^2 \approx \SI{1.53}{\GeV^2}$ used in Ref.~\cite{Czarnecki:2002nt}

There are many operators at dimension $5$ and $6$, which produce terms of order $q^{-6}$ and higher. In the chiral limit, these provide the first non-perturbative corrections and, in agreement with the non-renormalization of the anomaly, contribute only to $\tilde{\W}_2$. Reference~\cite{Czarnecki:2002nt} estimates in the chiral limit
\begin{equation}
	\label{eq:W2Q6OPE}
	\tilde{\W}_2(q^2) = - \frac{N_c \tr(\Q_5 \Q^2)}{4 \pi^2 q^2} \left[ 1 -\frac{(\SI{0.71}{\GeV})^4}{q^4} + \Order\left( \frac{1}{q^6} \right) \right]\,,
\end{equation}
assuming a simple factorization model in terms of quark condensates. We adopt the same number as central value, assigning a $100\%$ uncertainty to the $q^{-6}$ coefficient.

\mbox{}

\subsection{Implementation of SDCs}
\label{sec:VVA_SDC_Implementation}

In the results of the dispersion relations for fixed photon virtualities, Eqs.~\eqref{eq:PSPoleFixedVirt} and \eqref{eq:APoleFixedVirt}, the contribution from each intermediate state asymptotically falls off as $\O(q^{-4})$ or faster, which is a consequence of the asymptotic behavior of the pseudoscalar and axial-vector TFFs. Similarly to the case of HLbL, the OPE constraint cannot get saturated by a finite number of pseudoscalar or axial-vector intermediate states. An infinite tower of heavier intermediate states could be described in terms of hadronic models as discussed in Sect.~\ref{sec:VVAModels}.

In contrast, in the dispersion relation in $g-2$ kinematics, each contribution to the two scalar functions in Eq.~\eqref{eq:ResultScalarFunctionsgm2TwoPionExplicit} asymptotically behaves as $\O(q^{-2})$, suggesting that the OPE constraints could be directly saturated. However, even though the overall asymptotic scaling agrees with the OPE, there is no reason to expect that the coefficients of the leading and sub-leading terms in the high-energy expansion will match the OPE constraint. Indeed, the comparison of the two dispersive approaches in Sect.~\ref{sec:Comparison} has shown that the two results are identical up to effects due to intermediate states that are not fully taken into account. In any case, the truncation of the infinite sum of intermediate states implies that each dispersive representation must be regarded as a low-energy description with a limited range of validity.

In the following, we will restrict the discussion to the isovector channel. Instead of modeling an infinite tower of higher states, we follow a similar strategy as for the sub-processes (in particular the pion TFF~\cite{Hoferichter:2018kwz}) and we add to the dispersive result a small number of effective poles
\begin{equation}
	\label{eq:EffectivePoleVVA}
	\tilde{\W}_i^{(3),\text{eff}}(q^2) = \sum_j \frac{1}{4\pi^2} \frac{n_{i,j}^{(3),\text{eff}}}{q^2 - \big( m_{i,j}^{(3),\text{eff}} \big)^2} \,,
\end{equation}
which are used to model the resummation of all heavier intermediate states that are not included explicitly. The effective-pole masses $m_{ij}^{(3),\text{eff}}$ should be chosen sufficiently large, i.e., in the region where we expect additional heavier intermediate states. The residues $n_{i,j}^{(3),\text{eff}}$ are then adjusted by requiring that the sum of dispersive result and effective poles match both low-energy and OPE constraints~\eqref{eq:OPEresNLOIsovector}. Obviously, the effective poles also contribute at low energies. This model-dependent effect is beyond the control of the dispersive approach, but it is constrained by both the low-energy expansion and the OPE.

As an alternative to effective poles, one could follow the approach of Ref.~\cite{Ludtke:2020moa} to smoothly interpolate between the dispersively described low-energy region and the asymptotic region where SDCs are valid. However, in this approach one does not obtain an estimate of the contribution of neglected intermediate states to the low-energy region below the matching point.

In order to perform the asymptotic matching, we require the asymptotic expansions of the dispersion integrals in the two-pion contributions to Eq.~\eqref{eq:ResultScalarFunctionsgm2TwoPionExplicit}. At large $s$, additional intermediate states become important so that the integrand ceases to provide a good approximation. Therefore, we replace the upper integration limit by a finite cutoff $s_\text{cut}$. In our numerical study, we choose $s_\text{cut} = (\SI{1.5 \pm 0.5}{\GeV})^2$ and the cutoff dependence will enter our uncertainty estimate. After cutting off the dispersion integrals, the integrand can be expanded before integration for large $q^2$ in order to obtain the coefficients of the $1/q^2$ expansion and perform the matching to the SDCs.

%% file: sections/Numerics.tex

\section{Numerical implementation for the isovector channel}
\label{sec:Numerics}

In this section, we present numerical implementations of both dispersive representations in the isovector channel of the axial current. The two dispersive approaches only differ by effects due to neglected intermediate states. In order to enable a direct comparison, we implement the common building blocks in an identical way in both dispersive approaches.

\subsection{Phase shift, Omn\`es function, and pion VFF}
\label{sec:PhasesOmnes}

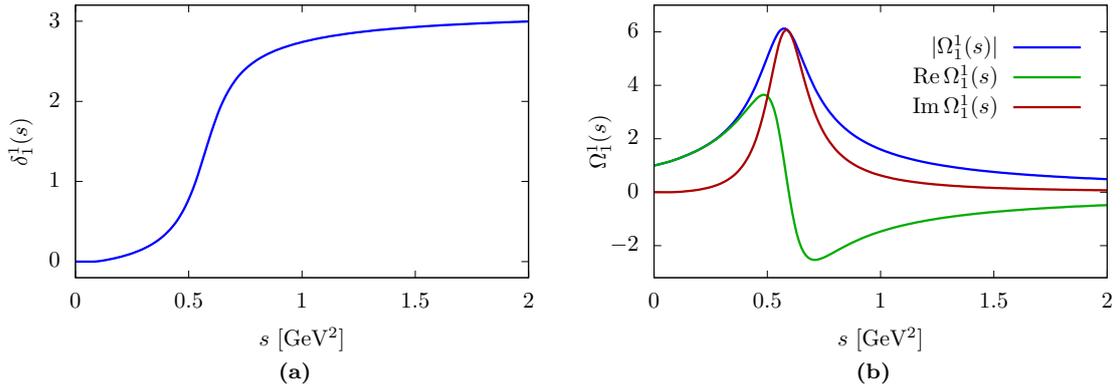
\begin{figure}[t]
	\qquad
	\begin{subfigure}[t]{0.45\textwidth}
		\scalebox{0.85}{\input{plots/delta11}}
		\caption{}
		\label{fig:PwavePhase}
	\end{subfigure} \qquad
	\begin{subfigure}[t]{0.45\textwidth}
		\scalebox{0.85}{\input{plots/Omega11}}
		\caption{}
		\label{fig:Omnes}
	\end{subfigure}
	\caption{{\bf (a)} Elastic $I=1$ $P$-wave phase shift from Refs.~\cite{Colangelo:2001df,Colangelo:2018mtw}. {\bf (b)} Omn\`es function obtained from this phase shift.}
\end{figure}

For the isovector elastic $\pi\pi$ $P$-wave phase shift $\delta_1^1(s)$, we use the results of Ref.~\cite{Colangelo:2018mtw}, based on the Roy-equation~\cite{Roy:1971tc} analysis of Refs.~\cite{Colangelo:2001df,Caprini:2011ky}. We guide it to the asymptotic value $\pi$ by switching to the function
\begin{equation}
	\delta_\text{asymp}(s) = \pi + \frac{a}{s + b}
\end{equation}
above $s = \SI{1}{\GeV^2}$, with the parameters $a$ and $b$ chosen such that the phase and its first derivative are continuous at the matching point. We have checked that the exact form of the asymptotic matching and the matching point have a negligible effect on our results for VVA. The phase shift is displayed in Fig.~\ref{fig:PwavePhase}. The resulting Omn\`es function Eq.~\eqref{eq:Omnes} is shown in Fig.~\ref{fig:Omnes}, exhibiting the strong $\rho$-resonance peak. In the pion VFF, we neglect isospin-breaking $\rho$--$\omega$ mixing~\cite{Colangelo:2018mtw} and approximate higher inelastic effects by a linear polynomial, which is set to a constant above $1\GeV$~\cite{Hoferichter:2018kwz}
\begin{equation}
	F_\pi^V(s) = \Omega_1^1(s) \left( 1 + \alpha_V s \right) \, .
\end{equation}
We fix the parameter $\alpha_V$ by matching the charge radius of the pion determined in Ref.~\cite{Colangelo:2018mtw}.\footnote{Enlarging the uncertainty on the charge radius in order to cover also the fit~\cite{Stoffer:2023gba} to the CMD-3 data set~\cite{CMD-3:2023alj,CMD-3:2023rfe} has no impact on our results.} In the following calculations, we propagate the uncertainties on the phase shift $\delta_1^1$ as determined in Ref.~\cite{Colangelo:2018mtw}.

\subsection[Pion TFF and $\gamma \to 3 \pi$]{\boldmath Pion TFF and $\gamma \to 3 \pi$}

\begin{figure}[t]
	\centering
	\scalebox{0.85}{\input{plots/TFF}}
	\caption{Singly-virtual pion TFF in the space-like region, compared to the dispersive evaluation of Ref.~\cite{Hoferichter:2018kwz}. We separately show the isoscalar and isovector components as well as the effective-pole contribution.}
	\label{fig:TFF}
\end{figure}
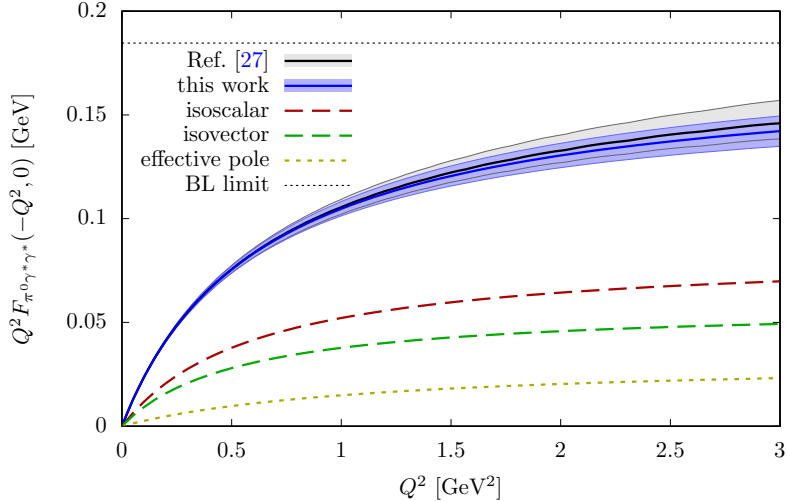

For the singly-virtual pion TFF, we use the simplified dispersive representation given in Eq.~\eqref{eq:PTFFsinglyVirtual} and we add an effective pole~\eqref{eq:PionTFFEffectivePole} to account for neglected higher intermediate states. The isovector contribution depends on the $\gamma\to3\pi$ partial wave $f_\pi^1(s,0)$, which we reconstruct by solving the Khuri--Treiman equations~\cite{Khuri:1960zz} for $\F_\pi(s, t, u; 0)$~\cite{Niecknig:2012sj, Hoferichter:2012pm, Hoferichter:2018kwz} with the $\pi\pi$-phase input discussed in Sect.~\ref{sec:PhasesOmnes}. In the isoscalar channel, we only include the $\omega$ meson, see Sect.~\ref{sec:VectorMesonInput}. The parameters of the effective pole are determined by matching the experimental normalization~\cite{PrimEx-II:2020jwd} and the slope of the TFF from Ref.~\cite{Hoferichter:2018kwz}. Numerically, our solution for the TFF agrees well with the state-of-the-art dispersive representation of Refs.~\cite{Hoferichter:2018kwz, Hoferichter:2018dmo}, see Fig.~\ref{fig:TFF}.

In the formalism in $g-2$ kinematics, we need the singly-virtual pion TFF evaluated at the time-like value $m_\pi^2$. We obtain
\begin{equation}
	F_{\pi^0 \gamma^* \gamma^*}(m_{\pi^0}^2, 0) = 0.2844(21)\GeV^{-1} \,,
\end{equation}
which agrees with a quadratic extrapolation from the real-photon point using the normalization from Ref.~\cite{PrimEx-II:2020jwd} and the slope and curvature parameters from Ref.~\cite{Hoferichter:2018kwz}.
For the dispersion relations in $g-2$ kinematics, also the solution of the Khuri--Treiman equations for $\gamma\to3\pi$ is required only at a single point. We find
\begin{equation}
	\F_\pi(m_{\pi^0}^2, m_\pi^2, m_\pi^2; 0) = f_\pi^1(m_{\pi^0}^2, 0) =  \frac{\Omega_1^1(m_{\pi^0}^2)}{4 \pi^2 F_\pi^3} \times 1.072(10) \, ,
\end{equation}
where quark-mass effects in the normalization~\cite{Bijnens:1989ff, Hoferichter:2012pm, Hoferichter:2018kwz} are responsible for the largest part of the difference to the chiral limit in Eq.~\eqref{eq:ChPT_g3pi}.

The two-pion contribution saturates the isovector sum rule Eq.~\eqref{eq:pionTFFIVNorm} at the level of $80\%$, which lies somewhat below previous findings~\cite{Hoferichter:2012pm} and is largely explained by a smaller value of the pion charge radius~\cite{Colangelo:2018mtw}. The isoscalar sum rule~\eqref{eq:sumRulePionTFFIsoscalar} is saturated by the $\omega$ contribution alone to 100\%, confirming the assumed dominance of the $\omega$ over further isoscalar contributions. The neglected contribution of the $\phi$ meson would lead to a correction to the isoscalar part of the pion TFF normalization of about $7\%$ and has to be compensated by even heavier or non-resonant contributions.

\subsection{Vector-meson couplings}
\label{sec:VectorMesonInput}

For the vector-meson poles in the dispersive framework in $g-2$ kinematics, we need the parameters $\tilde{\V}_i^{(3)}(m_V^2)$. According to Eq.~\eqref{eq:Vtildei}, these can be written as
\begin{align}
	\tilde{\V}_1^{(3)}(m_V^2) &=  - \frac{F_\pi F_{V\pi}}{m_V^2 - m_\pi^2}  + \frac{F_{a_1}^{(3)}}{m_{a_1}} \left( F_{V{a_1}}^1 - F_{V{a_1}}^2 - F_{V{a_1}}^3 \right) \,, \nn
	\tilde{\V}_2^{(3)}(m_V^2) &= - F_{a_1}^{(3)} m_{a_1} \frac{F_{V{a_1}}^2 + F_{V{a_1}}^3}{m_V^2 - m_{a_1}^2} \,,
\end{align}
where we have taken into account $\pi^0$ and $a_1$ poles. The coupling constants in this equation are related to decay widths by
\begin{align}
	\label{eq:WidthVPgAndAVg}
	\Gamma_{V \to P \gamma} &= \frac{\alpha}{24} \left( \frac{m_V^2 - m_P^2}{m_V} \right)^3 (F_{VP})^2 \,, \nn
	\Gamma_{A \to V \gamma} &= \frac{\alpha}{24} \frac{(m_A^2 - m_V^2)^3}{m_A^5} \bigg[ (m_A^2 - m_V^2)^2 \left( F_{VA}^1 \right)^2 + m_V^2 (m_A^2 + m_V^2) \left( F_{VA}^2 + F_{VA}^3 \right)^2 \nn
					&\qquad \qquad \qquad \qquad \quad + 2 m_V^2 (m_A^2 - m_V^2) F_{VA}^1 \left( F_{VA}^2 + F_{VA}^3 \right) \bigg] \,.
\end{align}
In the following, we will explicitly include only the lightest isoscalar vector resonance, the $\omega$-meson, for the following reasons. First, the parameters of the $\phi$-meson are related to those of the $\omega$ through the approximate $U(3)$-symmetry relation
\begin{equation}
	\frac{F_{\phi \pi}}{F_{\omega \pi}} = \frac{\tilde{F}_{\phi a_1}^i}{\tilde{F}_{\omega a_1}^i} = \frac{\cot \theta_V - \sqrt{2}}{1 + \sqrt{2} \cot \theta_V} \,.
\end{equation}
The vector-meson mixing angle $\theta_V$ is close to the ideal mixing angle $\theta_V^\text{ideal} = \tan^{-1}(1/\sqrt{2})$, which implies $F_{\phi \pi} \ll F_{\omega \pi}$, 
$\tilde{F}_{\phi a_1}^i \ll \tilde{F}_{\omega a_1}^i$ and thus $\tilde{\V}_i^{\phi, (3)}(m_\phi^2) \ll \tilde{\V}_i^{\omega, (3)}(m_\omega^2)$. Second, the masses of all heavier vector mesons are significantly larger than \SI{1}{\GeV} and there are other intermediate states in this mass range that we do not include explicitly, such as the $\pi(1300)$. The sum of all these contributions thus has to be described through the matching to SDCs.

The partial decay width of $\omega \to e^+ e^-$~\cite{ParticleDataGroup:2022pth}
\begin{align}
	\Gamma_{\omega \to e^+ e^-} = \frac{4 \pi \alpha^2 f_\omega^2}{3 m_\omega} \, ,
\end{align}
where we neglect the electron mass, determines the $\omega$ decay constant $f_\omega = \SI{45.9 \pm 0.8}{\MeV}$. The experimental value for the width $\Gamma_{\omega \to \pi \gamma}$~\cite{ParticleDataGroup:2022pth} together with Eq.~\eqref{eq:WidthVPgAndAVg} leads to
\begin{equation}
	F_{\omega \pi} = \SI{2.33 \pm 0.04}{\GeV^{-1}} \, ,
\end{equation}
where the sign ambiguity can be fixed with the sum rule for the TFF, Eq.~\eqref{eq:sumRulePionTFFIsoscalar}.

For the determination of the vector--axial-vector--photon couplings, we would need the decay width $a_1 \to \omega \gamma$, which has not yet been measured.
However, some information can be obtained by assuming $U(3)$ symmetry, which relates the decay width for $a_1 \to \omega\gamma$ to the partial decay width $\Gamma_{f_1 \to \rho \gamma}$, with $\mathrm{BR}(f_1\to\rho\gamma) = 4.2(1.0)\%$ taken from Ref.~\cite{Hoferichter:2023tgp}. For ideal vector-meson mixing, we find
\begin{equation}
	\frac{\Gamma_{a_1\to\omega\gamma}}{\Gamma_{f_1\to\rho\gamma}} = \frac{3}{\left( \sin\theta_A + \sqrt{2} \cos\theta_A \right)^2} \, ,
\end{equation}
with the mixing angle $\theta_A = 62(5)^\circ$ as determined by the L3 collaboration~\cite{L3:2007obw}. The two relevant linear combinations of couplings, $F_{\omega a_1}^1$ and $F_{\omega a_1}^2 + F_{\omega a_1}^3$, are constrained to an ellipse by Eq.~\eqref{eq:WidthVPgAndAVg}, shown in Fig.~\ref{fig:FVA}. The uncertainty band is given by the parametric errors and a generic 30\% $U(3)$ uncertainty, summed in quadrature.

The anomaly constraint implies that in the chiral limit the $\omega$-pole residue in $\tilde\W_1$ proportional to $\tilde\V_1^{(3)}(m_\omega^2)$ has to vanish. The linear combination of couplings that leads to a vanishing value of $\tilde\V_1^{(3)}(m_\omega^2)$ is indicated by the dashed line in Fig.~\ref{fig:FVA}. Assuming only small chiral corrections, we expect the values of the couplings to lie close to one of the two intersection points of this line with the ellipse obtained using $U(3)$ symmetry.

\subsection[Longitudinal VVA function in $g-2$ kinematics]{\boldmath Longitudinal VVA function in $g-2$ kinematics}
\label{sec:VVA_W1Numerics}

\begin{figure}[t]
	\centering
	\scalebox{0.8}{\input{plots/FVAellipse}}
	\caption{Vector--axial-vector meson couplings as determined from $\Gamma_{a_1\to\omega\gamma}$ and $U(3)$ symmetry and compared to the constraints obtained from the longitudinal VVA scalar function.}
	\label{fig:FVA}
\end{figure}
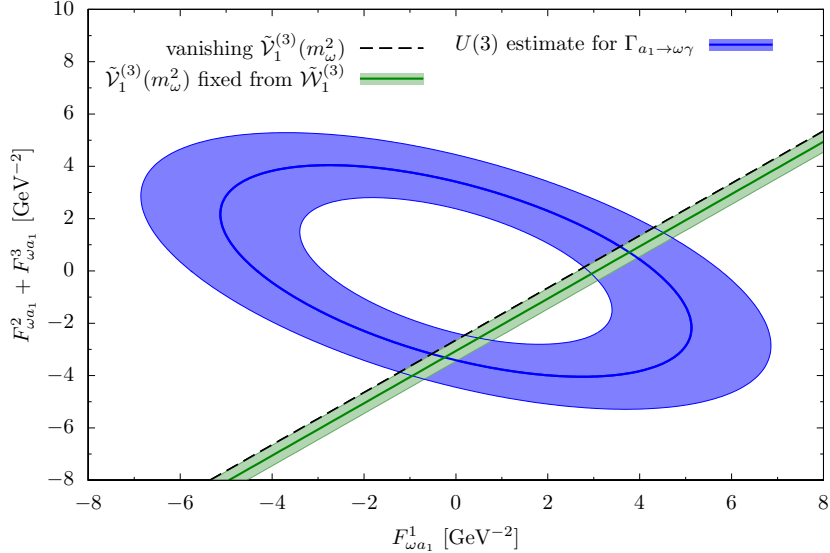

With the input discussed in the previous sections, we determine the isovector longitudinal function $\tilde\W_1^{(3)}(q^2)$ based on the dispersion relation in $g-2$ kinematics~\eqref{eq:ResultScalarFunctionsgm2TwoPionExplicit}. Including only the lowest intermediate states, we arrive at
\begin{align}
	\label{eq:W1Numerics}
	\tilde{\W}_1^{(3)}(q^2) &= F_\pi \frac{F_{\pi^0 \gamma^* \gamma^*}(m_{\pi^0}^2, 0)}{q^2 - m_{\pi^0}^2} + \frac{m_\omega f_\omega \tilde{\V}_1^{(3)}(m_\omega^2)}{q^2 - m_\omega^2} - \frac{1}{48 \pi^2} \int_{4m_\pi^2}^{\Lambda^2} \dif s \frac{\sigma_\pi^3(s)  F_\pi^V(s) \Omega_1^1(s)^* \, \tilde\U_1^{(3)}(0)}{s - q^2 - \imag \epsilon} \nn*
		&\quad + \frac{1}{96\pi^2} \int_{4m_\pi^2}^{\Lambda^2} \dif s \frac{s \sigma_\pi^3(s)  F_\pi^V(s) \Omega_1^1(s)^*}{(s - q^2 - \imag \epsilon)(s - m_{\pi^0}^2 - \imag \epsilon)} \frac{F_\pi \F_{\pi^0}}{\Omega_1^1(m_{\pi^0}^2)} \,,
\end{align}
where we vary the cutoff in the range ${\Lambda^2} = (1\ldots2\GeV)^2$. We determine the $\omega$-pole residue by imposing the leading $1/q^2$ asymptotic behavior in Eq.~\eqref{eq:OPEresNLOIsovector}.
This leads to
\begin{equation}
	m_\omega f_\omega \tilde{\V}_1^{(3)}(m_\omega^2) = 2.3(1.5) \times 10^{-3} \,,
\end{equation}
and determines one linear combination of vector--axial-vector--photon couplings
\begin{equation}
	\label{eq:FVA1m2m3fromW1}
	F_{\omega a_1}^1 - F_{\omega a_1}^2 - F_{\omega a_1}^3 = 3.1(4) \GeV^{-2} \, ,
\end{equation}
where the uncertainties are dominated by the $5\%$ error that we assign to the constraint~\eqref{eq:AVPiPiSubtractionChPT} on the subtraction constant $\tilde\U_1^{(3)}(0)$ due to higher-order corrections in $SU(2)$ \chpt{}.
As expected, there is a significant cancellation in $\tilde{\V}_1^{(3)}(m_\omega^2)$ between the $\pi^0$ and $a_1$ contributions. The constraint~\eqref{eq:FVA1m2m3fromW1} on the couplings is shown by the green band in Fig.~\ref{fig:FVA}. The combination with the $U(3)$ estimate for the partial decay width leaves two possible solutions: either the coupling $F_{\omega a_1}^1$ is close to zero and $F_{\omega a_1}^2 + F_{\omega a_1}^3$ is negative, or $F_{\omega a_1}^1$ is larger than $F_{\omega a_1}^2 + F_{\omega a_1}^3$ but both are positive. We will discuss in Sect.~\ref{sec:VVA_a1TFF} how to resolve this ambiguity.

The resulting normalization of the longitudinal function is
\begin{equation}
	\tilde\W_1^{(3)}(0) = -1.439(10) \GeV^{-2}
\end{equation}
and exactly matches the \chpt{} constraint~\eqref{eq:NormChPT}. The subleading $1/q^4$ asymptotic behavior does not automatically agree with the OPE constraint~\eqref{eq:OPEresNLOIsovector} and requires an additional asymptotic matching. As explained in Sect.~\ref{sec:VVA_SDC_Implementation}, we add an effective pole~\eqref{eq:EffectivePoleVVA}, which comes with two free parameters. Fixing them with the OPE constraint as well as the normalization leads to unstable results, as the normalization already perfectly agrees with the \chpt{} constraint without the addition of an effective pole. Hence, we vary the effective-pole mass in the typical scale of neglected intermediate states, $m_{1,1}^{(3),\mathrm{eff}} = (1.4\ldots2.0)\GeV$. The pole residue is small,
\begin{equation}
	n_{1,1}^{(3),\mathrm{eff}} = 1.1(7) \times 10^{-2} \, ,
\end{equation}
the impact on the $\omega$ parameters is relatively mild,
\begin{align}
	\label{eq:W1constraintOmegaParametersWithEffPole}
	m_\omega f_\omega \tilde{\V}_1^{(3)}(m_\omega^2)\Big|_\text{eff. pole} &= 2.0(1.5) \times 10^{-3} \,, \nn
	F_{\omega a_1}^1 - F_{\omega a_1}^2 - F_{\omega a_1}^3 \Big|_\text{eff. pole} &= 3.1(4) \GeV^{-2} \,,
\end{align}
and the normalization remains unaffected,
\begin{equation}
	\label{eq:W1NormDispersive}
	\tilde\W_1^{(3)}(0) \Big|_\text{eff. pole} = -1.439(10) \GeV^{-2} \, .
\end{equation}

\subsection[Transverse VVA function in $g-2$ kinematics]{\boldmath Transverse VVA function in $g-2$ kinematics}
\label{sec:VVA_W2Numerics}

Including the $\omega$ and $a_1(1260)$ mesons as the lowest intermediate states, the dispersive representation of the isovector transverse VVA function in $g-2$ kinematics~\eqref{eq:ResultScalarFunctionsgm2TwoPionExplicit} reads
\begin{align}
	\label{eq:W2gm2Kinematics}
	\tilde{\W}_2^{(3)}(q^2) &= - \frac{F_{a_1}^{(3)}}{m_{a_1}} \frac{\F_2^{a_1}(m_{a_1}^2, 0)}{q^2 - m_{a_1}^2} + \frac{m_\omega f_\omega \tilde{\V}_2^{(3)}(m_\omega^2)}{q^2 - m_\omega^2} \nn
		&\quad + \frac{1}{48 \pi^2} \int_{4m_\pi^2}^{\Lambda^2} \dif s \frac{\sigma_\pi^3(s) F_\pi^V(s) \Omega_1^1(s)^* \, \tilde\U_1^{(3)}(0)}{s - q^2 - \imag \epsilon} \nn
		&\quad + \frac{1}{48\pi^2} \int_{4m_\pi^2}^{\Lambda^2} \dif s \frac{s \sigma_\pi^3(s) F_\pi^V(s) \Omega_1^1(s)^*}{(s - q^2 - \imag \epsilon)(s - m_{a_1}^2 - \imag \epsilon)} \frac{F_{a_1}^{(3)}}{m_{a_1}} \frac{\tilde\F_{{a_1},1}}{\Omega_1^1(m_{a_1}^2)} \,.
\end{align}
This representation depends on the $a_1$ TFF $\F_2^{a_1}(m_{a_1}^2, 0)$ and on the $a_1\gamma\to\pi\pi$ amplitude in one kinematic point. Both quantities contain an imaginary part that arises due to two-pion rescattering effects. For their determination, we rely on the dispersive representation of the axial TFFs given in App.~\ref{sec:axialTFF} and the simplified Omn\`es representation for $a_1\gamma\to\pi\pi$ introduced in App.~\ref{sec:AgpipiParametrization}. As will be discussed in more detail in Sect.~\ref{sec:VVA_a1TFF}, these dispersive representations contain a single free parameter. In particular, the complex phase of $\tilde\F_{{a_1},1}$ is equal to $\delta_1^1(m_{a_1}^2)$ and the imaginary part of $\F_2^{a_1}(m_{a_1}^2, 0)$ follows from the requirement that $\tilde{\W}_2^{(3)}(q^2)$ be real below the two-pion threshold. The resulting representation~\eqref{eq:W2gm2Kinematics} does not fulfill both the constraints from \chpt{} at low energies and the asymptotic constraints from the OPE. However, this can be amended by introducing effective poles in $\tilde\W_2^{(3)}(q^2)$ as described in Sect.~\ref{sec:VVA_SDC_Implementation}. If we introduce a single effective pole with a mass around the $\rho'$ resonance and impose the leading and sub-leading OPE behavior, we obtain a normalization of the transverse function of
\begin{equation}
	\tilde\W_2^{(3)}(0) \Big|_\text{1 eff. pole} = 0.042(20) \GeV^{-2} \, ,
\end{equation}
which is perfectly compatible with the \chpt{} constraint~\eqref{eq:NormChPT}. For our central solution, we include an additional effective contribution to $\F_2^{a_1}(m_{a_1}^2,0)$ due to neglected higher intermediate states in the TFF, as well as two effective poles in $\tilde\W_2^{(3)}(q^2)$ itself. We vary the mass of the lighter effective pole in the range $m_{2,1}^{(3),\mathrm{eff}} = (1.4\ldots2.0)\GeV$ and take a splitting to the heavier effective pole of $m_{2,2}^{(3),\mathrm{eff}} - m_{2,1}^{(3),\mathrm{eff}} = (250\ldots500)\MeV$, i.e., the variations of the effective poles cover the energy region of the excited vector-meson resonances $\rho'/\omega'$ and $\rho''/\omega''$. These contributions gives us enough freedom to impose the \chpt{} constraint on the normalization, as well as the leading and sub-leading OPE constraints, including a constraint on the $1/q^{6}$ behavior as estimated in Ref.~\cite{Czarnecki:2002nt}, see Eq.~\eqref{eq:W2Q6OPE}.

For the modulus of the $a_1\gamma\to\pi\pi$ amplitude, we obtain
\begin{equation}
	|\tilde \F_{a_1,1}(m_{a_1}^2,m_\pi^2,m_\pi^2) | = 6.3(2.4) \GeV^{-2} \, .
\end{equation}
The $a_1$ TFF including the contribution of higher intermediate states amounts to
\begin{equation}
	\label{eq:F2TFFatma12}
	\F_2^{a_1}(m_{a_1}^2,0) = -0.44(28) - 0.05(2) i \, .
\end{equation}
For comparison, the hQCD model reviewed in Sect.~\ref{sec:VVAModels} leads to $\F_2^{a_1}(m_{a_1}^2,0) = -0.35$.

Overall, the combination of the dispersive representation with low-energy and asymptotic constraints leads to a rather robust representation of the transverse function, despite the scarce experimental information on the $a_1$ axial-vector meson. In order to test the validity of our approximations and to improve our representation, more information on the $a_1$--$\omega$--$\gamma$ couplings, the $a_1\to\pi\pi\gamma$ amplitude, and the $a_1$ TFFs would be desirable. Ideally, one should perform a full dispersive analysis of $a_1\gamma\to\pi\pi$ as proposed in App.~\ref{sec:AgpipiParametrization}.

\subsection[Singly-virtual $a_1$ TFFs]{\boldmath Singly-virtual $a_1$ TFFs}
\label{sec:VVA_a1TFF}

The VVA dispersion relations in $g-2$ kinematics determine the $a_1$ TFF $\F_2^{a_1}$ in one kinematic point, Eq.~\eqref{eq:F2TFFatma12}. For the dispersion relations for fixed photon virtualities, we require the TFFs in the space-like region.
In the case of the $f_1$ axial-vector resonance, Refs.~\cite{Zanke:2021wiq,Hoferichter:2023tgp} performed a comprehensive data analysis based on a VMD parametrization of the axial TFFs. Using $U(3)$ symmetry, one could relate the $a_1$ to the $f_1$ TFFs.

In the following, we discuss the implementation of the dispersive representation of the singly-virtual $a_1$ TFFs and the simplified Omn\`es representation for $a_1\gamma\to\pi\pi$, see App.~\ref{sec:axialTFF} and App.~\ref{sec:AgpipiParametrization}. The parameters in these representations can be constrained by the requirement that the axial-vector contributions agree between the two dispersive approaches to VVA. Including only the $a_1$, we impose the second of the conditions~\eqref{eq:AxialDRCondition}, which amounts to
\begin{equation}
	\bar\F_{a_1,1} - m_{a_1}^2 \bar\F_{a_1,2} = - \frac{m_{a_1}}{F_{a_1}^{(3)}} \, \tilde\U_1^{(3)}(0) = - 10.9(8) \GeV^{-2} \, ,
\end{equation}
with the estimate $F_{a_1}^{(3)} = 168(7)\MeV$ from light-cone sum rules~\cite{Yang:2007zt} and the \chpt{} constraint~\eqref{eq:AVPiPiSubtractionChPT} on the subtraction constant $\tilde\U_1^{(3)}(0)$.

The first of the conditions~\eqref{eq:AxialDRCondition} implies $\bar\F_{a_1,1}' = 0$, but its validity relies on the assumption of a sufficiently restrictive asymptotic behavior of the $a_1\gamma\to\pi\pi$ sub-process. While this condition is of no direct consequence for the VVA scalar functions in $g-2$ kinematics, it would lead to a very large two-pion contribution to the TFF $\F_1^{a_1}$, making it difficult to implement asymptotic constraints. As we already neglect crossed-channel effects in $a_1\gamma\to\pi\pi$, we refrain from making the parametrization in App.~\ref{sec:AgpipiParametrization} even more rigid.

In the dispersive representation of VVA in $g-2$ kinematics, the amplitude $a_1\gamma\to\pi\pi$ is required in one kinematic point, which within our approximations is given by
\begin{align}
	\tilde\F_{a_1,1}(m_{a_1}^2,m_\pi^2,m_\pi^2) &= \Omega_1^1(m_{a_1}^2) \left( \bar\F_{{a_1},1} + m_{a_1}^2 \bar\F_{{a_1},1}' \right) \nn
		&= - \tilde\U_1^{(3)}(0) \Omega_1^1(m_{a_1}^2)  \frac{m_{a_1}}{F_{a_1}^{(3)}} + \Omega_1^1(m_{a_1}^2) m_{a_1}^2 ( \bar\F_{{a_1},1}' + \bar\F_{a_1,2} ) \, ,
\end{align}
see Eq.~\eqref{eq:SimplifiedOmnesFA1}. Within the same approximations, we obtain the following representation for the two axial-vector TFFs:
\begin{align}
	\label{eq:a1TFFs}
	\F_1^{a_1}(q^2,0) &= \frac{q^2}{48 \pi^2} \int_{4 m_\pi^2}^\infty \dif s \frac{m_{a_1}^2 \sigma_\pi^3(s)}{s - q^2}  F_\pi^V(s)^* \Omega_1^1(s) \bar\F_{{a_1},2} - \frac{m_{a_1}^2 q^2}{q^2 - m_\omega^2} \frac{f_\omega}{m_\omega} F_{\omega{a_1}}^1 \,, \nn
	\F_2^{a_1}(q^2,0) &= \frac{m_{a_1}^2}{48 \pi^2} \int_{4 m_\pi^2}^\infty \dif s \frac{\sigma_\pi^3(s) F_\pi^V(s)^* \Omega_1^1(s)}{s - q^2} \left( \frac{m_{a_1}}{F_{a_1}^{(3)}} \tilde\U_1^{(3)}(0) - s (\bar\F_{a_1,1}' + \bar\F_{{a_1},2}) \right) \nn
		&\quad + \frac{m_{a_1}^2 m_\omega f_\omega}{q^2 - m_\omega^2}  ( F_{\omega a_1}^2 + F_{\omega a_1}^3) \,.
\end{align}
In the global fit of Ref.~\cite{Hoferichter:2023tgp}, the antisymmetric TFF $\F_1^{f_1}(q_1^2,q_2^2)$ turned out to be compatible with zero. Assuming $U(3)$ symmetry, we neglect the axial-vector TFF $\F_1^{a_1}(q^2,0)$, which amounts to setting $\bar\F_{{a_1},2} = 0$ and $F_{\omega{a_1}}^1 = 0$. This fixes the $\omega$ residue in the transverse VVA function~\eqref{eq:W2gm2Kinematics} to
\begin{equation}
	m_\omega f_\omega \tilde{\V}_2^{(3)}(m_\omega^2) = -2.5(3) \times 10^{-2} \,.
\end{equation}

As an alternative to neglecting the antisymmetric TFF $\F_1^{a_1}$, we could determine the vector--axial-vector--photon couplings with the constraint from the longitudinal function and $U(3)$ symmetry, as shown in Fig.~\ref{fig:FVA}. One of the two possible solutions is compatible with the assumption of a vanishing antisymmetric $a_1$ TFF. We separately considered the alternative solution, where $F_{\omega a_1}^2 + F_{\omega a_1}^3$ is positive and smaller than $F_{\omega a_1}^1$. It turns out that it leads to almost identical results for the transverse VVA function. However, the relative contributions of isoscalar and isovector components are very different, since the $\omega$-pole residues in both the axial-vector TFF $\F_2^{a_1}(q^2,0)$ and the transverse VVA function $\tilde\W_2^{(3)}(q^2)$ are proportional to $F_{\omega a_1}^2 + F_{\omega a_1}^3$. $U(3)$ symmetry predicts that the isoscalar and isovector contributions to $\F_2^{a_1}(0,0)$ are equal. For the alternative solution, we find that the $\omega$ contribution to $F_2^{a_1}(0,0)$ is only about $-0.1$ times the two-pion contribution. In contrast, if we neglect the antisymmetric $a_1$ TFF and set $F_{\omega a_1}^1 = 0$, we find an $\omega$ contribution to $F_2^{a_1}(0,0)$ that is about $40\%$ larger than the two-pion contribution, roughly compatible with expected $U(3)$ uncertainties. Therefore, we consider the alternative solution unphysical and we neglect $\F_1^{a_1}$ in the following.

The dispersive representation of $\F_2^{a_1}(q^2,0)$ given in Eq.~\eqref{eq:a1TFFs} asymptotically behaves as $\O(q^{-2})$. In order to match the asymptotic behavior derived in Ref.~\cite{Hoferichter:2020lap},
\begin{equation}
	\label{eq:AxialTFFBLLimit}
	\F_{2}^{a_1}(q^2,0) = \frac{2 F_{a_1}^{(3)} m_{a_1}^3}{q^4} + \O(q^{-6}) \, ,
\end{equation}
we introduce a sum of effective poles
\begin{equation}
	\F_2^{a_1,\mathrm{eff}}(q^2,0) = \sum_j \frac{g_j^\mathrm{eff}}{4\pi} \frac{(m_j^\mathrm{eff})^2}{q^2 - (m_j^\mathrm{eff})^2} \, ,
\end{equation}
which are added to Eq.~\eqref{eq:a1TFFs}. The masses of the effective poles are chosen above the $a_1$ mass, and varied in the region of the excited vector mesons $\rho'/\omega'$ and $\rho''/\omega''$. With three effective poles, the residues can be adjusted to implement the correct asymptotic behavior and to match the TFF value in Eq.~\eqref{eq:F2TFFatma12}.

The normalization of the TFF is related to the so-called equivalent two-photon decay width
\begin{equation}
	\tilde{\Gamma}_{a_1 \to \gamma \gamma} = \lim_{q^2 \to 0} \frac{m_{a_1}^2}{q^2} \frac{1}{2} \Gamma(a_1 \to \gamma_L^* \gamma_T) \,,
\end{equation}
where $q^2$ is the virtuality of the longitudinal off-shell photon. In Ref.~\cite{Hoferichter:2020lap}, it was estimated using $U(3)$ symmetry and the measurements for $f_1(1285)$ and $f_1'(1420)$~\cite{Achard:2001uu, L3:2007obw}, leading to
\begin{equation}
	\tilde{\Gamma}_{a_1 \to \gamma \gamma} = \SI{2.0 \pm 0.7}{\keV} \,.
\end{equation}
The relation to the form factor~\cite{Hoferichter:2020lap}
\begin{equation}
	\tilde{\Gamma}_{a_1 \to \gamma \gamma} = \frac{\pi \alpha^2 m_{a_1}}{12} |\F_2^{a_1}(0, 0)|^2
\end{equation}
determines the normalization up to a sign,
\begin{equation}
	\label{eq:a1_TFF_Norm_U3}
	|\F_2^{a_1}(0, 0)| = \num{0.34 \pm 0.06} \,.
\end{equation}
In our representation with three effective poles that are matched to the asymptotic behavior~\eqref{eq:AxialTFFBLLimit} and the value~\eqref{eq:F2TFFatma12} extracted from the transverse VVA function, we find for the TFF normalization
\begin{equation}
	\label{eq:a1_TFF_Norm}
	\F_2^{a_1}(0, 0) = 0.38(5) \,,
\end{equation}
in perfect agreement with the $U(3)$ prediction~\eqref{eq:a1_TFF_Norm_U3}.

\begin{figure}[t]
	\centering
	\scalebox{0.85}{\input{plots/axialTFF}}
	\caption{Singly-virtual axial TFF $\F_2^{a_1}$ in the space-like region. The VMD parametrization of Ref.~\cite{Hoferichter:2023tgp} is shown without an asymptotic contribution that can be used to match the BL limit~\cite{Hoferichter:2024bae}.}
	\label{fig:axialTFF}
\end{figure}
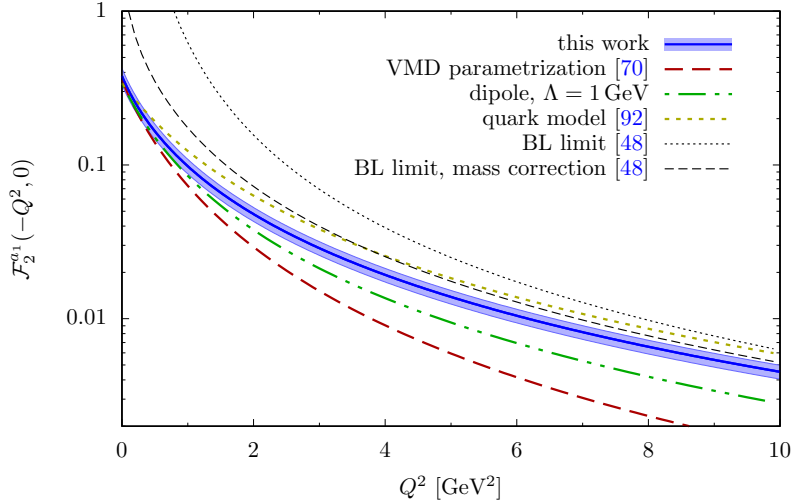

In Fig.~\ref{fig:axialTFF}, we compare our result for the space-like $a_1$ TFF $\F_2^{a_1}(q^2,0)$ to the VMD parametrization of Ref.~\cite{Hoferichter:2023tgp}, to the asymptotic limit~\eqref{eq:AxialTFFBLLimit} and its variant including higher mass corrections~\cite{Hoferichter:2020lap}, as well as to a dipole parametrization
\begin{equation}
	\F_2^{a_1}(q^2, 0) = \F_2^{a_1}(0, 0) \left( 1 - \frac{q^2}{\Lambda^2} \right)^{-2} \,.
\end{equation}
In the case of the $f_1(1285)$ and $f_1'(1420)$ resonances, the mass scale $\Lambda$ was found to be around $\SI{1}{\GeV}$~\cite{Achard:2001uu, L3:2007obw}. For $\Lambda = m_{a_1}$, the dipole form agrees with the singly-virtual limit of the quark-model of Ref.~\cite{Schuler:1997yw}. Although it has already the correct $\O(q^{-4})$ scaling, the VMD parametrization of Ref.~\cite{Hoferichter:2023tgp} is designed to extract information on all three form factors from the low-energy region and it still needs to be supplemented with an additional asymptotic contribution in order to match the coefficient~\eqref{eq:AxialTFFBLLimit} of the singly-virtual asymptotic limit~\cite{Hoferichter:2024bae}.

%% file: plots/delta11.tex
\begingroup
  \makeatletter
  \providecommand\color[2][]{%
    \GenericError{(gnuplot) \space\space\space\@spaces}{%
      Package color not loaded in conjunction with
      terminal option `colourtext'%
    }{See the gnuplot documentation for explanation.%
    }{Either use 'blacktext' in gnuplot or load the package
      color.sty in LaTeX.}%
    \renewcommand\color[2][]{}%
  }%
  \providecommand\includegraphics[2][]{%
    \GenericError{(gnuplot) \space\space\space\@spaces}{%
      Package graphicx or graphics not loaded%
    }{See the gnuplot documentation for explanation.%
    }{The gnuplot epslatex terminal needs graphicx.sty or graphics.sty.}%
    \renewcommand\includegraphics[2][]{}%
  }%
  \providecommand\rotatebox[2]{#2}%
  \@ifundefined{ifGPcolor}{%
    \newif\ifGPcolor
    \GPcolorfalse
  }{}%
  \@ifundefined{ifGPblacktext}{%
    \newif\ifGPblacktext
    \GPblacktexttrue
  }{}%
  \let\gplgaddtomacro\g@addto@macro
  \gdef\gplbacktext{}%
  \gdef\gplfronttext{}%
  \makeatother
  \ifGPblacktext
    \def\colorrgb#1{}%
    \def\colorgray#1{}%
  \else
    \ifGPcolor
      \def\colorrgb#1{\color[rgb]{#1}}%
      \def\colorgray#1{\color[gray]{#1}}%
      \expandafter\def\csname LTw\endcsname{\color{white}}%
      \expandafter\def\csname LTb\endcsname{\color{black}}%
      \expandafter\def\csname LTa\endcsname{\color{black}}%
      \expandafter\def\csname LT0\endcsname{\color[rgb]{1,0,0}}%
      \expandafter\def\csname LT1\endcsname{\color[rgb]{0,1,0}}%
      \expandafter\def\csname LT2\endcsname{\color[rgb]{0,0,1}}%
      \expandafter\def\csname LT3\endcsname{\color[rgb]{1,0,1}}%
      \expandafter\def\csname LT4\endcsname{\color[rgb]{0,1,1}}%
      \expandafter\def\csname LT5\endcsname{\color[rgb]{1,1,0}}%
      \expandafter\def\csname LT6\endcsname{\color[rgb]{0,0,0}}%
      \expandafter\def\csname LT7\endcsname{\color[rgb]{1,0.3,0}}%
      \expandafter\def\csname LT8\endcsname{\color[rgb]{0.5,0.5,0.5}}%
    \else
      \def\colorrgb#1{\color{black}}%
      \def\colorgray#1{\color[gray]{#1}}%
      \expandafter\def\csname LTw\endcsname{\color{white}}%
      \expandafter\def\csname LTb\endcsname{\color{black}}%
      \expandafter\def\csname LTa\endcsname{\color{black}}%
      \expandafter\def\csname LT0\endcsname{\color{black}}%
      \expandafter\def\csname LT1\endcsname{\color{black}}%
      \expandafter\def\csname LT2\endcsname{\color{black}}%
      \expandafter\def\csname LT3\endcsname{\color{black}}%
      \expandafter\def\csname LT4\endcsname{\color{black}}%
      \expandafter\def\csname LT5\endcsname{\color{black}}%
      \expandafter\def\csname LT6\endcsname{\color{black}}%
      \expandafter\def\csname LT7\endcsname{\color{black}}%
      \expandafter\def\csname LT8\endcsname{\color{black}}%
    \fi
  \fi
    \setlength{\unitlength}{0.0500bp}%
    \ifx\gptboxheight\undefined%
      \newlength{\gptboxheight}%
      \newlength{\gptboxwidth}%
      \newsavebox{\gptboxtext}%
    \fi%
    \setlength{\fboxrule}{0.5pt}%
    \setlength{\fboxsep}{1pt}%
    \definecolor{tbcol}{rgb}{1,1,1}%
\begin{picture}(4320.00,3024.00)%
    \gplgaddtomacro\gplbacktext{%
      \csname LTb\endcsname
      \put(198,736){\makebox(0,0)[r]{\strut{}$0$}}%
      \put(198,1444){\makebox(0,0)[r]{\strut{}$1$}}%
      \put(198,2151){\makebox(0,0)[r]{\strut{}$2$}}%
      \put(198,2859){\makebox(0,0)[r]{\strut{}$3$}}%
      \put(330,374){\makebox(0,0){\strut{}$0$}}%
      \put(1324,374){\makebox(0,0){\strut{}$0.5$}}%
      \put(2318,374){\makebox(0,0){\strut{}$1$}}%
      \put(3312,374){\makebox(0,0){\strut{}$1.5$}}%
      \put(4306,374){\makebox(0,0){\strut{}$2$}}%
    }%
    \gplgaddtomacro\gplfronttext{%
      \csname LTb\endcsname
      \put(-143,1797){\rotatebox{-270}{\makebox(0,0){\strut{}$\delta_1^1(s)$}}}%
      \put(2318,44){\makebox(0,0){\strut{}$s$ [GeV${}^2$]}}%
    }%
    \gplbacktext
    \put(0,0){\includegraphics[width={216.00bp},height={151.20bp}]{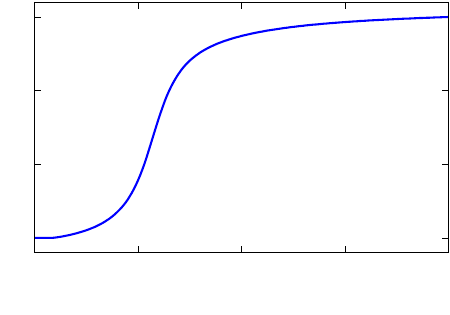}}%
    \gplfronttext
  \end{picture}%
\endgroup

%% file: plots/Omega11.tex
\begingroup
  \makeatletter
  \providecommand\color[2][]{%
    \GenericError{(gnuplot) \space\space\space\@spaces}{%
      Package color not loaded in conjunction with
      terminal option `colourtext'%
    }{See the gnuplot documentation for explanation.%
    }{Either use 'blacktext' in gnuplot or load the package
      color.sty in LaTeX.}%
    \renewcommand\color[2][]{}%
  }%
  \providecommand\includegraphics[2][]{%
    \GenericError{(gnuplot) \space\space\space\@spaces}{%
      Package graphicx or graphics not loaded%
    }{See the gnuplot documentation for explanation.%
    }{The gnuplot epslatex terminal needs graphicx.sty or graphics.sty.}%
    \renewcommand\includegraphics[2][]{}%
  }%
  \providecommand\rotatebox[2]{#2}%
  \@ifundefined{ifGPcolor}{%
    \newif\ifGPcolor
    \GPcolorfalse
  }{}%
  \@ifundefined{ifGPblacktext}{%
    \newif\ifGPblacktext
    \GPblacktexttrue
  }{}%
  \let\gplgaddtomacro\g@addto@macro
  \gdef\gplbacktext{}%
  \gdef\gplfronttext{}%
  \makeatother
  \ifGPblacktext
    \def\colorrgb#1{}%
    \def\colorgray#1{}%
  \else
    \ifGPcolor
      \def\colorrgb#1{\color[rgb]{#1}}%
      \def\colorgray#1{\color[gray]{#1}}%
      \expandafter\def\csname LTw\endcsname{\color{white}}%
      \expandafter\def\csname LTb\endcsname{\color{black}}%
      \expandafter\def\csname LTa\endcsname{\color{black}}%
      \expandafter\def\csname LT0\endcsname{\color[rgb]{1,0,0}}%
      \expandafter\def\csname LT1\endcsname{\color[rgb]{0,1,0}}%
      \expandafter\def\csname LT2\endcsname{\color[rgb]{0,0,1}}%
      \expandafter\def\csname LT3\endcsname{\color[rgb]{1,0,1}}%
      \expandafter\def\csname LT4\endcsname{\color[rgb]{0,1,1}}%
      \expandafter\def\csname LT5\endcsname{\color[rgb]{1,1,0}}%
      \expandafter\def\csname LT6\endcsname{\color[rgb]{0,0,0}}%
      \expandafter\def\csname LT7\endcsname{\color[rgb]{1,0.3,0}}%
      \expandafter\def\csname LT8\endcsname{\color[rgb]{0.5,0.5,0.5}}%
    \else
      \def\colorrgb#1{\color{black}}%
      \def\colorgray#1{\color[gray]{#1}}%
      \expandafter\def\csname LTw\endcsname{\color{white}}%
      \expandafter\def\csname LTb\endcsname{\color{black}}%
      \expandafter\def\csname LTa\endcsname{\color{black}}%
      \expandafter\def\csname LT0\endcsname{\color{black}}%
      \expandafter\def\csname LT1\endcsname{\color{black}}%
      \expandafter\def\csname LT2\endcsname{\color{black}}%
      \expandafter\def\csname LT3\endcsname{\color{black}}%
      \expandafter\def\csname LT4\endcsname{\color{black}}%
      \expandafter\def\csname LT5\endcsname{\color{black}}%
      \expandafter\def\csname LT6\endcsname{\color{black}}%
      \expandafter\def\csname LT7\endcsname{\color{black}}%
      \expandafter\def\csname LT8\endcsname{\color{black}}%
    \fi
  \fi
    \setlength{\unitlength}{0.0500bp}%
    \ifx\gptboxheight\undefined%
      \newlength{\gptboxheight}%
      \newlength{\gptboxwidth}%
      \newsavebox{\gptboxtext}%
    \fi%
    \setlength{\fboxrule}{0.5pt}%
    \setlength{\fboxsep}{1pt}%
    \definecolor{tbcol}{rgb}{1,1,1}%
\begin{picture}(4320.00,3024.00)%
    \gplgaddtomacro\gplbacktext{%
      \csname LTb\endcsname
      \put(198,877){\makebox(0,0)[r]{\strut{}$-2$}}%
      \put(198,1349){\makebox(0,0)[r]{\strut{}$0$}}%
      \put(198,1821){\makebox(0,0)[r]{\strut{}$2$}}%
      \put(198,2293){\makebox(0,0)[r]{\strut{}$4$}}%
      \put(198,2765){\makebox(0,0)[r]{\strut{}$6$}}%
      \put(330,374){\makebox(0,0){\strut{}$0$}}%
      \put(1324,374){\makebox(0,0){\strut{}$0.5$}}%
      \put(2318,374){\makebox(0,0){\strut{}$1$}}%
      \put(3312,374){\makebox(0,0){\strut{}$1.5$}}%
      \put(4306,374){\makebox(0,0){\strut{}$2$}}%
    }%
    \gplgaddtomacro\gplfronttext{%
      \csname LTb\endcsname
      \put(-143,1797){\rotatebox{-270}{\makebox(0,0){\strut{}$\Omega_1^1(s)$}}}%
      \put(2318,44){\makebox(0,0){\strut{}$s$ [GeV${}^2$]}}%
      \csname LTb\endcsname
      \put(3352,2628){\makebox(0,0)[r]{\strut{}$|\Omega_1^1(s)|$}}%
      \csname LTb\endcsname
      \put(3352,2364){\makebox(0,0)[r]{\strut{}$\Re\Omega_1^1(s)$}}%
      \csname LTb\endcsname
      \put(3352,2100){\makebox(0,0)[r]{\strut{}$\Im\Omega_1^1(s)$}}%
    }%
    \gplbacktext
    \put(0,0){\includegraphics[width={216.00bp},height={151.20bp}]{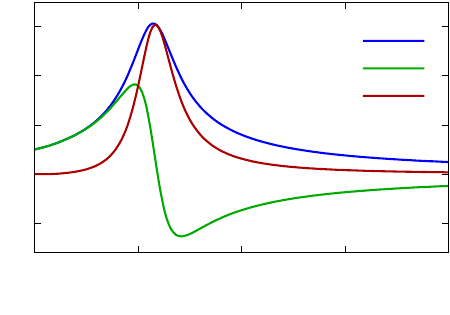}}%
    \gplfronttext
  \end{picture}%
\endgroup

%% file: plots/TFF.tex
\begingroup
  \makeatletter
  \providecommand\color[2][]{%
    \GenericError{(gnuplot) \space\space\space\@spaces}{%
      Package color not loaded in conjunction with
      terminal option `colourtext'%
    }{See the gnuplot documentation for explanation.%
    }{Either use 'blacktext' in gnuplot or load the package
      color.sty in LaTeX.}%
    \renewcommand\color[2][]{}%
  }%
  \providecommand\includegraphics[2][]{%
    \GenericError{(gnuplot) \space\space\space\@spaces}{%
      Package graphicx or graphics not loaded%
    }{See the gnuplot documentation for explanation.%
    }{The gnuplot epslatex terminal needs graphicx.sty or graphics.sty.}%
    \renewcommand\includegraphics[2][]{}%
  }%
  \providecommand\rotatebox[2]{#2}%
  \@ifundefined{ifGPcolor}{%
    \newif\ifGPcolor
    \GPcolorfalse
  }{}%
  \@ifundefined{ifGPblacktext}{%
    \newif\ifGPblacktext
    \GPblacktexttrue
  }{}%
  \let\gplgaddtomacro\g@addto@macro
  \gdef\gplbacktext{}%
  \gdef\gplfronttext{}%
  \makeatother
  \ifGPblacktext
    \def\colorrgb#1{}%
    \def\colorgray#1{}%
  \else
    \ifGPcolor
      \def\colorrgb#1{\color[rgb]{#1}}%
      \def\colorgray#1{\color[gray]{#1}}%
      \expandafter\def\csname LTw\endcsname{\color{white}}%
      \expandafter\def\csname LTb\endcsname{\color{black}}%
      \expandafter\def\csname LTa\endcsname{\color{black}}%
      \expandafter\def\csname LT0\endcsname{\color[rgb]{1,0,0}}%
      \expandafter\def\csname LT1\endcsname{\color[rgb]{0,1,0}}%
      \expandafter\def\csname LT2\endcsname{\color[rgb]{0,0,1}}%
      \expandafter\def\csname LT3\endcsname{\color[rgb]{1,0,1}}%
      \expandafter\def\csname LT4\endcsname{\color[rgb]{0,1,1}}%
      \expandafter\def\csname LT5\endcsname{\color[rgb]{1,1,0}}%
      \expandafter\def\csname LT6\endcsname{\color[rgb]{0,0,0}}%
      \expandafter\def\csname LT7\endcsname{\color[rgb]{1,0.3,0}}%
      \expandafter\def\csname LT8\endcsname{\color[rgb]{0.5,0.5,0.5}}%
    \else
      \def\colorrgb#1{\color{black}}%
      \def\colorgray#1{\color[gray]{#1}}%
      \expandafter\def\csname LTw\endcsname{\color{white}}%
      \expandafter\def\csname LTb\endcsname{\color{black}}%
      \expandafter\def\csname LTa\endcsname{\color{black}}%
      \expandafter\def\csname LT0\endcsname{\color{black}}%
      \expandafter\def\csname LT1\endcsname{\color{black}}%
      \expandafter\def\csname LT2\endcsname{\color{black}}%
      \expandafter\def\csname LT3\endcsname{\color{black}}%
      \expandafter\def\csname LT4\endcsname{\color{black}}%
      \expandafter\def\csname LT5\endcsname{\color{black}}%
      \expandafter\def\csname LT6\endcsname{\color{black}}%
      \expandafter\def\csname LT7\endcsname{\color{black}}%
      \expandafter\def\csname LT8\endcsname{\color{black}}%
    \fi
  \fi
    \setlength{\unitlength}{0.0500bp}%
    \ifx\gptboxheight\undefined%
      \newlength{\gptboxheight}%
      \newlength{\gptboxwidth}%
      \newsavebox{\gptboxtext}%
    \fi%
    \setlength{\fboxrule}{0.5pt}%
    \setlength{\fboxsep}{1pt}%
    \definecolor{tbcol}{rgb}{1,1,1}%
\begin{picture}(6120.00,4284.00)%
    \gplgaddtomacro\gplbacktext{%
      \csname LTb\endcsname
      \put(198,594){\makebox(0,0)[r]{\strut{}$0$}}%
      \put(198,1511){\makebox(0,0)[r]{\strut{}$0.05$}}%
      \put(198,2428){\makebox(0,0)[r]{\strut{}$0.1$}}%
      \put(198,3344){\makebox(0,0)[r]{\strut{}$0.15$}}%
      \put(198,4261){\makebox(0,0)[r]{\strut{}$0.2$}}%
      \put(330,374){\makebox(0,0){\strut{}$0$}}%
      \put(1293,374){\makebox(0,0){\strut{}$0.5$}}%
      \put(2255,374){\makebox(0,0){\strut{}$1$}}%
      \put(3218,374){\makebox(0,0){\strut{}$1.5$}}%
      \put(4180,374){\makebox(0,0){\strut{}$2$}}%
      \put(5143,374){\makebox(0,0){\strut{}$2.5$}}%
      \put(6105,374){\makebox(0,0){\strut{}$3$}}%
    }%
    \gplgaddtomacro\gplfronttext{%
      \csname LTb\endcsname
      \put(-539,2427){\rotatebox{-270}{\makebox(0,0){\strut{}$Q^2 F_{\pi^0\gamma^*\gamma^*}(-Q^2,0)$ [GeV]}}}%
      \put(3217,44){\makebox(0,0){\strut{}$Q^2$ [GeV${}^2$]}}%
      \csname LTb\endcsname
      \put(1626,3821){\makebox(0,0)[r]{\strut{} }}%
      \csname LTb\endcsname
      \put(1626,3601){\makebox(0,0)[r]{\strut{} }}%
    }%
    \gplgaddtomacro\gplbacktext{%
    }%
    \gplgaddtomacro\gplfronttext{%
      \csname LTb\endcsname
      \put(-11,2427){\rotatebox{-270}{\makebox(0,0){\strut{} }}}%
      \put(3217,374){\makebox(0,0){\strut{} }}%
      \csname LTb\endcsname
      \put(1626,3821){\makebox(0,0)[r]{\strut{}Ref.~\cite{Hoferichter:2018kwz}}}%
      \csname LTb\endcsname
      \put(1626,3601){\makebox(0,0)[r]{\strut{}this work}}%
      \csname LTb\endcsname
      \put(1626,3381){\makebox(0,0)[r]{\strut{}isoscalar}}%
      \csname LTb\endcsname
      \put(1626,3161){\makebox(0,0)[r]{\strut{}isovector}}%
      \csname LTb\endcsname
      \put(1626,2941){\makebox(0,0)[r]{\strut{}effective pole}}%
      \csname LTb\endcsname
      \put(1626,2721){\makebox(0,0)[r]{\strut{}BL limit}}%
    }%
    \gplbacktext
    \put(0,0){\includegraphics[width={306.00bp},height={214.20bp}]{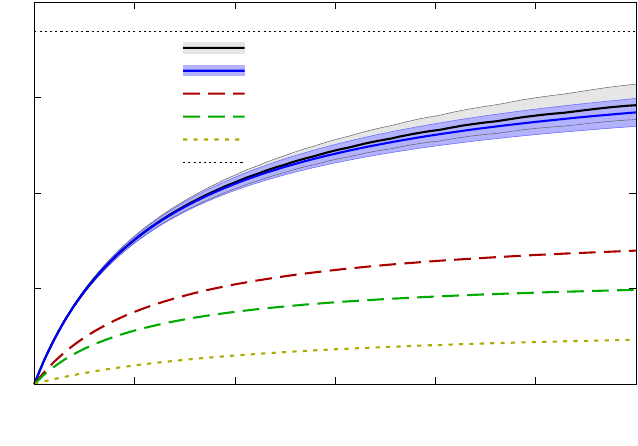}}%
    \gplfronttext
  \end{picture}%
\endgroup

%% file: plots/FVAellipse.tex
\begingroup
  \makeatletter
  \providecommand\color[2][]{%
    \GenericError{(gnuplot) \space\space\space\@spaces}{%
      Package color not loaded in conjunction with
      terminal option `colourtext'%
    }{See the gnuplot documentation for explanation.%
    }{Either use 'blacktext' in gnuplot or load the package
      color.sty in LaTeX.}%
    \renewcommand\color[2][]{}%
  }%
  \providecommand\includegraphics[2][]{%
    \GenericError{(gnuplot) \space\space\space\@spaces}{%
      Package graphicx or graphics not loaded%
    }{See the gnuplot documentation for explanation.%
    }{The gnuplot epslatex terminal needs graphicx.sty or graphics.sty.}%
    \renewcommand\includegraphics[2][]{}%
  }%
  \providecommand\rotatebox[2]{#2}%
  \@ifundefined{ifGPcolor}{%
    \newif\ifGPcolor
    \GPcolorfalse
  }{}%
  \@ifundefined{ifGPblacktext}{%
    \newif\ifGPblacktext
    \GPblacktexttrue
  }{}%
  \let\gplgaddtomacro\g@addto@macro
  \gdef\gplbacktext{}%
  \gdef\gplfronttext{}%
  \makeatother
  \ifGPblacktext
    \def\colorrgb#1{}%
    \def\colorgray#1{}%
  \else
    \ifGPcolor
      \def\colorrgb#1{\color[rgb]{#1}}%
      \def\colorgray#1{\color[gray]{#1}}%
      \expandafter\def\csname LTw\endcsname{\color{white}}%
      \expandafter\def\csname LTb\endcsname{\color{black}}%
      \expandafter\def\csname LTa\endcsname{\color{black}}%
      \expandafter\def\csname LT0\endcsname{\color[rgb]{1,0,0}}%
      \expandafter\def\csname LT1\endcsname{\color[rgb]{0,1,0}}%
      \expandafter\def\csname LT2\endcsname{\color[rgb]{0,0,1}}%
      \expandafter\def\csname LT3\endcsname{\color[rgb]{1,0,1}}%
      \expandafter\def\csname LT4\endcsname{\color[rgb]{0,1,1}}%
      \expandafter\def\csname LT5\endcsname{\color[rgb]{1,1,0}}%
      \expandafter\def\csname LT6\endcsname{\color[rgb]{0,0,0}}%
      \expandafter\def\csname LT7\endcsname{\color[rgb]{1,0.3,0}}%
      \expandafter\def\csname LT8\endcsname{\color[rgb]{0.5,0.5,0.5}}%
    \else
      \def\colorrgb#1{\color{black}}%
      \def\colorgray#1{\color[gray]{#1}}%
      \expandafter\def\csname LTw\endcsname{\color{white}}%
      \expandafter\def\csname LTb\endcsname{\color{black}}%
      \expandafter\def\csname LTa\endcsname{\color{black}}%
      \expandafter\def\csname LT0\endcsname{\color{black}}%
      \expandafter\def\csname LT1\endcsname{\color{black}}%
      \expandafter\def\csname LT2\endcsname{\color{black}}%
      \expandafter\def\csname LT3\endcsname{\color{black}}%
      \expandafter\def\csname LT4\endcsname{\color{black}}%
      \expandafter\def\csname LT5\endcsname{\color{black}}%
      \expandafter\def\csname LT6\endcsname{\color{black}}%
      \expandafter\def\csname LT7\endcsname{\color{black}}%
      \expandafter\def\csname LT8\endcsname{\color{black}}%
    \fi
  \fi
    \setlength{\unitlength}{0.0500bp}%
    \ifx\gptboxheight\undefined%
      \newlength{\gptboxheight}%
      \newlength{\gptboxwidth}%
      \newsavebox{\gptboxtext}%
    \fi%
    \setlength{\fboxrule}{0.5pt}%
    \setlength{\fboxsep}{1pt}%
    \definecolor{tbcol}{rgb}{1,1,1}%
\begin{picture}(7200.00,5040.00)%
    \gplgaddtomacro\gplbacktext{%
      \csname LTb\endcsname
      \put(198,594){\makebox(0,0)[r]{\strut{}$-8$}}%
      \put(198,1085){\makebox(0,0)[r]{\strut{}$-6$}}%
      \put(198,1577){\makebox(0,0)[r]{\strut{}$-4$}}%
      \put(198,2068){\makebox(0,0)[r]{\strut{}$-2$}}%
      \put(198,2560){\makebox(0,0)[r]{\strut{}$0$}}%
      \put(198,3051){\makebox(0,0)[r]{\strut{}$2$}}%
      \put(198,3543){\makebox(0,0)[r]{\strut{}$4$}}%
      \put(198,4034){\makebox(0,0)[r]{\strut{}$6$}}%
      \put(198,4526){\makebox(0,0)[r]{\strut{}$8$}}%
      \put(198,5017){\makebox(0,0)[r]{\strut{}$10$}}%
      \put(330,374){\makebox(0,0){\strut{}$-8$}}%
      \put(1187,374){\makebox(0,0){\strut{}$-6$}}%
      \put(2044,374){\makebox(0,0){\strut{}$-4$}}%
      \put(2901,374){\makebox(0,0){\strut{}$-2$}}%
      \put(3758,374){\makebox(0,0){\strut{}$0$}}%
      \put(4614,374){\makebox(0,0){\strut{}$2$}}%
      \put(5471,374){\makebox(0,0){\strut{}$4$}}%
      \put(6328,374){\makebox(0,0){\strut{}$6$}}%
      \put(7185,374){\makebox(0,0){\strut{}$8$}}%
    }%
    \gplgaddtomacro\gplfronttext{%
      \csname LTb\endcsname
      \put(-275,2805){\rotatebox{-270}{\makebox(0,0){\strut{}$F_{\omega a_1}^2 + F_{\omega a_1}^3$ [GeV${}^{-2}$]}}}%
      \put(3757,44){\makebox(0,0){\strut{}$F_{\omega a_1}^1$ [GeV${}^{-2}$]}}%
      \csname LTb\endcsname
      \put(5987,4686){\makebox(0,0)[r]{\strut{} }}%
    }%
    \gplgaddtomacro\gplbacktext{%
    }%
    \gplgaddtomacro\gplfronttext{%
      \csname LTb\endcsname
      \put(5987,4686){\makebox(0,0)[r]{\strut{}$U(3)$ estimate for $\Gamma_{a_1\to\omega\gamma}$}}%
    }%
    \gplgaddtomacro\gplbacktext{%
    }%
    \gplgaddtomacro\gplfronttext{%
      \csname LTb\endcsname
      \put(2731,4653){\makebox(0,0)[r]{\strut{} }}%
      \csname LTb\endcsname
      \put(2731,4367){\makebox(0,0)[r]{\strut{} }}%
    }%
    \gplgaddtomacro\gplbacktext{%
    }%
    \gplgaddtomacro\gplfronttext{%
      \csname LTb\endcsname
      \put(2731,4653){\makebox(0,0)[r]{\strut{}vanishing $\tilde\V_1^{(3)}(m_\omega^2)$}}%
      \csname LTb\endcsname
      \put(2731,4367){\makebox(0,0)[r]{\strut{}$\tilde\V_1^{(3)}(m_\omega^2)$ fixed from $\tilde\W_1^{(3)}$}}%
    }%
    \gplbacktext
    \put(0,0){\includegraphics[width={360.00bp},height={252.00bp}]{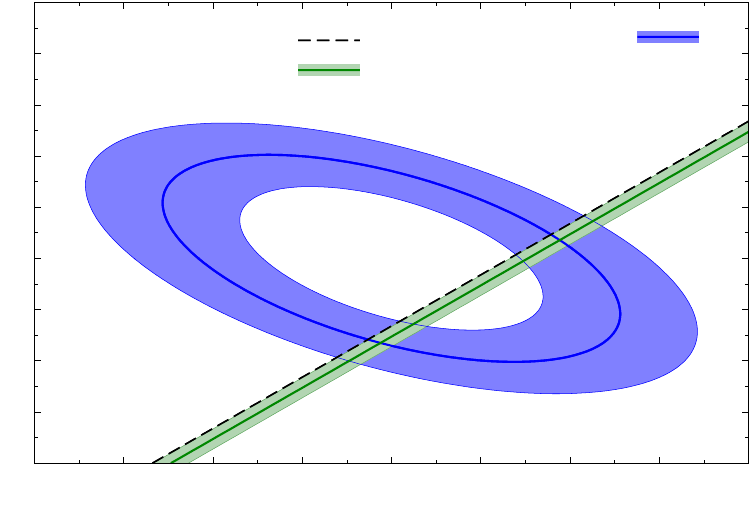}}%
    \gplfronttext
  \end{picture}%
\endgroup

%% file: plots/axialTFF.tex
\begingroup
  \makeatletter
  \providecommand\color[2][]{%
    \GenericError{(gnuplot) \space\space\space\@spaces}{%
      Package color not loaded in conjunction with
      terminal option `colourtext'%
    }{See the gnuplot documentation for explanation.%
    }{Either use 'blacktext' in gnuplot or load the package
      color.sty in LaTeX.}%
    \renewcommand\color[2][]{}%
  }%
  \providecommand\includegraphics[2][]{%
    \GenericError{(gnuplot) \space\space\space\@spaces}{%
      Package graphicx or graphics not loaded%
    }{See the gnuplot documentation for explanation.%
    }{The gnuplot epslatex terminal needs graphicx.sty or graphics.sty.}%
    \renewcommand\includegraphics[2][]{}%
  }%
  \providecommand\rotatebox[2]{#2}%
  \@ifundefined{ifGPcolor}{%
    \newif\ifGPcolor
    \GPcolorfalse
  }{}%
  \@ifundefined{ifGPblacktext}{%
    \newif\ifGPblacktext
    \GPblacktexttrue
  }{}%
  \let\gplgaddtomacro\g@addto@macro
  \gdef\gplbacktext{}%
  \gdef\gplfronttext{}%
  \makeatother
  \ifGPblacktext
    \def\colorrgb#1{}%
    \def\colorgray#1{}%
  \else
    \ifGPcolor
      \def\colorrgb#1{\color[rgb]{#1}}%
      \def\colorgray#1{\color[gray]{#1}}%
      \expandafter\def\csname LTw\endcsname{\color{white}}%
      \expandafter\def\csname LTb\endcsname{\color{black}}%
      \expandafter\def\csname LTa\endcsname{\color{black}}%
      \expandafter\def\csname LT0\endcsname{\color[rgb]{1,0,0}}%
      \expandafter\def\csname LT1\endcsname{\color[rgb]{0,1,0}}%
      \expandafter\def\csname LT2\endcsname{\color[rgb]{0,0,1}}%
      \expandafter\def\csname LT3\endcsname{\color[rgb]{1,0,1}}%
      \expandafter\def\csname LT4\endcsname{\color[rgb]{0,1,1}}%
      \expandafter\def\csname LT5\endcsname{\color[rgb]{1,1,0}}%
      \expandafter\def\csname LT6\endcsname{\color[rgb]{0,0,0}}%
      \expandafter\def\csname LT7\endcsname{\color[rgb]{1,0.3,0}}%
      \expandafter\def\csname LT8\endcsname{\color[rgb]{0.5,0.5,0.5}}%
    \else
      \def\colorrgb#1{\color{black}}%
      \def\colorgray#1{\color[gray]{#1}}%
      \expandafter\def\csname LTw\endcsname{\color{white}}%
      \expandafter\def\csname LTb\endcsname{\color{black}}%
      \expandafter\def\csname LTa\endcsname{\color{black}}%
      \expandafter\def\csname LT0\endcsname{\color{black}}%
      \expandafter\def\csname LT1\endcsname{\color{black}}%
      \expandafter\def\csname LT2\endcsname{\color{black}}%
      \expandafter\def\csname LT3\endcsname{\color{black}}%
      \expandafter\def\csname LT4\endcsname{\color{black}}%
      \expandafter\def\csname LT5\endcsname{\color{black}}%
      \expandafter\def\csname LT6\endcsname{\color{black}}%
      \expandafter\def\csname LT7\endcsname{\color{black}}%
      \expandafter\def\csname LT8\endcsname{\color{black}}%
    \fi
  \fi
    \setlength{\unitlength}{0.0500bp}%
    \ifx\gptboxheight\undefined%
      \newlength{\gptboxheight}%
      \newlength{\gptboxwidth}%
      \newsavebox{\gptboxtext}%
    \fi%
    \setlength{\fboxrule}{0.5pt}%
    \setlength{\fboxsep}{1pt}%
    \definecolor{tbcol}{rgb}{1,1,1}%
\begin{picture}(6120.00,4284.00)%
    \gplgaddtomacro\gplbacktext{%
      \csname LTb\endcsname
      \put(198,1544){\makebox(0,0)[r]{\strut{}$0.01$}}%
      \put(198,2902){\makebox(0,0)[r]{\strut{}$0.1$}}%
      \put(198,4261){\makebox(0,0)[r]{\strut{}$1$}}%
      \put(330,374){\makebox(0,0){\strut{}$0$}}%
      \put(1485,374){\makebox(0,0){\strut{}$2$}}%
      \put(2640,374){\makebox(0,0){\strut{}$4$}}%
      \put(3795,374){\makebox(0,0){\strut{}$6$}}%
      \put(4950,374){\makebox(0,0){\strut{}$8$}}%
      \put(6105,374){\makebox(0,0){\strut{}$10$}}%
    }%
    \gplgaddtomacro\gplfronttext{%
      \csname LTb\endcsname
      \put(-539,2427){\rotatebox{-270}{\makebox(0,0){\strut{}$\F_2^{a_1}(-Q^2,0)$}}}%
      \put(3217,44){\makebox(0,0){\strut{}$Q^2$ [GeV${}^2$]}}%
      \csname LTb\endcsname
      \put(4961,3968){\makebox(0,0)[r]{\strut{} }}%
    }%
    \gplgaddtomacro\gplbacktext{%
    }%
    \gplgaddtomacro\gplfronttext{%
      \csname LTb\endcsname
      \put(-11,2427){\rotatebox{-270}{\makebox(0,0){\strut{} }}}%
      \put(3217,374){\makebox(0,0){\strut{} }}%
      \csname LTb\endcsname
      \put(4961,3968){\makebox(0,0)[r]{\strut{}this work}}%
      \csname LTb\endcsname
      \put(4961,3748){\makebox(0,0)[r]{\strut{}VMD parametrization~\cite{Hoferichter:2023tgp}}}%
      \csname LTb\endcsname
      \put(4961,3528){\makebox(0,0)[r]{\strut{}dipole, $\Lambda = 1\GeV$}}%
      \csname LTb\endcsname
      \put(4961,3308){\makebox(0,0)[r]{\strut{}quark model~\cite{Schuler:1997yw}}}%
      \csname LTb\endcsname
      \put(4961,3088){\makebox(0,0)[r]{\strut{}BL limit~\cite{Hoferichter:2020lap}}}%
      \csname LTb\endcsname
      \put(4961,2868){\makebox(0,0)[r]{\strut{}BL limit, mass correction~\cite{Hoferichter:2020lap}}}%
    }%
    \gplbacktext
    \put(0,0){\includegraphics[width={306.00bp},height={214.20bp}]{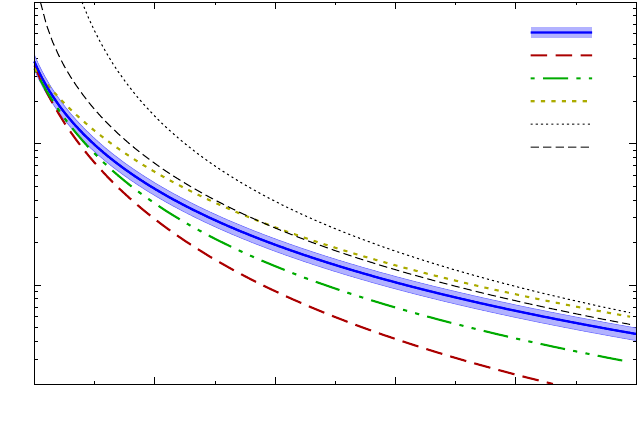}}%
    \gplfronttext
  \end{picture}%
\endgroup

%% file: sections/Results.tex

\section{\boldmath Results and first-family VVA contribution to $a_\mu$}
\label{sec:Results}

In the following, we compare the results of the two dispersive approaches, in particular the difference related to truncation effects as well as the low-energy and asymptotic constraints on VVA and on the sub-amplitudes. We also compare the dispersive results to the hadronic models discussed in Sect.~\ref{sec:VVAModels} and provide results for $a_\mu^\text{VVA}$.

\subsection[Dispersive results in $g-2$ kinematics]{\boldmath Dispersive results in $g-2$ kinematics}

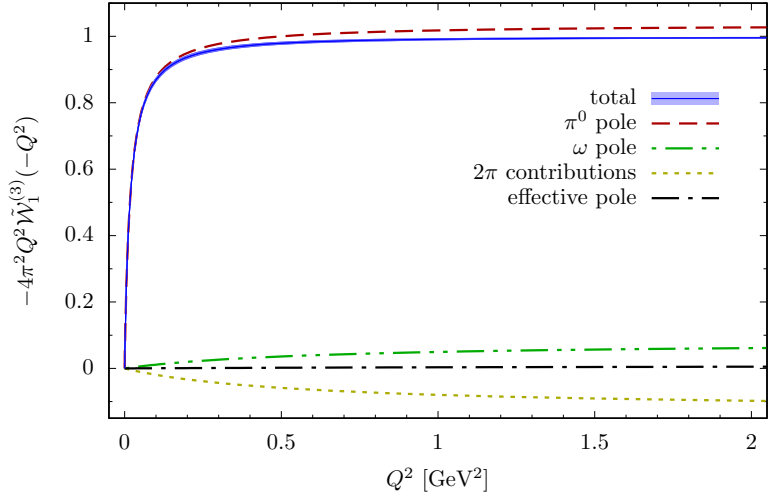
\begin{figure}
	\centering
	\scalebox{0.85}{\input{plots/W1gm2}}
	\caption{Individual contributions to $\tilde{\W}_1^{(3)}$ in the dispersive framework in $g-2$ kinematics. The blue band shows the uncertainty of the total result.}
	\label{Fig:VVA_ContW1}
\end{figure}

In Fig.~\ref{Fig:VVA_ContW1}, we show our central result for the longitudinal VVA function $\tilde\W_1^{(3)}$, obtained in the dispersive representation in $g-2$ kinematics. As discussed in Sect.~\ref{sec:VVA_W1Numerics}, the representation automatically fulfills the low-energy \chpt{} constraint and we implement the asymptotic behavior up to $\O(q^{-4})$ by including one effective pole. In Fig.~\ref{Fig:VVA_ContW1}, we split the contributions to $\tilde\W_1^{(3)}$ according to the different rows in Tab.~\ref{tab:Reshuffling}. The pure pion pole dominates the whole function. We observe a partial cancellation between the small contributions of the isoscalar vector-meson $\omega$ and the isovector two-pion state. The effective-pole contribution that is required to match the asymptotic behavior is tiny. As shown in the plot, the resulting uncertainties are very small.

\begin{figure}
	\centering
	\scalebox{0.85}{\input{plots/W2gm2}}
	\caption{Individual contributions to $\tilde{\W}_2^{(3)}$ in the dispersive framework in $g-2$ kinematics. The blue band shows the uncertainty of the total result.}
	\label{Fig:VVA_ContW2}
\end{figure}
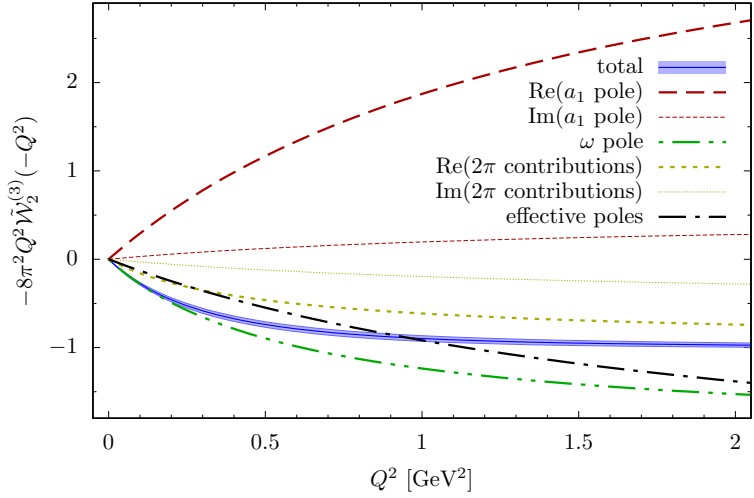

\begin{figure}
	\centering
	\scalebox{0.85}{\input{plots/W2gm2Constr}}
	\caption{Effect of the low- and high-energy constraints on the uncertainty on $\tilde{\W}_2^{(3)}$.}
	\label{fig:VVAW2Constr}
\end{figure}
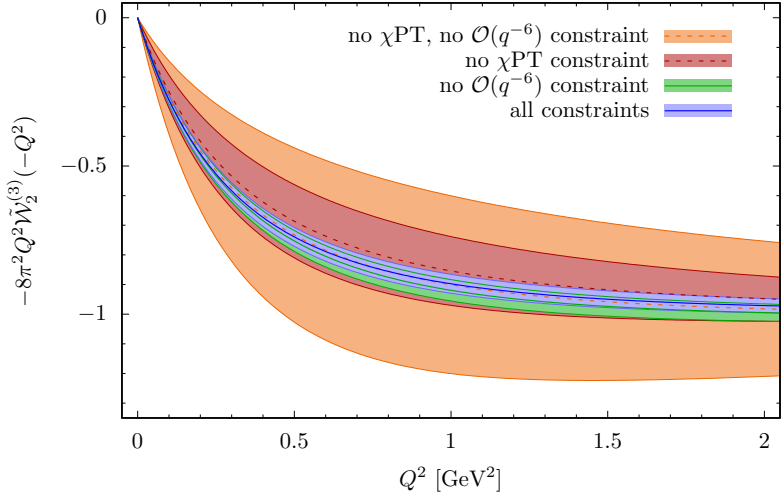

In Fig.~\ref{Fig:VVA_ContW2}, we show our central result for the transverse VVA function $\tilde\W_2^{(3)}$ in the dispersive representation in $g-2$ kinematics, as well as the different contributions according to the rows in Tab.~\ref{tab:Reshuffling}. We impose both the low-energy \chpt{} constraint as well as the asymptotic behavior up to $\O(q^{-6})$, which requires effective-pole contributions to the axial-vector TFF and two effective poles in $\tilde\W_2^{(3)}$ itself. The dispersive approach leads to imaginary parts of the axial-vector and two-pion contributions, which cancel each other by construction.

As discussed previously, in the case of $\tilde\W_2^{(3)}$ the fulfillment of the asymptotic constraints requires significant cancellations between the different hadronic contributions. While the $a_1$ contribution is the largest in size, it gets overcompensated by the contributions of the $\omega$ meson, the two-pion state, as well as the effective poles. Despite these large cancellations, the combination of the dispersive representation with low- and high-energy constraints results in a small overall uncertainty. It is instructive to investigate the effect of the different constraints on the uncertainty in the final result. In Fig.~\ref{fig:VVAW2Constr} we show our solution for different matching strategies: the orange band shows the result that we obtain if we do not impose the \chpt{} constraint and implement the asymptotic behavior only up to $\O(q^{-4})$ by including one effective pole in $\tilde\W_2^{(3)}$. The red band is obtained if we implement in addition the $\O(q^{-6})$ behavior by including a second effective pole, whereas the green band instead implements the \chpt{} constraint. The blue band is our central result and implements both the \chpt{} and the $\O(q^{-6})$ behavior by including an effective contribution to the $a_1$ TFF. We find that for $\tilde\W_2^{(3)}$, the dispersive representation alone comes with large uncertainties due to the intricate cancellation effects between different hadronic contributions, which however can be well controlled by making use of low- and high-energy constraints.

In combination with the electron contribution, we obtain the first-family longitudinal and transverse VVA contributions to $a_\mu$
\begin{equation}
	a_{\mu}^{\text{VVA}, L}[u, d, e] = -0.892(10) \times 10^{-11} \, , \qquad a_{\mu}^{\text{VVA}, T}[u, d, e] = -1.192(29) \times 10^{-11} \, ,
\end{equation}
leading to a total first-family VVA contribution of
\begin{equation}
	\label{eq:VVA_res_gm2}
	a_{\mu}^\text{VVA}[u, d, e] = -2.08(3) \times 10^{-11} \,.
\end{equation}
Without implementing the \chpt{} and $\O(q^{-6})$ constraints in the transverse function, we would instead obtain a much larger uncertainty of $0.3 \times 10^{-11}$. Therefore, it is the combination of the dispersive analysis with low- and high-energy constraints that allows us to obtain the result~\eqref{eq:VVA_res_gm2} with uncertainties improved by roughly an order of magnitude compared to the model of Ref.~\cite{Czarnecki:2002nt}.

\subsection{Dispersive results for fixed photon virtualities}

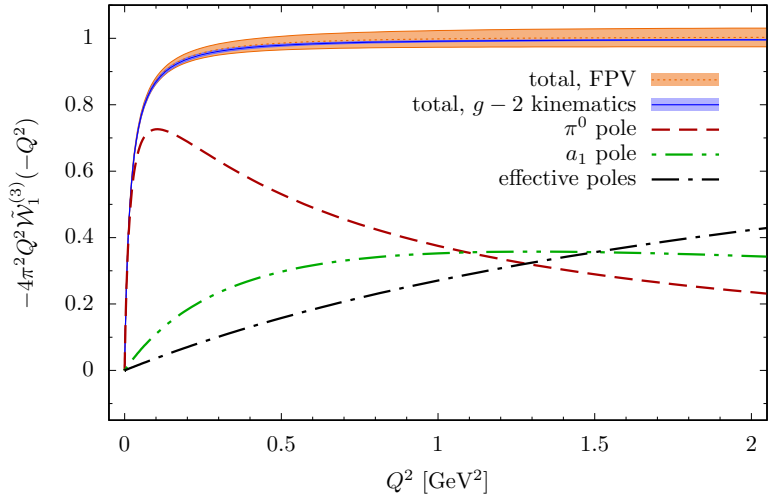
\begin{figure}
	\centering
	\scalebox{0.85}{\input{plots/W1FPV}}
	\caption{Individual contributions to $\tilde{\W}_1^{(3)}$ in the dispersive framework for fixed photon virtualities. The orange band shows the uncertainty of the total result, the blue band again shows the result in $g-2$ kinematics.}
	\label{fig:W1FPV}
\end{figure}
\begin{figure}
	\centering
	\scalebox{0.85}{\input{plots/W2FPV}}
	\caption{Individual contributions to $\tilde{\W}_2^{(3)}$ in the dispersive framework for fixed photon virtualities. The orange band shows the uncertainty of the total result, the blue band again shows the result in $g-2$ kinematics.}
	\label{fig:W2FPV}
\end{figure}
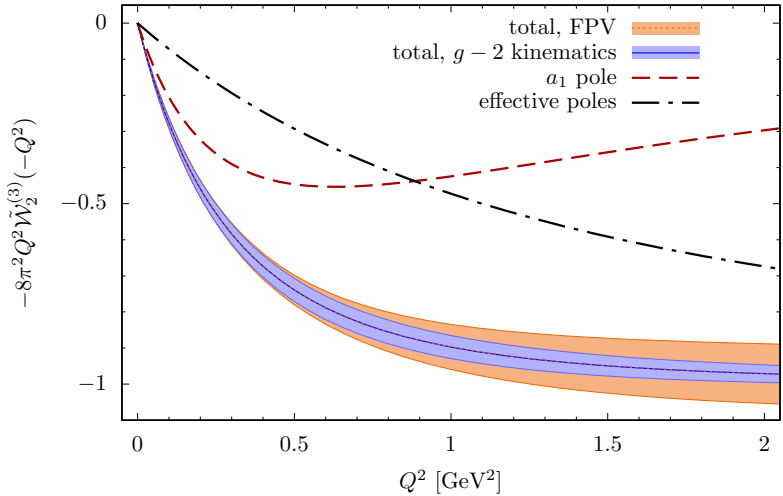

We next consider the dispersion relations for fixed photon virtualities (FPV). In Fig.~\ref{fig:W1FPV}, we show the result for the longitudinal function $\tilde\W_1^{(3)}$. We now split the total into individual contributions according to the columns of Tab.~\ref{tab:Reshuffling}, i.e., the pion pole, axial pole, and the asymptotic matching, which requires three effective poles in order to match both the \chpt{} constraint and the $\O(q^{-4})$ behavior. Without imposing the \chpt{} constraint, the normalization comes out smaller in magnitude, but still compatible with Eq.~\eqref{eq:NormChPT}. We observe a few important differences to the dispersion relations in $g-2$ kinematics. First, the pion pole alone is no longer a good approximation of the full longitudinal VVA function, since the $q^2$-dependence TFF leads to a $\O(q^{-4})$ behavior. In addition to pion and axial poles, we require a contribution from effective poles that is substantial already at small energies well below $1\GeV$. Consequently, the variation of the effective-pole parameters leads to a much larger overall uncertainty on the longitudinal VVA function (orange band) than the uncertainty obtained with the dispersive result in $g-2$ kinematics (blue band in Fig.~\ref{fig:W1FPV}).

The result for the transverse VVA function is shown in Fig.~\ref{fig:W2FPV}. As before, we implement both the \chpt{} constraint on the normalization and the asymptotic constraints up to $\O(q^{-6})$. The central result coincides with the dispersive result in $g-2$ kinematics, but we find somewhat larger uncertainties than before. This can be traced back to the uncorrelated variation of the effective-pole parameters in the $a_1$ TFF and $\tilde\W_2^{(3)}$ itself, which are used to implement the asymptotic behavior in both of these functions. 

For the longitudinal and transverse VVA contributions to $a_\mu$, we obtain
\begin{equation}
	a_{\mu,\text{FPV}}^{\text{VVA}, L}[u, d, e] = -0.877(55) \times 10^{-11} \,, \qquad a_{\mu,\text{FPV}}^{\text{VVA}, T}[u, d, e] = -1.192(64) \times 10^{-11} \, ,
\end{equation}
resulting in a total VVA contribution of
\begin{equation}
	\label{eq:VVA_res_FPV}
	a_{\mu,\text{FPV}}^\text{VVA}[u, d, e] = -2.07(11) \times 10^{-11} \,.
\end{equation}

\subsection{Comparison to models}
\label{sec:VVAComp}

In Figs.~\ref{Fig:VVA_ComparisonW1} and \ref{Fig:VVA_ComparisonW2}, we compare the two dispersive representations to the models discussed in Sect.~\ref{sec:VVAModels} for the two scalar functions.\footnote{We thank the authors of Ref.~\cite{Masjuan:2020jsf} for providing us with their results.} For $\tilde{\W}_1$, the hQCD model~\cite{Leutgeb:2019gbz, Cappiello:2019hwh, Hirn:2005nr} is very close to the dispersive representation in $g-2$ kinematics over the whole plotted $Q^2$ range. The model~\eqref{eq:VVA_CMVModel} of Ref.~\cite{Czarnecki:2002nt} is also close, but lies outside our uncertainty band. This can be understood as a consequence of the model consisting only of a pion pole with the residue being adjusted to the SDC. With the leading-order expression for the pion TFF normalization, the model agrees with the dispersively defined pion pole up to a shift of the argument from $m_\pi^2$ to $0$. This shift slightly reduces the residue, which influences the scalar function at small $Q^2$. The pion-pole residue in the holographic model is closer to our value, which explains why this model stays closer to our dispersive result. The Regge model of Refs.~\cite{Colangelo:2019lpu, Colangelo:2019uex}, which consists of a tower of excited pseudoscalars, provides a decent approximation of $\tilde\W_1$ only at low energies, where the pion pole dominates, as well as at very high energies, where the behavior is dictated by the SDC, but in the intermediate-energy region one finds large deviations from the dispersive result and the other models, for the reasons explained in Ref.~\cite{Colangelo:2021nkr}. The differences at very low energies in the longitudinal function are illustrated best by comparing the normalization obtained in the different models with the dispersive result~\eqref{eq:W1NormDispersive}, which agrees with the \chpt{} constraint~\eqref{eq:NormChPT}:
\begin{align}
	\label{eq:VVA_Wt1Normalization}
	\tilde{\W}_{1,\text{CMV}}^{(3)}(0) &= \SI{-1.390}{\GeV^{-2}} \, , \qquad \tilde{\W}_{1,\text{hQCD}}^{(3)}(0) = \SI{-1.418}{\GeV^{-2}} \, , \nn
	\tilde{\W}_{1,\text{PS-Regge}}^{(3)}(0) &= \SI{-1.395}{\GeV^{-2}} \,.
\end{align}
The normalization in the holographic model is closer to the \chpt{} constraint, but still outside the uncertainty. In contrast, the Regge-like model of Ref.~\cite{Masjuan:2020jsf} fulfills the \chpt{} constraint.

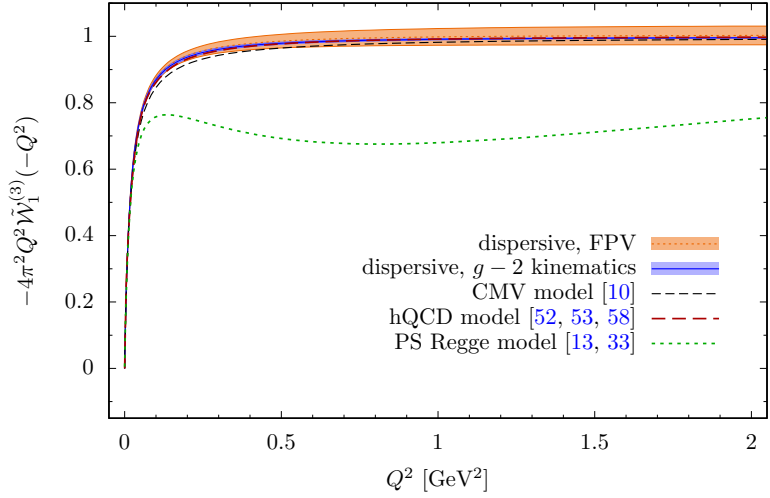
\begin{figure}
	\centering
	\scalebox{0.85}{\input{plots/W1comparisonNew}}
	\caption{Comparison of the two dispersive approaches to VVA with the models~\cite{Colangelo:2019lpu, Colangelo:2019uex, Czarnecki:2002nt, Leutgeb:2019gbz, Cappiello:2019hwh, Hirn:2005nr, Masjuan:2020jsf} described in Sect.~\ref{sec:VVAModels} for $\tilde{\W}_1$.}
	\label{Fig:VVA_ComparisonW1}
\end{figure}
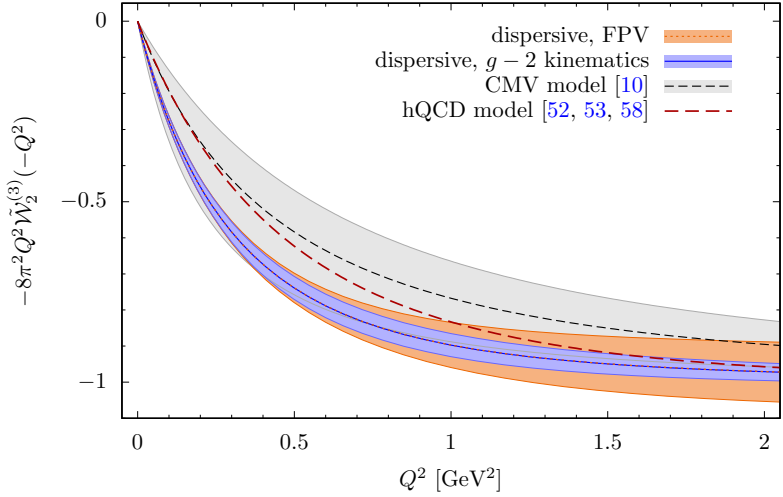
\begin{figure}
	\centering
	\scalebox{0.85}{\input{plots/W2comparisonNew}}
	\caption{Comparison of the two dispersive approaches to VVA with the models~\cite{Czarnecki:2002nt, Leutgeb:2019gbz, Cappiello:2019hwh, Hirn:2005nr, Masjuan:2020jsf} described in Sect.~\ref{sec:VVAModels} for $\tilde{\W}_2$. The gray band corresponds to the model uncertainty according to the prescription given in Ref.~\cite{Czarnecki:2002nt}.}
	\label{Fig:VVA_ComparisonW2}
\end{figure}

For the transverse function, the CMV model is compatible with our dispersive results due to its large uncertainties, obtained by varying the hadron masses in the range given in Ref.~\cite{Czarnecki:2002nt}. The holographic model shows better agreement with the dispersive result at higher energies, whereas at low energies it lies outside the uncertainty band of our dispersive result, which is reflected in the normalization of the transverse function,
\begin{equation}
	\label{eq:VVA_Wt2Normalization}
	\tilde{\W}_{2,\text{CMV}}^{(3)}(0) = 0.029^{+0.030}_{-0.010}\si{\GeV^{-2}} \,, \qquad \tilde{\W}_{2,\text{hQCD}}^{(3)}(0) = \SI{0.028}{\GeV^{-2}} \,,
\end{equation}
whereas the \chpt{} constraint~\eqref{eq:NormChPT} is implemented in the dispersive result as well as in the Regge-like model of Ref.~\cite{Masjuan:2020jsf}.

\subsection{Intermediate-state reshuffling}
\label{sec:Reshuffling}

\begin{figure}
	\centering
	\qquad\scalebox{0.575}{\input{plots/W1Cells}}
	\qquad\quad
	\scalebox{0.575}{\input{plots/W2Cells}}
	\\[0.75cm]
	\caption{Comparison of the different contributions to $\tilde\W_1^{(3)}$ and  $\tilde\W_2^{(3)}$ between the two dispersive approaches to VVA, according to Tab.~\ref{tab:Reshuffling}. We show our central values without uncertainties. The blue line is the total result for the dispersion relations in $g-2$ kinematics. Ellipses denote further contributions that are modeled by effective poles, whereas `all' denotes the sum over the columns.}
	\label{fig:W12Cells}
\end{figure}
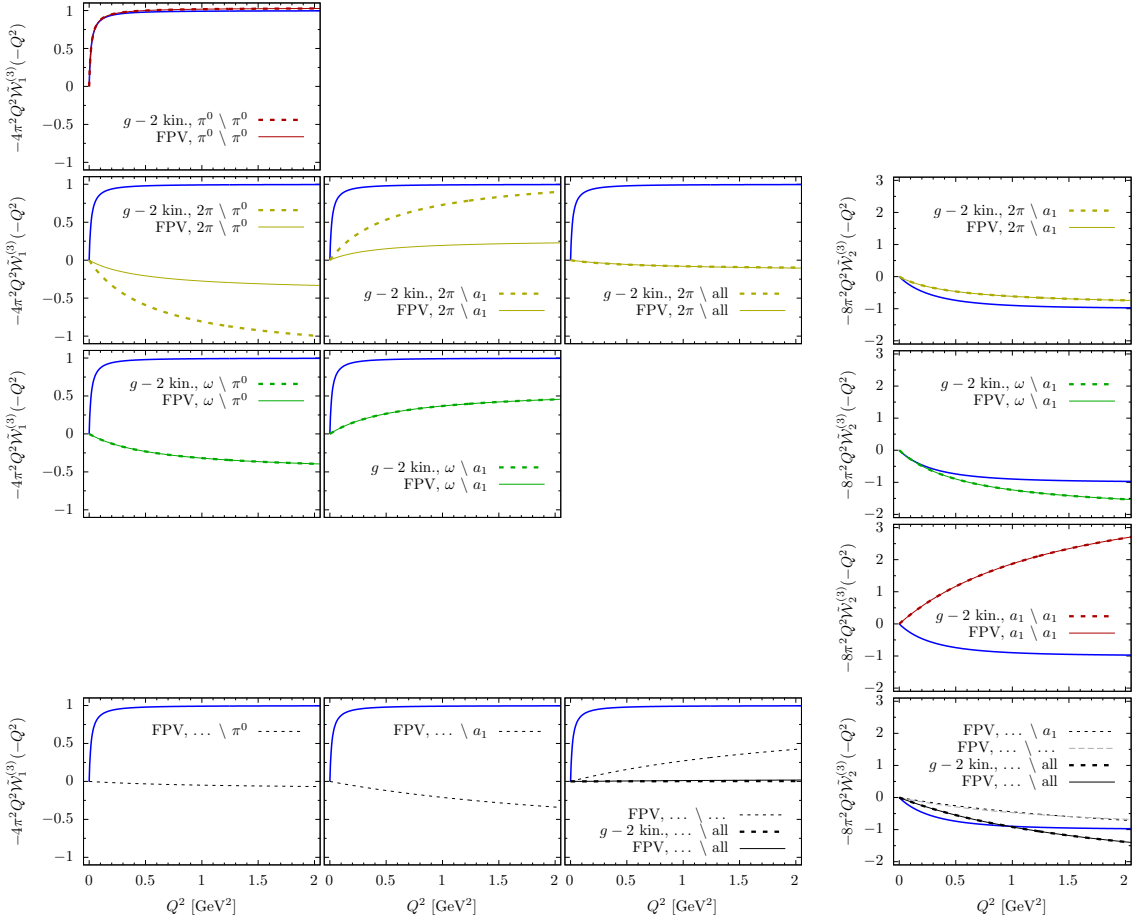

\begin{figure}
	\centering
	\scalebox{0.84}{\input{plots/W1Diff}}
	\caption{Difference between the total two-pion and effective-pole contributions to $\tilde\W_1^{(3)}$ in the two dispersive approaches, including uncertainties. Due to the low-energy and asymptotic constraints on $\tilde\W_1^{(3)}$, the two-pion and effective-pole differences are anticorrelated, so that the total difference between the two dispersive approaches has a smaller uncertainty, see Fig.~\ref{fig:W1FPV}.}
	\label{fig:W1Differences}
\end{figure}
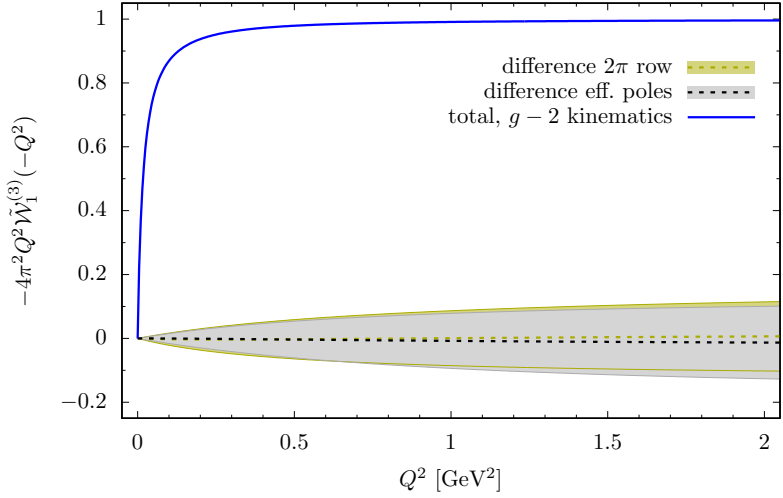

In this section, we compare the two dispersive results in more detail by disentangling the individual contributions according to the primary unitarity cuts in each of the dispersion relations. In Fig.~\ref{fig:W12Cells}, we show the longitudinal function in the first three columns and the transverse function in the rightmost column, according to the individual cells of Tab.~\ref{tab:Reshuffling}. We find that all contributions exactly agree between the two dispersive approaches, with the exception of the two-pion contributions to $\tilde\W_1^{(3)}$. The observed differences are due to two effects. On the one hand, in the contributions that derive from the pion pole for fixed photon virtualities, a difference arises due to the three-pion cut shown in Fig.~\ref{fig:ThreeCuts}, see Sect.~\ref{sec:ComparisonPSPole}. On the other hand, in the contributions from the axial pole for fixed photon virtualities, another difference appears due to the fact that we do not implement the first of the conditions~\eqref{eq:AxialDRCondition}, which would assume a more restrictive asymptotic behavior for $a_1\gamma\to\pi\pi$. Instead, the parameter $\bar\F_{a_1,1}'$ is adjusted together with the effective-pole parameters to fulfill the \chpt{} constraint and the SDCs on $\tilde\W_2^{(3)}$. This has the surprising effect that an almost perfect cancellation happens between the shifts in the $2\pi$ row. This cancellation of the shifts is given by the approximate equality
\begin{equation}
	\label{eq:2piCancellations}
	\frac{1}{96\pi^2} \int_{4 m_\pi^2}^{\Lambda^2} \dif s \frac{s \sigma_\pi^3(s) |\Omega_1^1(s)|^2}{s - q^2} F_\pi \frac{\frac{f_P^1(s, 0)}{\Omega_1^1(s)} - \frac{f_P^1(m_P^2, 0)}{\Omega_1^1(m_P^2)}}{s - m_P^2} \approx -\frac{1}{96 \pi^2} \int_{4 m_\pi^2}^{\Lambda^2} \dif s \frac{s \sigma_\pi^3(s) |\Omega_1^1(s)|^2}{s - q^2} \frac{2F_{a_1}^{(3)}}{m_{a_1}}  \bar\F_{A,1}' \, ,
\end{equation}
which for our central solution holds to a few percent, illustrated in the third column and second row of Fig.~\ref{fig:W12Cells}. Although these shifts are due to higher intermediate states that are not fully taken into account in the dispersion relations, both sides of Eq.~\eqref{eq:2piCancellations} are large (corresponding to the differences between dispersive results in $g-2$ kinematics and fixed photon virtualities in the second row of Fig.~\ref{fig:W12Cells}). Their cancellation implies that also the total contribution of the effective poles, shown in the last row in Fig.~\ref{fig:W12Cells}, agrees to good approximation between the two dispersive approaches. Therefore, the implementation of all low-energy and high-energy constraints leads to the result that the two dispersive approaches largely agree, even at the level of individual rows in Tab.~\ref{tab:Reshuffling} or Fig.~\ref{fig:W12Cells}. To our understanding, the level of the cancellation due to Eq.~\eqref{eq:2piCancellations} is accidental: the error analysis shows that a reshuffling between two-pion and effective-pole contributions between the two dispersive approaches can be expected to happen at the level of $10\%$, as shown by the error bands on the differences in Fig.~\ref{fig:W1Differences}.

%% file: plots/W1gm2.tex
\begingroup
  \makeatletter
  \providecommand\color[2][]{%
    \GenericError{(gnuplot) \space\space\space\@spaces}{%
      Package color not loaded in conjunction with
      terminal option `colourtext'%
    }{See the gnuplot documentation for explanation.%
    }{Either use 'blacktext' in gnuplot or load the package
      color.sty in LaTeX.}%
    \renewcommand\color[2][]{}%
  }%
  \providecommand\includegraphics[2][]{%
    \GenericError{(gnuplot) \space\space\space\@spaces}{%
      Package graphicx or graphics not loaded%
    }{See the gnuplot documentation for explanation.%
    }{The gnuplot epslatex terminal needs graphicx.sty or graphics.sty.}%
    \renewcommand\includegraphics[2][]{}%
  }%
  \providecommand\rotatebox[2]{#2}%
  \@ifundefined{ifGPcolor}{%
    \newif\ifGPcolor
    \GPcolorfalse
  }{}%
  \@ifundefined{ifGPblacktext}{%
    \newif\ifGPblacktext
    \GPblacktexttrue
  }{}%
  \let\gplgaddtomacro\g@addto@macro
  \gdef\gplbacktext{}%
  \gdef\gplfronttext{}%
  \makeatother
  \ifGPblacktext
    \def\colorrgb#1{}%
    \def\colorgray#1{}%
  \else
    \ifGPcolor
      \def\colorrgb#1{\color[rgb]{#1}}%
      \def\colorgray#1{\color[gray]{#1}}%
      \expandafter\def\csname LTw\endcsname{\color{white}}%
      \expandafter\def\csname LTb\endcsname{\color{black}}%
      \expandafter\def\csname LTa\endcsname{\color{black}}%
      \expandafter\def\csname LT0\endcsname{\color[rgb]{1,0,0}}%
      \expandafter\def\csname LT1\endcsname{\color[rgb]{0,1,0}}%
      \expandafter\def\csname LT2\endcsname{\color[rgb]{0,0,1}}%
      \expandafter\def\csname LT3\endcsname{\color[rgb]{1,0,1}}%
      \expandafter\def\csname LT4\endcsname{\color[rgb]{0,1,1}}%
      \expandafter\def\csname LT5\endcsname{\color[rgb]{1,1,0}}%
      \expandafter\def\csname LT6\endcsname{\color[rgb]{0,0,0}}%
      \expandafter\def\csname LT7\endcsname{\color[rgb]{1,0.3,0}}%
      \expandafter\def\csname LT8\endcsname{\color[rgb]{0.5,0.5,0.5}}%
    \else
      \def\colorrgb#1{\color{black}}%
      \def\colorgray#1{\color[gray]{#1}}%
      \expandafter\def\csname LTw\endcsname{\color{white}}%
      \expandafter\def\csname LTb\endcsname{\color{black}}%
      \expandafter\def\csname LTa\endcsname{\color{black}}%
      \expandafter\def\csname LT0\endcsname{\color{black}}%
      \expandafter\def\csname LT1\endcsname{\color{black}}%
      \expandafter\def\csname LT2\endcsname{\color{black}}%
      \expandafter\def\csname LT3\endcsname{\color{black}}%
      \expandafter\def\csname LT4\endcsname{\color{black}}%
      \expandafter\def\csname LT5\endcsname{\color{black}}%
      \expandafter\def\csname LT6\endcsname{\color{black}}%
      \expandafter\def\csname LT7\endcsname{\color{black}}%
      \expandafter\def\csname LT8\endcsname{\color{black}}%
    \fi
  \fi
    \setlength{\unitlength}{0.0500bp}%
    \ifx\gptboxheight\undefined%
      \newlength{\gptboxheight}%
      \newlength{\gptboxwidth}%
      \newsavebox{\gptboxtext}%
    \fi%
    \setlength{\fboxrule}{0.5pt}%
    \setlength{\fboxsep}{1pt}%
    \definecolor{tbcol}{rgb}{1,1,1}%
\begin{picture}(6120.00,4284.00)%
    \gplgaddtomacro\gplbacktext{%
      \csname LTb\endcsname
      \put(198,1034){\makebox(0,0)[r]{\strut{}$0$}}%
      \put(198,1621){\makebox(0,0)[r]{\strut{}$0.2$}}%
      \put(198,2207){\makebox(0,0)[r]{\strut{}$0.4$}}%
      \put(198,2794){\makebox(0,0)[r]{\strut{}$0.6$}}%
      \put(198,3381){\makebox(0,0)[r]{\strut{}$0.8$}}%
      \put(198,3968){\makebox(0,0)[r]{\strut{}$1$}}%
      \put(468,374){\makebox(0,0){\strut{}$0$}}%
      \put(1843,374){\makebox(0,0){\strut{}$0.5$}}%
      \put(3218,374){\makebox(0,0){\strut{}$1$}}%
      \put(4593,374){\makebox(0,0){\strut{}$1.5$}}%
      \put(5968,374){\makebox(0,0){\strut{}$2$}}%
    }%
    \gplgaddtomacro\gplfronttext{%
      \csname LTb\endcsname
      \put(-407,2427){\rotatebox{-270}{\makebox(0,0){\strut{}$-4\pi^2 Q^2 \tilde\W_1^{(3)}(-Q^2)$}}}%
      \put(3217,44){\makebox(0,0){\strut{}$Q^2$ [GeV${}^2$]}}%
      \csname LTb\endcsname
      \put(4961,3418){\makebox(0,0)[r]{\strut{}total}}%
    }%
    \gplgaddtomacro\gplbacktext{%
    }%
    \gplgaddtomacro\gplfronttext{%
      \csname LTb\endcsname
      \put(-11,2427){\rotatebox{-270}{\makebox(0,0){\strut{} }}}%
      \put(3217,374){\makebox(0,0){\strut{} }}%
      \csname LTb\endcsname
      \put(4961,3418){\makebox(0,0)[r]{\strut{} }}%
      \csname LTb\endcsname
      \put(4961,3198){\makebox(0,0)[r]{\strut{}$\pi^0$ pole}}%
      \csname LTb\endcsname
      \put(4961,2978){\makebox(0,0)[r]{\strut{}$\omega$ pole}}%
      \csname LTb\endcsname
      \put(4961,2758){\makebox(0,0)[r]{\strut{}$2\pi$ contributions}}%
      \csname LTb\endcsname
      \put(4961,2538){\makebox(0,0)[r]{\strut{}effective pole}}%
    }%
    \gplbacktext
    \put(0,0){\includegraphics[width={306.00bp},height={214.20bp}]{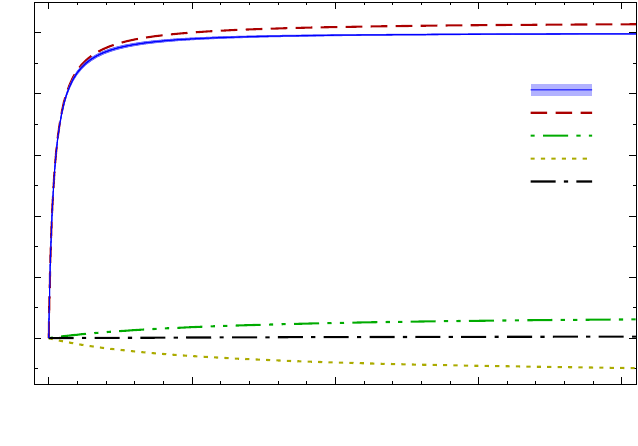}}%
    \gplfronttext
  \end{picture}%
\endgroup

%% file: plots/W2gm2.tex
\begingroup
  \makeatletter
  \providecommand\color[2][]{%
    \GenericError{(gnuplot) \space\space\space\@spaces}{%
      Package color not loaded in conjunction with
      terminal option `colourtext'%
    }{See the gnuplot documentation for explanation.%
    }{Either use 'blacktext' in gnuplot or load the package
      color.sty in LaTeX.}%
    \renewcommand\color[2][]{}%
  }%
  \providecommand\includegraphics[2][]{%
    \GenericError{(gnuplot) \space\space\space\@spaces}{%
      Package graphicx or graphics not loaded%
    }{See the gnuplot documentation for explanation.%
    }{The gnuplot epslatex terminal needs graphicx.sty or graphics.sty.}%
    \renewcommand\includegraphics[2][]{}%
  }%
  \providecommand\rotatebox[2]{#2}%
  \@ifundefined{ifGPcolor}{%
    \newif\ifGPcolor
    \GPcolorfalse
  }{}%
  \@ifundefined{ifGPblacktext}{%
    \newif\ifGPblacktext
    \GPblacktexttrue
  }{}%
  \let\gplgaddtomacro\g@addto@macro
  \gdef\gplbacktext{}%
  \gdef\gplfronttext{}%
  \makeatother
  \ifGPblacktext
    \def\colorrgb#1{}%
    \def\colorgray#1{}%
  \else
    \ifGPcolor
      \def\colorrgb#1{\color[rgb]{#1}}%
      \def\colorgray#1{\color[gray]{#1}}%
      \expandafter\def\csname LTw\endcsname{\color{white}}%
      \expandafter\def\csname LTb\endcsname{\color{black}}%
      \expandafter\def\csname LTa\endcsname{\color{black}}%
      \expandafter\def\csname LT0\endcsname{\color[rgb]{1,0,0}}%
      \expandafter\def\csname LT1\endcsname{\color[rgb]{0,1,0}}%
      \expandafter\def\csname LT2\endcsname{\color[rgb]{0,0,1}}%
      \expandafter\def\csname LT3\endcsname{\color[rgb]{1,0,1}}%
      \expandafter\def\csname LT4\endcsname{\color[rgb]{0,1,1}}%
      \expandafter\def\csname LT5\endcsname{\color[rgb]{1,1,0}}%
      \expandafter\def\csname LT6\endcsname{\color[rgb]{0,0,0}}%
      \expandafter\def\csname LT7\endcsname{\color[rgb]{1,0.3,0}}%
      \expandafter\def\csname LT8\endcsname{\color[rgb]{0.5,0.5,0.5}}%
    \else
      \def\colorrgb#1{\color{black}}%
      \def\colorgray#1{\color[gray]{#1}}%
      \expandafter\def\csname LTw\endcsname{\color{white}}%
      \expandafter\def\csname LTb\endcsname{\color{black}}%
      \expandafter\def\csname LTa\endcsname{\color{black}}%
      \expandafter\def\csname LT0\endcsname{\color{black}}%
      \expandafter\def\csname LT1\endcsname{\color{black}}%
      \expandafter\def\csname LT2\endcsname{\color{black}}%
      \expandafter\def\csname LT3\endcsname{\color{black}}%
      \expandafter\def\csname LT4\endcsname{\color{black}}%
      \expandafter\def\csname LT5\endcsname{\color{black}}%
      \expandafter\def\csname LT6\endcsname{\color{black}}%
      \expandafter\def\csname LT7\endcsname{\color{black}}%
      \expandafter\def\csname LT8\endcsname{\color{black}}%
    \fi
  \fi
    \setlength{\unitlength}{0.0500bp}%
    \ifx\gptboxheight\undefined%
      \newlength{\gptboxheight}%
      \newlength{\gptboxwidth}%
      \newsavebox{\gptboxtext}%
    \fi%
    \setlength{\fboxrule}{0.5pt}%
    \setlength{\fboxsep}{1pt}%
    \definecolor{tbcol}{rgb}{1,1,1}%
\begin{picture}(6120.00,4284.00)%
    \gplgaddtomacro\gplbacktext{%
      \csname LTb\endcsname
      \put(198,1218){\makebox(0,0)[r]{\strut{}$-1$}}%
      \put(198,1998){\makebox(0,0)[r]{\strut{}$0$}}%
      \put(198,2779){\makebox(0,0)[r]{\strut{}$1$}}%
      \put(198,3559){\makebox(0,0)[r]{\strut{}$2$}}%
      \put(468,374){\makebox(0,0){\strut{}$0$}}%
      \put(1843,374){\makebox(0,0){\strut{}$0.5$}}%
      \put(3218,374){\makebox(0,0){\strut{}$1$}}%
      \put(4593,374){\makebox(0,0){\strut{}$1.5$}}%
      \put(5968,374){\makebox(0,0){\strut{}$2$}}%
    }%
    \gplgaddtomacro\gplfronttext{%
      \csname LTb\endcsname
      \put(-275,2427){\rotatebox{-270}{\makebox(0,0){\strut{}$-8\pi^2 Q^2 \tilde\W_2^{(3)}(-Q^2)$}}}%
      \put(3217,44){\makebox(0,0){\strut{}$Q^2$ [GeV${}^2$]}}%
      \csname LTb\endcsname
      \put(5163,3693){\makebox(0,0)[r]{\strut{}total}}%
    }%
    \gplgaddtomacro\gplbacktext{%
    }%
    \gplgaddtomacro\gplfronttext{%
      \csname LTb\endcsname
      \put(-11,2427){\rotatebox{-270}{\makebox(0,0){\strut{} }}}%
      \put(3217,374){\makebox(0,0){\strut{} }}%
      \csname LTb\endcsname
      \put(5163,3693){\makebox(0,0)[r]{\strut{} }}%
      \csname LTb\endcsname
      \put(5163,3473){\makebox(0,0)[r]{\strut{}Re($a_1$ pole)}}%
      \csname LTb\endcsname
      \put(5163,3253){\makebox(0,0)[r]{\strut{}Im($a_1$ pole)}}%
      \csname LTb\endcsname
      \put(5163,3033){\makebox(0,0)[r]{\strut{}$\omega$ pole}}%
      \csname LTb\endcsname
      \put(5163,2813){\makebox(0,0)[r]{\strut{}Re($2\pi$ contributions)}}%
      \csname LTb\endcsname
      \put(5163,2593){\makebox(0,0)[r]{\strut{}Im($2\pi$ contributions)}}%
      \csname LTb\endcsname
      \put(5163,2373){\makebox(0,0)[r]{\strut{}effective poles}}%
    }%
    \gplbacktext
    \put(0,0){\includegraphics[width={306.00bp},height={214.20bp}]{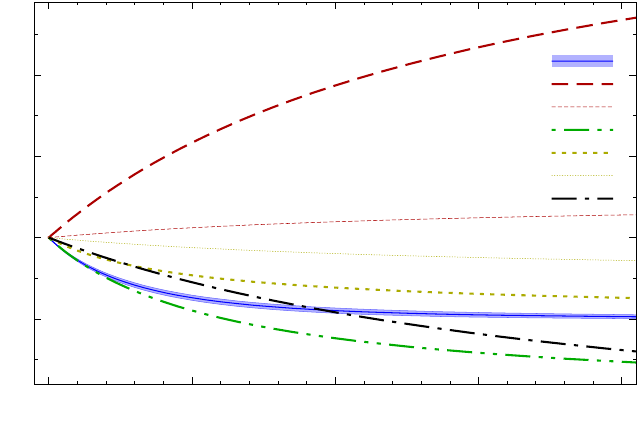}}%
    \gplfronttext
  \end{picture}%
\endgroup

%% file: plots/W2gm2Constr.tex
\begingroup
  \makeatletter
  \providecommand\color[2][]{%
    \GenericError{(gnuplot) \space\space\space\@spaces}{%
      Package color not loaded in conjunction with
      terminal option `colourtext'%
    }{See the gnuplot documentation for explanation.%
    }{Either use 'blacktext' in gnuplot or load the package
      color.sty in LaTeX.}%
    \renewcommand\color[2][]{}%
  }%
  \providecommand\includegraphics[2][]{%
    \GenericError{(gnuplot) \space\space\space\@spaces}{%
      Package graphicx or graphics not loaded%
    }{See the gnuplot documentation for explanation.%
    }{The gnuplot epslatex terminal needs graphicx.sty or graphics.sty.}%
    \renewcommand\includegraphics[2][]{}%
  }%
  \providecommand\rotatebox[2]{#2}%
  \@ifundefined{ifGPcolor}{%
    \newif\ifGPcolor
    \GPcolorfalse
  }{}%
  \@ifundefined{ifGPblacktext}{%
    \newif\ifGPblacktext
    \GPblacktexttrue
  }{}%
  \let\gplgaddtomacro\g@addto@macro
  \gdef\gplbacktext{}%
  \gdef\gplfronttext{}%
  \makeatother
  \ifGPblacktext
    \def\colorrgb#1{}%
    \def\colorgray#1{}%
  \else
    \ifGPcolor
      \def\colorrgb#1{\color[rgb]{#1}}%
      \def\colorgray#1{\color[gray]{#1}}%
      \expandafter\def\csname LTw\endcsname{\color{white}}%
      \expandafter\def\csname LTb\endcsname{\color{black}}%
      \expandafter\def\csname LTa\endcsname{\color{black}}%
      \expandafter\def\csname LT0\endcsname{\color[rgb]{1,0,0}}%
      \expandafter\def\csname LT1\endcsname{\color[rgb]{0,1,0}}%
      \expandafter\def\csname LT2\endcsname{\color[rgb]{0,0,1}}%
      \expandafter\def\csname LT3\endcsname{\color[rgb]{1,0,1}}%
      \expandafter\def\csname LT4\endcsname{\color[rgb]{0,1,1}}%
      \expandafter\def\csname LT5\endcsname{\color[rgb]{1,1,0}}%
      \expandafter\def\csname LT6\endcsname{\color[rgb]{0,0,0}}%
      \expandafter\def\csname LT7\endcsname{\color[rgb]{1,0.3,0}}%
      \expandafter\def\csname LT8\endcsname{\color[rgb]{0.5,0.5,0.5}}%
    \else
      \def\colorrgb#1{\color{black}}%
      \def\colorgray#1{\color[gray]{#1}}%
      \expandafter\def\csname LTw\endcsname{\color{white}}%
      \expandafter\def\csname LTb\endcsname{\color{black}}%
      \expandafter\def\csname LTa\endcsname{\color{black}}%
      \expandafter\def\csname LT0\endcsname{\color{black}}%
      \expandafter\def\csname LT1\endcsname{\color{black}}%
      \expandafter\def\csname LT2\endcsname{\color{black}}%
      \expandafter\def\csname LT3\endcsname{\color{black}}%
      \expandafter\def\csname LT4\endcsname{\color{black}}%
      \expandafter\def\csname LT5\endcsname{\color{black}}%
      \expandafter\def\csname LT6\endcsname{\color{black}}%
      \expandafter\def\csname LT7\endcsname{\color{black}}%
      \expandafter\def\csname LT8\endcsname{\color{black}}%
    \fi
  \fi
    \setlength{\unitlength}{0.0500bp}%
    \ifx\gptboxheight\undefined%
      \newlength{\gptboxheight}%
      \newlength{\gptboxwidth}%
      \newsavebox{\gptboxtext}%
    \fi%
    \setlength{\fboxrule}{0.5pt}%
    \setlength{\fboxsep}{1pt}%
    \definecolor{tbcol}{rgb}{1,1,1}%
\begin{picture}(6120.00,4284.00)%
    \gplgaddtomacro\gplbacktext{%
      \csname LTb\endcsname
      \put(198,1511){\makebox(0,0)[r]{\strut{}$-1$}}%
      \put(198,2820){\makebox(0,0)[r]{\strut{}$-0.5$}}%
      \put(198,4130){\makebox(0,0)[r]{\strut{}$0$}}%
      \put(468,374){\makebox(0,0){\strut{}$0$}}%
      \put(1843,374){\makebox(0,0){\strut{}$0.5$}}%
      \put(3218,374){\makebox(0,0){\strut{}$1$}}%
      \put(4593,374){\makebox(0,0){\strut{}$1.5$}}%
      \put(5968,374){\makebox(0,0){\strut{}$2$}}%
    }%
    \gplgaddtomacro\gplfronttext{%
      \csname LTb\endcsname
      \put(-539,2427){\rotatebox{-270}{\makebox(0,0){\strut{}$-8\pi^2 Q^2 \tilde\W_2^{(3)}(-Q^2)$}}}%
      \put(3217,44){\makebox(0,0){\strut{}$Q^2$ [GeV${}^2$]}}%
      \csname LTb\endcsname
      \put(4961,3968){\makebox(0,0)[r]{\strut{}no \chpt{}, no $\O(q^{-6})$ constraint}}%
      \csname LTb\endcsname
      \put(4961,3748){\makebox(0,0)[r]{\strut{}no \chpt{} constraint}}%
      \csname LTb\endcsname
      \put(4961,3528){\makebox(0,0)[r]{\strut{}no $\O(q^{-6})$ constraint}}%
      \csname LTb\endcsname
      \put(4961,3308){\makebox(0,0)[r]{\strut{}all constraints}}%
    }%
    \gplgaddtomacro\gplbacktext{%
    }%
    \gplgaddtomacro\gplfronttext{%
      \csname LTb\endcsname
      \put(-11,2427){\rotatebox{-270}{\makebox(0,0){\strut{} }}}%
      \put(3217,374){\makebox(0,0){\strut{} }}%
      \csname LTb\endcsname
      \put(4961,3968){\makebox(0,0)[r]{\strut{} }}%
      \csname LTb\endcsname
      \put(4961,3748){\makebox(0,0)[r]{\strut{} }}%
      \csname LTb\endcsname
      \put(4961,3528){\makebox(0,0)[r]{\strut{} }}%
      \csname LTb\endcsname
      \put(4961,3308){\makebox(0,0)[r]{\strut{} }}%
    }%
    \gplbacktext
    \put(0,0){\includegraphics[width={306.00bp},height={214.20bp}]{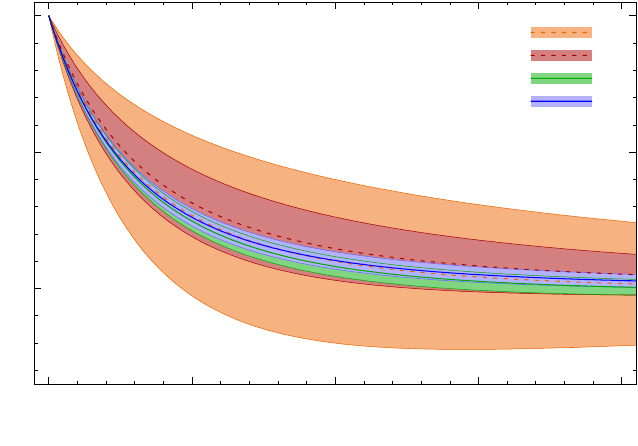}}%
    \gplfronttext
  \end{picture}%
\endgroup

%% file: plots/W1FPV.tex
\begingroup
  \makeatletter
  \providecommand\color[2][]{%
    \GenericError{(gnuplot) \space\space\space\@spaces}{%
      Package color not loaded in conjunction with
      terminal option `colourtext'%
    }{See the gnuplot documentation for explanation.%
    }{Either use 'blacktext' in gnuplot or load the package
      color.sty in LaTeX.}%
    \renewcommand\color[2][]{}%
  }%
  \providecommand\includegraphics[2][]{%
    \GenericError{(gnuplot) \space\space\space\@spaces}{%
      Package graphicx or graphics not loaded%
    }{See the gnuplot documentation for explanation.%
    }{The gnuplot epslatex terminal needs graphicx.sty or graphics.sty.}%
    \renewcommand\includegraphics[2][]{}%
  }%
  \providecommand\rotatebox[2]{#2}%
  \@ifundefined{ifGPcolor}{%
    \newif\ifGPcolor
    \GPcolorfalse
  }{}%
  \@ifundefined{ifGPblacktext}{%
    \newif\ifGPblacktext
    \GPblacktexttrue
  }{}%
  \let\gplgaddtomacro\g@addto@macro
  \gdef\gplbacktext{}%
  \gdef\gplfronttext{}%
  \makeatother
  \ifGPblacktext
    \def\colorrgb#1{}%
    \def\colorgray#1{}%
  \else
    \ifGPcolor
      \def\colorrgb#1{\color[rgb]{#1}}%
      \def\colorgray#1{\color[gray]{#1}}%
      \expandafter\def\csname LTw\endcsname{\color{white}}%
      \expandafter\def\csname LTb\endcsname{\color{black}}%
      \expandafter\def\csname LTa\endcsname{\color{black}}%
      \expandafter\def\csname LT0\endcsname{\color[rgb]{1,0,0}}%
      \expandafter\def\csname LT1\endcsname{\color[rgb]{0,1,0}}%
      \expandafter\def\csname LT2\endcsname{\color[rgb]{0,0,1}}%
      \expandafter\def\csname LT3\endcsname{\color[rgb]{1,0,1}}%
      \expandafter\def\csname LT4\endcsname{\color[rgb]{0,1,1}}%
      \expandafter\def\csname LT5\endcsname{\color[rgb]{1,1,0}}%
      \expandafter\def\csname LT6\endcsname{\color[rgb]{0,0,0}}%
      \expandafter\def\csname LT7\endcsname{\color[rgb]{1,0.3,0}}%
      \expandafter\def\csname LT8\endcsname{\color[rgb]{0.5,0.5,0.5}}%
    \else
      \def\colorrgb#1{\color{black}}%
      \def\colorgray#1{\color[gray]{#1}}%
      \expandafter\def\csname LTw\endcsname{\color{white}}%
      \expandafter\def\csname LTb\endcsname{\color{black}}%
      \expandafter\def\csname LTa\endcsname{\color{black}}%
      \expandafter\def\csname LT0\endcsname{\color{black}}%
      \expandafter\def\csname LT1\endcsname{\color{black}}%
      \expandafter\def\csname LT2\endcsname{\color{black}}%
      \expandafter\def\csname LT3\endcsname{\color{black}}%
      \expandafter\def\csname LT4\endcsname{\color{black}}%
      \expandafter\def\csname LT5\endcsname{\color{black}}%
      \expandafter\def\csname LT6\endcsname{\color{black}}%
      \expandafter\def\csname LT7\endcsname{\color{black}}%
      \expandafter\def\csname LT8\endcsname{\color{black}}%
    \fi
  \fi
    \setlength{\unitlength}{0.0500bp}%
    \ifx\gptboxheight\undefined%
      \newlength{\gptboxheight}%
      \newlength{\gptboxwidth}%
      \newsavebox{\gptboxtext}%
    \fi%
    \setlength{\fboxrule}{0.5pt}%
    \setlength{\fboxsep}{1pt}%
    \definecolor{tbcol}{rgb}{1,1,1}%
\begin{picture}(6120.00,4284.00)%
    \gplgaddtomacro\gplbacktext{%
      \csname LTb\endcsname
      \put(198,1034){\makebox(0,0)[r]{\strut{}$0$}}%
      \put(198,1621){\makebox(0,0)[r]{\strut{}$0.2$}}%
      \put(198,2207){\makebox(0,0)[r]{\strut{}$0.4$}}%
      \put(198,2794){\makebox(0,0)[r]{\strut{}$0.6$}}%
      \put(198,3381){\makebox(0,0)[r]{\strut{}$0.8$}}%
      \put(198,3968){\makebox(0,0)[r]{\strut{}$1$}}%
      \put(468,374){\makebox(0,0){\strut{}$0$}}%
      \put(1843,374){\makebox(0,0){\strut{}$0.5$}}%
      \put(3218,374){\makebox(0,0){\strut{}$1$}}%
      \put(4593,374){\makebox(0,0){\strut{}$1.5$}}%
      \put(5968,374){\makebox(0,0){\strut{}$2$}}%
    }%
    \gplgaddtomacro\gplfronttext{%
      \csname LTb\endcsname
      \put(-407,2427){\rotatebox{-270}{\makebox(0,0){\strut{}$-4\pi^2 Q^2 \tilde\W_1^{(3)}(-Q^2)$}}}%
      \put(3217,44){\makebox(0,0){\strut{}$Q^2$ [GeV${}^2$]}}%
      \csname LTb\endcsname
      \put(4961,3601){\makebox(0,0)[r]{\strut{}total, FPV}}%
      \csname LTb\endcsname
      \put(4961,3381){\makebox(0,0)[r]{\strut{}total, $g-2$ kinematics}}%
    }%
    \gplgaddtomacro\gplbacktext{%
    }%
    \gplgaddtomacro\gplfronttext{%
      \csname LTb\endcsname
      \put(-11,2427){\rotatebox{-270}{\makebox(0,0){\strut{} }}}%
      \put(3217,374){\makebox(0,0){\strut{} }}%
      \csname LTb\endcsname
      \put(4961,3601){\makebox(0,0)[r]{\strut{} }}%
      \csname LTb\endcsname
      \put(4961,3381){\makebox(0,0)[r]{\strut{} }}%
      \csname LTb\endcsname
      \put(4961,3161){\makebox(0,0)[r]{\strut{}$\pi^0$ pole}}%
      \csname LTb\endcsname
      \put(4961,2941){\makebox(0,0)[r]{\strut{}$a_1$ pole}}%
      \csname LTb\endcsname
      \put(4961,2721){\makebox(0,0)[r]{\strut{}effective poles}}%
    }%
    \gplbacktext
    \put(0,0){\includegraphics[width={306.00bp},height={214.20bp}]{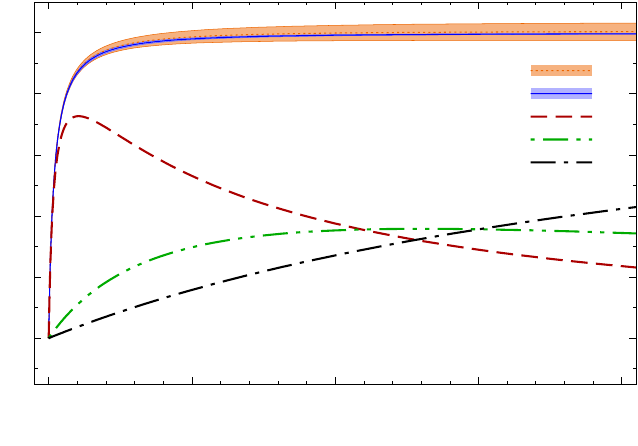}}%
    \gplfronttext
  \end{picture}%
\endgroup

%% file: plots/W2FPV.tex
\begingroup
  \makeatletter
  \providecommand\color[2][]{%
    \GenericError{(gnuplot) \space\space\space\@spaces}{%
      Package color not loaded in conjunction with
      terminal option `colourtext'%
    }{See the gnuplot documentation for explanation.%
    }{Either use 'blacktext' in gnuplot or load the package
      color.sty in LaTeX.}%
    \renewcommand\color[2][]{}%
  }%
  \providecommand\includegraphics[2][]{%
    \GenericError{(gnuplot) \space\space\space\@spaces}{%
      Package graphicx or graphics not loaded%
    }{See the gnuplot documentation for explanation.%
    }{The gnuplot epslatex terminal needs graphicx.sty or graphics.sty.}%
    \renewcommand\includegraphics[2][]{}%
  }%
  \providecommand\rotatebox[2]{#2}%
  \@ifundefined{ifGPcolor}{%
    \newif\ifGPcolor
    \GPcolorfalse
  }{}%
  \@ifundefined{ifGPblacktext}{%
    \newif\ifGPblacktext
    \GPblacktexttrue
  }{}%
  \let\gplgaddtomacro\g@addto@macro
  \gdef\gplbacktext{}%
  \gdef\gplfronttext{}%
  \makeatother
  \ifGPblacktext
    \def\colorrgb#1{}%
    \def\colorgray#1{}%
  \else
    \ifGPcolor
      \def\colorrgb#1{\color[rgb]{#1}}%
      \def\colorgray#1{\color[gray]{#1}}%
      \expandafter\def\csname LTw\endcsname{\color{white}}%
      \expandafter\def\csname LTb\endcsname{\color{black}}%
      \expandafter\def\csname LTa\endcsname{\color{black}}%
      \expandafter\def\csname LT0\endcsname{\color[rgb]{1,0,0}}%
      \expandafter\def\csname LT1\endcsname{\color[rgb]{0,1,0}}%
      \expandafter\def\csname LT2\endcsname{\color[rgb]{0,0,1}}%
      \expandafter\def\csname LT3\endcsname{\color[rgb]{1,0,1}}%
      \expandafter\def\csname LT4\endcsname{\color[rgb]{0,1,1}}%
      \expandafter\def\csname LT5\endcsname{\color[rgb]{1,1,0}}%
      \expandafter\def\csname LT6\endcsname{\color[rgb]{0,0,0}}%
      \expandafter\def\csname LT7\endcsname{\color[rgb]{1,0.3,0}}%
      \expandafter\def\csname LT8\endcsname{\color[rgb]{0.5,0.5,0.5}}%
    \else
      \def\colorrgb#1{\color{black}}%
      \def\colorgray#1{\color[gray]{#1}}%
      \expandafter\def\csname LTw\endcsname{\color{white}}%
      \expandafter\def\csname LTb\endcsname{\color{black}}%
      \expandafter\def\csname LTa\endcsname{\color{black}}%
      \expandafter\def\csname LT0\endcsname{\color{black}}%
      \expandafter\def\csname LT1\endcsname{\color{black}}%
      \expandafter\def\csname LT2\endcsname{\color{black}}%
      \expandafter\def\csname LT3\endcsname{\color{black}}%
      \expandafter\def\csname LT4\endcsname{\color{black}}%
      \expandafter\def\csname LT5\endcsname{\color{black}}%
      \expandafter\def\csname LT6\endcsname{\color{black}}%
      \expandafter\def\csname LT7\endcsname{\color{black}}%
      \expandafter\def\csname LT8\endcsname{\color{black}}%
    \fi
  \fi
    \setlength{\unitlength}{0.0500bp}%
    \ifx\gptboxheight\undefined%
      \newlength{\gptboxheight}%
      \newlength{\gptboxwidth}%
      \newsavebox{\gptboxtext}%
    \fi%
    \setlength{\fboxrule}{0.5pt}%
    \setlength{\fboxsep}{1pt}%
    \definecolor{tbcol}{rgb}{1,1,1}%
\begin{picture}(6120.00,4284.00)%
    \gplgaddtomacro\gplbacktext{%
      \csname LTb\endcsname
      \put(198,913){\makebox(0,0)[r]{\strut{}$-1$}}%
      \put(198,2507){\makebox(0,0)[r]{\strut{}$-0.5$}}%
      \put(198,4102){\makebox(0,0)[r]{\strut{}$0$}}%
      \put(468,374){\makebox(0,0){\strut{}$0$}}%
      \put(1843,374){\makebox(0,0){\strut{}$0.5$}}%
      \put(3218,374){\makebox(0,0){\strut{}$1$}}%
      \put(4593,374){\makebox(0,0){\strut{}$1.5$}}%
      \put(5968,374){\makebox(0,0){\strut{}$2$}}%
    }%
    \gplgaddtomacro\gplfronttext{%
      \csname LTb\endcsname
      \put(-539,2427){\rotatebox{-270}{\makebox(0,0){\strut{}$-8\pi^2 Q^2 \tilde\W_2^{(3)}(-Q^2)$}}}%
      \put(3217,44){\makebox(0,0){\strut{}$Q^2$ [GeV${}^2$]}}%
      \csname LTb\endcsname
      \put(4673,4059){\makebox(0,0)[r]{\strut{}total, FPV}}%
      \csname LTb\endcsname
      \put(4673,3839){\makebox(0,0)[r]{\strut{}total, $g-2$ kinematics}}%
    }%
    \gplgaddtomacro\gplbacktext{%
    }%
    \gplgaddtomacro\gplfronttext{%
      \csname LTb\endcsname
      \put(-11,2427){\rotatebox{-270}{\makebox(0,0){\strut{} }}}%
      \put(3217,374){\makebox(0,0){\strut{} }}%
      \csname LTb\endcsname
      \put(4673,4059){\makebox(0,0)[r]{\strut{} }}%
      \csname LTb\endcsname
      \put(4673,3839){\makebox(0,0)[r]{\strut{} }}%
      \csname LTb\endcsname
      \put(4673,3619){\makebox(0,0)[r]{\strut{}$a_1$ pole}}%
      \csname LTb\endcsname
      \put(4673,3399){\makebox(0,0)[r]{\strut{}effective poles}}%
    }%
    \gplbacktext
    \put(0,0){\includegraphics[width={306.00bp},height={214.20bp}]{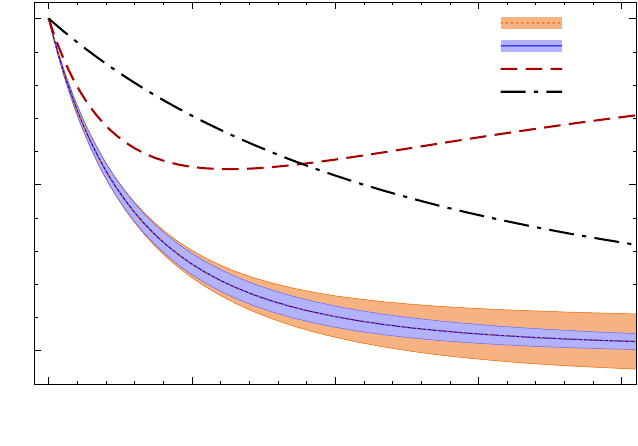}}%
    \gplfronttext
  \end{picture}%
\endgroup

%% file: plots/W1comparisonNew.tex
\begingroup
  \makeatletter
  \providecommand\color[2][]{%
    \GenericError{(gnuplot) \space\space\space\@spaces}{%
      Package color not loaded in conjunction with
      terminal option `colourtext'%
    }{See the gnuplot documentation for explanation.%
    }{Either use 'blacktext' in gnuplot or load the package
      color.sty in LaTeX.}%
    \renewcommand\color[2][]{}%
  }%
  \providecommand\includegraphics[2][]{%
    \GenericError{(gnuplot) \space\space\space\@spaces}{%
      Package graphicx or graphics not loaded%
    }{See the gnuplot documentation for explanation.%
    }{The gnuplot epslatex terminal needs graphicx.sty or graphics.sty.}%
    \renewcommand\includegraphics[2][]{}%
  }%
  \providecommand\rotatebox[2]{#2}%
  \@ifundefined{ifGPcolor}{%
    \newif\ifGPcolor
    \GPcolorfalse
  }{}%
  \@ifundefined{ifGPblacktext}{%
    \newif\ifGPblacktext
    \GPblacktexttrue
  }{}%
  \let\gplgaddtomacro\g@addto@macro
  \gdef\gplbacktext{}%
  \gdef\gplfronttext{}%
  \makeatother
  \ifGPblacktext
    \def\colorrgb#1{}%
    \def\colorgray#1{}%
  \else
    \ifGPcolor
      \def\colorrgb#1{\color[rgb]{#1}}%
      \def\colorgray#1{\color[gray]{#1}}%
      \expandafter\def\csname LTw\endcsname{\color{white}}%
      \expandafter\def\csname LTb\endcsname{\color{black}}%
      \expandafter\def\csname LTa\endcsname{\color{black}}%
      \expandafter\def\csname LT0\endcsname{\color[rgb]{1,0,0}}%
      \expandafter\def\csname LT1\endcsname{\color[rgb]{0,1,0}}%
      \expandafter\def\csname LT2\endcsname{\color[rgb]{0,0,1}}%
      \expandafter\def\csname LT3\endcsname{\color[rgb]{1,0,1}}%
      \expandafter\def\csname LT4\endcsname{\color[rgb]{0,1,1}}%
      \expandafter\def\csname LT5\endcsname{\color[rgb]{1,1,0}}%
      \expandafter\def\csname LT6\endcsname{\color[rgb]{0,0,0}}%
      \expandafter\def\csname LT7\endcsname{\color[rgb]{1,0.3,0}}%
      \expandafter\def\csname LT8\endcsname{\color[rgb]{0.5,0.5,0.5}}%
    \else
      \def\colorrgb#1{\color{black}}%
      \def\colorgray#1{\color[gray]{#1}}%
      \expandafter\def\csname LTw\endcsname{\color{white}}%
      \expandafter\def\csname LTb\endcsname{\color{black}}%
      \expandafter\def\csname LTa\endcsname{\color{black}}%
      \expandafter\def\csname LT0\endcsname{\color{black}}%
      \expandafter\def\csname LT1\endcsname{\color{black}}%
      \expandafter\def\csname LT2\endcsname{\color{black}}%
      \expandafter\def\csname LT3\endcsname{\color{black}}%
      \expandafter\def\csname LT4\endcsname{\color{black}}%
      \expandafter\def\csname LT5\endcsname{\color{black}}%
      \expandafter\def\csname LT6\endcsname{\color{black}}%
      \expandafter\def\csname LT7\endcsname{\color{black}}%
      \expandafter\def\csname LT8\endcsname{\color{black}}%
    \fi
  \fi
    \setlength{\unitlength}{0.0500bp}%
    \ifx\gptboxheight\undefined%
      \newlength{\gptboxheight}%
      \newlength{\gptboxwidth}%
      \newsavebox{\gptboxtext}%
    \fi%
    \setlength{\fboxrule}{0.5pt}%
    \setlength{\fboxsep}{1pt}%
    \definecolor{tbcol}{rgb}{1,1,1}%
\begin{picture}(6120.00,4284.00)%
    \gplgaddtomacro\gplbacktext{%
      \csname LTb\endcsname
      \put(198,1034){\makebox(0,0)[r]{\strut{}$0$}}%
      \put(198,1621){\makebox(0,0)[r]{\strut{}$0.2$}}%
      \put(198,2207){\makebox(0,0)[r]{\strut{}$0.4$}}%
      \put(198,2794){\makebox(0,0)[r]{\strut{}$0.6$}}%
      \put(198,3381){\makebox(0,0)[r]{\strut{}$0.8$}}%
      \put(198,3968){\makebox(0,0)[r]{\strut{}$1$}}%
      \put(468,374){\makebox(0,0){\strut{}$0$}}%
      \put(1843,374){\makebox(0,0){\strut{}$0.5$}}%
      \put(3218,374){\makebox(0,0){\strut{}$1$}}%
      \put(4593,374){\makebox(0,0){\strut{}$1.5$}}%
      \put(5968,374){\makebox(0,0){\strut{}$2$}}%
    }%
    \gplgaddtomacro\gplfronttext{%
      \csname LTb\endcsname
      \put(-407,2427){\rotatebox{-270}{\makebox(0,0){\strut{}$-4\pi^2 Q^2 \tilde\W_1^{(3)}(-Q^2)$}}}%
      \put(3217,44){\makebox(0,0){\strut{}$Q^2$ [GeV${}^2$]}}%
      \csname LTb\endcsname
      \put(4961,2134){\makebox(0,0)[r]{\strut{}dispersive, FPV}}%
      \csname LTb\endcsname
      \put(4961,1914){\makebox(0,0)[r]{\strut{}dispersive, $g-2$ kinematics}}%
    }%
    \gplgaddtomacro\gplbacktext{%
    }%
    \gplgaddtomacro\gplfronttext{%
      \csname LTb\endcsname
      \put(-11,2427){\rotatebox{-270}{\makebox(0,0){\strut{} }}}%
      \put(3217,374){\makebox(0,0){\strut{} }}%
      \csname LTb\endcsname
      \put(4961,2134){\makebox(0,0)[r]{\strut{} }}%
      \csname LTb\endcsname
      \put(4961,1914){\makebox(0,0)[r]{\strut{} }}%
      \csname LTb\endcsname
      \put(4961,1694){\makebox(0,0)[r]{\strut{}CMV model~\cite{Czarnecki:2002nt}}}%
      \csname LTb\endcsname
      \put(4961,1474){\makebox(0,0)[r]{\strut{}hQCD model~\cite{Leutgeb:2019gbz, Cappiello:2019hwh, Hirn:2005nr}}}%
      \csname LTb\endcsname
      \put(4961,1254){\makebox(0,0)[r]{\strut{}PS Regge model~\cite{Colangelo:2019lpu, Colangelo:2019uex}}}%
      \csname LTb\endcsname
      \put(4961,1034){\makebox(0,0)[r]{\strut{}Regge-like model~\cite{Masjuan:2020jsf}}}%
    }%
    \gplbacktext
    \put(0,0){\includegraphics[width={306.00bp},height={214.20bp}]{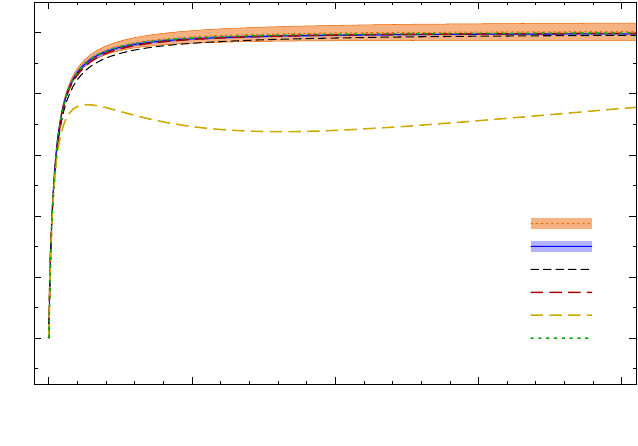}}%
    \gplfronttext
  \end{picture}%
\endgroup

%% file: plots/W2comparisonNew.tex
\begingroup
  \makeatletter
  \providecommand\color[2][]{%
    \GenericError{(gnuplot) \space\space\space\@spaces}{%
      Package color not loaded in conjunction with
      terminal option `colourtext'%
    }{See the gnuplot documentation for explanation.%
    }{Either use 'blacktext' in gnuplot or load the package
      color.sty in LaTeX.}%
    \renewcommand\color[2][]{}%
  }%
  \providecommand\includegraphics[2][]{%
    \GenericError{(gnuplot) \space\space\space\@spaces}{%
      Package graphicx or graphics not loaded%
    }{See the gnuplot documentation for explanation.%
    }{The gnuplot epslatex terminal needs graphicx.sty or graphics.sty.}%
    \renewcommand\includegraphics[2][]{}%
  }%
  \providecommand\rotatebox[2]{#2}%
  \@ifundefined{ifGPcolor}{%
    \newif\ifGPcolor
    \GPcolorfalse
  }{}%
  \@ifundefined{ifGPblacktext}{%
    \newif\ifGPblacktext
    \GPblacktexttrue
  }{}%
  \let\gplgaddtomacro\g@addto@macro
  \gdef\gplbacktext{}%
  \gdef\gplfronttext{}%
  \makeatother
  \ifGPblacktext
    \def\colorrgb#1{}%
    \def\colorgray#1{}%
  \else
    \ifGPcolor
      \def\colorrgb#1{\color[rgb]{#1}}%
      \def\colorgray#1{\color[gray]{#1}}%
      \expandafter\def\csname LTw\endcsname{\color{white}}%
      \expandafter\def\csname LTb\endcsname{\color{black}}%
      \expandafter\def\csname LTa\endcsname{\color{black}}%
      \expandafter\def\csname LT0\endcsname{\color[rgb]{1,0,0}}%
      \expandafter\def\csname LT1\endcsname{\color[rgb]{0,1,0}}%
      \expandafter\def\csname LT2\endcsname{\color[rgb]{0,0,1}}%
      \expandafter\def\csname LT3\endcsname{\color[rgb]{1,0,1}}%
      \expandafter\def\csname LT4\endcsname{\color[rgb]{0,1,1}}%
      \expandafter\def\csname LT5\endcsname{\color[rgb]{1,1,0}}%
      \expandafter\def\csname LT6\endcsname{\color[rgb]{0,0,0}}%
      \expandafter\def\csname LT7\endcsname{\color[rgb]{1,0.3,0}}%
      \expandafter\def\csname LT8\endcsname{\color[rgb]{0.5,0.5,0.5}}%
    \else
      \def\colorrgb#1{\color{black}}%
      \def\colorgray#1{\color[gray]{#1}}%
      \expandafter\def\csname LTw\endcsname{\color{white}}%
      \expandafter\def\csname LTb\endcsname{\color{black}}%
      \expandafter\def\csname LTa\endcsname{\color{black}}%
      \expandafter\def\csname LT0\endcsname{\color{black}}%
      \expandafter\def\csname LT1\endcsname{\color{black}}%
      \expandafter\def\csname LT2\endcsname{\color{black}}%
      \expandafter\def\csname LT3\endcsname{\color{black}}%
      \expandafter\def\csname LT4\endcsname{\color{black}}%
      \expandafter\def\csname LT5\endcsname{\color{black}}%
      \expandafter\def\csname LT6\endcsname{\color{black}}%
      \expandafter\def\csname LT7\endcsname{\color{black}}%
      \expandafter\def\csname LT8\endcsname{\color{black}}%
    \fi
  \fi
    \setlength{\unitlength}{0.0500bp}%
    \ifx\gptboxheight\undefined%
      \newlength{\gptboxheight}%
      \newlength{\gptboxwidth}%
      \newsavebox{\gptboxtext}%
    \fi%
    \setlength{\fboxrule}{0.5pt}%
    \setlength{\fboxsep}{1pt}%
    \definecolor{tbcol}{rgb}{1,1,1}%
\begin{picture}(6120.00,4284.00)%
    \gplgaddtomacro\gplbacktext{%
      \csname LTb\endcsname
      \put(198,1181){\makebox(0,0)[r]{\strut{}$-1$}}%
      \put(198,2648){\makebox(0,0)[r]{\strut{}$-0.5$}}%
      \put(198,4114){\makebox(0,0)[r]{\strut{}$0$}}%
      \put(468,374){\makebox(0,0){\strut{}$0$}}%
      \put(1843,374){\makebox(0,0){\strut{}$0.5$}}%
      \put(3218,374){\makebox(0,0){\strut{}$1$}}%
      \put(4593,374){\makebox(0,0){\strut{}$1.5$}}%
      \put(5968,374){\makebox(0,0){\strut{}$2$}}%
    }%
    \gplgaddtomacro\gplfronttext{%
      \csname LTb\endcsname
      \put(-539,2427){\rotatebox{-270}{\makebox(0,0){\strut{}$-8\pi^2 Q^2 \tilde\W_2^{(3)}(-Q^2)$}}}%
      \put(3217,44){\makebox(0,0){\strut{}$Q^2$ [GeV${}^2$]}}%
      \csname LTb\endcsname
      \put(4961,3968){\makebox(0,0)[r]{\strut{}dispersive, FPV}}%
      \csname LTb\endcsname
      \put(4961,3748){\makebox(0,0)[r]{\strut{}dispersive, $g-2$ kinematics}}%
      \csname LTb\endcsname
      \put(4961,3528){\makebox(0,0)[r]{\strut{}CMV model~\cite{Czarnecki:2002nt}}}%
      \csname LTb\endcsname
      \put(4961,3308){\makebox(0,0)[r]{\strut{} }}%
      \csname LTb\endcsname
      \put(4961,3088){\makebox(0,0)[r]{\strut{} }}%
    }%
    \gplgaddtomacro\gplbacktext{%
    }%
    \gplgaddtomacro\gplfronttext{%
      \csname LTb\endcsname
      \put(-11,2427){\rotatebox{-270}{\makebox(0,0){\strut{} }}}%
      \put(3217,374){\makebox(0,0){\strut{} }}%
      \csname LTb\endcsname
      \put(4961,3968){\makebox(0,0)[r]{\strut{} }}%
      \csname LTb\endcsname
      \put(4961,3748){\makebox(0,0)[r]{\strut{} }}%
      \csname LTb\endcsname
      \put(4961,3528){\makebox(0,0)[r]{\strut{} }}%
      \csname LTb\endcsname
      \put(4961,3308){\makebox(0,0)[r]{\strut{}hQCD model~\cite{Leutgeb:2019gbz, Cappiello:2019hwh, Hirn:2005nr}}}%
      \csname LTb\endcsname
      \put(4961,3088){\makebox(0,0)[r]{\strut{}Regge-like model~\cite{Masjuan:2020jsf}}}%
    }%
    \gplbacktext
    \put(0,0){\includegraphics[width={306.00bp},height={214.20bp}]{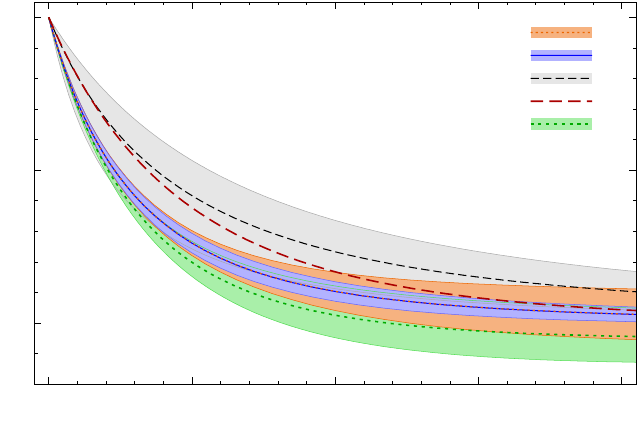}}%
    \gplfronttext
  \end{picture}%
\endgroup

%% file: plots/W1Cells.tex
\begingroup
  \makeatletter
  \providecommand\color[2][]{%
    \GenericError{(gnuplot) \space\space\space\@spaces}{%
      Package color not loaded in conjunction with
      terminal option `colourtext'%
    }{See the gnuplot documentation for explanation.%
    }{Either use 'blacktext' in gnuplot or load the package
      color.sty in LaTeX.}%
    \renewcommand\color[2][]{}%
  }%
  \providecommand\includegraphics[2][]{%
    \GenericError{(gnuplot) \space\space\space\@spaces}{%
      Package graphicx or graphics not loaded%
    }{See the gnuplot documentation for explanation.%
    }{The gnuplot epslatex terminal needs graphicx.sty or graphics.sty.}%
    \renewcommand\includegraphics[2][]{}%
  }%
  \providecommand\rotatebox[2]{#2}%
  \@ifundefined{ifGPcolor}{%
    \newif\ifGPcolor
    \GPcolorfalse
  }{}%
  \@ifundefined{ifGPblacktext}{%
    \newif\ifGPblacktext
    \GPblacktexttrue
  }{}%
  \let\gplgaddtomacro\g@addto@macro
  \gdef\gplbacktext{}%
  \gdef\gplfronttext{}%
  \makeatother
  \ifGPblacktext
    \def\colorrgb#1{}%
    \def\colorgray#1{}%
  \else
    \ifGPcolor
      \def\colorrgb#1{\color[rgb]{#1}}%
      \def\colorgray#1{\color[gray]{#1}}%
      \expandafter\def\csname LTw\endcsname{\color{white}}%
      \expandafter\def\csname LTb\endcsname{\color{black}}%
      \expandafter\def\csname LTa\endcsname{\color{black}}%
      \expandafter\def\csname LT0\endcsname{\color[rgb]{1,0,0}}%
      \expandafter\def\csname LT1\endcsname{\color[rgb]{0,1,0}}%
      \expandafter\def\csname LT2\endcsname{\color[rgb]{0,0,1}}%
      \expandafter\def\csname LT3\endcsname{\color[rgb]{1,0,1}}%
      \expandafter\def\csname LT4\endcsname{\color[rgb]{0,1,1}}%
      \expandafter\def\csname LT5\endcsname{\color[rgb]{1,1,0}}%
      \expandafter\def\csname LT6\endcsname{\color[rgb]{0,0,0}}%
      \expandafter\def\csname LT7\endcsname{\color[rgb]{1,0.3,0}}%
      \expandafter\def\csname LT8\endcsname{\color[rgb]{0.5,0.5,0.5}}%
    \else
      \def\colorrgb#1{\color{black}}%
      \def\colorgray#1{\color[gray]{#1}}%
      \expandafter\def\csname LTw\endcsname{\color{white}}%
      \expandafter\def\csname LTb\endcsname{\color{black}}%
      \expandafter\def\csname LTa\endcsname{\color{black}}%
      \expandafter\def\csname LT0\endcsname{\color{black}}%
      \expandafter\def\csname LT1\endcsname{\color{black}}%
      \expandafter\def\csname LT2\endcsname{\color{black}}%
      \expandafter\def\csname LT3\endcsname{\color{black}}%
      \expandafter\def\csname LT4\endcsname{\color{black}}%
      \expandafter\def\csname LT5\endcsname{\color{black}}%
      \expandafter\def\csname LT6\endcsname{\color{black}}%
      \expandafter\def\csname LT7\endcsname{\color{black}}%
      \expandafter\def\csname LT8\endcsname{\color{black}}%
    \fi
  \fi
    \setlength{\unitlength}{0.0500bp}%
    \ifx\gptboxheight\undefined%
      \newlength{\gptboxheight}%
      \newlength{\gptboxwidth}%
      \newsavebox{\gptboxtext}%
    \fi%
    \setlength{\fboxrule}{0.5pt}%
    \setlength{\fboxsep}{1pt}%
    \definecolor{tbcol}{rgb}{1,1,1}%
\begin{picture}(9354.00,11338.00)%
    \gplgaddtomacro\gplbacktext{%
      \csname LTb\endcsname
      \put(-106,9213){\makebox(0,0)[r]{\strut{}$-1$}}%
      \put(-106,9708){\makebox(0,0)[r]{\strut{}$-0.5$}}%
      \put(-106,10204){\makebox(0,0)[r]{\strut{}$0$}}%
      \put(-106,10699){\makebox(0,0)[r]{\strut{}$0.5$}}%
      \put(-106,11194){\makebox(0,0)[r]{\strut{}$1$}}%
      \put(99,8894){\makebox(0,0){\strut{}}}%
      \put(829,8894){\makebox(0,0){\strut{}}}%
      \put(1559,8894){\makebox(0,0){\strut{}}}%
      \put(2288,8894){\makebox(0,0){\strut{}}}%
      \put(3018,8894){\makebox(0,0){\strut{}}}%
    }%
    \gplgaddtomacro\gplfronttext{%
      \csname LTb\endcsname
      \put(-843,10203){\rotatebox{-270}{\makebox(0,0){\strut{}$-4\pi^2 Q^2 \tilde\W_1^{(3)}(-Q^2)$}}}%
      \csname LTb\endcsname
      \put(2159,9767){\makebox(0,0)[r]{\strut{}$g-2$ kin., $\pi^0$ \textbackslash{} $\pi^0$}}%
      \csname LTb\endcsname
      \put(2159,9547){\makebox(0,0)[r]{\strut{}FPV, $\pi^0$ \textbackslash{} $\pi^0$}}%
    }%
    \gplgaddtomacro\gplbacktext{%
      \csname LTb\endcsname
      \put(-106,6945){\makebox(0,0)[r]{\strut{}$-1$}}%
      \put(-106,7441){\makebox(0,0)[r]{\strut{}$-0.5$}}%
      \put(-106,7936){\makebox(0,0)[r]{\strut{}$0$}}%
      \put(-106,8431){\makebox(0,0)[r]{\strut{}$0.5$}}%
      \put(-106,8927){\makebox(0,0)[r]{\strut{}$1$}}%
      \put(99,6626){\makebox(0,0){\strut{}}}%
      \put(829,6626){\makebox(0,0){\strut{}}}%
      \put(1559,6626){\makebox(0,0){\strut{}}}%
      \put(2288,6626){\makebox(0,0){\strut{}}}%
      \put(3018,6626){\makebox(0,0){\strut{}}}%
    }%
    \gplgaddtomacro\gplfronttext{%
      \csname LTb\endcsname
      \put(-843,7936){\rotatebox{-270}{\makebox(0,0){\strut{}$-4\pi^2 Q^2 \tilde\W_1^{(3)}(-Q^2)$}}}%
      \csname LTb\endcsname
      \put(2159,8589){\makebox(0,0)[r]{\strut{}$g-2$ kin., $2\pi$ \textbackslash{} $\pi^0$}}%
      \csname LTb\endcsname
      \put(2159,8369){\makebox(0,0)[r]{\strut{}FPV, $2\pi$ \textbackslash{} $\pi^0$}}%
    }%
    \gplgaddtomacro\gplbacktext{%
      \csname LTb\endcsname
      \put(3012,6945){\makebox(0,0)[r]{\strut{}}}%
      \put(3012,7441){\makebox(0,0)[r]{\strut{}}}%
      \put(3012,7936){\makebox(0,0)[r]{\strut{}}}%
      \put(3012,8431){\makebox(0,0)[r]{\strut{}}}%
      \put(3012,8927){\makebox(0,0)[r]{\strut{}}}%
      \put(3217,6626){\makebox(0,0){\strut{}}}%
      \put(3946,6626){\makebox(0,0){\strut{}}}%
      \put(4676,6626){\makebox(0,0){\strut{}}}%
      \put(5406,6626){\makebox(0,0){\strut{}}}%
      \put(6135,6626){\makebox(0,0){\strut{}}}%
    }%
    \gplgaddtomacro\gplfronttext{%
      \csname LTb\endcsname
      \put(5276,7499){\makebox(0,0)[r]{\strut{}$g-2$ kin., $2\pi$ \textbackslash{} $a_1$}}%
      \csname LTb\endcsname
      \put(5276,7279){\makebox(0,0)[r]{\strut{}FPV, $2\pi$ \textbackslash{} $a_1$}}%
    }%
    \gplgaddtomacro\gplbacktext{%
      \csname LTb\endcsname
      \put(6130,6945){\makebox(0,0)[r]{\strut{}}}%
      \put(6130,7441){\makebox(0,0)[r]{\strut{}}}%
      \put(6130,7936){\makebox(0,0)[r]{\strut{}}}%
      \put(6130,8431){\makebox(0,0)[r]{\strut{}}}%
      \put(6130,8927){\makebox(0,0)[r]{\strut{}}}%
      \put(6335,6626){\makebox(0,0){\strut{}}}%
      \put(7064,6626){\makebox(0,0){\strut{}}}%
      \put(7794,6626){\makebox(0,0){\strut{}}}%
      \put(8524,6626){\makebox(0,0){\strut{}}}%
      \put(9253,6626){\makebox(0,0){\strut{}}}%
    }%
    \gplgaddtomacro\gplfronttext{%
      \csname LTb\endcsname
      \put(8394,7499){\makebox(0,0)[r]{\strut{}$g-2$ kin., $2\pi$ \textbackslash{} all}}%
      \csname LTb\endcsname
      \put(8394,7279){\makebox(0,0)[r]{\strut{}FPV, $2\pi$ \textbackslash{} all}}%
    }%
    \gplgaddtomacro\gplbacktext{%
      \csname LTb\endcsname
      \put(-106,4678){\makebox(0,0)[r]{\strut{}$-1$}}%
      \put(-106,5173){\makebox(0,0)[r]{\strut{}$-0.5$}}%
      \put(-106,5669){\makebox(0,0)[r]{\strut{}$0$}}%
      \put(-106,6164){\makebox(0,0)[r]{\strut{}$0.5$}}%
      \put(-106,6659){\makebox(0,0)[r]{\strut{}$1$}}%
      \put(99,4359){\makebox(0,0){\strut{}}}%
      \put(829,4359){\makebox(0,0){\strut{}}}%
      \put(1559,4359){\makebox(0,0){\strut{}}}%
      \put(2288,4359){\makebox(0,0){\strut{}}}%
      \put(3018,4359){\makebox(0,0){\strut{}}}%
    }%
    \gplgaddtomacro\gplfronttext{%
      \csname LTb\endcsname
      \put(-843,5668){\rotatebox{-270}{\makebox(0,0){\strut{}$-4\pi^2 Q^2 \tilde\W_1^{(3)}(-Q^2)$}}}%
      \csname LTb\endcsname
      \put(2159,6321){\makebox(0,0)[r]{\strut{}$g-2$ kin., $\omega$ \textbackslash{} $\pi^0$}}%
      \csname LTb\endcsname
      \put(2159,6101){\makebox(0,0)[r]{\strut{}FPV, $\omega$ \textbackslash{} $\pi^0$}}%
    }%
    \gplgaddtomacro\gplbacktext{%
      \csname LTb\endcsname
      \put(3012,4678){\makebox(0,0)[r]{\strut{}}}%
      \put(3012,5173){\makebox(0,0)[r]{\strut{}}}%
      \put(3012,5669){\makebox(0,0)[r]{\strut{}}}%
      \put(3012,6164){\makebox(0,0)[r]{\strut{}}}%
      \put(3012,6659){\makebox(0,0)[r]{\strut{}}}%
      \put(3217,4359){\makebox(0,0){\strut{}}}%
      \put(3946,4359){\makebox(0,0){\strut{}}}%
      \put(4676,4359){\makebox(0,0){\strut{}}}%
      \put(5406,4359){\makebox(0,0){\strut{}}}%
      \put(6135,4359){\makebox(0,0){\strut{}}}%
    }%
    \gplgaddtomacro\gplfronttext{%
      \csname LTb\endcsname
      \put(5276,5232){\makebox(0,0)[r]{\strut{}$g-2$ kin., $\omega$ \textbackslash{} $a_1$}}%
      \csname LTb\endcsname
      \put(5276,5012){\makebox(0,0)[r]{\strut{}FPV, $\omega$ \textbackslash{} $a_1$}}%
    }%
    \gplgaddtomacro\gplbacktext{%
      \csname LTb\endcsname
      \put(-106,143){\makebox(0,0)[r]{\strut{}$-1$}}%
      \put(-106,638){\makebox(0,0)[r]{\strut{}$-0.5$}}%
      \put(-106,1134){\makebox(0,0)[r]{\strut{}$0$}}%
      \put(-106,1629){\makebox(0,0)[r]{\strut{}$0.5$}}%
      \put(-106,2124){\makebox(0,0)[r]{\strut{}$1$}}%
      \put(99,-176){\makebox(0,0){\strut{}$0$}}%
      \put(829,-176){\makebox(0,0){\strut{}$0.5$}}%
      \put(1559,-176){\makebox(0,0){\strut{}$1$}}%
      \put(2288,-176){\makebox(0,0){\strut{}$1.5$}}%
      \put(3018,-176){\makebox(0,0){\strut{}$2$}}%
    }%
    \gplgaddtomacro\gplfronttext{%
      \csname LTb\endcsname
      \put(-843,1133){\rotatebox{-270}{\makebox(0,0){\strut{}$-4\pi^2 Q^2 \tilde\W_1^{(3)}(-Q^2)$}}}%
      \put(1558,-506){\makebox(0,0){\strut{}$Q^2$ [GeV${}^2$]}}%
      \csname LTb\endcsname
      \put(2159,1786){\makebox(0,0)[r]{\strut{}FPV, \ldots{} \textbackslash{} $\pi^0$}}%
    }%
    \gplgaddtomacro\gplbacktext{%
      \csname LTb\endcsname
      \put(3012,143){\makebox(0,0)[r]{\strut{}}}%
      \put(3012,638){\makebox(0,0)[r]{\strut{}}}%
      \put(3012,1134){\makebox(0,0)[r]{\strut{}}}%
      \put(3012,1629){\makebox(0,0)[r]{\strut{}}}%
      \put(3012,2124){\makebox(0,0)[r]{\strut{}}}%
      \put(3217,-176){\makebox(0,0){\strut{}$0$}}%
      \put(3946,-176){\makebox(0,0){\strut{}$0.5$}}%
      \put(4676,-176){\makebox(0,0){\strut{}$1$}}%
      \put(5406,-176){\makebox(0,0){\strut{}$1.5$}}%
      \put(6135,-176){\makebox(0,0){\strut{}$2$}}%
    }%
    \gplgaddtomacro\gplfronttext{%
      \csname LTb\endcsname
      \put(4676,-506){\makebox(0,0){\strut{}$Q^2$ [GeV${}^2$]}}%
      \csname LTb\endcsname
      \put(5276,1786){\makebox(0,0)[r]{\strut{}FPV, \ldots{} \textbackslash{} $a_1$}}%
    }%
    \gplgaddtomacro\gplbacktext{%
      \csname LTb\endcsname
      \put(6130,143){\makebox(0,0)[r]{\strut{}}}%
      \put(6130,638){\makebox(0,0)[r]{\strut{}}}%
      \put(6130,1134){\makebox(0,0)[r]{\strut{}}}%
      \put(6130,1629){\makebox(0,0)[r]{\strut{}}}%
      \put(6130,2124){\makebox(0,0)[r]{\strut{}}}%
      \put(6335,-176){\makebox(0,0){\strut{}$0$}}%
      \put(7064,-176){\makebox(0,0){\strut{}$0.5$}}%
      \put(7794,-176){\makebox(0,0){\strut{}$1$}}%
      \put(8524,-176){\makebox(0,0){\strut{}$1.5$}}%
      \put(9253,-176){\makebox(0,0){\strut{}$2$}}%
    }%
    \gplgaddtomacro\gplfronttext{%
      \csname LTb\endcsname
      \put(7794,-506){\makebox(0,0){\strut{}$Q^2$ [GeV${}^2$]}}%
      \csname LTb\endcsname
      \put(8394,697){\makebox(0,0)[r]{\strut{}FPV, \ldots{} \textbackslash{} \ldots{}}}%
      \csname LTb\endcsname
      \put(8394,477){\makebox(0,0)[r]{\strut{}$g-2$ kin., \ldots{} \textbackslash{} all}}%
      \csname LTb\endcsname
      \put(8394,257){\makebox(0,0)[r]{\strut{}FPV, \ldots{} \textbackslash{} all}}%
    }%
    \gplbacktext
    \put(0,0){\includegraphics[width={467.70bp},height={566.90bp}]{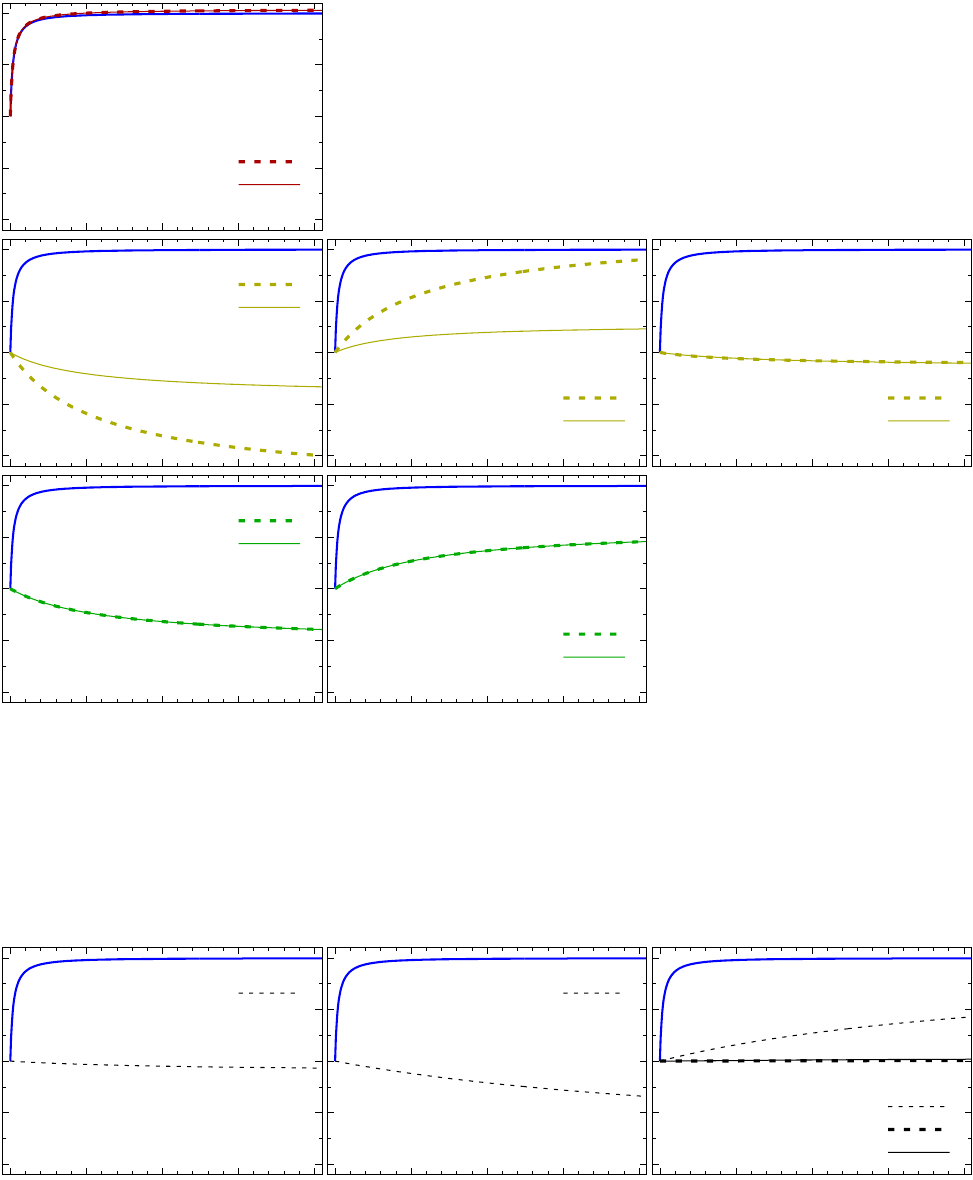}}%
    \gplfronttext
  \end{picture}%
\endgroup

%% file: plots/W2Cells.tex
\begingroup
  \makeatletter
  \providecommand\color[2][]{%
    \GenericError{(gnuplot) \space\space\space\@spaces}{%
      Package color not loaded in conjunction with
      terminal option `colourtext'%
    }{See the gnuplot documentation for explanation.%
    }{Either use 'blacktext' in gnuplot or load the package
      color.sty in LaTeX.}%
    \renewcommand\color[2][]{}%
  }%
  \providecommand\includegraphics[2][]{%
    \GenericError{(gnuplot) \space\space\space\@spaces}{%
      Package graphicx or graphics not loaded%
    }{See the gnuplot documentation for explanation.%
    }{The gnuplot epslatex terminal needs graphicx.sty or graphics.sty.}%
    \renewcommand\includegraphics[2][]{}%
  }%
  \providecommand\rotatebox[2]{#2}%
  \@ifundefined{ifGPcolor}{%
    \newif\ifGPcolor
    \GPcolorfalse
  }{}%
  \@ifundefined{ifGPblacktext}{%
    \newif\ifGPblacktext
    \GPblacktexttrue
  }{}%
  \let\gplgaddtomacro\g@addto@macro
  \gdef\gplbacktext{}%
  \gdef\gplfronttext{}%
  \makeatother
  \ifGPblacktext
    \def\colorrgb#1{}%
    \def\colorgray#1{}%
  \else
    \ifGPcolor
      \def\colorrgb#1{\color[rgb]{#1}}%
      \def\colorgray#1{\color[gray]{#1}}%
      \expandafter\def\csname LTw\endcsname{\color{white}}%
      \expandafter\def\csname LTb\endcsname{\color{black}}%
      \expandafter\def\csname LTa\endcsname{\color{black}}%
      \expandafter\def\csname LT0\endcsname{\color[rgb]{1,0,0}}%
      \expandafter\def\csname LT1\endcsname{\color[rgb]{0,1,0}}%
      \expandafter\def\csname LT2\endcsname{\color[rgb]{0,0,1}}%
      \expandafter\def\csname LT3\endcsname{\color[rgb]{1,0,1}}%
      \expandafter\def\csname LT4\endcsname{\color[rgb]{0,1,1}}%
      \expandafter\def\csname LT5\endcsname{\color[rgb]{1,1,0}}%
      \expandafter\def\csname LT6\endcsname{\color[rgb]{0,0,0}}%
      \expandafter\def\csname LT7\endcsname{\color[rgb]{1,0.3,0}}%
      \expandafter\def\csname LT8\endcsname{\color[rgb]{0.5,0.5,0.5}}%
    \else
      \def\colorrgb#1{\color{black}}%
      \def\colorgray#1{\color[gray]{#1}}%
      \expandafter\def\csname LTw\endcsname{\color{white}}%
      \expandafter\def\csname LTb\endcsname{\color{black}}%
      \expandafter\def\csname LTa\endcsname{\color{black}}%
      \expandafter\def\csname LT0\endcsname{\color{black}}%
      \expandafter\def\csname LT1\endcsname{\color{black}}%
      \expandafter\def\csname LT2\endcsname{\color{black}}%
      \expandafter\def\csname LT3\endcsname{\color{black}}%
      \expandafter\def\csname LT4\endcsname{\color{black}}%
      \expandafter\def\csname LT5\endcsname{\color{black}}%
      \expandafter\def\csname LT6\endcsname{\color{black}}%
      \expandafter\def\csname LT7\endcsname{\color{black}}%
      \expandafter\def\csname LT8\endcsname{\color{black}}%
    \fi
  \fi
    \setlength{\unitlength}{0.0500bp}%
    \ifx\gptboxheight\undefined%
      \newlength{\gptboxheight}%
      \newlength{\gptboxwidth}%
      \newsavebox{\gptboxtext}%
    \fi%
    \setlength{\fboxrule}{0.5pt}%
    \setlength{\fboxsep}{1pt}%
    \definecolor{tbcol}{rgb}{1,1,1}%
\begin{picture}(3118.00,9070.00)%
    \gplgaddtomacro\gplbacktext{%
      \csname LTb\endcsname
      \put(-106,6888){\makebox(0,0)[r]{\strut{}$-2$}}%
      \put(-106,7307){\makebox(0,0)[r]{\strut{}$-1$}}%
      \put(-106,7726){\makebox(0,0)[r]{\strut{}$0$}}%
      \put(-106,8145){\makebox(0,0)[r]{\strut{}$1$}}%
      \put(-106,8564){\makebox(0,0)[r]{\strut{}$2$}}%
      \put(-106,8983){\makebox(0,0)[r]{\strut{}$3$}}%
      \put(99,6626){\makebox(0,0){\strut{}}}%
      \put(828,6626){\makebox(0,0){\strut{}}}%
      \put(1558,6626){\makebox(0,0){\strut{}}}%
      \put(2288,6626){\makebox(0,0){\strut{}}}%
      \put(3017,6626){\makebox(0,0){\strut{}}}%
    }%
    \gplgaddtomacro\gplfronttext{%
      \csname LTb\endcsname
      \put(-579,7935){\rotatebox{-270}{\makebox(0,0){\strut{}$-8\pi^2 Q^2 \tilde\W_2^{(3)}(-Q^2)$}}}%
      \csname LTb\endcsname
      \put(2158,8588){\makebox(0,0)[r]{\strut{}$g-2$ kin., $2\pi$ \textbackslash{} $a_1$}}%
      \csname LTb\endcsname
      \put(2158,8368){\makebox(0,0)[r]{\strut{}FPV, $2\pi$ \textbackslash{} $a_1$}}%
    }%
    \gplgaddtomacro\gplbacktext{%
      \csname LTb\endcsname
      \put(-106,4621){\makebox(0,0)[r]{\strut{}$-2$}}%
      \put(-106,5040){\makebox(0,0)[r]{\strut{}$-1$}}%
      \put(-106,5459){\makebox(0,0)[r]{\strut{}$0$}}%
      \put(-106,5878){\makebox(0,0)[r]{\strut{}$1$}}%
      \put(-106,6297){\makebox(0,0)[r]{\strut{}$2$}}%
      \put(-106,6716){\makebox(0,0)[r]{\strut{}$3$}}%
      \put(99,4359){\makebox(0,0){\strut{}}}%
      \put(828,4359){\makebox(0,0){\strut{}}}%
      \put(1558,4359){\makebox(0,0){\strut{}}}%
      \put(2288,4359){\makebox(0,0){\strut{}}}%
      \put(3017,4359){\makebox(0,0){\strut{}}}%
    }%
    \gplgaddtomacro\gplfronttext{%
      \csname LTb\endcsname
      \put(-579,5668){\rotatebox{-270}{\makebox(0,0){\strut{}$-8\pi^2 Q^2 \tilde\W_2^{(3)}(-Q^2)$}}}%
      \csname LTb\endcsname
      \put(2158,6321){\makebox(0,0)[r]{\strut{}$g-2$ kin., $\omega$ \textbackslash{} $a_1$}}%
      \csname LTb\endcsname
      \put(2158,6101){\makebox(0,0)[r]{\strut{}FPV, $\omega$ \textbackslash{} $a_1$}}%
    }%
    \gplgaddtomacro\gplbacktext{%
      \csname LTb\endcsname
      \put(-106,2353){\makebox(0,0)[r]{\strut{}$-2$}}%
      \put(-106,2772){\makebox(0,0)[r]{\strut{}$-1$}}%
      \put(-106,3191){\makebox(0,0)[r]{\strut{}$0$}}%
      \put(-106,3611){\makebox(0,0)[r]{\strut{}$1$}}%
      \put(-106,4030){\makebox(0,0)[r]{\strut{}$2$}}%
      \put(-106,4449){\makebox(0,0)[r]{\strut{}$3$}}%
      \put(99,2091){\makebox(0,0){\strut{}}}%
      \put(828,2091){\makebox(0,0){\strut{}}}%
      \put(1558,2091){\makebox(0,0){\strut{}}}%
      \put(2288,2091){\makebox(0,0){\strut{}}}%
      \put(3017,2091){\makebox(0,0){\strut{}}}%
    }%
    \gplgaddtomacro\gplfronttext{%
      \csname LTb\endcsname
      \put(-579,3401){\rotatebox{-270}{\makebox(0,0){\strut{}$-8\pi^2 Q^2 \tilde\W_2^{(3)}(-Q^2)$}}}%
      \csname LTb\endcsname
      \put(2158,3291){\makebox(0,0)[r]{\strut{}$g-2$ kin., $a_1$ \textbackslash{} $a_1$}}%
      \csname LTb\endcsname
      \put(2158,3071){\makebox(0,0)[r]{\strut{}FPV, $a_1$ \textbackslash{} $a_1$}}%
    }%
    \gplgaddtomacro\gplbacktext{%
      \csname LTb\endcsname
      \put(-106,86){\makebox(0,0)[r]{\strut{}$-2$}}%
      \put(-106,505){\makebox(0,0)[r]{\strut{}$-1$}}%
      \put(-106,924){\makebox(0,0)[r]{\strut{}$0$}}%
      \put(-106,1343){\makebox(0,0)[r]{\strut{}$1$}}%
      \put(-106,1762){\makebox(0,0)[r]{\strut{}$2$}}%
      \put(-106,2181){\makebox(0,0)[r]{\strut{}$3$}}%
      \put(99,-176){\makebox(0,0){\strut{}$0$}}%
      \put(828,-176){\makebox(0,0){\strut{}$0.5$}}%
      \put(1558,-176){\makebox(0,0){\strut{}$1$}}%
      \put(2288,-176){\makebox(0,0){\strut{}$1.5$}}%
      \put(3017,-176){\makebox(0,0){\strut{}$2$}}%
    }%
    \gplgaddtomacro\gplfronttext{%
      \csname LTb\endcsname
      \put(-579,1133){\rotatebox{-270}{\makebox(0,0){\strut{}$-8\pi^2 Q^2 \tilde\W_2^{(3)}(-Q^2)$}}}%
      \put(1558,-506){\makebox(0,0){\strut{}$Q^2$ [GeV${}^2$]}}%
      \csname LTb\endcsname
      \put(2158,1786){\makebox(0,0)[r]{\strut{}FPV, \ldots{} \textbackslash{} $a_1$}}%
      \csname LTb\endcsname
      \put(2158,1566){\makebox(0,0)[r]{\strut{}FPV, \ldots{} \textbackslash{} \ldots{}}}%
      \csname LTb\endcsname
      \put(2158,1346){\makebox(0,0)[r]{\strut{}$g-2$ kin., \ldots{} \textbackslash{} all}}%
      \csname LTb\endcsname
      \put(2158,1126){\makebox(0,0)[r]{\strut{}FPV, \ldots{} \textbackslash{} all}}%
    }%
    \gplbacktext
    \put(0,0){\includegraphics[width={155.90bp},height={453.50bp}]{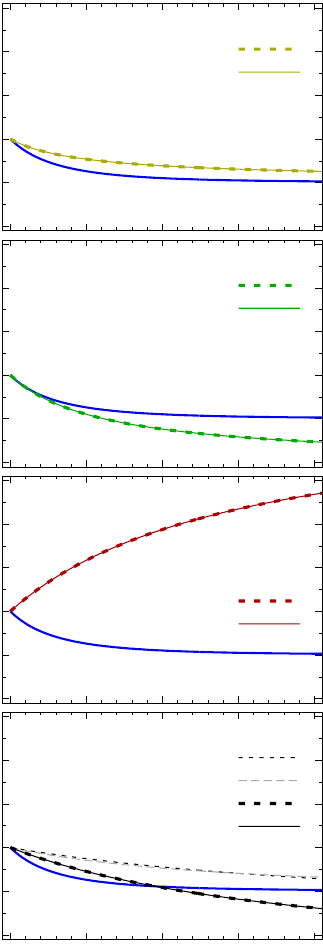}}%
    \gplfronttext
  \end{picture}%
\endgroup

%% file: plots/W1Diff.tex
\begingroup
  \makeatletter
  \providecommand\color[2][]{%
    \GenericError{(gnuplot) \space\space\space\@spaces}{%
      Package color not loaded in conjunction with
      terminal option `colourtext'%
    }{See the gnuplot documentation for explanation.%
    }{Either use 'blacktext' in gnuplot or load the package
      color.sty in LaTeX.}%
    \renewcommand\color[2][]{}%
  }%
  \providecommand\includegraphics[2][]{%
    \GenericError{(gnuplot) \space\space\space\@spaces}{%
      Package graphicx or graphics not loaded%
    }{See the gnuplot documentation for explanation.%
    }{The gnuplot epslatex terminal needs graphicx.sty or graphics.sty.}%
    \renewcommand\includegraphics[2][]{}%
  }%
  \providecommand\rotatebox[2]{#2}%
  \@ifundefined{ifGPcolor}{%
    \newif\ifGPcolor
    \GPcolorfalse
  }{}%
  \@ifundefined{ifGPblacktext}{%
    \newif\ifGPblacktext
    \GPblacktexttrue
  }{}%
  \let\gplgaddtomacro\g@addto@macro
  \gdef\gplbacktext{}%
  \gdef\gplfronttext{}%
  \makeatother
  \ifGPblacktext
    \def\colorrgb#1{}%
    \def\colorgray#1{}%
  \else
    \ifGPcolor
      \def\colorrgb#1{\color[rgb]{#1}}%
      \def\colorgray#1{\color[gray]{#1}}%
      \expandafter\def\csname LTw\endcsname{\color{white}}%
      \expandafter\def\csname LTb\endcsname{\color{black}}%
      \expandafter\def\csname LTa\endcsname{\color{black}}%
      \expandafter\def\csname LT0\endcsname{\color[rgb]{1,0,0}}%
      \expandafter\def\csname LT1\endcsname{\color[rgb]{0,1,0}}%
      \expandafter\def\csname LT2\endcsname{\color[rgb]{0,0,1}}%
      \expandafter\def\csname LT3\endcsname{\color[rgb]{1,0,1}}%
      \expandafter\def\csname LT4\endcsname{\color[rgb]{0,1,1}}%
      \expandafter\def\csname LT5\endcsname{\color[rgb]{1,1,0}}%
      \expandafter\def\csname LT6\endcsname{\color[rgb]{0,0,0}}%
      \expandafter\def\csname LT7\endcsname{\color[rgb]{1,0.3,0}}%
      \expandafter\def\csname LT8\endcsname{\color[rgb]{0.5,0.5,0.5}}%
    \else
      \def\colorrgb#1{\color{black}}%
      \def\colorgray#1{\color[gray]{#1}}%
      \expandafter\def\csname LTw\endcsname{\color{white}}%
      \expandafter\def\csname LTb\endcsname{\color{black}}%
      \expandafter\def\csname LTa\endcsname{\color{black}}%
      \expandafter\def\csname LT0\endcsname{\color{black}}%
      \expandafter\def\csname LT1\endcsname{\color{black}}%
      \expandafter\def\csname LT2\endcsname{\color{black}}%
      \expandafter\def\csname LT3\endcsname{\color{black}}%
      \expandafter\def\csname LT4\endcsname{\color{black}}%
      \expandafter\def\csname LT5\endcsname{\color{black}}%
      \expandafter\def\csname LT6\endcsname{\color{black}}%
      \expandafter\def\csname LT7\endcsname{\color{black}}%
      \expandafter\def\csname LT8\endcsname{\color{black}}%
    \fi
  \fi
    \setlength{\unitlength}{0.0500bp}%
    \ifx\gptboxheight\undefined%
      \newlength{\gptboxheight}%
      \newlength{\gptboxwidth}%
      \newsavebox{\gptboxtext}%
    \fi%
    \setlength{\fboxrule}{0.5pt}%
    \setlength{\fboxsep}{1pt}%
    \definecolor{tbcol}{rgb}{1,1,1}%
\begin{picture}(6120.00,4284.00)%
    \gplgaddtomacro\gplbacktext{%
      \csname LTb\endcsname
      \put(198,735){\makebox(0,0)[r]{\strut{}$-0.2$}}%
      \put(198,1299){\makebox(0,0)[r]{\strut{}$0$}}%
      \put(198,1863){\makebox(0,0)[r]{\strut{}$0.2$}}%
      \put(198,2428){\makebox(0,0)[r]{\strut{}$0.4$}}%
      \put(198,2992){\makebox(0,0)[r]{\strut{}$0.6$}}%
      \put(198,3556){\makebox(0,0)[r]{\strut{}$0.8$}}%
      \put(198,4120){\makebox(0,0)[r]{\strut{}$1$}}%
      \put(468,374){\makebox(0,0){\strut{}$0$}}%
      \put(1843,374){\makebox(0,0){\strut{}$0.5$}}%
      \put(3218,374){\makebox(0,0){\strut{}$1$}}%
      \put(4593,374){\makebox(0,0){\strut{}$1.5$}}%
      \put(5968,374){\makebox(0,0){\strut{}$2$}}%
    }%
    \gplgaddtomacro\gplfronttext{%
      \csname LTb\endcsname
      \put(-539,2427){\rotatebox{-270}{\makebox(0,0){\strut{}$-4\pi^2 Q^2 \tilde\W_1^{(3)}(-Q^2)$}}}%
      \put(3217,44){\makebox(0,0){\strut{}$Q^2$ [GeV${}^2$]}}%
      \csname LTb\endcsname
      \put(5163,3693){\makebox(0,0)[r]{\strut{}difference $2\pi$ row}}%
      \csname LTb\endcsname
      \put(5163,3473){\makebox(0,0)[r]{\strut{}difference eff.\ poles}}%
    }%
    \gplgaddtomacro\gplbacktext{%
    }%
    \gplgaddtomacro\gplfronttext{%
      \csname LTb\endcsname
      \put(-11,2427){\rotatebox{-270}{\makebox(0,0){\strut{} }}}%
      \put(3217,374){\makebox(0,0){\strut{} }}%
      \csname LTb\endcsname
      \put(5163,3693){\makebox(0,0)[r]{\strut{} }}%
      \csname LTb\endcsname
      \put(5163,3473){\makebox(0,0)[r]{\strut{} }}%
      \csname LTb\endcsname
      \put(5163,3253){\makebox(0,0)[r]{\strut{}total, $g-2$ kinematics}}%
    }%
    \gplbacktext
    \put(0,0){\includegraphics[width={306.00bp},height={214.20bp}]{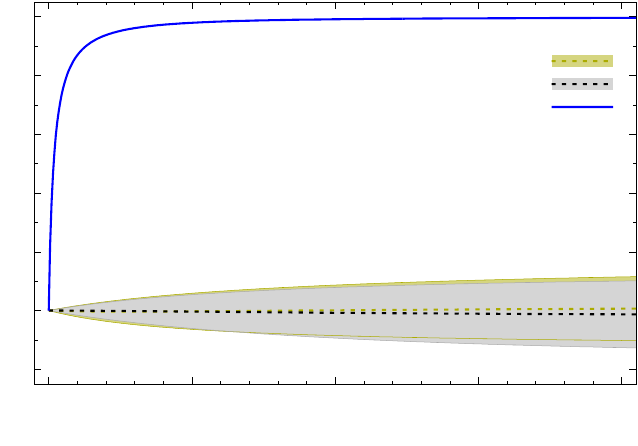}}%
    \gplfronttext
  \end{picture}%
\endgroup

%% file: sections/Conclusions.tex

\section{Conclusions}
\label{sec:Conclusions}

In the present work, we have analyzed the hadronic VVA correlator and its role in the EW contribution to the anomalous magnetic moment of the muon, introducing two dispersive approaches. In the first variant, we use dispersion relations for fixed photon virtualities, employing unitarity in the channel of the axial-current virtuality. This approach is analogous to the established dispersive approach to HLbL~\cite{Colangelo:2015ama,Colangelo:2017fiz}. In the case of HLbL, the more complicated tensor structure and redundancies in its decomposition lead to the appearance of spurious kinematic singularities that prevent an unambiguous evaluation of all contributions beyond spin 1. We proposed a solution to this problem in Ref.~\cite{Ludtke:2023hvz} by introducing a dispersive approach directly in the kinematic limit of a soft external photon. The comparison and combination of the two dispersive approaches requires a careful study of the reshuffling of the contribution of different intermediate states. In order to better understand the relation between the two dispersion relations, here we also establish the second dispersive approach for VVA, writing dispersion relations directly in $g-2$ kinematics, i.e., for one soft photon. We have performed a detailed analysis of the reshuffling, both analytically in Sect.~\ref{sec:Comparison} and numerically in Sect.~\ref{sec:Reshuffling}. As consistency requires, the two dispersive approaches agree on all contributions that are included explicitly in the unitarity relations, whereas we find differences that are due to only partially included higher intermediate states. In Sect.~\ref{sec:Comparison}, we have provided arguments why we expect the dispersion relation for fixed photon virtualities to be affected more severely by the truncation of intermediate states. In the numerical analysis, we find that a partial cancellation happens between these effects in the $\pi^0$-pole and $a_1$-pole contributions, so that within errors both dispersion relations give essentially identical results. In order to fulfill the \chpt{} constraints as well as the SDCs, the dispersion relations for fixed photon virtualities require the introduction of more effective-pole parameters than the dispersion relations in $g-2$ kinematics. These modeled matching contributions are degenerate with the ones in the sub-processes, so that an uncorrelated variation of all parameters leads to a larger uncertainty estimate than what we obtain from the dispersion relations in $g-2$ kinematics.

We quote again our central result for the first-family VVA contribution to $a_\mu$,
\begin{equation}
	a_{\mu}^\text{VVA}[u, d, e] = -2.08(3) \times 10^{-11} \,, \tag{\ref{eq:VVA_res_gm2}}
\end{equation}
which reduces the uncertainty in comparison to Ref.~\cite{Czarnecki:2002nt} by almost an order of magnitude and motivates an overall update of the EW contribution to $a_\mu$~\cite{Hoferichter:2025yih}. Of course, in the context of the present uncertainties in the HVP and HLbL contributions, this improvement is irrelevant for the comparison of the SM prediction of $a_\mu$ with the experiment. However, the present analysis paves the way for a similar complete analysis of HLbL, on the one hand by elucidating the relation between the two dispersive approaches, on the other hand by providing input for the asymptotic matching of HLbL, where the VVA correlator is the leading contribution to the mixed-region OPE~\cite{Melnikov:2003xd}. Importantly, a side-product of our analysis is a dispersive treatment of the singly-virtual $a_1$ TFF $\F_2^{a_1}$, which is constrained by VVA itself. The normalization of the TFF is an output of the analysis and perfectly agrees with the $U(3)$ estimate. Since experimental information on the axial-vector TFFs is scarce, these results are very valuable input for the evaluation of axial-vector contributions to HLbL.

%% file: sections/Nomenclature.tex

\section{Overview of the nomenclature and conventions}
\label{sec:Conventions}

\begin{table}[t]
	\centering
	\small
	\begin{tabular}{l l l l l l l}
	\toprule
	process & definitions & amplitude & tensors & scalar func.
	\\
	\midrule \\[-0.52cm] \midrule
	VVA & \eqref{eq:defW}, \eqref{eq:TensorDecompVVA} & $\W^{\mu\nu\rho}$ & $\tau_i^{\mu\nu\rho}$ & $\W_i$ \\
	VVA, soft-photon limit & \eqref{eq:decompositiongm2} & $\W^{\mu\nu\rho}$ & $q_{2\sigma} \tau_i^{\mu\nu\rho\sigma}$ & $\tilde\W_i$ \\
	\midrule
	$V\to\gamma^*a^*$ & \eqref{eq:gammaAVMatrixElement} & $\V^{\mu\nu\rho}$ & $\tau_i^{\mu\nu\rho}$ & $\V_i$ \\
	$V\to\gamma a^*$, soft-photon limit & \eqref{eq:VtildeDef} & $\V^{\mu\nu\rho}$ & $q_{2\sigma} \tau_i^{\mu\nu\rho\sigma}$ & $\tilde\V_i$ \\
	$V$ decay constant, $V\to \gamma^*$ & \eqref{eq:VectorDecayConst} & & & $f_V$ \\
	\midrule \\[-0.52cm] \midrule
	pion VFF & \eqref{eq:pionVFF} & & & $F_\pi^V$ \\
	\midrule \\[-0.52cm] \midrule
	$P$ TFF, $P\to\gamma^*\gamma^*$ & \eqref{eq:pionTFF} & & & $F_{P\gamma^*\gamma^*}$ \\
	singly-virtual isovector $P$ TFF & \eqref{eq:pseudoscalarTFFIVIS} & & & $F_v^P$ \\
	singly-virtual isoscalar $P$ TFF & \eqref{eq:pseudoscalarTFFIVIS} & & & $F_s^P$ \\
	$P$ decay constant, $P\to a^*$ & \eqref{eq:AxialPion} & & & $F_P^a$ \\
	\midrule
	$A$ TFF, $A\to\gamma^*\gamma^*$ & \eqref{eq:axialGammaGamma}, \eqref{eq:AxialTFFDefinition} & $\M^{\mu\nu\alpha}$ & $T_i^{\mu\nu\alpha}$ & $\F_i^A$ \\
	$A$ decay constant, $A\to a^*$ & \eqref{eq:AxialMatEl} & & & $F_A^a$ \\
	\midrule \\[-0.52cm] \midrule
	$P\gamma^*\to\pi\pi$ & \eqref{eq:gammaPiPiPScFun} & & & $\F_P$ \\
	\midrule
	$a^*\gamma^*\to\pi\pi$ & \eqref{eq:defU}, \eqref{eq:tensorDecompU} & $\U^{\nu\rho}$ & $t_i^{\nu\rho}$ & $\U_i$ \\
	$a^*\gamma\to\pi\pi$, soft-photon limit & \eqref{eq:tensorDecompUSoftLimit} & $\U^{\nu\rho}$ & $q_{2\sigma} t_i^{\nu\rho\sigma}$ & $\tilde\U_i$ \\
	$A\gamma^*\to\pi\pi$ & \eqref{eq:AgammaPiPiAmpliutude}, \eqref{eq:AgammaPiPiDecomposition} & $\A^{\nu\rho}$ & $t_i^{\nu\rho}$ & $\F_{A,i}$ \\
	$A\gamma\to\pi\pi$, soft-photon limit & \eqref{eq:FAtilde}, \eqref{eq:AgammaPiPiDecompositionSoftPhoton} & & $q_{2\sigma} t_1^{\nu\rho\sigma}$ & $\tilde\F_{A,1}$ \\
	\midrule \\[-0.52cm] \midrule
	$V\to\gamma P$ & \eqref{eq:GammaVPLorentz} & & $$ & $F_{VP}$ \\
	\midrule
	$V\to\gamma A$ & \eqref{eq:gammaAmesonVmesonMatrixElement} & $\V_A^{\mu\nu\rho}$ & $\tau_i^{\mu\nu\rho}$ & $F_{VA}^i$ \\
	\bottomrule
	\end{tabular}
	\caption{Nomenclature for the different amplitudes used in this work. $\gamma^*$ ($a^*$) refers to an off-shell vector (axial-vector) current, $\gamma$ to an on-shell photon, whereas $P$, $V$, and $A$ denote a pseudoscalar, vector, and axial-vector mesons.}
	\label{tab:Nomenclature}
\end{table}

In the following, we list our main conventions. The QED and QCD Lagrangian for the light quarks $q = (u,d,s)^T$ is
\begin{align}
	\mathcal{L}_\mathrm{QED+QCD} = - \frac{1}{4} F_{\mu\nu} F^{\mu\nu} - \frac{1}{4} G_{\mu\nu}^A G^{A,\mu\nu} + \bar q ( i \slashed D - \M ) q \, ,
\end{align}
where the covariant derivative is $D_\mu = \p_\mu + i e \Q A_\mu + i g T^A G_\mu^A$ with $SU(3)_c$ generators $T^A$, and the quark mass and charge matrices in flavor space are given by
\begin{equation}
	\M = \mathrm{diag}(m_u, m_d, m_s) \, , \quad \Q = \mathrm{diag}\left(\frac{2}{3}, -\frac{1}{3}, -\frac{1}{3}\right) \, .
\end{equation}
The photon--quark interaction Lagrangian is $\mathcal{L} = - e A_\mu j_\mathrm{em}^\mu$, with the electromagnetic current
\begin{equation}
	j_\mathrm{em}^\mu(x) = \bar q(x) \Q \gamma^\mu q(x) \, ,
\end{equation}
and the photon equations of motion are
\begin{equation}
	\p_\mu F^{\mu\nu}(x) = e j_\mathrm{em}^\mu(x) \, .
\end{equation}
The LSZ reduction for incoming and outgoing photons reads
\begin{align}
	\< \mathrm{out} | \gamma(p,\lambda) , \mathrm{in} \> &= -i e \, \epsilon_\mu^\lambda(p) \int d^4x \, e^{-i p\cdot x} \< \mathrm{out} | j_\mathrm{em}^\mu(x) | \mathrm{in} \> \, , \nn
	\< \gamma(p,\lambda) , \mathrm{out} | \mathrm{in} \> &= -i e \, \epsilon_\mu^\lambda(p)^* \int d^4x \, e^{i p\cdot x} \< \mathrm{out} | j_\mathrm{em}^\mu(x) | \mathrm{in} \> \, .
\end{align}
In order to work out unitarity relations, it is useful to employ an auxiliary axial-vector field $a_\mu$ with an interaction Lagrangian $\mathcal{L} = -g_A a^\mu(x) j_\mu^5(x)$, where the axial-vector current is defined in Eq.~\eqref{eq:VVA_axialCurrent}. For $\gamma_5$ and the Levi-Civita symbol, we are using the sign convention
\begin{equation}
	\label{eq:Gamma5LCConvention}
	\gamma_5 = - \frac{i}{4!} \epsilon_{\mu\nu\lambda\sigma} \gamma^\mu \gamma^\nu \gamma^\lambda \gamma^\sigma = i \gamma^0 \gamma^1 \gamma^2 \gamma^3 \, , \quad \epsilon^{0123} = 1 \, .
\end{equation}
The LSZ reduction for the auxiliary axial-vector particle is given by
\begin{align}
	\label{eq:LSZAuxiliaryAxial}
	\< \mathrm{out} | a(p,\lambda) , \mathrm{in} \> &= - i g_A \, \epsilon_\mu^\lambda(p) \int d^4x \, e^{-i p\cdot x} \< \mathrm{out} | j_5^\mu(x) | \mathrm{in} \> \, , \nn
	\< a(p,\lambda) , \mathrm{out} | \mathrm{in} \> &= - i g_A \, \epsilon_\mu^\lambda(p)^* \int d^4x \, e^{i p\cdot x} \< \mathrm{out} | j_5^\mu(x) | \mathrm{in} \> \, .
\end{align}

Throughout the paper, we encounter a rather large number of (tensor) matrix elements: our naming conventions for the amplitudes, their tensor decomposition, and the scalar coefficient functions is provided in Tab.~\ref{tab:Nomenclature}. For helicity amplitudes, we follow the same convention as Refs.~\cite{Colangelo:2015ama,Colangelo:2017fiz}, understanding them as reduced matrix elements without explicit factors of the electromagnetic coupling, i.e., we relate them to the $S$-matrix elements according to
\begin{equation}
	S = 1 + i T \, , \quad T = \text{(factors of $e$)} \times (2\pi)^4 \delta^{(4)}(p_\mathrm{in} - p_\mathrm{out}) \times H \, .
\end{equation}

%% file: sections/TFFs.tex

\section{Dispersive representation of transition form factors}
\label{sec:TFFs}

In this section, we discuss dispersive reconstructions of the pseudoscalar and axial-vector TFFs. The TFFs are an input in both dispersive approaches to VVA, and their dispersive reconstruction is required for a detailed comparison.

As in the dispersion relations for VVA in $g-2$ kinematics, we explicitly include two-pion and single-vector intermediate states in the unitarity relation for the TFFs. The contribution of additional intermediate states will be modeled by effective poles. In the case of the pseudoscalars, our goal is not to compete with the sophisticated dispersive analyses available in the literature~\cite{Hoferichter:2014vra, Hoferichter:2018dmo, Hoferichter:2018kwz, Hanhart:2013vba, Holz:2022hwz, Holz:2015tcg, Kubis:2018bej}, but to derive a simplified representation that is suitable for the comparison of the two dispersive approaches to VVA.

\subsection{Pseudoscalar transition form factors}
\label{sec:DispForPSTFFs}

The pion TFF can be decomposed into components where either the first or the second vector current is an isovector~\cite{Hoferichter:2018kwz}
\begin{equation}
	\label{eq:IsospinDecompositionPionTFF}
	F_{\pi^0 \gamma^*\gamma^*}(q_1^2, q_2^2) = F_{vs}^{\pi^0}(q_1^2, q_2^2) + F_{vs}^{\pi^0}(q_2^2, q_1^2)\,.
\end{equation}
Similarly, for the $\eta$ and $\eta'$ TFFs the decomposition reads~\cite{Holz:2015tcg}
\begin{equation}
	F_{\eta^{(\prime)} \gamma^*\gamma^*}(q_1^2, q_2^2) = F_{vv}^{\eta^{(\prime)}}(q_1^2, q_2^2) + F_{ss}^{\eta^{(\prime)}}(q_1^2, q_2^2)\,.
\end{equation}
In both cases, the singly-virtual TFF takes the form
\begin{equation}
	\label{eq:pseudoscalarTFFIVIS}
	F_{P \gamma^*\gamma^*}(q_1^2, 0) = F_v^P(q_1^2) + F_s^P(q_1^2)\,,
\end{equation}
where we only specify the isospin component of the current coupling to the virtual photon. Explicitly, the definitions of the isovector and isoscalar form factors read
\begin{equation}
	\label{eq:pseudoscalarTFFDefinition}
	\< \gamma(q_2, \lambda_2) | j_\mu^{v/s}(0) | P(q_1 + q_2) \> = - e \, \epsilon_{\lambda_2}^{\nu}(q_2)^* \epsilon_{\mu \nu \rho \sigma} q_1^\rho q_2^\sigma F_{v/s}^P(q_1^2) \, ,
\end{equation}
with $j_\mu^{v/s}(x)$ the isovector and isoscalar components of the electromagnetic current,
\begin{equation}
	\label{eq:currentIVIS}
	j^\mu_\mathrm{em} = j^\mu_v + j^\mu_s \,, \qquad j^\mu_v = j^\mu_3 \,, \qquad j^\mu_s = \frac{1}{\sqrt{3}} j^\mu_8 \,, \qquad j^\mu_a = \bar q \frac{\lambda_a}{2} \gamma^\mu q \, ,
\end{equation}
and $q_2^2 = 0$. We consider two-pion intermediate states in the isovector channel and narrow vector resonances in the isoscalar channel.

\subsubsection{Two-pion intermediate state}

The unitarity relation in the isovector-photon channel reads
\begin{align}
	\Delta \left[ -\imag \braket{\gamma_v^*(q_1, \lambda_1) | P(p) \gamma(q_2, \lambda_2)} \right] = \sum_n \frac{1}{2 S_n} \left(\prod_{i=1}^{N_n} \int \widetilde{\dif p_i} \right) & \braket{n;\{p_i\} | P(p) \gamma(q_2, \lambda_2)} \nn
	&\quad \times \braket{n;\{p_i\} | \gamma_v^*(q_1, \lambda_1)}^* \,.
\end{align}
Taking the two-pion contribution from the sum over intermediate states, one obtains
\begin{align}
	\Delta^{\pi\pi} & \left[ -\imag \braket{\gamma_v^*(q_1, \lambda_1) | P(p) \gamma(q_2, \lambda_2)} \right] \nn
	& \quad = \frac{1}{2} \int \widetilde{\dif p_1} \widetilde{\dif p_2} {\phantom{\big|}}_{I=1}\!\!\braket{\pi^+(p_1) \pi^-(p_2) | P(p) \gamma(q_2, \lambda_2) } \! {\phantom{\big|}}_{I=1}\!\!\braket{\pi^+(p_1) \pi^-(p_2) | \gamma_v^*(q_1, \lambda_1)}^* \,,
\end{align}
which leads to
\begin{align}
	\epsilon^{\mu \nu \rho \sigma} q_{1 \rho} q_{2 \sigma} \Delta^{\pi\pi} F_v^P(q_1^2) &= -\frac{1}{2} \int \widetilde{\dif p_1} \widetilde{\dif p_2} (2\pi)^4 \delta^{(4)}(q_1 - p_1 - p_2) \epsilon^{\nu \alpha \beta \gamma} p_{1\alpha} p_{2\beta} (q_1 - q_2)_\gamma (p_1 - p_2)^\mu  \nn
		&\qquad \times \F_P(q_1^2, (q_2 - p_1)^2, (q_2 - p_2)^2; 0) F_\pi^V(q_1^2)^*\,,
\end{align}
where the function $\F_P$ is the only scalar amplitude for $\gamma^* \to \pi^+ \pi^- P$ and defined by~\cite{Hoferichter:2012pm, Hoferichter:2017ftn, Hoferichter:2018kwz, Hoferichter:2019mqg, Holz:2015tcg}
\begin{align}
	\label{eq:gammaPiPiPScFun}
	\< 0 | j_\mathrm{em}^\mu(0) & | \pi^+(p_+) \pi^-(p_-) P(p_0) \> \nn
	&\quad = \epsilon^{\mu \nu \rho \sigma} p_{+\nu} p_{-\rho} p_{0\sigma} \, \F_P \left( (p_+ + p_-)^2, (p_- + p_0)^2, (p_+ + p_0)^2; q^2 \right)
\end{align}
and the pion VFF is defined in Eq.~\eqref{eq:pionVFF}.
Contraction with $\epsilon_{\mu \nu \rho \sigma} q_1^\rho q_2^\sigma / (2 (q_1 \cdot q_2)^2)$ simplifies this expression to
\begin{align}
	&\Delta^{\pi\pi} F_v^P(q_1^2) \nn
	&\quad = \frac{1}{2} \int \widetilde{\dif p_1} \widetilde{\dif p_2} (2\pi)^4 \delta^{(4)}(q_1 - p_1 - p_2) \frac{q_1^2 \sigma_\pi^2(q_1^2)}{4} (1 - z^2) F_\pi^V(q_1^2)^* \nn
	&\qquad \times \F_P\left( q_1^2, m_\pi^2 - \frac{1}{2} (q_1^2 - m_P^2)(1 - z \sigma_\pi(q_1^2)), m_\pi^2 - \frac{1}{2} (q_1^2 - m_P^2)(1 + z \sigma_\pi(q_1^2)); 0 \right) \,,
\end{align}
where $z = ((q_2 - p_1)^2 - (q_2 - p_2)^2) / (\sigma_\pi(q_1^2) (q_1^2 - m_P^2))$ is the cosine of the scattering angle in the $\pi\pi$ center-of-mass frame.
We perform a partial-wave expansion~\cite{Hoferichter:2018kwz}
\begin{equation}
	\label{eq:PWEPppg}
	\F_P(s, t, u; q^2) = \sum_{l \text{ odd}} f_P^l(s, q^2) P_l^\prime(z)\,,
\end{equation}
where the restriction to odd $l$ is due to the two pions being in an isovector state. 
The phase-space integrals project onto the $P$-wave
\begin{equation}
	\label{eq:2piDiscPSTFF}
	\Delta^{\pi\pi} F_v^P(q_1^2) = \frac{1}{96 \pi} q_1^2 \sigma_\pi^3(q_1^2) f_P^1(q_1^2, 0) F_\pi^V(q_1^2)^*\,,
\end{equation}
and the dispersion relation gives~\cite{Hoferichter:2018kwz}
\begin{align}
	\label{eq:2piContributionPSTFF}
	F_v^P(q_1^2) = \frac{1}{96 \pi^2} \int_{4m_\pi^2}^\infty \dif s\frac{s \sigma_\pi^3(s) f_P^1(s, 0) F_\pi^V(s)^*}{s - q_1^2} \,,
\end{align}
where we neglected contributions beyond the two-pion intermediate state.

\subsubsection{Vector-meson poles}

While isovector vector mesons will be included in the two-pion intermediate state through the $\pi\pi$-scattering phase shift, isoscalar vector mesons couple in the isospin limit only to heavier intermediate states starting from three pions. We will include the very narrow $\omega$ and $\phi$ mesons using a NWA.

We start again with the unitarity relation, this time for the isoscalar part of the TFF
\begin{align}
	&\Delta \left[ -\imag \braket{\gamma_s^*(q_1, \lambda_1) | P(p) \gamma(q_2, \lambda_2)} \right] \nn
	&\quad = \sum_n \frac{1}{2 S_n} \left(\prod_{i=1}^{N_n} \int \widetilde{\dif p_i} \right) \braket{P(p) \gamma(q_2, \lambda_2) | n;\{p_i\}}^* \braket{\gamma_s^*(q_1, \lambda_1) | n;\{p_i\}} \, ,
\end{align}
and we single out a vector meson $V$ from the sum over intermediate states
\begin{align}
	&\Delta^V \left[ -\imag \braket{\gamma_s^*(q_1, \lambda_1) | P(p) \gamma(q_2, \lambda_2)} \right] \nn
	&\qquad = \sum_{\lambda_V} \frac{1}{2} \int \widetilde{\dif p_V} \braket{P(p) \gamma(q_2, \lambda_2) | V(p_V, \lambda_V)}^* \braket{\gamma_s^*(q_1, \lambda_1) | V(p_V, \lambda_V)}\,.
\end{align}
The second matrix element on the right-hand side is related to the vector-meson decay constant defined in Eq.~\eqref{eq:VectorDecayConst}.
For the other matrix element, we reduce the photon,
\begin{equation}
	\label{eq:GammaVPMatrixElement}
	\braket{\gamma(-q_2, \lambda_2) V(p_V, \lambda_V) | P(p)}= -\imag e \epsilon_\nu^{\lambda_2}(-q_2)^* (2 \pi)^4 \delta^{(4)}(q_2 + p - p_V) \bra{V(p_V, \lambda_V)} j^\nu_\mathrm{em}(0) \ket{P(p)} \, ,
\end{equation}
and use that there is only a single tensor structure
\begin{equation}
	\label{eq:GammaVPLorentz}
	\bra{V(p_V, \lambda_V)} j_\mathrm{em}^\nu(0) \ket{P(p)} = \epsilon_\rho^{\lambda_V}(p_V)^* \epsilon^{\rho \nu \alpha \beta} p_{V\alpha} q_{2\beta} F_{VP} \, ,
\end{equation}
with $q_2 = p_V - p$. Due to $q_2^2 = 0$, $F_{VP}$ does not depend on any free kinematic variable. Inserting the matrix elements and performing the polarization sum and phase-space integral, we obtain
\begin{equation}
	\Delta^V F_s^P(q_1^2) = \pi \delta(q_1^2 - m_V^2) m_V f_V F_{VP}\,.
\end{equation}
Thus, summing over all contributing vector mesons and neglecting other intermediate states in the isoscalar channel, we arrive at
\begin{equation}
	F_s^P(q_1^2) = -\sum_V \frac{m_V f_V F_{VP}}{q_1^2 - m_V^2}\,.
\end{equation}

\subsubsection{Sum rule and effective pole}

For the pion TFF, the isospin decomposition Eq.~\eqref{eq:IsospinDecompositionPionTFF} shows that the normalization is distributed equally between isovector and isoscalar components. If we only include the two-pion intermediate state in the isovector channel and vector-meson intermediate states in the isoscalar channel, this provides the sum rules (up to small chiral corrections)
\begin{equation}
	\label{eq:pionTFFIVNorm}
	F_v^{\pi^0}(0) = \frac{1}{96 \pi^2} \int_{4m_\pi^2}^\infty \dif s\, \sigma_\pi^3(s) f_{\pi^0}^1(s, 0) F_\pi^V(s)^* = \frac{1}{2} F_{\pi^0 \gamma^* \gamma^*}(0, 0) = \frac{1}{8 \pi^2 F_\pi} =: \frac{F_{\pi\gamma\gamma}}{2} \,,
\end{equation}
as well as
\begin{equation}
	\label{eq:sumRulePionTFFIsoscalar}
	F_s^{\pi^0}(0) = \sum_V \frac{f_V F_{V\pi^0}}{m_V} = \frac{F_{\pi\gamma\gamma}}{2} \,.
\end{equation}
In the case of $\eta$ and $\eta'$,  due to the different isospin decomposition only the sum of isovector and isoscalar normalizations is constrained
\begin{align}
	\sum_{P =  \eta, \eta'} F_P^a \left( F_v^P(0) + F_s^P(0) \right) &= \sum_{P = \eta, \eta'} F_P^a \left( \frac{1}{96 \pi^2} \int_{4m_\pi^2}^\infty \dif s\, \sigma_\pi^3(s) f_P^1(s, 0) F_\pi^V(s)^* + \sum_V \frac{f_V F_{V P}}{m_V} \right) \nn
		&= \frac{3}{4 \pi^2} \tr(\Q^2 \lambda^a)\,,
\end{align}
where we have accounted for $\eta$--$\eta'$ mixing.

Summing isovector and isoscalar contributions, the total singly-virtual pseudoscalar TFFs read
\begin{equation}
	\label{eq:PTFFsinglyVirtual}
	F_{P\gamma^*\gamma^*}(q_1^2, 0) = \frac{1}{96 \pi^2} \int_{4m_\pi^2}^\infty \dif s\frac{s \sigma_\pi^3(s) f_P^1(s, 0) F_\pi^V(s)^*}{s - q_1^2} - \sum_V \frac{m_V f_V F_{VP}}{q_1^2 - m_V^2}\,.
\end{equation}

As observed in Refs.~\cite{Schneider:2012ez,Hoferichter:2012pm,Hoferichter:2014vra,Hoferichter:2018kwz}, the two-pion intermediate state saturates the sum rule~\eqref{eq:pionTFFIVNorm} for the $\pi^0$ TFF at the level of $90\%$. We follow Ref.~\cite{Hoferichter:2018kwz} and parametrize the contribution of further intermediate states by adding to Eq.~\eqref{eq:PTFFsinglyVirtual} an effective pole
\begin{equation}
	\label{eq:PionTFFEffectivePole}
	F_{\pi^0\gamma^*\gamma^*}^\mathrm{eff}(q^2,0) = \frac{g_\mathrm{eff}}{4\pi^2 F_\pi} \frac{m_\mathrm{eff}^2}{m_\mathrm{eff}^2 - q^2} \, ,
\end{equation}
where the coupling $g_\mathrm{eff}$ and effective pole mass $m_\mathrm{eff}$ are determined by the normalization and slope parameter of the TFF
\begin{equation}
	a_\pi = \frac{m_{\pi^0}^2}{F_{\pi\gamma\gamma}} \frac{\p}{\p q^2} F_{\pi^0\gamma^*\gamma^*}(q^2,0) \big|_{q^2 = 0} \, .
\end{equation}

For our purpose, the NWA for the contribution of isoscalar vector mesons is an adequate approximation. In the more sophisticated analysis of Ref.~\cite{Hoferichter:2018kwz}, isoscalar vector mesons are included through a $q^2$-dependent subtraction function $a(q^2)$ in the Khuri--Treiman equations for the sub-process $\gamma^* \to 3 \pi$, where $q^2$ is the virtuality of the isoscalar photon. The doubly-virtual TFF fulfills a double-spectral representation: the discontinuities in the isoscalar virtuality arise on the one hand from three-pion intermediate states, reconstructed by the Khuri--Treiman equations, on the other hand from the imaginary part of the subtraction function $a(q^2)$, which is due to vector mesons as well as inelastic contributions described in Ref.~\cite{Hoferichter:2018kwz} in terms of a conformal polynomial, which is constrained to produce the correct asymptotic behavior. In the limit of vanishing vector-meson widths, we expect the singly-virtual limit of this representation to reproduce Eq.~\eqref{eq:PTFFsinglyVirtual}.

\subsection{Axial-vector transition form factors}
\label{sec:axialTFF}

The isospin decomposition of the axial-meson TFFs is similar to that of the pseudoscalar TFFs in the previous section with the difference that the TFFs are not symmetric in their two arguments. The crossing properties~\cite{Hoferichter:2020lap}
\begin{equation}
	\F_1^A(q_1^2,q_2^2) = -\F_1^A(q_2^2,q_1^2) \, , \quad
	\F_2^A(q_1^2,q_2^2) = -\F_3^A(q_2^2,q_1^2)
\end{equation}
imply for the isospin decompositions
\begin{align}
	\F_i^{a_1}(q_1^2, q_2^2) &= \F_{i, vs}^{a_1}(q_1^2, q_2^2) + \F_{i, sv}^{a_1}(q_1^2, q_2^2)\,, \nn
	\F_i^{f_1^{(\prime)}}(q_1^2, q_2^2) &= \F_{i, vv}^{f_1^{(\prime)}}(q_1^2, q_2^2) + \F_{i, ss}^{f_1^{(\prime)}}(q_1^2, q_2^2)\,
\end{align}
the following relations:
\begin{align}
	\F_{1,vs}^{a_1}(q_1^2,q_2^2) &= -\F_{1,sv}^{a_1}(q_2^2,q_1^2) \, , \quad
	\F_{2,vs}^{a_1}(q_1^2,q_2^2) = -\F_{3,sv}^{a_1}(q_2^2,q_1^2) \, , \quad
	\F_{3,vs}^{a_1}(q_1^2,q_2^2) = -\F_{2,sv}^{a_1}(q_2^2,q_1^2) \, , \nn
	\F_{1,vv}^{f_1^{(\prime)}}(q_1^2,q_2^2) &= -\F_{1,vv}^{f_1^{(\prime)}}(q_2^2,q_1^2) \, , \quad
	\F_{2,vv}^{f_1^{(\prime)}}(q_1^2,q_2^2) = -\F_{3,vv}^{f_1^{(\prime)}}(q_2^2,q_1^2) \, , \nn
	\F_{1,ss}^{f_1^{(\prime)}}(q_1^2,q_2^2) &= -\F_{1,ss}^{f_1^{(\prime)}}(q_2^2,q_1^2) \, , \quad
	\F_{2,ss}^{f_1^{(\prime)}}(q_1^2,q_2^2) = -\F_{3,ss}^{f_1^{(\prime)}}(q_2^2,q_1^2) \, .
\end{align}
In the singly-virtual case, all the axial-meson TFFs can be split into two pieces, labeled by the isospin of the off-shell current
\begin{align}
\F_i^A(q_1^2, 0) = \F_{i, v}^A(q_1^2) + \F_{i, s}^A(q_1^2)\,.
\end{align}
In the following, we discuss two-pion and vector-meson intermediate states in these TFFs in analogy to the pseudoscalar TFFs in the previous section.

\subsubsection{Two-pion intermediate state}
\label{sec:axialTFFTwoPion}

We start with the unitarity relation for the isovector part of the TFFs,
\begin{align}
	&\Delta \left[ -\imag \braket{\gamma_v^*(q_1, \lambda_1) | A(p, \lambda_A) \gamma(q_2, \lambda_2)} \right] \nn
	&\quad= \sum_n \frac{1}{2 S_n} \left( \prod_{i=1}^{N_n} \int \widetilde{\dif p_i} \right) \braket{A(p, \lambda_A) \gamma(q_2, \lambda_2) | n; \{p_i\}}^* \braket{\gamma_v^*(q_1, \lambda_1) | n; \{p_i\}} \, ,
\end{align}
and single out the isovector two-pion intermediate state:
\begin{align}
	\Delta^{\pi\pi} &\left[ -\imag \braket{\gamma_v^*(q_1, \lambda_1) | A(p, \lambda_A) \gamma(q_2, \lambda_2)} \right] \nn
	&\quad= \frac{1}{2} \int \widetilde{\dif p_1} \widetilde{\dif p_2} \;{}_{I=1} \!\< \pi^+(p_1) \pi^-(p_2) | A(p, \lambda_A) \gamma(q_2, \lambda_2) \> \;{}_{I=1}\!\< \pi^+(p_1) \pi^-(p_2) | \gamma_v^*(q_1, \lambda_1) \>^* \,.
\end{align}
The matrix element for $A\gamma\to\pi\pi$ is discussed in detail in App.~\ref{sec:AGammaPiPi}, including the BTT tensor decomposition, which for $q_2^2 = 0$ leads to
\begin{align}
	&\Delta^{\pi\pi} \left[\sum_i T_i^{\mu\nu\alpha}(q_1, -q_2) \F_{i, v}^A(q_1^2)\right] = \frac{m_A^2}{2} F_\pi^V(q_1^2)^* \int \widetilde{\dif p_1} \widetilde{\dif p_2} (2\pi)^4 \delta^{(4)}(q_1 - p_1 - p_2) \nn
	& \qquad\times \sum_{i=1}^3 t_i^{\nu\alpha}(q_2, p_1, p_2) (p_1 - p_2)^\mu \tilde{\F}_{A, i}(q_1^2, (q_2 - p_2)^2, (q_2 - p_1)^2)\,.
\end{align}
We project the right-hand side onto the axial-vector TFFs\footnote{The right-hand side also gives a contribution to the unphysical TFF $\F_{3,v}^A(q_1^2)$, which has to be taken into account in the projection, but is of no further relevance.} and express the functions $\tilde{\F}_{A, i}$ in terms of the helicity partial-wave amplitudes via Eq.~\eqref{eq:AgammaPiPiHelicityPW}. This gives
\begin{align}
	\Delta^{\pi^+\pi^-} \F_{1, v}^A(s) &= \frac{m_A^2 \sigma_\pi(s) F_\pi^V(s)^*}{(s - m_A^2)^2} \int \widetilde{\dif p_1} \widetilde{\dif p_2} (2\pi)^4 \delta^{(4)}(q_1 - p_1 - p_2) \nn
& \qquad \times\sum_{l \text{ odd}}\left( s z P_l(z) h_{++}^l(s) - \frac{m_A}{\sqrt{2}} \sqrt{s} (1 - z^2) P_l'(z) h_{+0}^l(s) \right) \,,\nn
	\Delta^{\pi^+\pi^-} \F_{2, v}^A(s) &= -\frac{m_A^2 \sigma_\pi(s) F_\pi^V(s)^*}{(s-m_A^2)} \int \widetilde{\dif p_1} \widetilde{\dif p_2} (2\pi)^4 \delta^{(4)}(q_1 - p_1 - p_2) \sum_{l \text{ odd}} z P_l(z) h_{++}^l(s) \, ,
\end{align}
with $s = (p_1 + p_2)^2 = q_1^2$. Performing the phase-space integrals, we obtain
\begin{align}
	\Delta^{\pi^+\pi^-} \F_{1, v}^A(s) &= \frac{m_A^2 \sigma_\pi^3(s)}{24 \pi (s-m_A^2)} F_\pi^V(s)^* \left( s g_{++}^1(s) - \sqrt{2s} m_A g_{+0}^1(s) \right) \,,\nn
	\Delta^{\pi^+\pi^-} \F_{2, v}^A(s) &= -\frac{m_A^2 \sigma_\pi^3(s)}{24 \pi}  F_\pi^V(s)^*g_{++}^1(s) \, ,
\end{align}
in terms of the rescaled partial waves $g_{\lambda_2\lambda_A}^l$ defined in Eq.~\eqref{eq:AGammaPiPiThreshold}. The residue of the apparent pole of $\Delta^{\pi^+\pi^-} \F_{1, v}^A(s)$ at $s = m_A^2$ vanishes due to the threshold constraint Eq.~\eqref{eq:AGammaPiPiConstraint}.

Given the asymptotic behavior~\eqref{eq:AxialTFFBLLimit}, we assume the singly-virtual TFFs $\F_{1,2}^A(q_1^2,0)$ to fulfill unsubtracted dispersion relations. However, the antisymmetry property $\F_1^A(q_1^2,q_2^2) = - \F_1^A(q_2^2,q_1^2)$ implies the constraint $\F_1^A(0,0) = 0$, which we implement by writing an unsubtracted dispersion relation for $\F_1(q_1^2,0)/q_1^2$. For both TFFs, we expect a number of sum rules to hold that guarantee the correct asymptotic behavior.

Therefore, the dispersion relations for the isovector TFFs lead to the two-pion contributions
\begin{align}
	\label{eq:AxialTFFPiPi}
	\F_{1, v}^A(q_1^2) &= \frac{q_1^2}{24 \pi^2} \int_{4 m_\pi^2}^\infty \dif s \frac{m_A^2 \sigma_\pi^3(s)}{s(s - q_1^2)}  F_\pi^V(s)^* \left(\frac{ s g_{++}^1(s) - \sqrt{2s} m_A g_{+0}^1(s) }{s - m_A^2}\right) \,,\nn
	\F_{2, v}^A(q_1^2) &= -\frac{1}{24 \pi^2} \int_{4 m_\pi^2}^\infty \dif s \frac{m_A^2 \sigma_\pi^3(s)}{s - q_1^2}  F_\pi^V(s)^* g_{++}^1(s) \,.
\end{align}
If we use the simple $s$-channel Omn\`es parametrization of the $A\gamma\to\pi\pi$ partial waves described in App.~\ref{sec:AgpipiParametrization}, we obtain
\begin{align}
	\label{eq:AxialTFFPiPiOmnes}
	\F_{1, v}^A(q_1^2) &= \frac{q_1^2}{48 \pi^2} \int_{4 m_\pi^2}^\infty \dif s \frac{m_A^2 \sigma_\pi^3(s)}{s - q_1^2}  F_\pi^V(s)^* \Omega_1^1(s) \bar\F_{A,2} \,,\nn
	\F_{2, v}^A(q_1^2) &= -\frac{1}{48 \pi^2} \int_{4 m_\pi^2}^\infty \dif s \frac{m_A^2 \sigma_\pi^3(s)}{s - q_1^2}  F_\pi^V(s)^* \Omega_1^1(s) \left( \bar\F_{A,1} + s \bar\F_{A,1}' + (s-m_A^2) \bar\F_{A,2} \right) \,.
\end{align}

\subsubsection{Vector-meson poles}

The unitarity relation for the isoscalar part of the TFFs reads
\begin{align}
	&\Delta \left[ -\imag \braket{\gamma_s^*(q_1, \lambda_1) | A(p, \lambda_A) \gamma(q_2, \lambda_2)} \right] \nn
	&\qquad= \sum_n \frac{1}{2 S_n} \left( \prod_{i=1}^n \int \widetilde{\dif p_i} \right) \braket{A(p, \lambda_A) \gamma(q_2, \lambda_2) | n; \{p_i\}}^* \braket{\gamma_s^*(q_1, \lambda_1) | n; \{p_i\}}
\end{align}
and we again only consider contributions from single vector mesons
\begin{align}
	&\Delta^V \left[ -\imag \braket{\gamma_s^*(q_1, \lambda_1) | A(p, \lambda_A) \gamma(q_2, \lambda_2)} \right] \nn
	&\qquad= \sum_{\lambda_V} \frac{1}{2} \int \widetilde{\dif p_V} \braket{A(p, \lambda_A) \gamma(q_2, \lambda_2) | V(p_V, \lambda_V)}^* \braket{\gamma_s^*(q_1, \lambda_1) | V(p_V, \lambda_V)}\,.
\end{align}
The second matrix element on the right-hand side is given in Eq.~\eqref{eq:vectorGammaMatrixElement}. For the first one, we reduce the photon
\begin{align}
	\label{eq:gammaAmesonVmesonMatrixElement}
	\< A(p, \lambda_A) \gamma(q_2, \lambda_2) &| V(p_V, \lambda_V) \> = - \imag e \epsilon_\nu^{\lambda_2}(q_2)^* (2 \pi)^4 \delta^{(4)}(p + q_2 - p_V) \bra{A(p, \lambda_A)} j_\mathrm{em}^\nu(0) \ket{V(p_V, \lambda_V)} \nn
	&= - e \epsilon_\nu^{\lambda_2}(q_2)^* \epsilon_\rho^{\lambda_A}(p)^* \epsilon_\lambda^{\lambda_V}(p_V) (2 \pi)^4 \delta^{(4)}(p + q_2 - p_V) \V_A^{\lambda\nu\rho}(-p_V, q_2) \,.
\end{align}
The matrix element $\V_A^{\lambda\nu\rho}(-p_V, q_2)$ is reminiscent of the VVA correlator itself: the off-shell axial-vector current and one vector current are replaced by on-shell massive axial-vector and vector mesons. There is no $U(1)_\mathrm{em}$ Ward identity for the contraction with $p_{V\lambda}$, but every observable will be contracted with the vector-meson polarization sum, which still allows one to drop one tensor structure. The result of the BTT decomposition shows that the same tensor structures $\tau_i^{\lambda\nu\rho}$ as for VVA can be employed without introducing kinematic singularities.
Furthermore, since in every observable $\V_A^{\lambda\nu\rho}(-p_V, q_2)$ appears contracted with the axial-vector polarization sum, we can drop the tensor structure $\tau_0^{\lambda\nu\rho}(-p_V, q_2)$ and we thus obtain the decomposition
\begin{equation}
	\label{eq:BTTVMVectorMeson}
	\V_A^{\lambda\nu\rho}(-p_V, q_2) = \sum_{i=1}^3 \tau_i^{\lambda \nu \rho}(-p_V, q_2) F_{VA}^i \,.
\end{equation}
As in the pseudoscalar case, $F_{VA}^i$ are constants at $q_2^2 = 0$. Inserting this decomposition into the unitarity relation leads to
\begin{align}
	\Delta^V \F_{1,s}^A(q_1^2) &= \pi \delta(q_1^2 - m_V^2) m_A^2 m_V f_V F_{VA}^1 \,, \nn
	\Delta^V \F_{2,s}^A(q_1^2) &= -\pi \delta(q_1^2 - m_V^2) m_A^2 m_V f_V (F_{VA}^2 + F_{VA}^3) \,, \nn
	\Delta^V \F_{3,s}^A(q_1^2) &= \pi \delta(q_1^2 - m_V^2) m_A^2 m_V f_V (F_{VA}^2 - F_{VA}^3)
\end{align}
and the dispersion relations give
\begin{align}
	\label{eq:AxialTFFVpole}
	\F_{1,s}^A(q_1^2) &= -\sum_V \frac{m_A^2 q_1^2}{q_1^2 - m_V^2} \frac{f_V}{m_V} F_{VA}^1 \,, \nn
	\F_{2,s}^A(q_1^2) &= \sum_V \frac{m_A^2 m_V f_V}{q_1^2 - m_V^2} (F_{VA}^2 + F_{VA}^3) \,, \nn
	\F_{3,s}^A(q_1^2) &= -\sum_V \frac{m_A^2 m_V f_V}{q_1^2 - m_V^2} (F_{VA}^2 - F_{VA}^3) \,,
\end{align}
where the third TFF $\F_{3,s}^A(q_1^2)$ has no physical relevance in the singly-virtual limit.

In summary, we obtain for the axial-vector TFFs
\begin{align}
	\label{eq:AxialTFFSolution}
	\F_1^A(q_1^2,0) &= \frac{q_1^2}{24 \pi^2} \int_{4 m_\pi^2}^\infty \dif s \frac{m_A^2 \sigma_\pi^3(s)}{s(s - q_1^2)}  F_\pi^V(s)^* \left(\frac{ s g_{++}^1(s) - \sqrt{2s} m_A g_{+0}^1(s) }{s - m_A^2}\right) - \sum_V \frac{m_A^2 q_1^2}{q_1^2 - m_V^2} \frac{f_V}{m_V} F_{VA}^1 \,, \nn
	\F_2^A(q_1^2,0) &= -\frac{1}{24 \pi^2} \int_{4 m_\pi^2}^\infty \dif s \frac{m_A^2 \sigma_\pi^3(s)}{s - q_1^2}  F_\pi^V(s)^* g_{++}^1(s) + \sum_V \frac{m_A^2 m_V f_V}{q_1^2 - m_V^2}  (F_{VA}^2 + F_{VA}^3) \,.
\end{align}

%% file: sections/VgammaA.tex


\section{\boldmath Vector-meson transition amplitude $V\to\gamma a^*$}
\label{sec:VectorMesonTransitionAmplitude}

In the following, we consider the form factors $\tilde{\V}_i(q^2)$ defined in Eq.~\eqref{eq:VtildeDef}, which correspond to the transition amplitude $V\to\gamma a^*$ with an on-shell photon and an off-shell axial current with invariant mass $q^2$. In the soft-photon limit, the on-shell condition for the vector meson fixes $q^2 = m_V^2$, leaving no free kinematic invariant. Therefore, the dispersive reconstruction, which is presented in App.~\ref{eq:DispersionRelationVgammaA}, needs to be formulated away from the soft-photon limit.
In App.~\ref{sec:AsymptoticsVgammaA}, we work out the asymptotic scaling of the form factors $\tilde\V_i$ for large space-like virtualities using a light-cone expansion~\cite{Lepage:1979zb, Lepage:1980fj, Brodsky:1981rp}, in close analogy to the case of meson TFFs derived in Ref.~\cite{Hoferichter:2020lap}.

\subsection{Dispersive reconstruction}
\label{eq:DispersionRelationVgammaA}

Assuming that the asymptotic scaling of the form factors, $\tilde\V_i(q^2) \asymp 1/q^2$, translates into the time-like region, both form factors $\tilde\V_{1,2}$ can be described by an unsubtracted dispersion relation:
\begin{equation}
	\tilde\V_i(q^2) = \frac{1}{\pi} \int \dif s \frac{\Delta \tilde\V_i(s)}{s - q^2 - \imag \epsilon}\,.
\end{equation}
In the unitarity relation for $V\to\gamma a^*$,
\begin{align}
	\Delta \bigl[ -\imag \< &a^*(q_3, \lambda_a) | \gamma(-q_2, \lambda_2) V(p_V, \lambda_V) \> \bigr] \nn*
		&= \sum_n \frac{1}{2 S_n} \left( \prod_{i=1}^{N_n} \int \widetilde{\dif p}_i \right) \braket{\gamma(-q_2, \lambda_2) V(p_V, \lambda_V) | n; \{p_i\}}^* \braket{a^*(q_3, \lambda_a) | n; \{p_i\}} \, ,
\end{align}
we consider pseudoscalar and axial-meson intermediate states. The contribution of a pseudoscalar meson $P$ is given by
\begin{align}
	\Delta^P \bigl[ -\imag \< a^*(q_3, \lambda_a) &| \gamma(-q_2, \lambda_2) V(p_V, \lambda_V) \> \bigr] \nn
		&= \frac{1}{2} \int \widetilde{\dif p} \braket{\gamma(-q_2, \lambda_2) V(p_V, \lambda_V) | P(p)}^*\braket{a^*(q_3, \lambda_a) | P(p)} \, .
\end{align}
Using Eqs.~\eqref{eq:gammaAVMatrixElement}, \eqref{eq:GammaVPLorentz} and \eqref{eq:AxialPion}, we find
\begin{equation}
	\Delta^P \V^{\lambda \nu \rho}(-p_V, q_2) = 2 \pi \delta(q_3^2 - m_P^2) \sum_a C_a F_P^a F_{VP} \tau_0^{\lambda \nu \rho}(-p_V, q_2)
\end{equation}
and therefore
\begin{equation}
	\Delta^P \V_i(0, q_3^2) = 2 \pi \delta(q_3^2 - m_P^2) \sum_a C_a F_P^a F_{VP} \delta_{i0}\,.
\end{equation}
The dispersion relation then gives
\begin{align}
	\label{eq:VtildeiPpole}
	\tilde{\V}_1^\text{$P$-pole}(q^2) = -2 \sum_a C_a F_P^a \frac{F_{VP}}{q^2 - m_P^2}  \, , \quad \tilde\V_2^\text{$P$-pole}(q^2) = 0 \, .
\end{align}

An axial-vector meson $A$ contributes
\begin{align}
	\Delta^A \bigl[ -\imag \< a^*(q_3, \lambda_a) &| \gamma(-q_2, \lambda_2) V(p_V, \lambda_V) \> \bigr] \nn
		&= \frac{1}{2} \sum_{\lambda_A} \int \widetilde{\dif p} \braket{\gamma(-q_2, \lambda_2) V(p_V, \lambda_V) | A(p, \lambda_A)}^* \braket{a^*(q_3, \lambda_a) | A(p, \lambda_A)}
\end{align}
to the sum over intermediate states. Using Eqs.~\eqref{eq:gammaAmesonVmesonMatrixElement} and \eqref{eq:AxialMatEl}, we obtain
\begin{equation}
	\Delta^A \V^{\lambda \nu \rho}(-p_V, q_2) = 2 \pi \delta(q_3^2 - m_A^2) \left( -g_\alpha^\rho + \frac{q_{3\alpha} q_3^\rho}{m_A^2} \right) m_A \sum_{i=1}^3 \tau_i^{\lambda \nu \alpha}(p_V, -q_2) F_{VA}^i \sum_a C_a F_A^a
\end{equation}
and the projection onto the scalar functions $\V_i$ gives
\begin{align}
	\Delta^A \V_0(0, q_3^2) &= 2 \pi \delta(q_3^2 - m_A^2) \sum_a \frac{C_a F_A^a}{m_A} \left( m_V^2 ( -F_{VA}^1 +  F_{VA}^2 +  F_{VA}^3) - m_A^2 F_{VA}^2 \right) \,, \nn
	\Delta^A \V_1(0, q_3^2) &= 2 \pi \delta(q_3^2 - m_A^2) \sum_a C_a F_A^a m_A F_{VA}^1 \,, \nn
	\Delta^A \V_2(0, q_3^2) &= 2 \pi \delta(q_3^2 - m_A^2) \sum_a C_a F_A^a m_A F_{VA}^2 \,, \nn
	\Delta^A \V_3(0, q_3^2) &= 2 \pi \delta(q_3^2 - m_A^2) \sum_a C_a F_A^a m_A F_{VA}^3 \,,
\end{align}
and hence
\begin{align}
	\Delta^A \tilde{\V}_1(q^2) &= 2 \pi \delta(q^2 - m_A^2) \sum_a C_a F_A^a \frac{m_A^2 - m_V^2}{m_A} \left( F_{VA}^1 - F_{VA}^2 - F_{VA}^3 \right) \,, \nn
	\Delta^A \tilde{\V}_2(q^2) &= 2 \pi \delta(q^2 - m_A^2) \sum_a C_a F_A^a m_A \left(F_{VA}^2 + F_{VA}^3 \right) \,.
\end{align}
The dispersion relation then leads to
\begin{align}
	\label{eq:VtildeiApole}
	\tilde{\V}_1^\text{$A$-pole}(q^2) &= 2 \sum_a \frac{C_a F_A^a}{m_A} \frac{m_V^2 - m_A^2}{q^2 - m_A^2} \left( F_{VA}^1 - F_{VA}^2 - F_{VA}^3 \right) \,, \nn
	\tilde{\V}_2^\text{$A$-pole}(q^2) &= -2 \sum_a C_a F_A^a m_A \frac{F_{VA}^2 + F_{VA}^3}{q^2 - m_A^2} \,.
\end{align}
In total, we obtain
\begin{align}
	\label{eq:Vtildei}
	\tilde{\V}_1(q^2) &=  -2 \sum_P \sum_a C_a F_P^a \frac{F_{VP}}{q^2 - m_P^2}  + 2 \sum_A \sum_a \frac{C_a F_A^a}{m_A} \frac{m_V^2 - m_A^2}{q^2 - m_A^2} \left( F_{VA}^1 - F_{VA}^2 - F_{VA}^3 \right) + \ldots \,, \nn
	\tilde{\V}_2(q^2) &= -2 \sum_A \sum_a C_a F_A^a m_A \frac{F_{VA}^2 + F_{VA}^3}{q^2 - m_A^2} + \ldots \,,
\end{align}
where the ellipses denote higher intermediate states. The constants $F_{VP}$ and $F_{VA}^i$ can be constrained by measurements of the partial decay widths $\Gamma_{V \to P \gamma}$ and $\Gamma_{A \to V \gamma}$ if available, see Sect.~\ref{sec:Numerics}.

\subsection{Light-cone expansion and asymptotic scaling}
\label{sec:AsymptoticsVgammaA}

In the following, we derive the asymptotic behavior of the form factors of the vector-meson-to-vacuum matrix element of a vector and axial-vector current defined in Eq.~\eqref{eq:vectorMesonTFF}. The form factors of the BTT tensor decomposition in Eq.~\eqref{eq:tensorDecompositionVAVectorMeson} can be most easily obtained by constructing projectors $P^i_{\lambda\nu\rho}$ that satisfy
\begin{equation}
	P^i_{\lambda\nu\rho} \tau_j^{\lambda\nu\rho} = \delta_{ij} \, ,
\end{equation}
hence
\begin{equation}
	\V_i(q_2^2,q_3^2) = P^i_{\lambda\nu\rho} \V^{\lambda\nu\rho}(-p,q_2) \, .
\end{equation}

We define the $SU(3)$ vector-meson decay constants
\begin{equation}
	\< 0 | \bar q(0) \frac{\lambda^a}{2} \gamma_\mu q(0) | V(p,\lambda_V) \> = m_V f_V^a \epsilon_\mu^{\lambda_V}(p) \, ,
\end{equation}
where the electromagnetic decay constant~\eqref{eq:VectorDecayConst} is the linear combination
\begin{equation}
	f_V = f_V^3 + \frac{1}{\sqrt{3}} f_V^8 \, .
\end{equation}

In close analogy to the light-cone expansion of TFFs considered in Ref.~\cite{Hoferichter:2020lap}, we now derive the asymptotic behavior of the form factors $\V_i$. Consider the time-ordered product of a vector and axial-vector current:
\begin{equation}
	T\{ j_\mathrm{em}^\nu(x) j_5^\rho(0) \} = T\left\{ \bar q(x) \Q \gamma^\nu q(x) \bar q(0) \Q_5 \gamma^\rho \gamma_5 q(0) \right\} \, .
\end{equation}
By performing the contractions of quark fields with free propagators, one obtains
\begin{equation}
	T\{ j_\mathrm{em}^\nu(x) j_5^\rho(0) \} = \bar q(x) \Q \Q_5 \gamma^\nu \gamma^\alpha \gamma^\rho \gamma_5 q(0) S_\alpha^F(x) + \bar q(0) \Q_5 \Q \gamma^\rho \gamma^\alpha \gamma^\nu \gamma_5 q(x) S_\alpha^F(-x) \, ,
\end{equation}
where
\begin{equation}
	 S_\alpha^F(x) = \int \frac{d^4p}{(2\pi)^4} \frac{i p_\alpha e^{-i p\cdot x}}{p^2 + i\epsilon} = \frac{i x_\alpha}{2\pi^2(x^2-i\epsilon)^2} \, .
\end{equation}
Using translation invariance, we obtain
\begin{align}
	\epsilon_\lambda^{\lambda_V}(p) \V^{\lambda \nu \rho}(-p, q_2) = \int d^4 x \bigg[
		&e^{i q_2 \cdot x}  \< 0 | \bar q(x) \Q\Q_5 \gamma^\nu \gamma^\alpha \gamma^\rho \gamma_5 q(0) | V(p, \lambda_V) \>  \nn
		& + e^{i (p-q_2) \cdot x} \< 0 | \bar q(x) \Q_5\Q \gamma^\rho \gamma^\alpha \gamma^\nu \gamma_5 q(0) | V(p, \lambda_V) \> 
		\bigg] S_\alpha^F(x) \, .
\end{align}
The Dirac structure can be simplified with the Chisholm identity
\begin{equation}
	\gamma^\nu \gamma^\alpha \gamma^\rho \gamma_5 = g^{\alpha\nu} \gamma^\rho \gamma_5 + g^{\alpha\rho} \gamma^\nu \gamma_5 - g^{\nu\rho} \gamma^\alpha \gamma_5 + i \epsilon^{\nu\alpha\rho\beta} \gamma_\beta \, .
\end{equation}

The distribution amplitudes for vector mesons up to twist-4 contributions have been discussed in detail in Refs.~\cite{Ball:1998ff,Ball:1998sk}. We introduce a light-like vector
\begin{equation}
	k_\mu = p_\mu - z_\mu \frac{m_V^2}{2 p \cdot z} \, ,
\end{equation}
which for $z^2=0$ fulfills $k^2 = 0$. The vector $z_\mu$ is given by
\begin{equation}
	z_\mu = x_\mu - p_\mu \frac{1}{m_V^2}\left( x \cdot p - \sqrt{ (x\cdot p)^2 - x^2 m_V^2 } \right) = x_\mu - p_\mu \frac{x^2}{2 z\cdot k} + \Order(x^4) \, .
\end{equation}
The polarization vector of the vector meson is decomposed according to
\begin{equation}
	\epsilon_\mu^{\lambda_V} = \frac{\epsilon^{\lambda_V} \cdot z}{k\cdot z} k_\mu + \frac{\epsilon^{\lambda_V} \cdot k}{k \cdot z} z_\mu + \epsilon_{\perp\mu}^{\lambda_V} \, .
\end{equation}
We decompose the flavor structure as
\begin{equation}
	\Q \Q_5 = \Q_5 \Q = \sum_a c_a \frac{\lambda^a}{2} \, , \quad c_a = \tr[\Q \Q_5 \lambda^a] \, ,
\end{equation}
which leads to
\begin{align}
	\epsilon_\lambda^{\lambda_V}(p) \V^{\lambda \nu \rho}(-p, q_2) = \sum_a c_a \int d^4 x \bigg[
		& \big( e^{i q_2 \cdot x} + e^{i (p-q_2) \cdot x} \big) \, g^{\alpha\nu} V_5^{a,\rho}(x,p)  \nn
		&+ \big( e^{i q_2 \cdot x} + e^{i (p-q_2) \cdot x} \big) \, g^{\alpha\rho} V_5^{a,\nu}(x,p)  \nn
		&- \big( e^{i q_2 \cdot x} + e^{i (p-q_2) \cdot x} \big) \, g^{\nu\rho} V_5^{a,\alpha}(x,p)  \nn
		&+\big( e^{i q_2 \cdot x}  - e^{i (p-q_2) \cdot x} \big) \, i \epsilon^{\nu\alpha\rho\beta} V_\beta^a(x,p)
		\bigg] S_\alpha^F(x) \, ,
\end{align}
where
\begin{align}
	V_5^{a,\mu}(x,p) &:= \< 0 | \bar q(x) \frac{\lambda^a}{2} \gamma^\mu \gamma_5 q(0) | V(p, \lambda_V) \> \, , \nn
	V^a_\mu(x,p) &:= \< 0 | \bar q(x) \frac{\lambda^a}{2} \gamma_\mu q(0) | V(p, \lambda_V) \> \, .
\end{align}
Neglecting $x^2$ corrections and using the light-cone distribution amplitudes of Refs.~\cite{Ball:1998ff,Ball:1998sk}, these matrix elements can be written as\footnote{Note the different sign convention for the Levi-Civita symbol in Refs.~\cite{Ball:1998ff,Ball:1998sk}.}
\begin{align}
	V_5^{a,\mu}(x,p) &= e^{-i p\cdot x/2} \< 0 | \bar q\left(\frac{x}{2}\right) \frac{\lambda^a}{2} \gamma^\mu \gamma_5 q\left(-\frac{x}{2}\right) | V(p, \lambda_V) \> \nn
		&= -\frac{1}{4} f_V^a m_V \epsilon^{\mu\nu\alpha\beta} \epsilon_{\perp\nu}^{\lambda_V} p_\alpha x_\beta \int_0^1 du \, e^{i(u-1) p\cdot x} g_{\perp}^{(A)}(u,\mu^2) \, , \nn
	V_\mu^a(x,p) &= e^{-i p\cdot x/2} \< 0 | \bar q\left(\frac{x}{2}\right) \frac{\lambda^a}{2} \gamma_\mu q\left(-\frac{x}{2}\right) | V(p, \lambda_V) \> \nn
		&= f_V^a m_V \bigg[ k_\mu \frac{\epsilon^{\lambda_V} \cdot x}{p\cdot x} \int_0^1 du \, e^{i(u-1) p\cdot x} \phi_{\parallel}(u,\mu^2) + \epsilon^{\lambda_V}_{\perp\mu} \int_0^1 du \, e^{i(u-1)p\cdot x} g_\perp^{(V)}(u,\mu^2) \nn
			&\qquad\qquad - \frac{1}{2} x_\mu \frac{\epsilon^{\lambda_V} \cdot x}{(p \cdot x)^2} m_V^2 \int_0^1 du \, e^{i(u-1) p\cdot x} g_3(u,\mu^2) \bigg] \, .
\end{align}
We will neglect the contribution of $g_3$, which is of twist 4. The distribution amplitudes are normalized to
\begin{equation}
	\int_0^1 du \, \varphi(u,\mu^2) = 1 \, , \quad \varphi \in \{ \phi_\parallel, g_\perp^{(A)}, g_\perp^{(V)} \} \, ,
\end{equation}
and we will use their asymptotic form:
\begin{align}
	\phi_\parallel(u) &= g_{\perp}^{(A)}(u) = 6 u(1-u) =: \phi(u) \, , \nn
	g_{\perp}^{(V)}(u) &= \frac{3}{4} (1 + (2u-1)^2) \, ,
\end{align}
which also fulfill the Wandzura--Wilczek relations, where three-particle contributions are neglected~\cite{Ball:1998sk}:
\begin{align}
	g_{\perp}^{(V),WW}(u) &= \frac{1}{2} \left[ \int_0^u dv \frac{1}{1-v} \phi_\parallel(v) + \int_u^1 dv \frac{1}{v} \phi_\parallel(v) \right] \, , \nn
	g_{\perp}^{(A),WW}(u) &= 2(1-u) \int_0^u dv \frac{1}{1-v} \phi_\parallel(v) + 2u \int_u^1 dv \frac{1}{v} \phi_\parallel(v) \, .
\end{align}
Using integration by parts and
\begin{equation}
	\int d^4x \, S_F^\mu(x) e^{i q \cdot x} = i \frac{q^\mu}{q^2} \, , \quad \int d^4x \, x^\mu S_F^\nu(x) e^{i q \cdot x} = \frac{g^{\mu\nu}}{q^2} - \frac{2q^\mu q^\nu}{(q^2)^2} \, , \quad 
\end{equation}
the integral over $x$ can be performed and one obtains a result that is already gauge invariant. Therefore, we can perform the projection onto the BTT decomposition and obtain:
\begin{align}
	\V_0(q_2^2, q_3^2) &= \sum_a \frac{c_a f_V^a}{2 m_V} \int_0^1 du \frac{(u^2+u+2)\phi(u)m_V^2 - 6 u^2 q_3^2}{(u q_3^2 + (1-u)q_2^2 - u(1-u) m_V^2)^2} \, , \nn
	\V_1(q_2^2, q_3^2) &= 0 \, , \nn
	\V_2(q_2^2, q_3^2) &= \sum_a \frac{c_a f_V^a}{2 m_V} \int_0^1 du \frac{(1-u)(u-2)\phi(u) + 6 u^2 q_3^2}{(u q_3^2 + (1-u)q_2^2 - u(1-u) m_V^2)^2} \, , \nn
	\V_3(q_2^2, q_3^2) &= \sum_a \frac{c_a f_V^a}{2 m_V} \int_0^1 du \frac{(1-u)(u+2)\phi(u) + 6 u^2 q_3^2}{(u q_3^2 + (1-u)q_2^2 - u(1-u) m_V^2)^2} \, .
\end{align}
In terms of the average virtuality and asymmetry parameter
\begin{equation}
	Q^2 = \frac{q_2^2 + q_3^2}{2} \, , \quad w = \frac{q_3^2 - q_2^2}{q_2^2 + q_3^2} \, ,
\end{equation}
this can be expressed to leading order in $m_V$ as
\begin{equation}
	\V_0(q_2^2, q_3^2) = - \V_2(q_2^2, q_3^2) = - \V_3(q_2^2, q_3^2) = \sum_a \frac{c_a f_V^a}{2 m_V} \frac{a^V(w)}{Q^2} \, , \nn
\end{equation}
with the asymmetry function
\begin{equation}
	a^V(w) = -\frac{3}{2w^2} \left( 2 + \frac{1-w^2}{w} \log\left( \frac{1-w}{1+w} \right) \right) \, ,
\end{equation}
which has no end-point singularity for $w\to1$. In the limit of $g-2$ kinematics, this leads to
\begin{align}
	\tilde\V_1(q^2) &= - \tilde \V_2(q^2) = - \sum_a \frac{6 c_a f_V^a}{m_V q^2} + \Order\big(q^{-4}\big) \, .
\end{align}

%% file: sections/AgammaPiPi.tex

\color{black}

\section{Two-pion amplitudes}
\label{sec:TwoPionAmplitudes}

In the two dispersive approaches to the VVA correlator, we encounter two-pion sub-processes with one photon and one axial vector---either an off-shell axial-vector current or an on-shell massive axial-vector meson. In Sect.~\ref{sec:AxialCurrentGammaPiPi}, we first consider the matrix element with an axial-vector current, before discussing the case of an axial-vector meson in Sect.~\ref{sec:AGammaPiPi}.


\subsection[Reconstruction of $a^*\gamma\to\pi\pi$]{\boldmath Reconstruction of $a^*\gamma\to\pi\pi$}
\label{sec:AxialCurrentGammaPiPi}

In the following, we discuss the two-pion matrix element of a vector and an axial-vector current.
\begin{equation}
	\tag{\ref{eq:defU}}
	\U^{\nu \rho}(q_2, p_1, p_2) = \int \dif^4 x \, e^{\imag q_2 \cdot x} \bra{0} \T\{j_\mathrm{em}^\nu(x) j_5^\rho(0)\} \ket{\pi^+(p_1) \pi^-(p_2)}_{I=1} \,,
\end{equation}
which enters the dispersion relation for VVA in $g-2$ kinematics in Sect.~\ref{sec:VVA_2pi}.

\subsubsection{Tensor decomposition}
\label{sec:AxialCurrentGammaPiPiTensorDecomposition}

We start with the tensor decomposition of the function $\U^{\nu\rho}(q_2, p_1, p_2)$ in general kinematics, employing the BTT recipe~\cite{Bardeen:1968ebo,Tarrach:1975tu}. The tensor structures have to contain a Levi-Civita tensor due to conservation of parity and they depend on three independent momenta. Thus, there are nine tensor structures to start with
\begin{align}
\{
&\epsilon^{\nu\rho\alpha\beta}q_{2\alpha} p_{1\beta}\,,\ 
\epsilon^{\nu\rho\alpha\beta}q_{2\alpha} p_{2\beta}\,,\ 
\epsilon^{\nu\rho\alpha\beta}p_{1\alpha} p_{2\beta}\,,\ 
\epsilon^{\nu\alpha\beta\gamma}q_{2\alpha} p_{1\beta} p_{2\gamma} q_2^\rho\,,\ 
\epsilon^{\nu\alpha\beta\gamma}q_{2\alpha} p_{1\beta} p_{2\gamma} p_1^\rho\,,\nn
&\epsilon^{\nu\alpha\beta\gamma}q_{2\alpha} p_{1\beta} p_{2\gamma} p_2^\rho\,,\ 
\epsilon^{\rho\alpha\beta\gamma}q_{2\alpha} p_{1\beta} p_{2\gamma} q_2^\nu\,,\ 
\epsilon^{\rho\alpha\beta\gamma}q_{2\alpha} p_{1\beta} p_{2\gamma} p_1^\nu\,,\ 
\epsilon^{\rho\alpha\beta\gamma}q_{2\alpha} p_{1\beta} p_{2\gamma} p_2^\nu
\}\,.
\end{align}
The Ward identity $q_{2\nu} \U^{\nu\rho}(q_2, p_1, p_2) = 0$ is implemented by contracting the tensor structures with the gauge projector $g_{\mu\nu} - q_{2\mu} q_{2\nu}/q_2^2$ and then the BTT procedure is applied taking into account the Schouten identities. The final gauge-invariant tensor structures can be chosen as
\begin{align}
	\label{eq:structuresU}
	\{t_i^{\nu\rho}(q_2, p_1, p_2) \} = \{
		&\epsilon^{\nu\rho\alpha\beta}q_{2\alpha} (p_1-p_2)_\beta,
		\epsilon^{\nu\rho\alpha\beta}q_{2\alpha} (p_1+p_2)_\beta, \nn
		& \epsilon^{\nu\alpha\beta\gamma}q_{2\alpha} p_{1\beta} p_{2\gamma} (p_1-p_2)^\rho,
		\epsilon^{\nu\alpha\beta\gamma}q_{2\alpha} p_{1\beta} p_{2\gamma} q_2^\rho,
		\epsilon^{\nu\alpha\beta\gamma}q_{2\alpha} p_{1\beta} p_{2\gamma} q_3^\rho
	\} \, ,
\end{align}
where $q_3 = p_1 + p_2 - q_2$ is the momentum of the axial current. The matrix element $\U^{\nu\rho}$ can thus be written as
\begin{equation}
	\label{eq:tensorDecompU}
	\U^{\nu\rho}(q_2, p_1, p_2) = \sum_{i=1}^{5} \U_i\left(s, t; q_2^2, q_3^2 \right) t_i^{\nu\rho}(q_2, p_1, p_2) \,,
\end{equation}
where the scalar functions $\U_i$ are free from kinematic singularities and zeros. They depend on the two virtualities and the Mandelstam variables of the process,
\begin{equation}
	s = (p_1+p_2)^2 \,, \quad t = (p_1-q_2)^2 \,, \quad u = (p_2-q_2)^2 \,, \quad s + t + u = 2 m_\pi^2 + q_2^2 + q_3^2 \, .
\end{equation}
Crossing symmetry for the isospin $I=1$ pions implies 
\begin{align}
	\U_{1,4,5}(s, t; q_2^2, q_3^2) &= \U_{1,4,5}(s,u; q_2^2, q_3^2) \,, & \U_{2,3}(s, t; q_2^2, q_3^2) &= -\U_{2,3}(s,u; q_2^2, q_3^2) \, .
\end{align}

Due to the Ward identity, the tensor structures in Eq.~\eqref{eq:structuresU} vanish at least linearly with $q_2$. Therefore, in the soft-photon limit, Eq.~\eqref{eq:tensorDecompU} simplifies to
\begin{equation}
	\label{eq:tensorDecompUSoftLimit}
	\U^{\nu\rho}(q_2, p_1, p_2) = \sum_{i=1}^2 \tilde\U_i (q_3^2) t_i^{\nu\rho\sigma}(p_1, p_2) q_{2\sigma} + \Order(q_2^2) \, ,
\end{equation}
where
\begin{equation}
	t_1^{\nu\rho\sigma}(p_1, p_2) = \epsilon^{\nu\rho\sigma\alpha} (p_1 - p_2)_{\alpha} \, , \quad
	t_2^{\nu\rho\sigma}(p_1, p_2) = \epsilon^{\nu\sigma\alpha\beta} p_{1\alpha} p_{2\beta} (p_1+p_2)^\rho
\end{equation}
and we define $\tilde\U_i(q_3^2) = \tilde\U_i (q_3^2, m_\pi^2; 0, q_3^2)$ with
\begin{equation}
	\tilde\U_1(s, t; q_2^2, q_3^2) = \U_1(s, t; q_2^2, q_3^2) \,, \quad \tilde\U_2(s, t; q_2^2, q_3^2) = \U_5(s, t; q_2^2, q_3^2) \,.
\end{equation}

Next, we insert the tensor decomposition Eq.~\eqref{eq:tensorDecompUSoftLimit} into the two-pion contribution to the discontinuity Eq.~\eqref{eq:2PiDisc}. We express the result in terms of the tensor structures $\tau_i^{\mu\nu\rho\sigma}(q_1)$ and read off
\begin{align}
	\Delta^{\pi\pi} \tilde{\W}_1 (q^2) &= -\frac{1}{12} \int \widetilde{\dif p}_1 \widetilde{\dif p}_2 (2\pi)^4 \delta^{(4)}(q + p_1 + p_2) F_\pi^V(q^2) \sigma_\pi^2(q^2) \left[2 \tilde\U_1(q^2)^* + q^2 \tilde\U_2(q^2)^*\right] \,, \nn
	\Delta^{\pi\pi} \tilde{\W}_2 (q^2) &= \frac{1}{6} \int \widetilde{\dif p}_1 \widetilde{\dif p}_2 (2\pi)^4 \delta^{(4)}(q + p_1 + p_2) F_\pi^V(q^2) \sigma_\pi^2(q^2) \tilde\U_1(q^2)^* \,,
\end{align}
where $\sigma_\pi(q^2) = \sqrt{1 - 4m_\pi^2/q^2}$ and we took into account the complex conjugation in Eq.~\eqref{eq:DiscWtildei}. Performing the phase-space integrals, we obtain Eq.~\eqref{eq:WipipiDisc}.

\subsubsection{Unitarity relations}

\begin{figure}[t]
	\centering
	\begin{align*}
		\Delta \left[ \minidiagSize{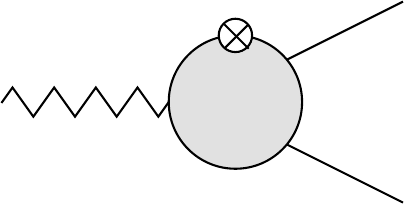}{2cm} \; \right]
		\quad = \quad 
		\left( \minidiagSize{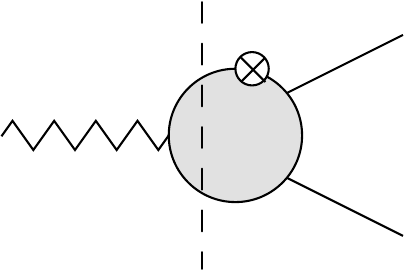}{2cm} \right)^*
		\quad + \quad
		\minidiagSize{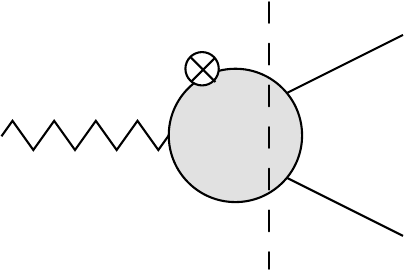}{2cm} \\[-0.75cm]
	\end{align*}
	\caption{Unitarity cuts contributing to the discontinuity of the $\tilde\U_i$ functions in the soft-photon limit. The axial current is denoted by a zigzag line and the static external electromagnetic field by a crossed circle.}
	\label{fig:aVpipiTwoCuts}
\end{figure}

The scalar functions $\tilde\U_i(q^2)$ are free of kinematic singularities and thus amenable to a dispersive treatment. The discontinuity can be calculated analogously to Eq.~\eqref{eq:DiscWtildei},
\begin{equation}
	\label{eq:DeltaUtildei}
	\Delta \tilde\U_i(q^2) = \lim_{s \to q^2} \left[\left( \Delta_3 \tilde\U_i(s + \imag \epsilon, m_\pi^2; 0, q^2)\right)^* + \Delta_s \tilde\U_i(s, m_\pi^2; 0, q^2 + \imag \epsilon) \right] \,,
\end{equation}
as illustrated in Fig.~\ref{fig:aVpipiTwoCuts}. Again, there is a freedom in choosing which of the contributions should be complex conjugated.

\begin{figure}[t]
	\centering
	\begin{align*}
		\minidiagSize{aVpipi-q3cut}{2cm}
		\quad = \quad 
		\minidiagSize{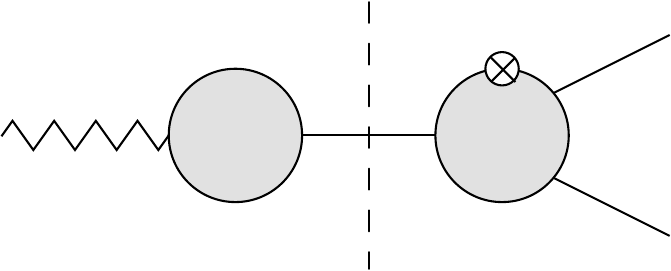}{3cm}
		\quad + \quad
		\minidiagSize{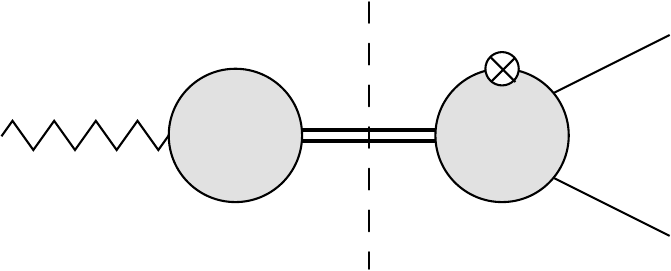}{3cm}
		\quad + \quad \ldots \\[-0.75cm]
	\end{align*}
	\caption{Pseudoscalar and axial-vector poles due to intermediate states in the unitarity relation Eq.~\eqref{eq:q3UnitarityU}, evaluated in the soft-photon limit.}
	\label{fig:aVpipi-q3Cut}
\end{figure}

\begin{figure}[t]
	\centering
	\begin{align*}
		\minidiagSize{aVpipi-scut}{2cm}
		\quad = \quad
		\minidiagSize{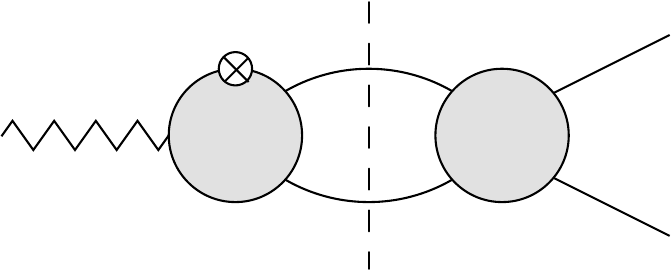}{3cm}
		\quad + \quad \ldots  \\[-0.75cm]
	\end{align*}
	\caption{Two-pion intermediate state in the unitarity relation Eq.~\eqref{eq:sUnitarityU}.}
	\label{fig:aVpipi-sCut}
\end{figure}

The two discontinuities in general kinematics follow from the unitarity relations for the $q_3^2$-channel process $\gamma^*(-q_2, \lambda_2) \pi^+(p_1) \pi^-(p_2) \to a^*(q_3, \lambda_a)$,
\begin{align}
	\label{eq:q3UnitarityU}
	&\Delta_3\left( -\imag \braket{a^*(q_3, \lambda_a) | \gamma^*(-q_2, \lambda_2) [\pi^+(p_1) \pi^-(p_2)]_{I=1}} \right) \nn
	&\quad = \sum_n \frac{1}{2 S_n} \left( \prod_{i=1}^{N_n} \int \widetilde{\dif p}_i \right) \braket{\gamma^*(-q_2, \lambda_2) [\pi^+(p_1) \pi^-(p_2)]_{I=1}| n;\{p_i\}}^* \braket{a^*(q_3, \lambda_a)| n;\{p_i\}} \,,
\end{align}
and the $s$-channel process $\pi^+(p_1) \pi^-(p_2) \to \gamma^*(q_2, \lambda_2) a^*(q_3, \lambda_a)$ with the two pions in an isospin $I = 1$ state,
\begin{align}
	\label{eq:sUnitarityU}
	&\Delta_s\left( - \imag \braket{\gamma^*(q_2, \lambda_2) a^*(q_3, \lambda_a) | \pi^+(p_1) \pi^-(p_2)}_{I=1} \right) \nn
	&\quad = \sum_n \frac{1}{2 S_n} \left(\prod_{i=1}^n \int \widetilde{\dif p}_i\right) {}_{I=1}\!\braket{\pi^+(p_1) \pi^-(p_2)| n;\{p_i\}}^* \braket{\gamma^*(q_2, \lambda_2) a^*(q_3, \lambda_A) | n;\{p_i\}}\,,
\end{align}
where we implicitly understand an analytic continuation to the upper rim of the cut in all kinematic variables except for the one in which the discontinuity is calculated.

As illustrated in Fig.~\ref{fig:aVpipi-q3Cut}, the discontinuity in $q_3^2$ includes pseudoscalar and axial poles, whereas the discontinuity in $s=(p_1+p_2)^2$ is dominated at low energies by the two-pion intermediate state, as shown in Fig.~\ref{fig:aVpipi-sCut}. We will discuss these intermediate states next.

\subsubsection{Pseudoscalar-pole contribution}

We single out the pseudoscalar poles in the $q_3^2$-channel unitarity relation:
\begin{align}
	&\Delta_3^P\left( -\imag \braket{a^*(q_3, \lambda_a) | \gamma^*(-q_2, \lambda_2) [\pi^+(p_1) \pi^-(p_2)]_{I=1}} \right) \nn
	&\quad = \frac{1}{2} \int \widetilde{\dif p} \braket{\gamma^*(-q_2, \lambda_2) \pi^+(p_1) \pi^-(p_2)| P(p)}^* \braket{a^*(q_3, \lambda_a)| P(p)}\,.
\end{align}
The amplitude for $\gamma^{(*)} \to \pi^+ \pi^- P$ is described by the function $\F_P$ defined in Eq.~\eqref{eq:gammaPiPiPScFun}. The second matrix element is related to the decay constant according to Eq.~\eqref{eq:AxialPion}. To linear order in $q_2$, we find\footnote{The analytic continuation applied to the unitarity relation removes the complex conjugation from the function $\F_P$. This works up to further cuts in the $\pi^+ \pi^- \gamma^*$-channel, which are neglected. The same applies to the axial pole below.}
\begin{equation}
	\Delta_3^{P} \U^{\nu\rho}(q_2, p_1, p_2) = 2 \pi \delta(q_3^2 - m_P^2) \sum_a C_a F_P^a \, \F_P(m_P^2, m_\pi^2, m_\pi^2; 0) t_2^{\nu\rho\sigma}(p_1, p_2) q_{2\sigma} + \Order(q_2^2)
\end{equation}
and thus
\begin{equation}
	\label{eq:Delta3PUtildei}
	\Delta^{P} \tilde\U_1(s) = 0\,, \quad \Delta^{P} \tilde\U_2(s) = 2 \pi \delta(s - m_P^2) \sum_a C_a F_P^a \F_P(m_P^2, m_\pi^2, m_\pi^2; 0)\,.
\end{equation}
For $m_P^2 > 4m_\pi^2$, the kinematic point at which the function $\F_{P}$ is needed corresponds to the threshold of the process $\gamma P \to \pi^+ \pi^-$. In the partial-wave expansion Eq.~\eqref{eq:PWEPppg}, the partial-wave amplitudes behave at threshold according to
\begin{equation}
	f_P^l(s, 0) \sim (s - m_P^2)^{l-1}\,.
\end{equation}
Thus, only the $P$-wave amplitude can contribute and we have
\begin{equation}
	\F_P(m_P^2, m_\pi^2, m_\pi^2; 0) = f_P^1(m_P^2, 0)\,.
\end{equation}
This is consistent with the virtual photon coupling only to $P$-wave two-pion states. If $m_P^2 < 4 m_\pi^2$, the same result follows from analytic continuation in $m_P^2$. In the chiral limit (and for the $\eta'$ also in the large-$N_c$ limit), the value of the $\gamma P \pi^+\pi^-$ coupling is fixed by a low-energy theorem and the chiral anomaly~\cite{Wess:1971yu,Witten:1983tw, Bijnens:1989ff, Hoferichter:2012pm, Hoferichter:2018kwz, Gan:2020aco}
\begin{equation}
	\sum_{\mathclap{P=\pi^0, \eta, \eta'}} \; F_P^a \F_{P}(0, 0, 0; 0) = \frac{N^a}{4 \pi^2 F_\pi^2} \, ,
\end{equation}
with $N^3 = 1$, $N^8 = 1/\sqrt{3}$, and $N^0 = \sqrt{2/3}$, in particular
\begin{equation}
	f_{\pi^0}^1(0, 0) = \F_{\pi^0}(0, 0, 0; 0) = \frac{1}{4 \pi^2 F_\pi^3} \, .
\end{equation}

\subsubsection{Axial-pole contribution}
\label{sec:AxialCurrentGammaPiPiAxialPole}

Next, we consider axial-vector mesons in the sum over intermediate states of the $q_3^2$-channel unitarity relation
\begin{align}
	&\Delta_3^A\left( -\imag \braket{a^*(q_3, \lambda_a) | \gamma^*(-q_2, \lambda_2) [\pi^+(p_1) \pi^-(p_2)]_{I=1}} \right) \nn
	&\quad = \frac{1}{2} \int \widetilde{\dif p} \sum_{\lambda_A} \braket{\gamma^*(-q_2, \lambda_2) [\pi^+(p_1) \pi^-(p_2)]_{I=1} | A(p, \lambda_A)}^* \braket{a^*(q_3, \lambda_a)| A(p, \lambda_A)}\,.
\end{align}
The matrix element with an axial-vector meson will be discussed in more detail in Sect.~\ref{sec:AGammaPiPi}, see Eq.~\eqref{eq:axialPiPiGamma}, while the second matrix element on the right-hand side is given in Eq.~\eqref{eq:AxialMatEl}. This leads to
\begin{equation}
	\Delta_3^A \U^{\nu\rho}(q_2, p_1, p_2) = - 2 \pi \delta(q_3^2 - m_A^2) \sum_a C_a F_A^a m_A \left({g_\alpha}^\rho - \frac{q_{3\alpha} q_3^\rho}{m_A^2}\right) \A^{\nu\alpha}(q_2,p_1,p_2) \,.
\end{equation}
Projecting onto the scalar functions $\tilde\U_i(q^2)$ in $g-2$ kinematics gives
\begin{align}
	\label{eq:Delta3Utildei}
	\Delta^A \tilde\U_1(s) &= - 2 \pi \delta(s - m_A^2) \sum_a C_a F_A^a m_A \tilde\F_{A,1}(m_A^2, m_\pi^2, m_\pi^2) \,, \nn
	\Delta^A \tilde\U_2(s) &= 4 \pi \delta(s - m_A^2) \sum_a C_a \frac{F_A^a}{m_A} \tilde\F_{A,1}(m_A^2, m_\pi^2, m_\pi^2) \,,
\end{align}
where $\tilde\F_{A,1}$ describes $\gamma A\to\pi^+\pi^-$, which in the soft-photon limit does not depend on any kinematic variable anymore. As will be discussed in Sect.~\ref{sec:AgammaPiPiHelAmps}, it receives a contribution only from the leading partial wave at threshold, in analogy to the case of pseudoscalars.

\subsubsection{Two-pion rescattering}

The lowest-lying cut with respect to the variable $s$ is due to an isospin $I=1$ two-pion state. The contribution to the unitarity relation~\eqref{eq:sUnitarityU} reads
\begin{align}
	&\Delta_s^{\pi^+\pi^-}\left( - \imag \braket{\gamma^*(q_2, \lambda_2) a^*(q_3, \lambda_a) | \pi^+(p_1) \pi^-(p_2)}_{I=1} \right) \\
	&\quad = \frac{1}{2} \int \widetilde{\dif k}_1 \widetilde{\dif k}_2 \ {}_{I=1}\!\braket{\pi^+(p_1) \pi^-(p_2)| \pi^+(k_1) \pi^-(k_2)}_{I=1}^* \braket{\gamma^*(q_2, \lambda_2) a^*(q_3, \lambda_a)| \pi^+(k_1) \pi^-(k_2)}_{I=1} \, , \nonumber
\end{align}
which implies
\begin{align}
	\Delta_s^{\pi^+\pi^-} \U^{\nu\rho}(q_2, p_1, p_2) &= \frac{1}{4} \int \widetilde{\dif k}_1 \widetilde{\dif k}_2 (2\pi)^4 \delta^{(4)}(q_2 + q_3 - k_1 - k_2) \nn
	&\qquad \times \mathcal{T}^{(1)}(k_1, k_2 \to p_1, p_2)^* \U^{\nu\rho}(q_2, k_1, k_2)\,.
\end{align}
The partial-wave expansion of the $I=1$ $\pi\pi$-scattering amplitude is
\begin{equation}
	\mathcal{T}^{(1)}(k_1, k_2 \to p_1, p_2) = \sum_{l \text{ odd}} (2l + 1) P_l(\cos\theta') t_l^1(s) \,,
\end{equation}
where $\theta'$ is the angle between $\vec{p}_1$ and $\vec{k}_1$ in the $\pi\pi$ center-of-mass frame and the elastic partial waves are fixed in terms of the phase shifts
\begin{equation}
	t_l^1(s) = \frac{32\pi}{\sigma_\pi(s)} \sin \delta_l^1(s) e^{\imag \delta_l^1(s)} \, .
\end{equation}
After projecting onto the tensor structures in the soft-photon limit, one finds that only the $P$-wave contributes
\begin{equation}
	\label{eq:Utilde_2pi}
	\Delta^{\pi^+\pi^-} \tilde\U_i(s) = \sin \delta_1^1(s) e^{-\imag \delta_1^1(s)} \tilde\U_i(s)\,.
\end{equation}

\subsubsection{Solution of the dispersion relations}

The total discontinuity of $\tilde\U_i(s)$ can be written as
\begin{equation}
	\Delta \tilde\U_i(s) = \Delta^{\pi^+\pi^-} \tilde\U_i(s) + I_i(s) = \sin \delta_1^1(s) e^{-\imag \delta_1^1(s)} \tilde\U_i(s) + I_i(s) \,,
\end{equation}
which has the form of an inhomogeneous Muskhelishvili--Omn\`es (MO) problem~\cite{Muskhelishvili:1953,Omnes:1958hv} with inhomogeneity $I_i(s)$. The homogeneous MO problem is solved by a polynomial times the Omn\`es function~\eqref{eq:Omnes}.

The mass dimensions of the scalar functions are $[\tilde\U_1(q^2)] = -2$ and $[\tilde\U_1(q^2)] = -4$, hence we expect that asymptotically $\tilde\U_1(q^2)$ falls off faster by one power of $q^2$. We assume an asymptotic behavior $\tilde\U_1(q^2) \asymp 1/q^2$ and $\tilde\U_2(q^2) \asymp 1/q^4$. For the asymptotic behavior of the elastic phase shift, we assume $\delta_1^1(s) \asymp \pi$, hence the Omn\`es function behaves as $\Omega_1^1(q^2) \asymp 1/q^2$. This implies that the inhomogeneous MO solution for $\tilde\U_1$ involves one subtraction constant, while $\tilde\U_2$ fulfills an unsubtracted MO representation:
\begin{align}
	\tilde\U_1(q^2) &= \Omega_1^1(q^2) \left[ \tilde\U_1(0) + \frac{q^2}{\pi} \int_{4m_\pi^2}^\infty \dif s \frac{e^{\imag \delta_1^1(s)}I_1(s)}{|\Omega_1^1(s)| s (s - q^2 - \imag \epsilon)} \right] \,, \nn
	\tilde\U_2(q^2) &= \Omega_1^1(q^2) \frac{1}{\pi} \int_{4m_\pi^2}^\infty \dif s \frac{e^{\imag \delta_1^1(s)}I_2(s)}{|\Omega_1^1(s)| (s - q^2 - \imag \epsilon)} \,.
\end{align}
The inhomogeneities $I_i(s)$ are due to pseudoscalar and axial-vector poles and given by
\begin{align}
	I_1(s) &= - 2 \pi \sum_A  \delta(s - m_A^2) \sum_a C_a F_A^a m_A \tilde\F_{A,1}(m_A^2, m_\pi^2, m_\pi^2)^* \,,\nn
	I_2(s) &= 2 \pi \sum_P \delta(s - m_P^2) \sum_a C_a F_P^a \F_P(m_P^2, m_\pi^2, m_\pi^2; 0)^* \nn
	&\quad + 4 \pi \sum_A \delta(s - m_A^2) \sum_a \frac{C_a F_A^a}{m_A} \tilde\F_{A,1}(m_A^2, m_\pi^2, m_\pi^2)^* \,.
\end{align}
Therefore, we obtain
\begin{align}
	\label{eq:ResultsTildeUi}
	\tilde\U_1(q^2) &= \Omega_1^1(q^2) \left[ \tilde\U_1(0) + \sum_A \sum_a \frac{2 C_a F_A^a}{m_A} \frac{q^2 \tilde\F_{A,1}^*}{\Omega_1^1(m_A^2)^* (q^2 - m_A^2)} \right] \,, \nn
	\tilde\U_2(q^2) &= \Omega_1^1(q^2) \left[ - \sum_P \sum_a \frac{2 C_a F_P^a \F_{P}^*}{\Omega_1^1(m_P^2)^* (q^2 - m_P^2)} - \sum_A \sum_a \frac{4 C_a F_A^a}{m_A} \frac{\tilde\F_{A,1}^*}{\Omega_1^1(m_A^2)^* (q^2 - m_A^2)} \right] \, ,
\end{align}
with the abbreviations $\F_P = \F_P(m_P^2, m_\pi^2, m_\pi^2;0)$ and $\tilde\F_{A,1} = \tilde\F_{A,1}(m_A^2, m_\pi^2, m_\pi^2)$.


The subtraction constant $\tilde\U_1(0)$ can be fixed by matching the dispersion relation to \chpt{}. As in the numerical analysis, we focus on the isovector component and consider the $SU(2)$ chiral Lagrangian. We use the leading-order odd-intrinsic-parity Lagrangian coupled to external currents~\cite{Witten:1983tw}, in the conventions of Ref.~\cite{Bijnens:2001bb}. There are two tree-level diagrams: the $\gamma a^* \pi^+ \pi^-$ contact interaction and the $\gamma \pi^+ \pi^- \pi^0$ vertex with subsequent mixing of the neutral pion with the axial current. Each of them contributes to only one scalar function. In the soft-photon limit, we obtain
\begin{equation}
	\tilde\U_1^{(3)}(q^2) = \frac{N_c}{24 \pi^2 F^2} \,, \qquad \tilde\U_2^{(3)}(q^2) = -\frac{N_c}{12 \pi^2 F^2 (q^2 - m_\pi^2)} \,.
\end{equation}
At this order, we have
\begin{equation}
	\label{eq:ChPT_g3pi}
	\Omega_1^1(q^2) = 1 \,, \qquad \F_{\pi^0}(m_\pi^2, m_\pi^2, m_\pi^2; 0) = \frac{N_c}{12 \pi^2 F^3} \,,  \qquad F_\pi^{(3)} = F \, ,
\end{equation}
so that the pion-pole contribution to $\tilde\U_2^{(3)}$ in Eq.~\eqref{eq:ResultsTildeUi} is reproduced by the chiral expression. We thus conclude that the axial-pole and heavier pseudoscalar-pole contributions in $\tilde\U_2^{(3)}$ are of higher order in the chiral counting. The result for $\tilde\U_1^{(3)}(q^2)$ fixes the subtraction constant
\begin{equation}
	\label{eq:AVPiPiSubtractionChPT}
	\tilde\U_1^{(3)}(0) = \frac{N_c}{24 \pi^2 F^2} \,.
\end{equation}

\subsubsection{Implications for VVA}
\label{sec:TwoPionResultsInsertedIntoVVA}

The dispersive results Eq.~\eqref{eq:ResultsTildeUi} for the scalar functions $\tilde\U_i$ can now be inserted into the two-pion contribution to the VVA correlator in Eq.~\eqref{eq:Wipipi}:
\begin{align}
	\label{eq:respipig-2kin}
	\tilde{\W}_1^{\pi^+\pi^-}(q^2) &= - \frac{1}{48 \pi^2} \int_{4m_\pi^2}^\infty \dif s \frac{\sigma_\pi^3(s)  F_\pi^V(s) \Omega_1^1(s)^* \, \tilde\U_1(0)}{s - q^2 - \imag \epsilon} \nn
		&\quad + \frac{1}{48\pi^2} \sum_P \int_{4m_\pi^2}^\infty \dif s \frac{s \sigma_\pi^3(s)  F_\pi^V(s) \Omega_1^1(s)^*}{(s - q^2 - \imag \epsilon)(s - m_P^2 - \imag \epsilon)}  \sum_a \frac{C_a F_P^a \F_{P}}{\Omega_1^1(m_P^2)}\,, \nn
	\tilde{\W}_2^{\pi^+\pi^-}(q^2) &= \frac{1}{48 \pi^2} \int_{4m_\pi^2}^\infty \dif s \frac{\sigma_\pi^3(s) F_\pi^V(s) \Omega_1^1(s)^* \, \tilde\U_1(0)}{s - q^2 - \imag \epsilon} \nn
		&\quad + \frac{1}{24\pi^2} \sum_A \int_{4m_\pi^2}^\infty \dif s \frac{s \sigma_\pi^3(s) F_\pi^V(s) \Omega_1^1(s)^*}{(s - q^2 - \imag \epsilon)(s - m_A^2 - \imag \epsilon)} \sum_a \frac{C_a F_A^a}{m_A} \frac{\tilde\F_{A,1}}{\Omega_1^1(m_A^2)} \,,
\end{align}
where we have included the $\imag \epsilon$ terms explicitly to specify the integration contour for $m_{P,A}^2 > 4 m_\pi^2$. Interestingly, the axial-vector contribution to the longitudinal scalar function $\tilde\W_1$ cancels between $\tilde\U_1$ and $\tilde\U_2$.

Furthermore, we observe that the two-pion contribution to the isovector scalar function $\W_1^{(3)}$ vanishes in the chiral limit, as required by the anomaly constraint on $\tilde\W_1$: the decay constants of excited pseudoscalars vanish in the chiral limit, leaving only the $\pi^0$ contribution in the sum over pseudoscalars in Eq.~\eqref{eq:respipig-2kin}, which in the chiral limit $m_\pi^2\to0$ exactly cancels the first term proportional to the subtraction constant $\tilde\U_1^{(3)}(0)$.

\color{black}

\subsection[On-shell $A\gamma\to\pi\pi$ amplitude]{\boldmath On-shell $A\gamma\to\pi\pi$ amplitude}
\label{sec:AGammaPiPi}

In the description of two-pion intermediate states, the matrix element for $A\gamma\to\pi\pi$ appears in two places: on the one hand in the dispersive reconstruction of the axial-meson TFFs in App.~\ref{sec:axialTFFTwoPion}, on the other hand in the axial-meson pole contribution to $a^*\gamma\to\pi\pi$ discussed in App.~\ref{sec:AxialCurrentGammaPiPiAxialPole}, which is required for the dispersive approach to VVA in $g-2$ kinematics. The process $A\gamma\to\pi\pi$ is closely related to $a^*\gamma\to\pi\pi$ discussed in App.~\ref{sec:AxialCurrentGammaPiPi}. However, here we deal with an on-shell axial-vector meson instead of an axial current.

\subsubsection{Tensor decomposition}

We consider the matrix element
\begin{align}
	\label{eq:axialPiPiGamma}
	{}_{I=1} \!\< & \pi^+(p_1) \pi^-(p_2) | A(p, \lambda_A) \gamma(q_2, \lambda_2) \> \nn
	& = -\imag e \epsilon_\nu^{\lambda_2}(q_2) (2\pi)^4 \delta^{(4)}(p + q_2 - p_1 - p_2) \;{}_{I=1}\!\< \pi^+(p_1) \pi^-(p_2) | j_\mathrm{em}^\nu(0) | A(p, \lambda_A) \> \nn
	& = - e \epsilon_\nu^{\lambda_2}(q_2) \epsilon_\rho^{\lambda_A}(p) (2\pi)^4 \delta^{(4)}(p + q_2 - p_1 - p_2) \A^{\nu\rho}(q_2,p_1,p_2) \, ,
\end{align}
where
\begin{equation}
	\label{eq:AgammaPiPiAmpliutude}
	\epsilon_\rho^{\lambda_A}(p) \A^{\nu \rho}(q_2, p_1, p_2) =  i \; {}_{I=1}\!\bra{\pi^+(p_1) \pi^-(p_2)} j_\mathrm{em}^\nu(0) \ket{A(p, \lambda_A)} \, ,
\end{equation}
with $q_2 = p_1 + p_2 - p$. The BTT tensor decomposition of $\A^{\nu\rho}$ a priori leads to the same tensor structures $t^{\nu\rho}$ as for the case of $\U^{\nu\rho}$ with an axial-vector current in App.~\ref{sec:AxialCurrentGammaPiPiTensorDecomposition}. However, $\A^{\nu\rho}$ always appears contracted with the axial-vector polarization sum, which projects $t_5^{\nu\rho}$ to zero,
\begin{equation}
	\Big( g_{\sigma\rho} - \frac{p_\sigma p_\rho}{p^2} \Big) t_5^{\nu\rho}(q_2, p_1, p_2) = 0 \,.
\end{equation}
Therefore, we use the tensor decomposition
\begin{equation}
	\label{eq:AgammaPiPiDecomposition}
	\A^{\nu\rho}(q_2,p_1,p_2) = \sum_{i=1}^4 t_i^{\nu\rho}(q_2, p_1, p_2) \F_{A,i}(s, t, u; q_2^2) \, ,
\end{equation}
where the Mandelstam variables are $s = (p_1+p_2)^2$, $t = (p-p_1)^2$, $u = (p-p_2)^2$. Due to the restriction to the isospin $I=1$ component of the $\pi\pi$ state, pion-crossing symmetry implies the relations
\begin{align}
	\label{eq:crossingFM}
	\F_{A,1}(s,t,u;q_2^2) &= \F_{A,1}(s,u,t;q_2^2) \, , \quad &
	\F_{A,3}(s,t,u;q_2^2) &= - \F_{A,3}(s,u,t;q_2^2) \, , \nn
	\F_{A,2}(s,t,u;q_2^2) &= - \F_{A,2}(s,u,t;q_2^2) \, , \quad &
	\F_{A,4}(s,t,u;q_2^2) &= \F_{A,4}(s,u,t;q_2^2) \, ,
\end{align}
and in particular $\F_{A,2}(m_A^2,m_\pi^2,m_\pi^2;0) = \F_{A,3}(m_A^2,m_\pi^2,m_\pi^2;0) = 0$ in the soft-photon limit.

In the case of an on-shell photon, $q_2^2 = 0$, the tensor structure $t_4^{\nu\rho}$ becomes redundant. Using the Schouten identity, we find
\begin{align}
	t_4^{\nu\rho}(q_2,p_1,p_2) &- \frac{q_2 \cdot (p_1+p_2)}{2} t_1^{\nu\rho}(q_2,p_1,p_2) + \frac{q_2 \cdot (p_1-p_2)}{2} t_2^{\nu\rho}(q_2,p_1,p_2) \nn
		&= q_2^2 \epsilon^{\nu\rho\alpha\beta} {p_1}_\alpha {p_2}_\beta + q_2^\nu \epsilon^{\rho\alpha\beta\gamma} {p_1}_\alpha {p_2}_\beta {q_2}_\gamma \, .
\end{align}
Therefore, in the on-shell case we can use the decomposition
\begin{equation}
	\label{eq:AgammaPiPiDecompositionOnShell}
	\A^{\nu\rho}(q_2,p_1,p_2) = \sum_{i=1}^3 t_i^{\nu\rho}(q_2, p_1, p_2) \tilde\F_{A,i}(s, t, u) \, ,
\end{equation}
where
\begin{align}
	\label{eq:FAtilde}
	\tilde\F_{A,1}(s,t,u) &= \F_{A,1}(s,t,u;0) + \frac{s-m_A^2}{4} \F_{A,4}(s,t,u;0) \, , \nn
	\tilde\F_{A,2}(s,t,u) &= \F_{A,2}(s,t,u;0) - \frac{t-u}{4} \F_{A,4}(s,t,u;0) \, , \nn
	\tilde\F_{A,3}(s,t,u) &= \F_{A,3}(s,t,u;0) \, .
\end{align}
In the limit of a soft photon, one obtains
\begin{equation}
	\label{eq:AgammaPiPiDecompositionSoftPhoton}
	\A^{\nu\rho}(q_2,p_1,p_2) = t_1^{\nu\rho\sigma}(p_1, p_2) {q_2}_\sigma \, \tilde\F_{A,1}(m_A^2, m_\pi^2, m_\pi^2) + \Order(q_2^2) \, ,
\end{equation}
where $\tilde\F_{A,1}(m_A^2, m_\pi^2, m_\pi^2) = \F_{A,1}(m_A^2, m_\pi^2, m_\pi^2; 0)$.

\subsubsection{Helicity amplitudes and partial-wave expansion}
\label{sec:AgammaPiPiHelAmps}

Next, we derive the relation between the scalar functions $\tilde\F_{A,i}$ and $s$-channel helicity partial waves for the process $A\gamma\to\pi\pi$. We only consider the case of an on-shell photon, $q_2^2 = 0$. The momenta in the center-of-mass frame of the pion pair are
\begin{align}
	p &= \left(\frac{s + m_A^2}{2 \sqrt{s}}, 0, 0, \frac{s - m_A^2}{2 \sqrt{s}} \right) \,,\nn
	q_2 &= \left(\frac{s - m_A^2}{2 \sqrt{s}}, 0, 0, -\frac{s - m_A^2}{2 \sqrt{s}} \right) \,,\nn
	p_1 &= \left( \frac{\sqrt{s}}{2}, \frac{\sqrt{s} \sigma_\pi(s)}{2} \sin\theta \cos\phi, \frac{\sqrt{s} \sigma_\pi(s)}{2} \sin\theta \sin\phi, \frac{\sqrt{s} \sigma_\pi(s)}{2} \cos\theta \right) \,, \nn
	p_2 &= \left( \frac{\sqrt{s}}{2}, - \frac{\sqrt{s} \sigma_\pi(s)}{2} \sin\theta \cos\phi, - \frac{\sqrt{s} \sigma_\pi(s)}{2} \sin\theta \sin\phi, - \frac{\sqrt{s} \sigma_\pi(s)}{2} \cos\theta \right) \,.
\end{align}
The polarization vectors take the form
\begin{align}
	\quad \epsilon^\pm(p) &= \mp\frac{1}{\sqrt{2}}(0, 1, \pm \imag, 0) \,, \nn
	\epsilon^0(p) &= \left(\frac{s - m_A^2}{2 \sqrt{s} m_A}, 0, 0, \frac{s + m_A^2}{2 \sqrt{s} m_A}\right) \,, \nn
	\epsilon^\pm(q_2) &= \mp\frac{1}{\sqrt{2}}(0, 1, \mp \imag, 0) \,.
\end{align}
We define helicity amplitudes for $A(p) \gamma(q_2) \to \pi^+(p_1) \pi^-(p_2)$ as
\begin{align}
	\label{eq:helAmpsAgammaPiPi}
	H_{\lambda_2 \lambda_A}  &= - \epsilon_\nu^{\lambda_2}(q_2) \ {}_{I=1}\!\!\bra{\pi^+(p_1) \pi^-(p_2)} j_\mathrm{em}^\nu(0) \ket{A(p, \lambda_A)} =  i \epsilon_\nu^{\lambda_2}(q_2) \epsilon_\rho^{\lambda_A}(p)  \A^{\nu\rho}(q_2,p_1,p_2) \nn
		&= \sum_{i=1}^3 i \epsilon_\nu^{\lambda_2}(q_2) \epsilon_\rho^{\lambda_A}(p) t_i^{\nu\rho}(q_2,p_1,p_2) \tilde\F_{A,i}(s,t,u) \,.
\end{align}
Due to parity, there are only three independent helicity amplitudes. Inverting the system~\eqref{eq:helAmpsAgammaPiPi}, we can express the three scalar functions in terms of helicity amplitudes
\begin{align}
	\tilde\F_{A, 1}(s,t,u) &= \frac{2}{s - m_A^2} \left[  \frac{\sqrt{2} m_A}{\sqrt{s} \sigma_\pi(s) \sqrt{1 - z^2}} e^{\imag \phi} H_{+0} - \frac{(s + m_A^2) z}{s \sigma_\pi(s) (1 - z^2)} e^{2 \imag \phi} H_{+-} \right] \,,  \nn
	\tilde\F_{A, 2}(s,t,u) &= \frac{2}{s - m_A^2} \left[ H_{++} - \frac{\sqrt{2} m_A z}{\sqrt{s} \sqrt{1 - z^2}} e^{\imag \phi} H_{+0} + \frac{s + m_A^2 z^2}{s (1 - z^2)} e^{2 \imag \phi} H_{+-} \right] \,, \nn
	\tilde\F_{A,3}(s,t,u) &= -\frac{8}{s(s - m_A^2) \sigma_\pi^2(s)(1 - z^2)} e^{2 \imag \phi} H_{+-} \, ,
\end{align}
with the cosine of the scattering angle
\begin{align}
	z = \cos\theta = \frac{t - u}{\sigma_\pi(s)(s - m_A^2)} \, .
\end{align}
Next, we insert the partial-wave expansions for the helicity amplitudes, which take the form~\cite{Jacob:1959at}
\begin{align}
	H_{++} &= \sum_{l=0}^\infty D_{00}^l(\phi, \theta, -\phi) \tilde{h}_{++}^l(s) = \sum_{l=0}^\infty P_l(z) h_{++}^l(s) \,, \nn
	H_{+0} &= \sum_{l=1}^\infty D_{-1\,0}^l(\phi, \theta, -\phi) \tilde{h}_{+0}^l(s) = e^{-\imag \phi} \sqrt{1 - z^2} \sum_{l=1}^\infty P_l'(z) h_{+0}^l(s) \,, \nn
	H_{+-} &= \sum_{l=2}^\infty D_{-2\,0}^l(\phi, \theta, -\phi) \tilde{h}_{+-}^l(s) = e^{-2 \imag \phi} (1 - z^2) \sum_{l=2}^\infty P_l''(z) h_{+-}^l(s) \,,
\end{align}
where
\begin{align}
	h_{++}^l(s) &= \tilde{h}_{++}^l(s) \,, \nn
	h_{+0}^l(s) &= - \frac{\tilde{h}_{+0}^l(s)}{\sqrt{l(l+1)}} \,, \nn
	h_{+-}^l(s) &= \frac{\tilde{h}_{+-}^l(s)}{\sqrt{(l-1)l(l+1)(l+2)}}\,.
\end{align}
This leads to
\begin{align}
	\label{eq:AgammaPiPiHelicityPW}
	\tilde\F_{A, 1}(s,t,u) &= \frac{2}{s - m_A^2} \sum_{l \text{ odd}} \left[  \frac{\sqrt{2} m_A}{\sqrt{s} \sigma_\pi(s) } P_l'(z) h_{+0}^l(s) - \frac{(s + m_A^2) z}{s \sigma_\pi(s)} P_l''(z) h_{+-}^l(s) \right] \,,  \nn
	\tilde\F_{A, 2}(s,t,u) &= \frac{2}{s - m_A^2} \sum_{l \text{ odd}} \left[ P_l(z) h_{++}^l(s) - \frac{\sqrt{2} m_A z}{\sqrt{s}} P_l'(z) h_{+0}^l(s) + \frac{s + m_A^2 z^2}{s} P_l''(z) h_{+-}^l(s) \right] \,, \nn
	\tilde\F_{A,3}(s,t,u) &= -\frac{8}{s(s - m_A^2) \sigma_\pi^2(s)} \sum_{l \text{ odd}} P_l''(z) h_{+-}^l(s) \, ,
\end{align}
where the crossing relations for $I=1$ pions~\eqref{eq:crossingFM} forbid even partial waves.

Since the functions $\tilde{\F}_{A,i}$ are free of kinematic singularities, the poles at $s = m_A^2$ due to the explicit prefactors and the factors of $z$ have to be accompanied by zeros in the helicity partial-wave amplitudes $h_{\lambda_2\lambda_A}^l(s)$. This requires
\begin{equation}
	\label{eq:AGammaPiPiThreshold}
	h_{\lambda_2\lambda_A}^l(s) = \sigma_\pi(s)^l (s - m_A^2)^l g_{\lambda_2\lambda_A}^l(s) \, ,
\end{equation}
with $g_{\lambda_2\lambda_A}^l(s)$ regular at $s = m_A^2$. The absence of kinematic singularities in $\tilde\F_{A,2}$ requires one additional linear constraint for each value of $l$. For $l = 1$, it reads
\begin{equation}
	\label{eq:AGammaPiPiConstraint}
	g_{++}^1(m_A^2) = \sqrt{2} g_{+0}^1(m_A^2) \,.
\end{equation}
In the soft-photon limit, only the leading partial wave contributes and we obtain for the only non-vanishing scalar function
\begin{equation}
	\tilde\F_{A,1}(m_A^2,m_\pi^2,m_\pi^2) = 2\sqrt{2} g_{+0}^1(m_A^2) \, .
\end{equation}

\subsubsection{Simplified Omn\`es parametrization}
\label{sec:AgpipiParametrization}

In the soft-photon limit, all kinematic variables in the process $A\gamma\to\pi\pi$ are fixed by the on-shell conditions. Therefore, a dispersive reconstruction is only possible away from the soft-photon limit, i.e., in four-point kinematics. In the following, we only sketch the necessary steps for a dispersive reconstruction and we will make use of a simplified parametrization.

\begin{figure}[t]
	\centering
	\begin{align*}
		\minidiagSize{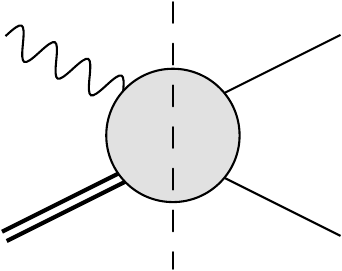}{2cm}
		\quad = \quad 
		\minidiagSize{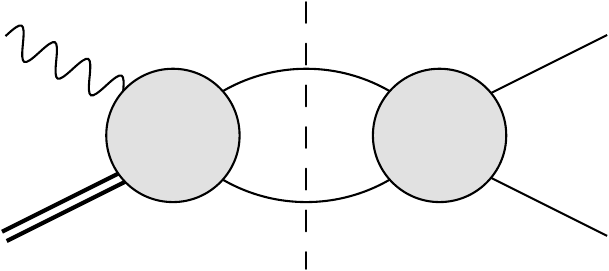}{3cm}
		\quad + \quad \ldots \\[-0.75cm]
	\end{align*}
	\caption{Two-pion intermediate state in the $s$-channel of $A\gamma\to\pi\pi$.}
	\label{fig:Agpipi-sCut}
\end{figure}

\begin{figure}[t]
	\centering
	\begin{align*}
		\minidiagSize{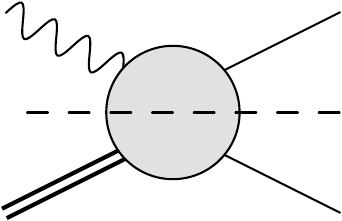}{2cm}
		\quad = \quad
		\minidiagSize{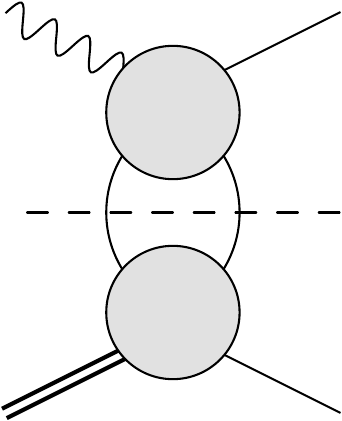}{1.75cm}
		\quad + \quad \ldots  \\[-0.75cm]
	\end{align*}
	\caption{Two-pion intermediate state in the $t$- and $u$-channels of $A\gamma\to\pi\pi$.}
	\label{fig:Agpipi-tCut}
\end{figure}

In principle, for the dispersive analysis of $A\gamma\to\pi\pi$ one needs to consider the three channels of the Mandelstam variables $s$, $t$, and $u$. The lowest intermediate state in each channel consists of two pions: in the $s$-channel, this amounts to elastic $\pi\pi$ final-state rescattering, shown in Fig.~\ref{fig:Agpipi-sCut}. In the $t$- and $u$-channels, unitarity relates the imaginary parts to the sub-processes $\gamma\to3\pi$ as well as $A\to3\pi$, as shown in Fig.~\ref{fig:Agpipi-tCut}. With suitable input for these sub-processes, one could set up a dispersive treatment in the form of a reconstruction theorem~\cite{Stern:1993rg,Ananthanarayan:2000cp,Zdrahal:2008bd,Niecknig:2012sj,Colangelo:2015kha}. To this end, another dispersive analysis of $A\to3\pi$ would be of interest, which could be constrained via $\tau\to3\pi\nu$~\cite{Lorenz:2016gsj}. However, this goes beyond the scope of the present work and we apply a simpler approximation, neglecting the effects of intermediate states in the $t$- and $u$-channels. Furthermore, in the $s$-channel we neglect intermediate states beyond $\pi\pi$ $P$-waves and we make a simple homogeneous Omn\`es ansatz for the form factors
\begin{equation}
	\tilde\F_{A,1}(s,t,u) = \Omega_1^1(s) (\bar\F_{A,1} + s \bar\F_{A,1}') \, , \quad
	\tilde\F_{A,2}(s,t,u) = (t-u) \Omega_1^1(s)  \bar\F_{A,2} \, , \quad
	\tilde\F_{A,3}(s,t,u) = 0  \, .
\end{equation}
In this approximation, the non-vanishing partial waves are given by
\begin{align}
	g_{++}^1(s) &= \frac{1}{2} \Omega_1^1(s) \left( \bar\F_{A,1} + s \bar\F_{A,1}' + (s-m_A^2) \bar\F_{A,2} \right) \, , \nn
	g_{+0}^1(s) &= \frac{\sqrt{s}}{2\sqrt{2} m_A} \Omega_1^1(s) \left( \bar\F_{A,1} + s  \bar\F_{A,1}' \right) \, ,
\end{align}
fulfilling the threshold constraint~\eqref{eq:AGammaPiPiConstraint}
\begin{equation}
	\label{eq:SimplifiedOmnesFA1}
	\tilde\F_{A,1}(m_A^2,m_\pi^2,m_\pi^2) = 2 g_{++}^1(m_A^2) = 2 \sqrt{2} g_{+0}^1(m_A^2) = \Omega_1^1(m_A^2) \left( \bar\F_{A,1} + m_A^2 \bar\F_{A,1}' \right) \, .
\end{equation}
The linear combinations of partial waves that enter the dispersion relation for the axial-vector TFFs are given by $g_{++}^1(s)$ as well as
\begin{equation}
	\frac{s g_{++}^1(s) - \sqrt{2s}m_A  g_{+0}^1(s)}{s-m_A^2} = \frac{s}{2} \Omega_1^1(s) \bar\F_{A,2} \, ,
\end{equation}
which in our approximation asymptotically both tend to a constant if $\Omega_1^1(s) \asymp s^{-1}$.